\documentclass[12pt]{book}
\usepackage{geometry}
\usepackage{amsmath}
\usepackage{multirow}
\usepackage{amsfonts}
\usepackage{amssymb,graphics,psfrag,float}
\usepackage{array,epsfig,multirow,graphicx}
\usepackage{comment}
\usepackage{slashed}
\usepackage{tensor}
\usepackage{booktabs}
\usepackage{colortbl}
\usepackage{hhline}
\usepackage{mathtools}
\usepackage{enumitem}
\usepackage{empheq}
\usepackage{xfrac}
\usepackage{tikz}
\usepackage[nosort]{cite}
\usetikzlibrary{decorations.markings}
\usetikzlibrary{shapes,arrows}
\usetikzlibrary{decorations.pathreplacing}
\usetikzlibrary{arrows,positioning}

\newcommand{\beq}{\begin{equation}}
\newcommand{\eeq}{\end{equation}}
\newcommand{\be}{\begin{equation}}
\newcommand{\ee}{\end{equation}}
\newcommand{\bea}{\begin{eqnarray}}
\newcommand{\eea}{\end{eqnarray}}
\newcommand{\ben}{\begin{eqnarray*}}
\newcommand{\een}{\end{eqnarray*}}               
\newcommand{\ba}{\begin{aligned}}
\newcommand{\ea}{\end{aligned}}
\newcommand{\bt}{\begin{tabular}}
\newcommand{\et}{\end{tabular}}
\newcommand{\bc}{\begin{center}}
\newcommand{\ec}{\end{center}}

\newcommand{\cO}{\mathcal{O}}

\newcommand{\cE}{\mathcal{E}}
\newcommand{\cP}{\mathcal{P}}
\newcommand{\cC}{\mathcal{C}}
\newcommand{\cD}{\mathcal{D}}
\newcommand{\cL}{\mathcal{L}}
\newcommand{\cS}{\mathcal{S}}
\newcommand{\cK}{\mathcal{K}}
\newcommand{\cN}{\mathcal{N}}
\newcommand{\cW}{\mathcal{W}}

\newcommand{\cA}{\mathcal{A}}
\newcommand{\cH}{\mathcal{H}}

\newcommand{\cI}{\mathcal{I}}
\newcommand{\cJ}{\mathcal{J}}
\newcommand{\cR}{\mathcal{R}}

\newcommand{\cV}{\mathcal{V}}

\newcommand{\cM}{\mathcal M}
\newcommand{\cQ}{\mathcal Q}

\newcommand{\I}{\text{Im}}
\newcommand{\R}{\text{Re}}

\newcommand{\fn}{\mathfrak{n}}

\newcommand{\bi}{{\bar \imath}}
\newcommand{\ib}{{\bar\imath }}
\newcommand{\jb}{{\bar\jmath }}
\newcommand{\bj}{{\bar\jmath}}

\DeclareMathOperator{\rk}{rank}
\DeclareMathOperator{\vol}{vol}
\DeclareMathOperator{\sign}{sign}

\newcommand{\bbZ}{\mathbb{Z}}

\newcommand{\nn}{\nonumber}

\newcommand{\cref}{{\bf [check ref]}}

\newcommand{\tr}{\mathrm{tr}}

\newcommand{\D}{\mathrm{D}}     

\newcommand{\tiv}{\tilde{v}}
\newcommand{\dd}{d}

\newcommand\Tstrut{\rule{0pt}{3.5ex}}     
\newcommand\Bstrut{\rule[-1.5ex]{0pt}{0pt}}
\newcommand\TTstrut{\rule{0pt}{2.3ex}}     
\newcommand\BBstrut{\rule[-0.9ex]{0pt}{0pt}}
\newcommand{\BB}{{\boldsymbol{B}}}

\newcommand{\Bpsi}{{\boldsymbol{\psi}}}
\newcommand{\Bphi}{{\boldsymbol{\phi}}}
\newcommand{\Blambda}{{\boldsymbol{\lambda}}}
\newcommand{\fe}{\sfrac{1}{2}}
\newcommand{\gr}{\sfrac{3}{2}}
\newcommand{\dynkinradius}{.05cm}
\newcommand{\dynkinstep}{.25cm}
\newcommand{\dynkindot}[2]{\fill (\dynkinstep*#1,\dynkinstep*#2) circle (\dynkinradius); }

\newcommand{\dynkinline}[4]{\draw[thin] (\dynkinstep*#1,\dynkinstep*#2) -- (\dynkinstep*#3,\dynkinstep*#4);}
\newcommand{\dynkindots}[4]{\draw[dotted] (\dynkinstep*#1,\dynkinstep*#2) -- (\dynkinstep*#3,\dynkinstep*#4);}
\newcommand{\dynkindoubleline}[4]{\draw[double,postaction={decorate}] (\dynkinstep*#1,\dynkinstep*#2) -- (\dynkinstep*#3,\dynkinstep*#4);}
\newcommand{\dynkintripleline}[4]{\draw[double,postaction={decorate}] (\dynkinstep*#1,\dynkinstep*#2) -- (\dynkinstep*#3,\dynkinstep*#4);
				  \draw (\dynkinstep*#1,\dynkinstep*#2) -- (\dynkinstep*#3,\dynkinstep*#4);}

\newenvironment{dynkin}{\begin{tikzpicture}[decoration={markings,mark=at position 0.6 with {\arrow[line width=0.15mm]{>}}}]}
{\end{tikzpicture}}
\definecolor{mppgreen}{RGB}{17,102,86}
\definecolor{mppgray}{RGB}{221,222,214}

\newcommand*\widefbox[1]{\fbox{\rule[-1.8cm]{0pt}{4cm}\hspace{2em}#1\hspace{2em}}}

\def\blfootnote{\xdef\@thefnmark{}\@footnotetext}
\long\def\symbolfootnote[#1]#2{\begingroup%
\def\thefootnote{\fnsymbol{footnote}}\footnote[#1]{#2}\endgroup}

\usepackage[utf8]{inputenc}
\usepackage{dtklogos}
\usetikzlibrary{mindmap,shadows,backgrounds}
\DeclareMathAlphabet{\mathpzc}{OT1}{pzc}{m}{it}

\newcommand{\tstar}[5]{
\pgfmathsetmacro{\starangle}{360/#3}
\draw[#5] (#4:#1)
\foreach \x in {1,...,#3}
{ -- (#4+\x*\starangle-\starangle/2:#2) -- (#4+\x*\starangle:#1)
}
-- cycle;
}

\newcommand{\ngram}[4]{
\pgfmathsetmacro{\starangle}{360/#2}
\pgfmathsetmacro{\innerradius}{#1*sin(98-\starangle)/sin(98+\starangle/2)}
\tstar{\innerradius}{#1}{#2}{#3}{#4}
}

\begin{document}

\begin{titlepage}

\newcommand{\HRule}{\rule{\linewidth}{0.7mm}} 

\center

\HRule \\[0.4cm]

\flushright { \huge \bfseries {Geometric Symmetries }}\\[0.2cm]

\flushright { \huge \bfseries {and Topological Terms}}\\[0.2cm]

\flushright { \huge \bfseries {in F-theory and Field Theory} }\\[0.8cm]

\flushright {\large \bfseries Andreas Kapfer}\\[0.4cm]

\HRule \\[1.5cm]
 
 \vspace*{\fill}

\centering \includegraphics[width=0.33\textwidth]{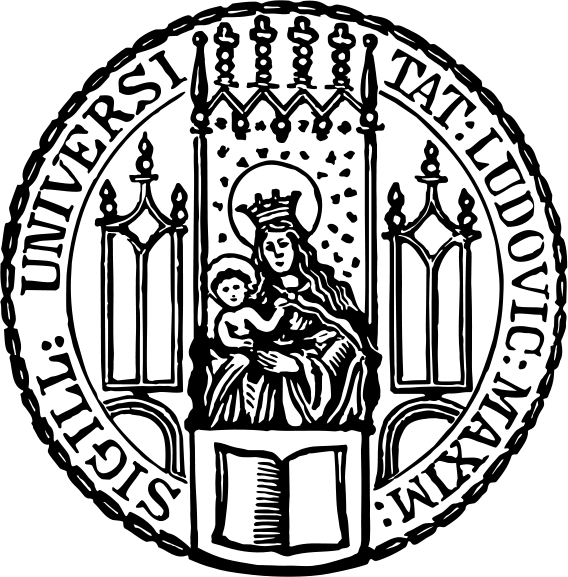}\\[1cm] 

\centering {\large M\"unchen, 2016}

\end{titlepage}

\newpage
\thispagestyle{empty}
\mbox{}


\newpage
\thispagestyle{empty}

\newcommand{\HRule}{\rule{\linewidth}{0.7mm}} 

{\center

\HRule \\[0.4cm]

{\flushright { \huge \bfseries {Geometric Symmetries }}\\[0.2cm]

\flushright { \huge \bfseries {and Topological Terms}}\\[0.2cm]

\flushright { \huge \bfseries {in F-theory and Field Theory} }\\[0.8cm]

\flushright {\large \bfseries Andreas Kapfer}\\[0.4cm]}

\HRule \\[1.5cm]

\textsc{\Large Dissertation\\ an der Fakult\"at f\"ur Physik\\ der Ludwig-Maximilians-Universit\"at\\ M\"unchen\\[1.4cm] vorgelegt von\\
Andreas Kapfer\\ aus Welden \\}
 
 \vspace*{\fill}

\large M\"unchen, 2016\\}


\newpage
\thispagestyle{empty}

 \vspace*{\fill}
 
 {\Large{\noindent Erstgutachter: Prof. Dr. Dieter L\"ust \\ Zweitgutachter: PD Dr. Ralph Blumenhagen \\ Tag der m\"undlichen Pr\"ufung: 6. Oktober 2016}}
 
 \newpage
\thispagestyle{empty}

{\flushright{\textsc{\Large Zusammenfassung}}}

\vspace*{1cm}
Die vorliegende Dissertation befasst sich mit topologischen Aspekten und a\-rith\-me\-ti\-schen Strukturen von Quantenfeldtheorie und String Theorie. Besonderes
Augenmerk wird hierbei auf konsistente Trunkierungen von Supergravitation und Kompaktifizierungen von F-Theorie gelegt.

Der erste Teil behandelt die Brechung von Supersymmetrie in f\"unf Dimensionen. Wir konzentrieren uns hierbei auf den 
\"Ubergang von $\cN=4$ auf $\cN=2$
in geeichter Supergravitation. F\"ur bestimmte Klassen von einbettenden Tensoren sind wir in der Lage, die Theorie um das Vakuum zu gro{\ss}en Teilen
zu analysieren. Es ist beachtlich, dass generisch Quantenkorrekturen zu Chern-Simons-Termen induziert werden, die unabh\"angig von der
Skala der Supersymmetriebrechung sind. Wir untersuchen konkrete Beispiele konsistenter Trunkierungen von Supergravitation und M-Theorie, die diese
Brechung von $\cN=4$ auf $\cN=2$ in f\"unf Dimensionen widerspiegeln. Insbesondere analysieren wir notwendige Bedingungen daf\"ur, dass diese konsistenten
Trunkierungen f\"ur ph\"anomenologische Zwecke herangezogen werden k\"onnen, indem wir fordern, dass sich die skaleninvarianten Korrekturen
zu den Chern-Simons-Kopplungen konsistent verhalten.

Im zweiten Teil untersuchen wir Anomalien und gro{\ss}e Eichtransformationen in kreisreduzierten Eichtheorien und F-Theorie. Wir setzen
vier- und sechsdimensionale Eichtheorien mit gekoppelter Materie auf einen Kreis und klassifizieren alle gro{\ss}en Eichtransformationen, die die 
Randbedingungen der Materiefelder erhalten. Die Forderung, dass diese Abbildungen konsistent auf Quantenkorrekturen zu Chern-Simons-Kopp\-lun\-gen
agieren sollen, liefert uns explizit alle h\"oherdimensionalen Anomaliebedingungen. Bezogen auf Kompaktifizierungen von F-Theorie
identifizieren wir die klassifizierten gro{\ss}en Eichtransformationen entlang des Kreises mit arithmetischen Strukturen auf
elliptisch-gefaserten Calabi-Yau-Mannigfaltigkeiten \"uber die duale Beschreibung mittels M-Theorie.
Integrale Abelsche gro{\ss}e Eichtransformationen entsprechen in der Tat 
freien Ver\-schie\-bun\-gen der Basis im Mordell-Weil-Gitter der rationalen Schnitte,
w\"ahrend spezielle nicht-ganzzahlige nicht-Abelsche gro{\ss}e Eichtransformationen zu torsionellen Verschiebungen in der Mordell-Weil-Gruppe geh\"oren.
F\"ur ganzzahlige nicht-Abelsche gro{\ss}e Eichtransformationen schlagen wir eine neue Gruppenstruktur auf aufgel\"osten elliptischen
Faserungen vor. Auf dieselbe Weise bringen wir eine neuartige Gruppenopperation f\"ur Mehrfachschnitte auf Faserungen von Geschlecht Eins
ohne echten Schnitt vor. Wir m\"ochten betonen, dass diese arithmetischen Strukturen die Aufhebung aller Eichanomalien in F-Theorie-Kompaktifizierungen
auf Calabi-Yau-Mannigfaltigkeiten sicherstellen.

\newpage
\thispagestyle{empty}
\mbox{}

\newpage
\thispagestyle{empty}

{\flushright{\textsc{\Large Abstract}}}

\vspace*{1cm}

In this thesis we investigate topological aspects and arithmetic structures in quantum field theory and string theory. Particular focus is put 
on consistent truncations of supergravity and compactifications of F-theory.

The first part treats settings of supersymmetry breaking in five dimensions. We focus on an $\cN =4$ to $\cN =2$ breaking in gauged
supergravity. For certain classes of embedding tensors we can analyze the theory around the vacuum to a great extent.
Importantly, one-loop corrections to
Chern-Simons terms are generically induced which are independent of the supersymmetry-breaking scale.
We investigate concrete examples of consistent truncations of supergravity and M-theory
which show this $\cN =4$ to $\cN =2$ breaking pattern in five dimensions.
In particular, we analyze necessary conditions for these consistent truncations to be used as effective theories
for phenomenology by demanding consistency of the scale-independent corrections to Chern-Simons couplings.

The second part is devoted to the study of anomalies and large gauge transformations in circle-reduced gauge theories and F-theory.
We consider four- and six-dimensional matter-coupled gauge theories on the circle and classify all large gauge transformations that preserve
the boundary conditions of the matter fields.
Enforcing that they act consistently on one-loop Chern-Simons
couplings in three and five dimensions explicitly yields all higher-dimensional gauge anomaly cancelation conditions. 
In the context of F-theory compactifications we identify the classified large gauge transformations along the circle
with arithmetic structures on elliptically-fibered Calabi-Yau manifolds via the dual M-theory setting. 
Integer Abelian large gauge transformations correspond to free basis shifts in the Mordell-Weil lattice of rational
sections while special fractional non-Abelian large gauge transformations are matched to torsional shifts in the Mordell-Weil group.
For integer non-Abelian large gauge transformations we suggest a new geometric group structure on resolved elliptic fibrations.
In the same way we also propose a novel group operation for multi-sections in genus-one fibrations without a proper section.
We stress that these arithmetic structures ensure the cancelation of all gauge anomalies in F-theory compactifications on Calabi-Yau manifolds.

\newpage
\thispagestyle{empty}
\mbox{}

\newpage
\thispagestyle{empty}

{\flushright{\textsc{\Large Acknowledgments}}}

\vspace*{1cm}

A lot of people contributed in all kinds of different ways to support me in writing this thesis over the
last three years.

First of all I am extremely grateful to Thomas Grimm who put confidence in my abilities and
provided working conditions which could not have been any better. He was willing to discuss whenever I desired to,
came up with projects and encouraged me to develop my own ideas. Besides his profound knowledge of the field
he was also always listening carefully when I came up with a maybe contrary point of view.

Many thanks also to my official supervisor Dieter L\"ust for gathering such a nice string theory group in Munich, which I'm very proud
to have been part of.

I am grateful to my collaborators Denis Klevers and Severin L\"ust, whom it was a pleasure to
work with and learn from. Special thanks goes to Jan Keitel, who introduced me to SAGE, and from whose deep geometric
knowledge (he would disclaim this characterization) I could always profit.

Many thanks to all (former)
members of the \textit{Grimm} group at the Max-Planck-Institut. Any single one of them helped me in solving problems
more than once: Federico Bonetti, Pierre Corvilain, Sebastian Greiner, Kilian Mayer, Tom Pugh, Diego Regalado,
Raffaele Savelli, Irene Valenzuela, Matthias Weissenbacher.
Thanks to my office mate Michael Fuchs and the members of the other string theory group, especially Ralph Blumenhagen,
I\~naki Garc\'ia Etxebarria, Daniela Herschmann, Rui Sun, Florian Wolf.

I'd like to express my deep gratitude to my family, who supported me from the start of my studies and took the load off me when needed.

Last but not least thanks to all of my friends, with whom I had a great time over the last three years, especially
Antonia, Bob, Viki, Valle, B\"arbel, Michi, Laura, Jana, Fabi, Julia, Leo, Matthle, Franzi, Peter, Dome, Alex, Maxl, Feldi, Kirsche, 
Anja, Mare,
and also Alina, Friederike and Simone.

\newpage
\thispagestyle{empty}

\tableofcontents

\newpage

\part{Introduction}\label{part:intro}

\chapter{Introduction}\label{ch:intro}

We start this thesis with a very general introduction which is directed mainly
towards non-string theorists and at the beginning to some extent even towards non-physicists.
In \autoref{ch:final_th} we review the general spirit of research in physics and the conceptually new approaches starting with the
beginning of the
20th century. On our way we try to convey how a final theory could look like, and why physicists expect that there exists a
unified description of all processes in nature. Our survey will be influenced by the historical proceedings as well as philosophical aspects.
We also state the current \textit{status quo} of high energy physics together with its open problems and possible resolutions.
After that we aim to an understandable summary of what string theory is about in \autoref{ch:intro_st}. We put some emphasis on
its original development, which was at first mainly influenced by considering string theory as a candidate
for describing nature. However,
we will also try to make clear that string theory is more than that.
In particular, besides the question if it is the correct theory for the
the world around us, it exists as an independent, presumably consistent
framework which has influenced both mathematics and physics beyond its original scope.
In this respect string theory is justified without reference to the outcome of any experiment.
To proceed further,
we find it useful to introduce the concept of symmetries and anomalies in quantum field theory for non-experts in \autoref{ch:popul_anom}.
These topics are very established in modern physics, and make up crucial parts of this thesis. In particular, we will explain why
anomalies render a theory inconsistent. Finally, we provide an outline of this thesis in
\autoref{sec:outline}. We shortly summarize the contents of the different parts there, and also state which kind of 
physical background is required
for the understanding of the individual parts.

\section{Towards a Final Theory}\label{ch:final_th}

      It seems to lie in our nature as human beings to be curios about the origin and the fate of the world we are living in.
      It is remarkable that we have always been asking questions and trying to get a picture of things which we are not directly confronted
      with in our everyday life.
      \textit{Where do we come from? Where do we go? What precisely} is \textit{our world?} 
      These are aspects which have bothered people in the past and still haven't
      lost any of their fascination today.
      Already with the most primitive cultures there had come some kind of religious dogma to explain what had been in the beginning,
      how the world looks like from a distant point of view, and
      what will happen when it all ends.
      Indeed, it is a common ground of nearly all cultures to possess a myth of creation and an idea of the apocalypse.
      
      While these two aspects had seemed to be far out of reach to be described via direct investigation,
      it has always been clear that at least some of the processes in nature follow certain rules.
      It is the
      subject of science to explain the latter based on a collection of fundamental principles which we would call a \textit{theory}.
      When years passed by, bit by bit phenomena of nature got uncovered or even downgraded from divine events to ones that could be
      explained logically although sometimes the underlying principles, \textit{i.e.~}science itself, were considered to be God-given. 
       However, the following categories had resisted a long time to be directly addressed by science (or still do):
      \begin{enumerate}
      \item \textbf{Space and time}
      
      Is space just empty or is it made out of some kind of 'substance' (aether)?
      
      Why are there three spatial dimensions and one time dimension, or are there even more?
      
      Is space curved, does it end somewhere, or is it infinite?
      
      Is it just a static stage for physical processes, or is it dynamical?

      \item \textbf{The evolution of the universe}
      
      Does the universe have a beginning and an end in time?
      
      If yes, how can they be described?
      
      Do there even exist several universes?
      
      \item \textbf{The origin of matter}
      
      What are the fundamental building blocks of matter in nature?
      
      How can they be described?
      
      Can matter be created, destroyed or converted?
      
      \item \textbf{The nature of forces and interactions}
      
      What is the origin of gravity?
      
      Which other forces do exist?
      
      How can they be described?
      
      \item \textbf{The origin of any theory or \textit{the} final theory itself}
      
      Why do there exist things at all?
      
      Where would a final theory come from, how is it 'selected' as \textit{the} final theory?
      \end{enumerate}
      It is the great achievement of the 20th century to directly accommodate for many questions in the
      (before inaccessible) categories 1-4,
      and it is the enormous task of the
      21st century to reconcile 1-4 in one final theory. I for myself cannot imagine how one could ever approach point 5.
      Of course, as many others I am convinced that 
      one can impose that a final theory should rely on simple principles, it should be aesthetically pleasing and maybe also unique
      concerning certain consistency conditions, \textit{e.g.~}as a consistent theory of quantum gravity.
      Although these guiding
      principles might lead us to finding the final theory in the end, they don't seem to explain
      the 'origin' of the latter. In contrast to mathematics, which exists as an
      abstract framework solely based on logic, physics is always subject to becoming reality.
      
      Let us now start reviewing how a lot of the questions which we raised under the points 1-4 got answered in the 20th century.
      The first big step was taken in the year 1915 by Einstein and his theory of general relativity.
     It is fully justified to say that the latter
      is one of the greatest feats in science at all.
      Namely for the first time in history (apart from special relativity to some extent) a theory was proposed
      which made predictions about space and time itself. In particular, matter and energy
      were conjectured to act as a source for the latter.
      Furthermore gravitational interactions got explained as an effect of matter curving spacetime.
      Einstein's perception
      is fundamentally different from basically all theories which had been formulated before. Formerly space
      and time had been considered as some kind of
      unalterable stage on which
      physical processes take place.
      General relativity in contrast completely changed our picture in this respect.
      Even more, at the time when the theory was formulated
      no limitations of Newtonian gravity were known which could be resolved by general relativity. Therefore from a practical perspective
      a more fundamental theory for gravity was not desirable. However, general relativity came along with new predictions which were confirmed
      only afterwards thereby establishing it as a
      solid theory in physics. It is absolutely essential to comprehend the novelty of this approach: Before Einstein physics was done
      in such a way to consider phenomena which could not be explained by current concepts, and then propose a new 
      theory which was able to explain
      these phenomena (perhaps in an aesthetically more pleasing way),
      and at best to predict new phenomena which serve as a test for the theory.
      Einstein simply omitted the first step and created a new theory out of his mind. This constitutes his outstanding achievement.
      Finally, it is worth pointing out that even
      another item in our list of topics was tackled by Einstein, namely for the first time a theory
      was able to make predictions about the origin and the future of the universe as a whole entity.
      This is clear since general relativity describes the dynamics of spacetime. Therefore, given the current status of the universe,
      one can (in principle at least) extrapolate the latter into the past and future using Einstein's field equations.  
      
      The second big revolution in physics of the 20th century took place with the arrival of quantum mechanics and later
      quantum field theory,
      whose development took many centuries starting with the early experiments of Max Planck in the year 1900 until the formulation of the
      \textit{Standard Model} of particle physics in the
      early 1970s.\footnote{Of course this is just a landmark. Quantum field theory remains an active area of research today. There are many
      aspects which are still not settled yet, especially concerning strongly coupled field theories.
      Of course the same holds also true for general relativity, however, developing the foundations of the latter was a more
      confined process than in quantum field theory.}
      Accordingly
      a lot of scientists have contributed to it over the years.
      Quantum mechanics first arose when scientists realized that at very tiny scales matter under some circumstances shows the
      behavior of particles and sometimes that of waves. This seemed to be an obvious contradiction to the at that time 
      separated notions of particles on the one hand and waves on the other hand.
      Even more severe, it turned out that seemingly it was conceptually not
      possible to predict the exact result of the outcome of a single experiment but rather only the probability to obtain a certain result.
      This point of view was confirmed in all following experiments with no contradiction up to this day. Moreover, the beautiful mathematical
      structure of quantum states in an abstract, so-called Hilbert space 
      was invoked in order to describe these processes and to calculate the probabilities.
      Loosely speaking, from that time on matter was understood as certain kind of waves,
      which in general 'contain' a superposition of many different possible values for observable quantities. When we
      carry out a measurement, we nevertheless obtain one unique value. As mentioned, only 
      the probability of the outcome can be calculated exactly. The macroscopic, heuristic illustration of this quantum effect is for instance provided by
      the famous \textit{Schr\"odinger's Cat} Gedankenexperiment.
      This is a (hopefully) fictional scenario where the outcome of an experiment at the quantum level 'decides' if a cat in a box
      gets killed by some kind of mechanism or not. The upshot is that the cat seems to be in a superposition
      of being alive and dead at the same time until one carries out a 'measurement' by opening the box and looking into it.
      Of course one should not take this illustration too seriously since the cat is far too big to be described by
      what physicists call a \textit{coherent quantum state} for which our statements concerning superpositions hold.
     In classical physics, \textit{i.e.~}before the arrival of quantum mechanics, the situation was different: As
     soon as one knows all properties of a physical system at some point in time,
     one can (at least in principle) predict every measurement in the future because classical physics is deterministic.
     In quantum mechanics this is fundamentally different. Although in case you are given the abstract quantum state of a 
     system at some point in time you
     can infer it for all other times, nevertheless the outcome of concrete measurements is only subject to probabilities.
     
     In the further development quantum mechanics was then reconciled with special relativity.
     This resulted in the fascinating and powerful quantum theory of fields.
     Let us highlight
     some important aspects of this theory. Particles were defined in a very precise sense
     as certain quantum states associated to so-called field operators.
     The latter were even classified into different types
     according to their spacetime symmetry properties called \textit{irreducible representations} of the \textit{Poincar\'e group.}
     It was discovered that
      for each particle there has to exist a partner called anti-particle and that particles
      can be created out of the vacuum, destroyed or converted. Interactions and forces (except for gravity)
      were nicely described by gauge theories, which are
      based on gauge symmetries as we will explain in \autoref{s:symmetries}.
      To put it in a nutshell, quantum field theory has provided us with a very clear picture of what matter is,
      how it behaves and interacts. Note that quantum field theory was even able to predict new particles 
      which were essential for keeping the theory consistent. Some of these were indeed discovered afterwards, like the famous Higgs boson.
      
      Some people might call
      the third and last big step in the 20th century the development of string theory.
      However, in contrast to general relativity
      and quantum field theory, which both constitute
      extremely well tested theories with astonishing precise descriptions of nature, string theory
      is still a work in progress with not a single of its inherent predictions confirmed yet.
      Partly this is due to the fact that many predictions of string theory are made for scales which are hard to access for us.
      We will comment on what points in our list at the beginning of this section string theory can accommodate for,
      and how it can unify quantum field theory and general relativity in \autoref{ch:intro_st}. Before we do so it might be fruitful to
      review what the current accepted and tested status of physics is, and what the basic puzzles are that still remain.

      \subsection{Where We Stand...}\label{sec:statusquo}
      In modern physics there are at the moment two in general accepted 
      theories which describe nature to extremely high accuracy at the fundamental level:
      the Standard Model
      of particle physics (supplemented by neutrino masses) and the $\Lambda$CDM Model of cosmology.
      The Standard Model is based on quantum field theory while the $\Lambda$CDM Model is mainly
      characterized by general relativity.
      Before we explain these models in some detail,
      let us stress that up to now
      there exists no satisfactory theory which unifies both theories into a single framework although string theory seems to be a 
      very promising candidate for achieving this. We will comment on the drawbacks of reconciling both theories in \autoref{sec:eff_th}.
      However, in most regimes of physical phenomena effects of either one or the other theory dominate. The Standard Model
      becomes important at small length scales, while the $\Lambda$CDM
      Model outweighs for settings with big masses. For example in particle colliders
      the length scales are usually very small and the masses also. That is why the effects of quantum field theory 
      dominate over general relativity.
      In contrast, when one observes the behavior of stars and galaxies, the masses are quite high but the length scales are 
      also very big. Therefore general
      relativity is trustworthy. However, there are settings where both theories 
      become important but make contradicting predictions.
      This is the regime of the mysterious theory of \textit{quantum gravity} 
      with small length scales and big masses. For instance effects of quantum gravity should become important
      for the description of black holes or the Big Bang.
      
      The Standard Model of particle physics describes all known interactions and forces
      (except for gravity), namely electromagnetism,
      the weak force and the strong force in terms of so-called gauge theories.
      Gauge theories always come with a certain number of spin-one fields which are called gauge
      bosons. They mediate forces between matter fields.
      As already mentioned, gauge theories also have associated gauge symmetry groups, and in the case of the Standard Model
      these are given by
      \begin{align}
       SU(3) \times SU(2)_L \times U(1)_Y \rightarrow SU(3) \times U(1)_{\textrm{em}} \, .
      \end{align}
      The arrow indicates that the gauge group is spontaneously broken down to a subgroup. This is analogous to ferromagnetism
      for which the underlying theory is in principle
      invariant under rotations,
      but below a certain temperature the spins (or 'elementary magnets')
      all spontaneously align into one direction thus breaking the rotational invariance in this way. 
      In the Standard Model a very similar effect is induced by the recently discovered Higgs field which
      mediates the symmetry breaking
      $SU(3) \times SU(2)_L \times U(1)_Y \rightarrow SU(3) \times U(1)_{\textrm{em}}$. In this way mass terms for
      the gauge bosons of the weak force, which are called W$^\pm$- and Z-bosons, are induced. 
      All other gauge bosons are massless, and for the strong force they are called
      gluons, and the one for electromagnetism is the well-known photon.
      The remaining fields of the theory have spin-$\sfrac{1}{2}$. These are called \textit{leptons} and \textit{quarks}
      and arrange in representations
      of the gauge groups, \textit{i.e.~}they in general carry charges under the gauge interactions.
      Their non-vanishing masses are also induced by the spontaneous symmetry breaking due to the Higgs field.
      
      Let us now only shortly comment on the current widely accepted
      description of cosmology, the $\Lambda$CDM (Lambda cold dark matter)
      Model. It is based on the framework of general relativity with a small positive cosmological constant, \textit{i.e.~}positive
      energy density in the vacuum.
      According to this theory our universe was created in a Big Bang around 13.8 billion years ago.
      It is able to explain the formation of stars and galaxies as
      well as the cosmic microwave background. In recent years physicists in addition proposed an
      early era of cosmic inflation. According to the latter there should have existed an epoch of exponential expansion of the universe
      just in a tiny fraction of a second after the Big Bang. As we have already noted, string theory provides an attempt of reconciling the 
      Standard Model with the $\Lambda$CDM Model. Indeed at the moment it becomes more and more attractive to try to realize the idea
      of inflation in string theory models. 
      
      Apart from the issue of unifying both theories there are in addition several puzzles concerning fine-tuning which might
      have to be addressed. In the following we comment on what is called the \textit{electroweak hierarchy problem}
      and the \textit{cosmological constant problem}:
      \begin{itemize}
       \item The electroweak hierarchy problem states that in the framework of the Standard Model one would generically expect the
       Higgs mass to be much bigger than the observed value due to quantum corrections like the ones we describe in \autoref{sec:eff_th}.
       The fact that it is so small requires an unimaginable fine-tuning of parameters over many scales. Thus it is solely a problem
       of naturalness but no inconsistency of the theory in principle. However, by imposing a new symmetry called \textit{supersymmetry}
       the measured value for the Higgs mass could lie in a range which seems to be more natural.
       \item The cosmological constant problem is in a similar spirit. In fact, the cosmological constant is unnaturally small,
       and it is extremely hard to come up with extensions of our theories which render this smallness more natural.
      \end{itemize}
    For both of these puzzles there exists a possible philosophical 
    resolution called the \textit{anthropic principle}, which is widely discussed. It states that if the Higgs mass and the cosmological
    constant would acquire the more generic big values, life of any form in such a universe would not be possible. Therefore in this case
    observers which could measure these quantities could not exist.
      Finally, there are further issues like the abundance of \textit{dark matter} or the \textit{strong CP problem} for which
    there do not exist satisfactory explanations by current theories. However, we will not go into the details here.
      
      \subsection{Effective Theories}\label{sec:eff_th}
      
      The notion of \textit{effective field theory} in quantum field theory is of particular importance for this thesis since it
      appears in the context of consistent truncations in \autoref{part:susybreak} and together with circle-compactified theories
      and F-theory in \autoref{part:fth}. We therefore dedicate this subsection to a short introduction into the main ideas.
      
      In principle one would naively expect a quantum field theory to be valid or even invariant
      at any energy scale. However, many quantities like
      the coupling constants to gauge interactions or the mass depend crucially on the energy scale. The extrapolation between the values for 
      these objects at different scales is governed by the renormalization group flow. It can now happen that this procedure only 
      yields good finite results
      as long as the energy is smaller than some maximal cut-off energy scale $E_{\rm max}$
      \begin{align}
       E < E_{\rm max} \, ,
      \end{align}
      and breaks down as $E \rightarrow E_{\rm max}$. This signals that the quantum field theory is
      only valid up to the scale $E_{\rm max}$, and one calls it non-renormalizable.
      For energies above this bound there have to be new degrees of freedom
      which become relevant, \textit{i.e.~}massive fields
      which didn't play a role at energies far below $E_{\rm max}$. The putative theory which includes also these degrees of freedom and is valid
      up to any energy scale is called an \textit{ultraviolet completion}. The original theory is called an
      \textit{effective theory} for the full, UV-complete theory. Since fields which have masses higher than
      the cut-off scale $E_{\rm max}$ do not constitute relevant degrees of freedom for $E<E_{\rm max}$, they can be removed from
      the description of the theory by a procedure which is called \textit{integrating out.} Via this process one obtains
      an effective description of the light degrees of freedom with $E<E_{\rm max}$. Note that in this so-obtained effective theory
      new couplings could have been generated or corrected. The latter usually scale with the energy and can therefore be neglected
      as one approaches $E \rightarrow 0$. In this thesis however we put special emphasis on corrections to Chern-Simons couplings in
      effective field theory. These have the crucial property of being \textit{independent} of the mass scale and therefore have to be
      included at arbitrarily low energies. For that reason they encode a lot of interesting physics as we will show in this
      thesis.
      
      Let us give a nice historical example of a non-renormalizable effective theory which was first proposed and later the more
      fundamental theory was found. Indeed, in 1933 Fermi described the beta decay by directly coupling the involved matter
      fields to each other. 
      This theory turned out to be non-renormalizable. Nevertheless much later when the more fundamental
      theory for the electroweak interactions
      was formulated, Fermi's theory could be understood in terms of an effective field theory with the massive gauge bosons
      integrated out. Put together, in the full theory massive gauge fields mediate the beta decay. By integrating them out 
      a non-renormalizable coupling between the involved matter fields is generated.
      
      We are now in a position to convey what goes wrong in trying to reconcile general relativity with quantum field theory.
      In fact, if we straightforwardly promote general relativity to a quantum field theory, we obtain
      a theory which is non-renormalizable. Therefore the latter can serve only as an effective theory which is valid
      at most up to the Planck scale around energies of $10^{19}$ GeV.
      At this scale we expect new degrees of freedom which guarantee
      a nice behavior in the ultraviolet. In string theory for instance these states are supposed to be provided by higher vibrational
      modes of the string.

      Finally let us close by mentioning also a very important issue related to strongly-coupled quantum field theory.
      In order to evaluate certain processes for a given quantum field theory one usually uses the technique of perturbation theory.
      Loosely speaking this corresponds to making a Taylor expansion in the coupling constant. It is clear that this procedure
      breaks down if one reaches the regime of strong coupling. In this case the description of the theory in general changes completely
      in the sense that one has to work with different fundamental degrees of freedom. For instance
      the fundamental degrees of freedom of matter for the strong interactions at weak coupling are the quarks.
      However, as we move to low energies, the theory becomes strongly coupled.
      That is why in our world
      of small energy scales we only observe bound states of quarks, namely protons, neutrons, pions and so on. These constitute
      the appropriate degrees of freedom for strong coupling. Note that nevertheless the precise description of the strong interactions
      at strong coupling is still far from being well-understood.

  \section{The Framework of String Theory}\label{ch:intro_st}
  
   \vspace{1cm}
  
  {\footnotesize\textit{
 ``}[\dots ]\textit{ From those incontrovertible premises, the librarian deduced that the Library is 'total'---perfect,
 complete, and whole---and that its bookshelves contain all possible combinations
 of the twenty-two orthographic symbols (a number which, though unimaginably vast, is not infinite)---that is, all that is able to be expressed,
 in every language.} All\textit{---the detailed history
 of the future, the autobiographies of the archangels, the faithful catalog of the Library, thousands and thousands of false catalogs, the
 proof of the falsity of those false catalogs, a proof of the falsity of the} true \textit{catalog, the gnostic gospel of Basilides,
 the commentary upon that gospel, the commentary on the commentary on that gospel, the true story of your death, the translation of every book
 into every language, the interpolations of every book into all books, the treatise Bede could have written (but did not) on the mythology of
 the Saxon people, the lost books of Tacitus.}
      
      \textit{When it was announced that the Library contained all books, the first reaction was unbounded joy.
      All men felt themselves the possessors of an intact and secret treasure. There was no personal problem, no world problem, whose eloquent
      solution did not exist---somewhere in some hexagon. The universe was justified; the universe suddenly became
      congruent with the unlimited width and breadth of humankind's hope.
      At that period there was much talk of The Vindications---books of} apologi\ae \textit{ and prophecies that would vindicate for all time
      the actions of
      every person in the universe and that held wondrous arcana for men's futures. Thousands of greedy
      individuals abandoned their sweet native hexagons
      and rushed downstairs, upstairs, spurred by the vain desire to find their Vindication. These pilgrims squabbled in the narrow corridors,
      muttered dark imprecations, strangled one another on the divine staircases, threw deceiving volumes down ventilation shafts,
      were themselves hurled to their deaths by men of distant regions. Others went insane\dots. The Vindications do exist
      (I have seen two of them, which refer to
      persons in the future, persons perhaps not imaginary), but those who went in quest of them failed to recall that the chance
      of a man's finding
      his own Vindication, or some perfidious version of his own, can be calculated to be zero.}
      
     \textit{At the same period there was also hope that the fundamental mysteries of mankind---the origin of the Library
      and of time---might be revealed. In all likelihood those profound mysteries can indeed be explained in words; if the language
      of the philosophers is not sufficient, then the multiform Library must surely have produced the extraordinary language
      that is required, together with
      the words and grammar of that language. For four centuries, men have been scouring the hexagons\dots.
      There are official searchers, the 'inquisitors'.
      I have seen them about their tasks: they arrive exhausted at some hexagon, they talk about a
      staircase that nearly killed them---some steps were missing---they speak with the librarian about galleries and
      staircases, and, once in a while, they take up the nearest book and
      leaf through it, searching for disgraceful or dishonorable words. Clearly, no one expects to discover anything }[\dots]\textit{''}\\}

      {\small\flushright{\textsc{La Biblioteca de Babel}, \textsc{J. L. Borges} (transl. Andrew Hurley, New York, Penguin)\\}}

  \vspace{1cm}

  In this section we give a short overview over the various aspects of string theory.
  We do not go into detail here but rather aim towards a pedagogical introduction for non-string-theorists.
  Our treatment follows the historical development, starting with the attempt of describing the strong interactions
  via strings, over the first and second
  string revolution until today with the numerous different branches of string theory.
  
  \subsection{The Beginning}
  
  String theory
  was first considered in the late 1960s as an attempt to describe the strong interactions which was however abandoned
  in the 1970s
  in favor of quantum chromodynamics, an ordinary quantum field theory based on point particles.
  Luckily, shortly after that physicists got interested in what is now called bosonic string theory.
  While it is only a predecessor of the more advanced
  superstring theories which constitute the theories of interest today, 
  it already showed many properties which excited people at that time and still do:
  String theory is about replacing point particles by extended one-dimensional objects called strings,
  which can be either open or closed. These fields can vibrate, and different vibrational
  modes correspond to different particles, like the different vibrational modes of a violin generate different tones. Importantly,
  in the spectrum of vibrational modes there is always an excitation, which describes the fluctuation of a background spacetime metric.
  This was considered as a hint that string theory could be a candidate for a consistent theory of quantum gravity. Indeed,
  it is astonishing how string theory deals with the bad non-renormalizable infinities in quantum field theory
  associated to gravitational interactions. The extended nature of the
  string delocalizes interaction vertices, and the problematic ultraviolet
   regime is mapped by a so-called \textit{duality} to the infrared regime which
   can be described easily. More precisely, this duality states that the physics of long strings at high energies is 
   the same as the physics
   of short strings at low energies. Via a precise 'dictionary' these regimes can be mapped to each other.
   All these nice properties have already appeared in the
   early version of bosonic string theory. However, the latter suffers from a couple of important drawbacks which makes it impossible
   to consider it as a theory of the world around us.
   First, it cannot account for spacetime fermions, which are the fundamental building blocks of our world.
   Second, in the spectrum of the theory one finds tachyons, \textit{i.e.}~modes of imaginary mass. These signal an instability of the theory.
   While the tachyon in the sector of open strings
   is quite well understood (we are sitting at the maximum of a potential, and rolling down corresponds to
   so-called D-brane condensation), the implications of the tachyon in the closed string sector
   are not clear but might most certainly render spacetime itself
   unstable. Both issues, the presence of tachyons and the absence of spacetime fermions, 
   soon got resolved by moving from bosonic string theory to superstring theory.
   By introducing a fermionic partner string for the bosonic string
   the theory acquires a new symmetry, namely two-dimensional supersymmetry. The latter is powerful enough to allow for stable solutions,
   and at the same time also leads to spacetime fermions, while keeping the nice properties in the ultraviolet regime.
   Indeed, it was found that there even exist five different superstring theories, which all require for consistency a total number of
   exactly ten spacetime dimensions. They are called \textit{type I}, \textit{type IIA},
    \textit{type IIB}, \textit{$SO(32)$ heterotic} and
    \textit{$E_8 \times E_8$ heterotic string theory}.

    \subsection{The Two Superstring Revolutions}
   It was in the mid-1980s when several discoveries (now called
   the \textit{first superstring revolution}), like for instance the anomaly cancelation of type I string theory via the Green-Schwarz
   mechanism, made physicists realize
   that superstring theories might be able to serve as fundamental theories of our world unifying quantum field theory and general relativity.
   The superfluous six dimensions out of the in total ten dimensions were argued to be
   very tiny in order to have escaped detection so far. Their description is captured
   by certain geometrical spaces which are solutions to the equations of motion of the theory.
   Importantly, much of the physics in the large four dimensions depends on the detailed shape of the six small dimensions.
   The process of rendering dimensions small is called compactification, and a big part of this thesis consists of considering
   the compactification of \textit{F-theory} on \textit{Calabi-Yau manifolds} to six and four large spacetime dimensions, respectively.
   We will have to say more about F-theory in a moment.
   
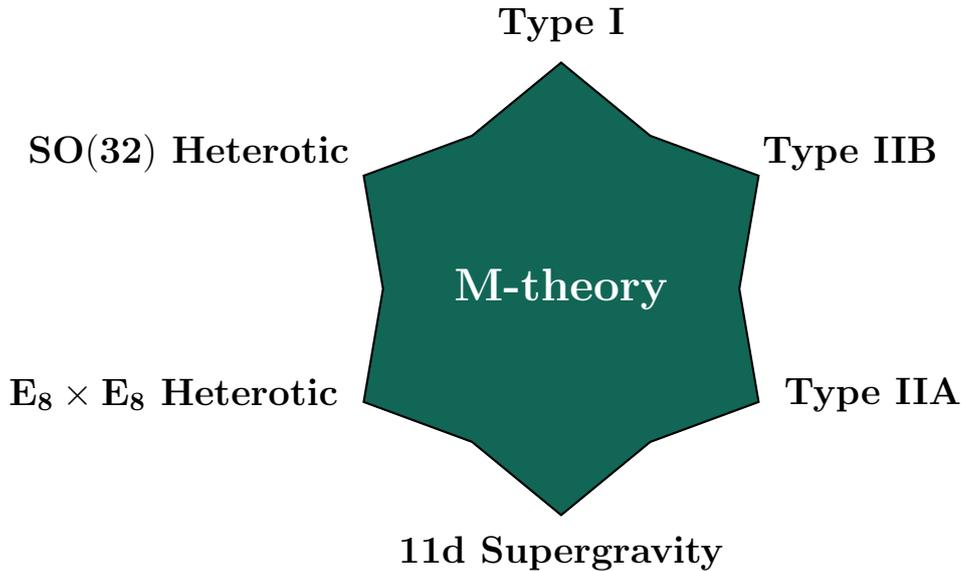
\begin{figure}
 \centering
   \begin{tikzpicture}
 \ngram{3}{6}{0}{thick,fill=mppgreen}
 \node[white] at (0cm,0cm) {\bf{\Large{M-theory}}};
 \node at (0cm,3.5cm) {\bf{\large{Type I}}};
 \node at (0cm,-3.5cm) {\bf{\large{11d Supergravity}}};
 \node at (3.8cm,1.8cm) {\bf{\large{Type IIB}}};
 \node at (4.1cm,-1.4cm) {\bf{\large{Type IIA}}};
 \node at (-4.9cm,1.8cm) {\bf{\large{$\bf SO(32)$ Heterotic}}};
  \node at (-5.1cm,-1.4cm) {\bf{\large{$\bf E_8 \times E_8$ Heterotic}}};
\end{tikzpicture}
\caption{We depict the famous M-theory star, which illustrates the unification of all five superstring theories and eleven-dimensional
supergravity. These constitute certain limits of the still not very well understood framework of M-theory.}
\label{fig:mstar}
\end{figure}

   Another ten years later in the mid-1990s Edward Witten and Joseph Polchinski initiated the \textit{second superstring revolution}.
   In fact, strong evidence
   was found that all five consistent superstring theories are linked together via dualities, and describe certain limits of a new
   conjectured eleven-dimensional theory, which Witten named M-theory, see \autoref{fig:mstar}.
   Finally it was shown that string theories
   are not theories of strings only, but for consistency also have to 
   allow for certain higher-dimensional, non-perturbative objects called branes.

   In 1997 Juan Maldacena found the first concrete realization of the \textit{holographic principle} in string theory,
   which in general conjectures that the properties of quantum gravity on some space is solely encoded by an ordinary quantum field
   theory on the lower-dimensional boundary of that space.
   Maldacena showed that
   the $\cN = 4$ super Yang-Mills theory in four dimensions (which is a quantum field theory) is dual to type IIB string theory compactified
   on a five-sphere. The remaining five dimensions span a space of negative curvature called anti-de Sitter space.
   The $\cN = 4$ super Yang-Mills theory is defined on the boundary of the latter.
   
   One year before, in 1996, Cumrun Vafa and David Morrison formulated F-theory, which makes up a considerable part of this thesis.
   Let us therefore spend some time in order to heuristically describe some of its properties. 
   We provide a much more detailed introduction in \autoref{ch:f_basics}.
   F-theory is a generalization of type IIB string theory. The latter possesses an inherent $SL(2, \mathbb Z)$-symmetry which is also
   precisely the reparametrization symmetry of a two-dimensional torus. The idea of F-theory is to take this fact seriously, and 
   extend the original ten spacetime dimensions to twelve by including an additional torus. Due to this property it is in F-theory
   even more crucial to analyze the compactification space in detail since most properties of physics in the remaining large dimensions
   can be directly read off from the geometry. In particular, the compactification
   spaces always have to include tori which in general collapse at some points.
   These encode gauge interactions and charged matter fields.
   
  \subsection{The Different Branches Today}
\begin{figure}
\centering
\scalebox{0.8}{
  \begin{tikzpicture}[mindmap,
  level 1 concept/.append style={level distance=130,sibling angle=30,font=\footnotesize},
  extra concept/.append style={color=blue!50,text=black}]


  \begin{scope}[mindmap, concept color=mppgreen!90, text=white]
    \node [concept] (st) at (-2.5,0) {String Theory, M-Theory}[clockwise from=210] 
       child {node [concept] (gen) {Fundamental Aspects}}
       child {node [concept] {Non-Perturbative Effects}}
       child {node [concept] (comp) {Compactifi-cation}}
              child {node [concept] (dualst) {Dualities}}
       child {node [concept] (amp) {Amplitudes}}
       child {node [concept] {Pheno-menology}}
       child {node [concept] {Cosmology}}
       child {node [concept] {Landscape}}
       child {node [concept] {Quantum Gravity}};
  \end{scope}

  \begin{scope}[mindmap, concept color=mppgray,text=black]
    \node [concept] (QFT) at (-6,-6) {Quantum Field Theory}[clockwise from=-75]
      child [level distance=130]
        {node [concept] (quan) {Quantization}}
      child [level distance=130] 
        {node [concept] (ampqft) {Amplitudes}}
        child [level distance=130] 
        {node [concept] (np) {Non-Perturbative Effects}}
      child [level distance=130] 
        {node [concept] (dualqft) {Dualities}}
        child [level distance=130] 
        {node [concept] (scft) {Conformal Field Theory}}
        child [level distance=130] 
        {node [concept] (cm) {Condensed Matter}};
  \end{scope}

  \begin{scope}[mindmap, concept color=blue!80,text=white]
    \node [concept] (math) at (1,-6) {Mathematics} [counterclockwise from=255]
      child [level distance=130] 
        {node [concept] (geo) {Geometry}}
      child [level distance=130] 
        {node [concept] (top) {Topology}}
child [level distance=130] 
        {node [concept] (num) {Number Theory}}
      child [level distance=130] 
        {node [concept] (group) {Group Theory}}
        child [level distance=130] 
        {node [concept] (sta) {Statistics}}
        child [level distance=130] 
        {node [concept] (func) {Functional Analysis}};
  \end{scope}

   \begin{pgfonlayer}{background}
   \path (st) to[circle connection bar] (math);
   \path (st) to[circle connection bar] (QFT);
   \path (QFT) to[circle connection bar] (math);
 \end{pgfonlayer}

\end{tikzpicture}
}
\caption{The different areas of string theory, quantum field theory and mathematics have strong influence on each other.}
\label{branches}
\end{figure}
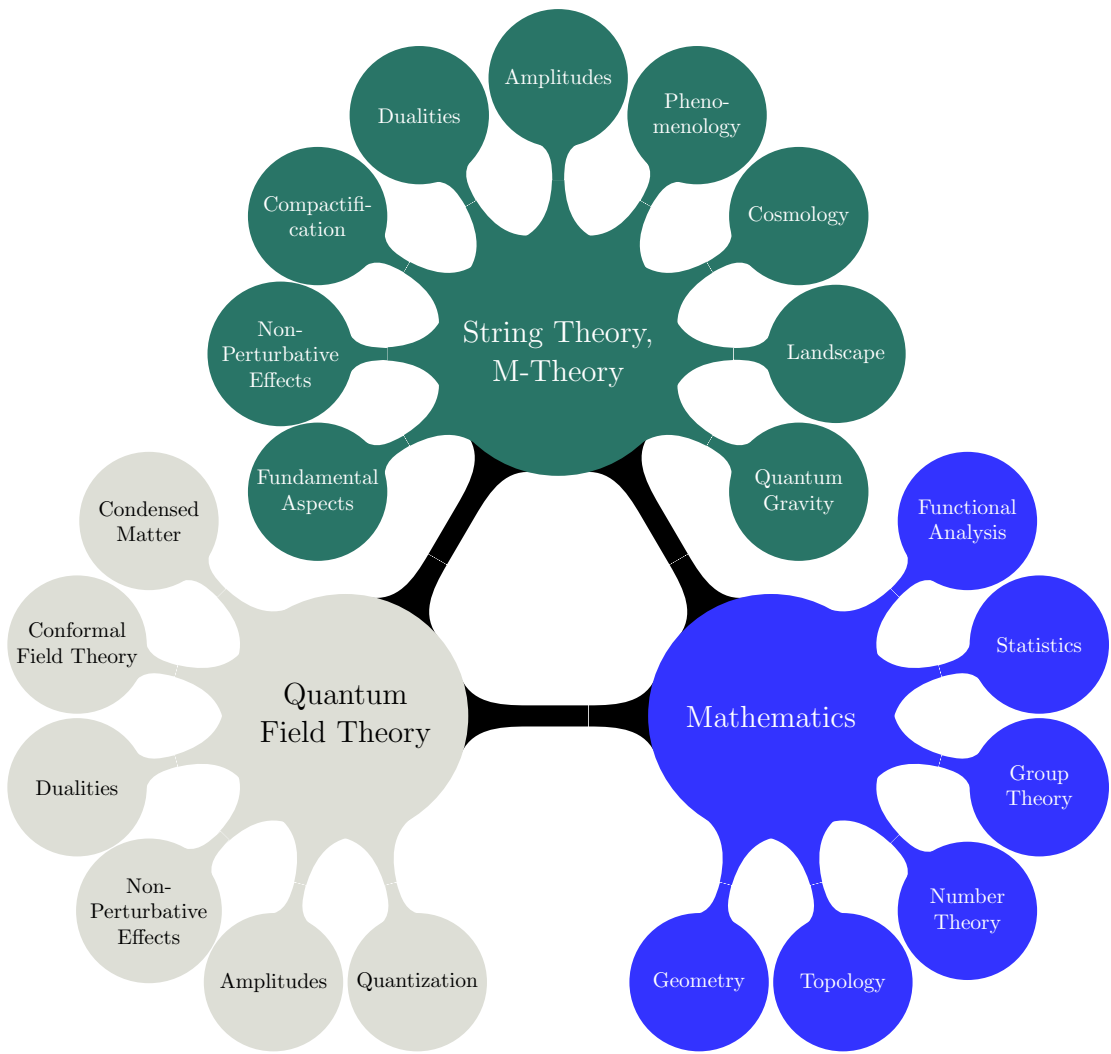

  We have seen that string theory, which had originally been developed 
  to describe the strong interactions, turned out to be a much more powerful framework
  than anyone could have imagined.
  Indeed, today there exist many different areas and aspects of this field on which theorists work actively,
  most of them having application and impact also outside of string theory.
  The present subsection 
  is meant as a short overview over the different branches of research within string theory. Note that we will not be able
  to do justice to
  all the branches but only focus on the most important ones.
  Moreover, we highlight their connection
  to quantum field theory and mathematics, which both profit a lot from developments in string theory,
  and conversely provide also crucial ingredients
  for the latter. The basic topics are also depicted in \autoref{branches}.
  
The oldest branch is certainly the study of the fundamental
aspects of string theory. It is however still one of the most mysterious and least understood ones despite the
intriguing connections and hints that have been found.
As already mentioned,
starting with the second superstring revolution it became clear that all consistent five superstring theories are connected via 
(non-perturbative)
dualities and constitute just different aspects of a more fundamental theory, namely M-theory. Nevertheless we still lack a full
(non-perturbative) formulation of string theory or M-theory, respectively. There has been at least
some progress in understanding the dynamics
of the fundamental objects in M-theory, M2-branes and M5-branes.
As we will explain in more detail in \autoref{sec:low_energy_m},
the former are described
in the context of ABJM theory, but we still seem to scratch just the tip of the iceberg of the theories describing M5-branes,
namely the 6d (2,0) superconformal field theories.
This is a nice example of how string theory broadened our knowledge of quantum field theory since interacting superconformal field theories
in six dimensions were long thought not to exist. From a string theory perspective however they are expected to be present as the low-energy
world-volume theories of multiple M5-branes. Furthermore, it is claimed that a certain class of F-theory compactifications exhausts the
full class of consistent 6d (2,0) superconformal field theories.
Apart from quantum field theory also for mathematics the fundamental aspects of string theory
yield new inspirations and insights.
For instance
non-commutative and non-associative structures, which appear naturally in string theory,
opened up new branches in group theory. Moreover, in this thesis we conjecture new mathematical structures on genus-one curves
from our intuition of F-theory.

 Another active area of intense research
 is compactification of string theories and M-theory, respectively. Also the much of the original work of this thesis in \autoref{part:susybreak}
 and \autoref{part:fth} falls into this category.
 In general, compactifications to any number of remaining large spacetime dimensions are interesting on their own and offer new insights,
 also to quantum field theory and mathematics.
 However, the case of four non-compact spacetime dimensions is particularly important and investigated heavily since this is the number of (large) dimensions
 we observe in nature. Indeed, in the field of \textit{string phenomenology} physicists try to reconcile the physics at the fundamental level of the world around us
 with string theory. This means that they in particular look for ways to realize the Standard Model of particle physics, which we introduced in \autoref{sec:statusquo},
 as a string compactification to four spacetime dimensions.
 In the same way physicists also try to embed the $\Lambda$CDM Model and in particular also the concept of inflation into string theory. This
 branch is usually referred to as \textit{string cosmology}.
 It is important to realize that the number of consistent string compactifications is unimaginably large.
 For this whole set people often use the term \textit{string landscape}. 
 There are some rough estimates on the size of the string landscape
 in the literature which are discussed heavily. The usual number of different vacua which people refer to is $10^{500}$.
 Even if one does not believe in the quantity $10^{500}$, it is unarguable that the number should be extremely big.
 At first it seems that we are in
 the same situation as the \textit{inquisitors}, mentioned in the citation at the beginning of this section,
 looking for a needle in a haystack. But there are some important differences: The string landscape is certainly not \textit{total} in the sense
 that every quantum field theory has a stringy realization, \textit{i.e.~}arises as a certain compactification of string theory. 
 Nevertheless it seems hopeless to go on the quest
 for a compactification geometry which exactly yields the Standard Model and the $\Lambda$CDM Model
 with all their details. In contrast, what people do is
 trying to find out if general features of these models can in principle be realized in string theory. 
 This means for instance: \textit{Is it possible to get the gauge group
 $SU(3) \times SU(2) \times U(1)$ in some way? Can the chiral spectrum and the observed mass hierarchies be generated?
 What are the restrictions on inflation in string theory?} Luckily, many questions of this kind can be answered very generally in string theory.
 Thus the aim is not to obtain the exact values for couplings and masses, but it is rather about getting the big picture right.
 We now close our heuristic discussion of string theory and proceed with a general treatment concerning symmetries and anomalies.
 
  \section{Symmetries and Anomalies in Field Theory}\label{ch:popul_anom}
  
  The appearance of anomalies in quantum field theory makes up a crucial part of this thesis,
  especially in \autoref{part:fth}.
 Therefore we heuristically discuss some basic background on symmetries in physics and quantum anomalies in general for the interested reader.
 We start with the introduction of global 
 and local symmetries in \autoref{s:symmetries}, and continue to explain in \autoref{s:anomaly_intro} how
 the breakdown of symmetries
 under quantization leads to anomalies in quantum field theory. For a more advanced recap of the concept of anomalies we refer
 to \autoref{ch:ano}.
  
  \subsection{Global Symmetries vs. Gauge Symmetries}\label{s:symmetries}
 
 In physics the notion of a symmetry refers to transformations of the configuration space which map solutions of 
 a particular theory again to solutions of the same theory.
 For instance consider the Poincar\'e group in the theory of electrodynamics. The continuous part of this group consists of translations
 in space and time, rotations in space
 and Lorentz boosts. Once we have found a solution in electrodynamics for some observer in spacetime, we can simply apply a Poincar\'e
 transformation and obtain in this way the solutions of the theory for all other observers related to the initial one by
 precisely this Poincar\'e transformation.
 Note that in contrast this does not work for observers with a relative acceleration, which are not related by Poincar\'e transformations.
 
 The role of the Poincar\'e group in electrodynamics is an example of what one calls an invariance under a \textit{global} symmetry group.
 We will find it important to distinguish between
 \textit{global} and \textit{gauge} (\textit{local})
 symmetries in this thesis (and also in general of course). Therefore let us explain the difference:
{ \allowdisplaybreaks \begin{itemize}
  \item For \textit{gauge} symmetries the symmetry parameter, \textit{e.g.}~the rotation angle, is allowed to depend
  non-trivially on the spacetime coordinates.
  This implies that gauge symmetries
 only describe redundancies of the theory since certain 'degrees of freedom' can be removed by a gauge transformation.
  One can equally well describe the theory after getting rid of all redundancies, this is called gauge-fixing. However, from a technical
  and mathematical point of view it is often more appealing to keep them.
  In a strict sense gauge symmetries are not even 'symmetries' at all for precisely that reason. As we have already mentioned,
  the weak, strong and electromagnetic interactions are based on gauge symmetries and therefore called gauge theories.
  \item Global symmetries
 only allow for a symmetry parameter which is constant over spacetime. They cannot be used to remove degrees of freedom, that's why they constitute actual symmetries.
 \end{itemize}}
It is absolutely essential to keep in mind the fundamentally different natures of \textit{global} and \textit{gauge} symmetries.
As we will explain in the next subsection,
this difference renders a theory with an anomalous gauge symmetry inconsistent while anomalous global symmetries are in general not 
problematic.
Finally, it is worth mentioning that it is believed that in a consistent full quantum gravity theory continuous global symmetries 
cannot exist, and all
symmetries are gauge symmetries.
  
  \subsection{Anomalies}\label{s:anomaly_intro}
  
 As already announced we now explain on a very basic level what quantum anomalies are, how they can 
 arise, as well as their implications for a theory. 
  Many quantum theories have an underlying classical theory, and the process of making the classical theory into a quantum theory is
  called quantization. It can happen that (global or gauge) symmetries break down under this quantization procedure,
  meaning that symmetries of the
  classical theory are not realized at the quantum level. In principle for global symmetries this constitutes no problem.
  On the one hand one might perhaps be interested in keeping certain global symmetries in the quantum theory, 
  but inconsistencies
  of the theory itself do not arise.
  For gauge symmetries the situation is completely different. Since they parametrize only redundant,
  \textit{i.e.~}unphysical 
  degrees of freedom in the theory, they should not disappear after quantization. In fact, an anomalous gauge symmetry
  renders the resulting quantum theory inconsistent.
  To put it in a nutshell, anomalies of global symmetries are in principle fine while gauge anomalies must always be canceled in order to
  retain a well-defined theory.
  
  There are a lot of equivalent definitions of an anomaly. The most intuitive one on a basic level is the one just described.
  Note that there are also a lot of examples of theories which do not have an underlying
  classical description but are only defined at the quantum level. These can also suffer from quantum 
  anomalies, however, at this heuristic stage we will not comment
  on such theories. Later in \autoref{ch:ano} we introduce 
  anomalies as non-conservation of certain currents, a definition which is independent
  of the existence of an underlying classical theory. 
  
    \section{Outline of the Thesis}\label{sec:outline}
  
  This thesis is divided into several different parts of which some can be read independently from each other.
  The introductory \autoref{part:intro}
  is first in \autoref{ch:intro}
  directed to readers without much background in string theory or quantum field theory.
  It conveys the general status of modern high energy physics as well as its historical origin and development.
  We give a short overview of the framework of string theory and its impact on other branches in mathematics and physics.
  Special emphasis is also put on effective field theories as well as the concept of anomalies in quantum field theory.
  Our aim is to put the work in this
  thesis in an understandable context for readers who are not experts in this field.
  Even non-physicists might be able
  to comprehend many of the points which are described in this chapter. 
  
  In the following \autoref{part:prel} we already require familiarity with quantum field theory at a very basic level.
  We review some important aspects and results which enter this thesis at several different stages.
  These are anomalies in quantum field theory in general, as well as Chern-Simons terms in three- and five-dimensional theories.
  Readers who are familiar with these concepts can safely skip this part or consult it if needed later.
  
  The presentation of original results of this thesis starts in \autoref{part:susybreak}.
  Fundamental knowledge of supergravity is needed in order to understand this part.
  We investigate (partial) supersymmetry
  breaking of five-dimensional $\cN=4$ gauged supergravity, and in particular derive crucial properties of the effective theory around vacua
  with special emphasis on $\cN=2$ vacua. Via a newly described tensorial Higgs mechanism tensors become massive by \textit{eating up} a vector, similar
  to the St\"uckelberg mechanism. We fully analyze the $\cN=2$ effective theory for purely Abelian magnetic gaugings.
  The set of modes which become massive in this breaking procedure are shown to induce
  one-loop corrections to Chern-Simons terms which are independent of the supersymmetry breaking scale.
  We find concrete
  realizations of the breaking $\cN=4 \rightarrow \cN=2$
  in consistent truncations of type IIB supergravity on five-dimensional (squashed)
  Sasaki-Einstein manifolds and M-theory on six-dimensional $SU(2)$-structure manifolds. The former truncation has already been described in the literature while the latter
  is derived here in detail.
  Exploiting the mentioned scale-invariant one-loop corrections we determine necessary
  conditions for consistent truncations to yield sensible effective theories. We test these constraints for our two examples of consistent
  truncations, and both times we obtain positive results.
  This part is based on the two publications \cite{Grimm:2014soa,Grimm:2014aha}:
  \begin{itemize}
   \item T. W. Grimm, and A. Kapfer, ``Self-Dual Tensors and Partial Supersymmetry Breaking in Five Dimensions,''
   \textit{JHEP} \textbf{1503} (2015) 008, \href{http://arxiv.org/abs/1402.3529}{1402.3529}
   \item T. W. Grimm, A. Kapfer, and S. L{\"u}st, ``Partial Supergravity Breaking and the Effective Action of Consistent Truncations,''
   \textit{JHEP} \textbf{1502} (2015) 093, \href{http://arxiv.org/abs/1409.0867}{1409.0867}
  \end{itemize}
  
  In \autoref{part:fth} we present original work in the context of circle-reduced gauge theories and F-theory.
  At first in \autoref{ch:circle_theories} and \autoref{ch:lgts} we assume only basic knowledge of
  quantum field theory, and we already obtain interesting results in these chapters. 
  In contrast, beginning with \autoref{ch:f_basics} familiarity with
  string theory is essential. Note that the reader is not at all required to be an expert in F-theory since we give a short introduction into the relevant topics of
  this subject.
  We start \autoref{part:fth} by
  reviewing the circle compactification of general four- and six-dimensional matter-coupled gauge theories along with
  the anomaly cancelation conditions of the uncompactified theories.
  Afterwards we classify large gauge transformations along the circle which preserve the boundary conditions of all matter fields. 
  Exploiting these maps we describe a
  procedure to extract the higher-dimensional anomaly cancelation conditions from the reduced theories on the circle.
  In order to do so we evaluate the large gauge transformations on one-loop Chern-Simons terms, and demand that they have to act in
  a way which is consistent with quantization. This procedure yields all gauge anomaly cancelation conditions in four and six
  dimensions as well as the mixed gauge-gravitational anomalies in six dimensions.
  In the context of F-theory
  compactifications we derive a precise dictionary
  which matches our classification of large gauge transformations along the circle
  to arithmetic structures on genus-one fibrations.
  We also comment on the implications of different choices for the zero-section of F-theory compactifications, which is a
  related topic.
  Some of these arithmetic structures
  are well-known (\textit{e.g.}~the Mordell-Weil group of rational sections), others
  are conjectured in this work since they have not been described in the mathematical literature yet.
  We find further evidence for the existence of our newly derived arithmetic structures by considering concrete geometric examples. In particular
  we investigate how novel group structures arise in Higgs transitions as remnants of the familiar Mordell-Weil group structure.
  Importantly our findings establish the cancelation of gauge anomalies in Calabi-Yau compactifications of F-theory.
  The results of this part are published in \cite{Grimm:2015zea,Grimm:2015wda}:
  \begin{itemize}
   \item T. W. Grimm, and A. Kapfer, ``Anomaly Cancelation in Field Theory and F-theory on a Circle,''
   \textit{JHEP} \textbf{1605} (2016) 102, \href{http://arxiv.org/abs/1502.05398}{1502.05398}
   \item T. W. Grimm, A. Kapfer, and D. Klevers, ``The Arithmetic of Elliptic Fibrations in Gauge Theories on a Circle,''
   \textit{JHEP} \textbf{1606} (2016) 112, \href{http://arxiv.org/abs/1510.04281}{1510.04281}
  \end{itemize}
   
   We stress that the two parts which contain the original work in this thesis, namely \autoref{part:susybreak} and \autoref{part:fth},
  treat two in principle very different topics and can therefore be read completely independently from each other. The connecting piece is only
  the usage of one-loop Chern-Simons terms in order to probe topological properties of gauge theories (at least as
  our considerations are concerned). As mentioned before,
  a review of one-loop Chern-Simons terms is provided at the beginning of this thesis in \autoref{part:prel}.
  
  Finally, in \autoref{part:add} we present further interesting ideas
  as well as first results which refer to \autoref{part:fth}
  but have
  not been published yet. In fact, some topics are more or less settled and just
have to be worked out in full detail, while others are still quite speculative.
  This part is on the one hand meant for the interested reader to demonstrate the universal applicability
  of the results in \autoref{part:fth} to
  other interesting areas of research, and on the other hand provides a concrete starting point for future investigation along these lines.
  
The main part of this thesis concludes with a summary of our results in \autoref{part:concl}
accompanied by a short outlook.
We complete
our work with additional material in \autoref{part:app} covering conventions, longish calculations and convenient tables
which would spoil the readability of the main text.

 \chapter{Preliminary Material}\label{part:prel}
    
 \section{Recap of Anomalies in Quantum Field Theory}\label{ch:ano}

 Anomalies in quantum field theory and string theory compactifications are very powerful objects. Because of their topological origin they are 
 very robust quantities which nevertheless carry a considerable amount of information about the chiral spectrum.
 In this thesis anomalies play a very prominent role and enter at several stages.
 Indeed, in \autoref{sec:anom_lgt} we discuss how one can obtain the gauge anomaly cancelation conditions of an arbitrary four- or six-dimensional
 theory after
 an additional circle compactification. These
 techniques are not only interesting on their own but also help us in
  better understanding the mechanisms underlying anomaly cancelation in F-theory compactifications. In particular combining the
  field theory results of \autoref{sec:anom_lgt} with the arithmetic structures on elliptic fibrations, which we introduce
  in \autoref{ch:arith},
  we are able to prove the cancelation of $U(1)$ gauge anomalies in F-theory explicitly, and find strong evidence that an
  arithmetic structure for blow-up divisors should also ensure the cancelation of all non-Abelian gauge anomalies.
  
 Since in the literature there are many excellent introductions into this well-known field,
 for example \cite{Nakahara:2003nw,Weinberg:1996kr,ZinnJustin:2002vj,Harvey:2005it,Bilal:2008qx}, we refrain from doing so in this thesis.
 We assume familiarity with the very basic facts about anomalies in quantum field theory.
 The interested reader is referred to these references for additional information. This chapter
 is far from being a survey of this vast topic,
 we rather recall some
 special aspects of anomalies, for instance anomaly polynomials,
 which are essential to understand the work of this thesis. We closely follow the notation and reasoning
 of \cite{Bilal:2008qx} in this short recap.
 
 In quantum field
 theory an anomaly $\cA_\Lambda$ of a symmetry appears when the conservation law for the symmetry current $j^\mu_\Lambda$ is violated
 \begin{align}
  \cA_\Lambda (x) = - D_\mu \langle j^\mu_\Lambda \rangle \, ,
 \end{align}
 where $\mu$ denotes the $d$-dimensional spacetime index, $D_\mu$ is the gauge covariant derivative, and
  the symmetry generators are given by $T_\Lambda$. The corresponding vector gauge field is denoted by $A_\mu = A_\mu^\Lambda T_\Lambda$.
  If the $T_\Lambda$ generate only a global symmetry, we assume in the following without loss of generality that the latter
  is gauged by coupling to a
  background gauge field $A_\mu$. Note that also (local) Lorentz transformations (or equivalently diffeomorphisms) can be anomalous.
  The results which we state here for genuine gauge symmetries hold in complete analogy also for
  local Lorentz transformations by simply replacing the gauge field $A$ by the spin connection $\omega$ and the field strength
  $F$ by the curvature two-form $\cR$.
 
  If the theory has a classical description
 in terms of an action, there are many different but equivalent ways of how to think of an anomaly.
 First it signals that the path-integral
 measure is not invariant under the respective symmetry although the classical action \textit{is} invariant.
 Or stated differently, the regularization scheme does not preserve the symmetry.
 Equivalently one can also think of an anomaly as a non-invariance of the quantum effective action $\Gamma$ under the
classical symmetry
  \begin{align}
   \delta_\epsilon \Gamma = \int d^d x \sqrt{-g} \ \epsilon^\Lambda (x) \cA_\Lambda (x)
  \end{align}
with $\epsilon = \epsilon^\Lambda (x) T_\Lambda$ the parameter of the variation.

The precise form of $\cA_\Lambda (x)$
can be determined for example by evaluating the transformation of the path-integral measure of the chiral fields in the theory
under a symmetry transformation, \textit{i.e.}~by calculating functional determinants. Recall that only chiral modes can induce anomalies.
Equivalently one can also determine the index of the corresponding chiral kinetic operators. These procedures
are nicely explained and carried out in full detail in \cite{Bilal:2008qx}. 
The result for the anomaly is 
 \begin{align}\label{e:structure_anomaly}
   \cA_\Lambda (x) = \sum_{\substack{\textrm{chiral}\\ \textrm{matter}}} c \ \epsilon^{\mu_1\nu_1\dots \mu_{\sfrac{d}{2}}\nu_{\sfrac{d}{2}}} \ 
   \tr_R \big( T_\Lambda\ \partial_{\mu_1} A_{\nu_1}\dots \partial_{\mu_{\sfrac{d}{2}}} A_{\nu_{\sfrac{d}{2}}} \big ) +
   \cO (A^{\sfrac{d}{2}+1}) \, ,
  \end{align}
 with $c$ a normalization constant depending on the type of fields,
 and the trace is taken in the representation $R$, in which the matter field which induces the anomaly
 transforms. When we treat anomalies for different fields in four and six dimensions later in this thesis, we will just state
 the values for $c$ and rather than calculating them explicitly.

 The structure \eqref{e:structure_anomaly} of $\cA_\Lambda (x)$ suggests to rewrite the anomalous variation of the effective action
 in the language of differential forms
 \begin{align}\label{e:var_eff_ac}
   \delta_\epsilon \Gamma = \int d^d x \sqrt{-g} \ \epsilon^\Lambda (x) \cA_\Lambda (x) = 
   \sum_{\substack{\textrm{chiral}\\ \textrm{matter}}}c\int q_d^1 (R) \, ,
  \end{align}
 where $q_d^1 (R)$ is a $d$-form 
 \begin{align}
  q_d^1 (R) :=  \tr_R \big( \epsilon\ (dA)^{\sfrac{d}{2}} \big ) + \cO (A^{\sfrac{d}{2}+1}) = 
   \tr_R \big( \epsilon\ F^{\sfrac{d}{2}} \big ) + \cO (A^{\sfrac{d}{2}+1}) \, .
 \end{align}
 It is important to notice that $q_d^1 (R)$ is not uniquely determined. In particular the integral in \eqref{e:var_eff_ac}
 is invariant under adding
 exact forms $d\psi_{d-1}$ which vanish at infinity. Furthermore one is always free to add local
 counterterms $c\int \phi_d$ to the action. This corresponds to adding the variation $\delta \phi_d$ to $q_d^1 (R)$.
 To put it in a nutshell, the ambiguity of determining $q_d^1 (R)$ is
 \begin{align}\label{e:eq_rel}
  q_d^1 (R) \sim q_d^1(R) + \delta \phi_d + d\psi_{d-1} \, .
 \end{align}
Nevertheless it would be desirable to find an unambiguous way of characterizing anomalies, and we will do so in the following
by using characteristic classes
thereby defining what is called the anomaly polynomial $I(R)$ which corresponds to $q_d^1 (R)$.

The first step is to formally extend our $d$-dimensional spacetime $\cM_d$ to $\cM_d \times \cD_2$, where $\cD_2$
is the two-dimensional disc. Note that also the gauge fields and their gauge transformations
are formally extended into the two new directions.
 Let us define the following  characteristic class on $\cM_d \times \cD_2$
 \begin{align}
  P_{d+2}(R) := \tr_R \big( F^{\sfrac{d}{2}+1} \big ) \, ,
 \end{align}
 which is a top degree form.
It is easy to see that $P_{d+2}(R)$ is closed and gauge-invariant
 \begin{align}
 dP_{d+2}(R) = \delta P_{d+2}(R) = 0 \, .
\end{align}
By the Poincar\'e lemma the closedness of $P_{d+2}(R)$ implies that it is locally exact
\begin{align}\label{e:des1}
 P_{d+2}(R) \overset{\textrm{locally}}{=} d Q^{\rm CS}_{d+1}(R) \, ,
\end{align}
where $Q^{\rm CS}_{d+1}(R)$ are locally defined $(d+1)$-forms which are called Chern-Simons forms.
It is important to notice that the Chern-Simons forms are not uniquely defined. In fact one can always add exact forms to them
\begin{align}\label{e:CS_ambi}
 Q^{\rm CS}_{d+1}(R) \sim Q^{\rm CS}_{d+1}(R) + d \Phi_d 
\end{align}
with $\Phi_d$ a form of degree $d$.
The fact that the variation of $P_{d+2}(R)$ vanishes implies
that the variations of the Chern-Simons forms are closed
\begin{align}
 0 = \delta P_{d+2}(R) \overset{\textrm{locally}}{=} \delta\big( d Q^{\rm CS}_{d+1} (R) \big ) = d \big( \delta Q^{\rm CS}_{d+1}(R) \big) \, .
\end{align}
From this we can once again conclude by the Poincar\'e lemma that the variations of the Chern-Simons forms, which are defined
on the individual simply connected patches, are exact
\begin{align}\label{e:des2}
 \delta Q^{\rm CS}_{d+1}(R) = d Q^1_{d}(R) \, ,
\end{align}
with $Q^1_{d} (R)$ differential forms of degree $d$. Again for a fixed choice of $\delta Q^{\rm CS}_{d+1}(R)$ the
corresponding $Q^1_{d}(R)$ is not uniquely defined since one can add exact forms $d \Psi_{d-1}$. Putting this together with the ambiguity
in defining the Chern-Simons forms \eqref{e:CS_ambi} we obtain the following equivalence relation
\begin{align}\label{e:ex_eq_rel}
 Q^1_{d}(R) \sim Q^1_{d}(R) + \delta \Phi_d + d \Psi_{d-1} \, .
\end{align}
The equations \eqref{e:des1}, \eqref{e:des2} together are called the \textit{descent equations}. 
Let us stress once again that these considerations were carried out in the
extended $(d+2)$-dimensional spacetime $\cM \times \cD_2$, and all differential forms $P_{d+2}(R)$, $Q^{\rm CS}_{d+1}(R)$, $Q^1_{d} (R)$,
$\Phi_d$, $\Psi_{d-1}$, as well as the exterior derivative
are defined on the latter. Also note that by '$\delta$' we mean a standard gauge variation and not a BRST transformation
in which the gauge parameter is replaced by a ghost field.
For a nice exposition on the beautiful connection between anomalies and BRST cohomology we again refer to \cite{Bilal:2008qx}.

We are now in the position to connect this rather formal treatment to anomalies.
Recall that we want to find a way to characterize anomalies in an unambiguous
way.
The crucial point is now that the unappealing equivalence relation of an anomaly \eqref{e:eq_rel} has precisely the same form as
\eqref{e:ex_eq_rel}. We just need map differential forms on $\cM_d$ to their formal extensions on $\cM_d \times \cD_2$
\begin{subequations}
\begin{align}
 q^1_{d}(R) & \mapsto Q^1_{d}(R)\, , \\
 \phi_d & \mapsto \Phi_d \, , \\
 \psi_{d-1} &\mapsto \Psi_{d-1} \, .
\end{align}
\end{subequations}
We can then characterize an anomaly $\cA_\Lambda$ by the anomaly polynomial $I(R)$ which is defined without
ambiguities on $\cM_d \times \cD_2$ via the characteristic class $P_{d+2}(R)$
\begin{align}
 I(R) := c P_{d+2}(R)
\end{align}
with the normalization constant $c$.
The vanishing of an anomaly is then equivalent to the vanishing of the total anomaly polynomial $I_{d+2}$
\begin{align}
 I_{d+2} := I^{\rm GS} +  \sum_{\substack{\textrm{chiral}\\ \textrm{matter}}} I(R) \, ,
\end{align}
where we included also an additional contribution $I^{\rm GS}$ coming from Green-Schwarz terms on which we shortly comment
at the end of this section. 
There is one subtlety which we didn't mention explicitly. In the preceding analysis we focused on a single gauge factor. 
If there are simultaneously several non-Abelian and Abelian gauge factors as well as
local Lorentz invariance present, there are also mixed anomalies, \textit{i.e.}~the field strengths of these three
categories appear with mixed products in the
$\cA_\Lambda$ and therefore also in the anomaly polynomial.
 
\begin{figure}[!h]
\centering
 \includegraphics{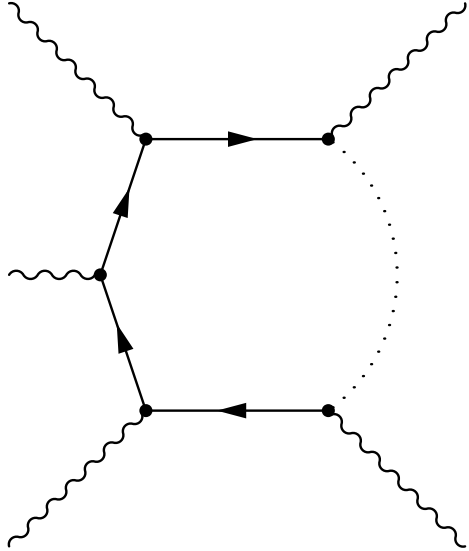}
 \caption{This is the general form of a one-loop contribution to the anomaly. The number of external legs is $\sfrac{d}{2} +1$, where $d$ is
 the number of spacetime dimensions.}
 \label{fig:anom_loop}
 \end{figure}

 One can show that a quantum anomaly is a one-loop effect, for example by introducing a loop
 counting parameter $S \rightarrow \frac{1}{\lambda}S$ in the calculation of the functional determinant.
More precisely, the fields which run in the loop are chiral modes, and the external legs 
are either gauge bosons or gravitons if we consider gravitational anomalies. The total number of external legs is given
by $\sfrac{d}{2} +1$, \textit{i.e.}~by the number of field strength tensors in the anomaly polynomial.
 Note that these constitute background fields
 for the case of anomalies of global symmetries. We depict the form of these loops in \autoref{fig:anom_loop}.
When the external legs are all gravitons,
 we face a pure a \textit{gravitational} anomaly, for solely gauge bosons a pure \textit{gauge}
 anomaly, and for mixed external legs (gravitons and
 gauge bosons) a \textit{mixed gravitational-gauge} anomaly.  

Note also that this characterization of anomalies in terms of polynomials in an extended space works also for theories without an underlying classical action, for example anomalies of six-dimensional non-Abelian tensor theories. It is not known if there
 exists an action for such theories. Nevertheless people determined
 the eight-dimensional anomaly polynomial, \textit{e.g.~}for the R-symmetry \cite{Bershadsky:1997sb,Freed:1998tg,Intriligator:2000eq,
 Yi:2001bz,Maxfield:2012aw,
 Monnier:2013kna,Monnier:2013rpa,Monnier:2014txa,Intriligator:2014eaa,Ohmori:2014pca,Ohmori:2014kda}.
 
Finally to close this section let us mention that there can also sometimes appear additional classical contributions to anomalies
which are crucial.
Indeed, axions or two-form fields might transform in a non-trivial way under gauge transformations and render the \textit{classical}
action non-gauge-invariant. This results in modified Bianchi identities and is called \textit{Green-Schwarz mechanism}.
Indeed, it first arose in type I string theory canceling its gauge anomalies \cite{Green:1984sg}.
In this thesis our focus is on four dimensions, where the Green-Schwarz mechanism is mediated by gauged axions, and six dimensions,
where two-forms are responsible for classical non-gauge-invariance \cite{Sagnotti:1992qw,Sadov:1996zm}.
 
 \section{One-Loop Chern-Simons Terms}\label{ch:CS}
 Throughout this thesis we frequently make use of one-loop corrections to Chern-Simons terms in three and five dimensions.
 They are induced from integrating out parity-violating massive modes, and by parity transformations we mean
 reflections of an odd number of spatial directions. Via this procedure
 the classical parity anomaly of the spectrum is transferred to the effective action
 since the Chern-Simons terms are not invariant under such transformations.
 Importantly, these loop-corrections are independent of the mass scale.
 We will see that this topological property makes them very robust quantities to encode crucial information about the spectrum.
 In \autoref{ch:cons_effth} we use them to formulate necessary conditions for a consistent truncation to be used as an effective theory
 for phenomenology. However, the presumably nicest and most important property is the relation of Chern-Simons terms
 to anomalies of theories in one dimension higher. While the precise relation was long unclear, we provide
 in \autoref{sec:anom_lgt} the procedure 
 of how to extract gauge anomalies in four and six dimensions from one-loop Chern-Simons terms in three and five
 dimensions, respectively,
 using large gauge transformations. In \autoref{ch:arith} we show that this mechanism has a natural implementation in the M-theory to 
 F-theory duality,
 and one can use geometric symmetries of the Weierstrass model in order to explicitly proof cancelation of gauge anomalies in F-theory.

 \subsection{Three Dimensions}
 
Let us introduce our conventions for Chern-Simons terms in three dimensions. For a general theory of Abelian vector fields 
$A^\Lambda$ with field strength $F^\Lambda = dA^\Lambda$ they take the form 
\begin{align}\label{e:def_CS_4d}
 S_{\textrm{CS}} = \int \Theta_{\Lambda \Sigma} A^\Lambda \wedge F^\Sigma \, ,
\end{align}
where $\Theta_{\Lambda \Sigma} $ are constants.

\begin{figure}[h]
\centering
 \includegraphics{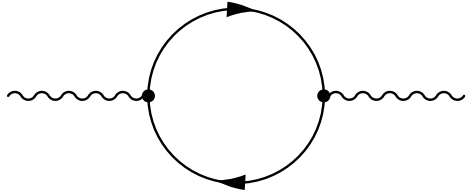}
 \caption{This is the loop which induces Chern-Simons terms in three dimensions. The external legs are Abelian gauge bosons
 under which the spin-$\sfrac{1}{2}$ fermions running in the loop are charged.}
 \label{fig:3dCS}
 \end{figure}

Importantly, these terms might not only appear at the classical level but can also receive quantum corrections which are one-loop exact.
The corresponding Feynman diagram is depicted in \autoref{fig:3dCS}.
It is well-known that a massive charged spin-$\sfrac12$-fermion $\Bpsi^{\sfrac12}$
contributes to $\Theta_{\Lambda\Sigma}$ as \cite{Niemi:1983rq,Redlich:1983dv,Aharony:1997bx}
\begin{align}\label{e:4d_single_loop_CS}
 \Theta_{\Lambda\Sigma}^{\textrm{loop}} = \frac{1}{2} q_\Lambda q_\Sigma \sign (m) \, ,
\end{align}
where $q_\Lambda$ is the charge of the fermion under the $U(1)$ gauge boson $A^\Lambda$ and
$\sign (m)$ is the sign of the mass $m$ in the Lagrangian
\begin{align}\label{e:lagr_3d}
 e^{-1} \cL = - \bar \Bpsi^{\sfrac12} \slashed\cD \Bpsi^{\sfrac12} + m \bar \Bpsi^{\sfrac12} \Bpsi^{\sfrac12} 
\end{align}
with
\begin{align}
 \cD_\mu \Bpsi^{\sfrac12} = \big(\nabla_\mu -i q_\Lambda A^\Lambda_\mu \big) \Bpsi^{\sfrac12} \, .
\end{align}
Let us shortly comment on the significance of $\sign (m)$. While the physical mass is of course positive semi-definite,
the meaning of $\sign (m)$ can be understood
as follows: The Lorentz group in three dimensions $SO(2,1)$ has only one (real) spinor representation.
However the Clifford algebra has two physically inequivalent representations related by $\gamma^\mu \rightarrow - \gamma^\mu$.
Similarly, the massive little group has two spinor representations. The sign of $m$ tells you under which spinor representation
of the little group the massive particle
transforms. However, for this to make sense you first have to fix a representation of the Clifford algebra since
by a change of the latter $\gamma^\mu \rightarrow - \gamma^\mu$ one has $m \rightarrow - m$. In this thesis we choose the basis of the Clifford
algebra in three dimensions in such a way that
the circle-compactification of a four-dimensional left-handed spinor yields 
the following Kaluza-Klein contribution for the mass of the $n$-th Kaluza-Klein mode
\begin{align}
 m = \dots + \frac{n}{r} \, ,
\end{align}
where $r$ is the radius of the circle.

Finally let us remark that the mass terms of the fermions violate parity. In particular, under such a transformation one finds
$m \rightarrow - m$. This classical property is also present in the quantum setting after integrating out these modes since
the Chern-Simons terms are also parity-odd.

  \subsection{Five Dimensions}\label{sec:5dcherns}
  Five-dimensional Chern-Simons terms for 
$U(1)$ gauge fields take the general form
\begin{subequations}\label{e:def_CS_6d}
\begin{align}
\label{5dgaugeCS} S^{\textrm{gauge}}_{\textrm{CS}} = - \frac{1}{12} \int k_{\Lambda\Sigma\Theta} A^{\Lambda} \wedge F^{\Sigma}
\wedge F^{\Theta} \, , \\
\label{5dgravCS} S^{\textrm{grav}}_{\textrm{CS}} = - \frac{1}{4} \int k_{\Lambda} A^{\Lambda} \wedge \tr ( \mathcal{R} \wedge \mathcal{R} ) \, ,
\end{align}
\end{subequations}
where $k_{\Lambda\Sigma\Theta} $ and $k_{\Lambda} $ are constants and  $\mathcal{R}$
is the five-dimensional curvature two-form. Although \eqref{5dgravCS} is higher-curvature, it plays an important role in our discussions.
 
\begin{figure}[h]
\centering
 \includegraphics{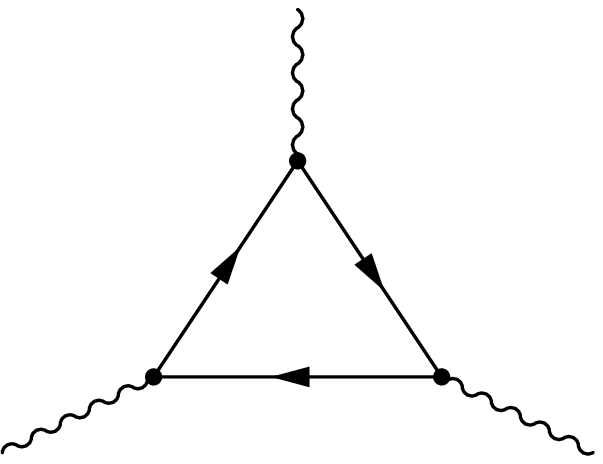}
 \caption{This type of loops induces the Chern-Simons couplings $k_{\Lambda\Sigma\Theta}$, $k_{\Lambda}$ in five dimensions.
 The external legs are either three Abelian gauge bosons
 or one Abelian gauge bosons plus two gravitons, respectively. The modes which run in the loop are
 spin-$\sfrac{1}{2}$ fermions, self-dual tensors and spin-$\sfrac{3}{2}$ fermions.}
 \label{fig:5dCS}
 \end{figure}

 In addition to Chern-Simons couplings which arise at the classical level the 
effective theory can admit one-loop induced Chern-Simons couplings from
integrating out massive charged spin-$\sfrac12$ fermions $\Bpsi^{\sfrac 12}$,
complex self-dual two-forms $\BB_{\mu\nu}$ in the sense of \cite{Townsend:1983xs}, 
and spin-$\sfrac32$ fermions $\Bpsi^{\sfrac 32}_{\mu}$
\cite{Witten:1996qb,Bonetti:2012fn,Bonetti:2013ela}. The form of the corresponding Feynman graphs is depicted in \autoref{fig:5dCS}.
The contributions of the individual fields are given by
\begin{align}
\label{e:6d_single_field_CS_1} k_{\Lambda\Sigma\Theta}^{\textrm{loop}} = c_{AFF}\, q_\Lambda q_\Sigma q_\Theta \, \sign (m) \, , \\
\label{e:6d_single_field_CS_2} k_{\Lambda}^{\textrm{loop}} = c_{A\cR\cR}\, q_\Lambda \, \sign (m) \, ,
\end{align}
where the $c_{AFF},c_{A\cR\cR}$ depend on the type of field and are given in \autoref{t:CS_correct}.
\begin{table}[h]
\begin{center}
\begin{tabular}{c|ccc}
 & Spin-$\sfrac{1}{2}$ fermion & Self-dual tensor & Spin-$\sfrac{3}{2}$ fermion\\
 \hline
\rule{0pt}{15pt} $c_{AFF}$ & $\ \ \frac{1}{2}$ & $-2$ & $\frac{5}{2}$\\
\rule{0pt}{15pt} $c_{A\cR\cR}$ & $-1$ & $-8$ & $19$\\
\end{tabular}
\end{center}
\caption{Normalization factors for one-loop Chern-Simons terms in five dimensions.}
\label{t:CS_correct}
\end{table}
The quantity $q_\Lambda$ is the charge under $A^\Lambda$ and $\sign (m)$ depends on the representation of the massive little group
$SO(4) \cong SU(2) \times SU(2)$ (locally). In analogy to the situation in three dimensions, 
which we discussed before, $SO(4)$ admits two spinor representations, and
the Clifford algebra in five dimensions has again two inequivalent representations related by $\gamma^\mu \rightarrow - \gamma^\mu$. The full
Lorentz group in five dimensions $SO(4,1)$ has only one (pseudo-real) spinor representation.
The quantity $\sign (m)$ is related to the representations of the massive little group $SO(4) \cong SU(2) \times SU(2)$ (locally) by
\begin{align}\label{def_sign_5d}
 \sign (m) = \begin{cases}
              +1 \quad \textrm{for } \big(\frac{1}{2},0\big),\big(1,0\big),\big(1,\frac{1}{2}\big)\, , \\
              -1 \quad \textrm{for } \big(0,\frac{1}{2}\big),\big(0,1\big),\big(\frac{1}{2},1\big) \, ,
             \end{cases}
\end{align}
where we labeled representations of $SU(2) \times SU(2)$ by their spins. Again we fix the representation of the Clifford algebra
by demanding that chiral
six-dimensional modes on a circle precisely yield \eqref{def_sign_5d} interpreting the representations of $SU(2) \times SU(2)$
as the six-dimensional helicity group. For convenience let us display the Lagrangians of the five-dimensional fields
\begin{subequations}\label{e:lagr_5d}
\begin{align}
\label{e:lagr_5d_1} e^{-1} \cL_{\sfrac 12} 
&= - \bar \Bpsi^{\sfrac 12} \slashed\cD \Bpsi^{\sfrac 12} + m \bar \Bpsi^{\sfrac 12} \Bpsi^{\sfrac 12} \, , \\
\label{e:lagr_5d_2} e^{-1}\cL_B &= -\frac{1}{4}i \sign (m) \, \epsilon^{\mu\nu\rho\sigma\tau}\bar \BB_{\mu\nu}\cD_\rho \BB_{\sigma\tau} 
 -\frac{1}{2} \vert m \vert \, \bar \BB_{\mu\nu} \BB^{\mu\nu}\, , \\
\label{e:lagr_5d_3} e^{-1} \cL_{\sfrac 32} &= - \bar \Bpsi_\mu^{\sfrac 32} \gamma^{\mu\nu\rho}  \cD_\nu \Bpsi_{\rho}^{\sfrac 32} + 
    m \, 
 \bar \Bpsi_\mu^{\sfrac 32} \gamma^{\mu\nu} \Bpsi_{\nu }^{\sfrac 32} \, ,
\end{align}
\end{subequations}
with
\begin{subequations}\label{e:cov_der_5d}
\begin{align}
 \cD_\mu \Bpsi^{\sfrac 12} &= \big(\nabla_\mu -i q_\Lambda A^\Lambda_\mu \big) \Bpsi^{\sfrac12}\, ,  \\
 \cD_{[\mu} \BB_{\nu\rho]} &= \big(\partial_{[\mu} -i q_\Lambda A^\Lambda_{[\mu} \big) \BB_{\nu\rho]}\, , \\
 \cD_{\lbrack\mu} \Bpsi_{\nu\rbrack}^{\sfrac 32} &= \big(\nabla_{[\mu} - i q_\Lambda A^\Lambda_{[\mu} \big) \Bpsi_{\nu]}^{\sfrac 32} \, .
\end{align}
\end{subequations}
Note that for the fermions again, as in three dimensions, the mass terms violate parity while for the massive tensor the
kinetic term is not invariant.

\part{Partial Supersymmetry Breaking and Consistent Truncations}\label{part:susybreak}

\chapter{Overview}

A systematic classification of supersymmetric vacua of supergravity theories in various dimensions 
has been a challenge since the first constructions of such theories. 
Supergravity theories with non-minimal supersymmetry can often admit 
Minkowski or anti-de Sitter ground states that preserve only a partial 
amount of supersymmetry. Finding such solutions is typically more involved 
than determining the fully supersymmetric solutions. 
For supergravity theories formulated in even spacetime dimensions various breaking patterns 
have been investigated in detail. For example the $\cN=2$ to $\cN=1$ breaking in four-dimensional 
supergravity theories has been investigated already in 
\cite{Cecotti:1984rk,Cecotti:1984wn,Cecotti:1985sf,Ferrara:1995gu,Antoniadis:1995vb,Fre:1996js,Kiritsis:1997ca,Andrianopoli:2002rm}. 
Recently
there has been a renewed interest in this direction \cite{Louis:2009xd,Louis:2010ui,Hansen:2013dda}, which was partially triggered by the application to flux 
compactifications of string theory \cite{Samtleben:2008pe}. The general analysis of \cite{Louis:2010ui} heavily employs 
the powerful techniques provided by the embedding tensor formalism \cite{deWit:2002vt,deWit:2005ub}.

The general study of partial supersymmetry breaking in 
odd-dimensional theories has attracted much less attention. 
Such theories however possess the interesting new possibility 
that the dynamics of some fields can arise from 
Chern-Simons-type couplings that are topological in nature. 
As was pointed out already for three-dimensional supergravity theories \cite{Hohm:2004rc}
such couplings can allow for special supersymmetry breaking patterns.
In this part we show that in five-dimensional supergravity theories 
with sixteen supercharges denoted as $\cN=4$, Chern-Simons-type 
couplings for two-form tensor fields can yield interesting new supersymmetry 
breaking patterns to vacua preserving eight supercharges denoted as $\cN=2$. 
Such tensor fields can have first-order kinetic terms
and become massive by a St\"uckelberg-like mechanism in which 
they \textit{eat} a dynamical vector field \cite{Townsend:1983xs,Dall'Agata:2001vb,Schon:2006kz}. 
The degrees of freedom of such tensors are counted by 
realizing that they have zero degrees of freedom before \textit{eating} the 
vector but admit three degrees of freedom as
massive fields. Hence they should be distinguished from 
tensors with standard kinetic and mass terms. They have been named \textit{self-dual tensors} in \cite{Townsend:1983xs},
and we introduced them in \autoref{sec:5dcherns}.
The mechanism rendering the tensor fields massive by eating a vector will be called 
\textit{tensorial Higgs mechanism} in the following.

We begin this part by studying general vacua of $\cN=4$ gauged supergravity in five dimensions using the embedding tensor formalism of
\cite{Schon:2006kz} which encodes the gauging of global symmetries in a very convenient way.
After assigning vacuum expectation values (VEVs) to the scalars
we calculate the gravitino masses, \textit{i.e.}~the number of broken supersymmetries, the cosmological constant, the bosonic spectrum including mass terms and charges,
as well as Chern-Simons terms. These quantities depend on the form of the embedding tensors contracted with the VEVs of the coset representatives
of the scalar manifold.
Once these objects are specified, one can fully analyze the theory around the vacuum. 
While such an analysis is possible for each considered vacuum, 
a classification of allowed vacua is beyond the scope of our work.

We continue by analyzing in full detail theories which are subject to a non-vanishing Abelian magnetic gauging only,
\textit{i.e.~}gauged by a constant anti-symmetric matrix $\xi_{MN}$.
The latter encodes
the couplings of the five-dimensional self-dual tensors to the vector fields and the form of 
the first order kinetic terms. 
A non-trivial $\xi_{MN}$ also induces vector gaugings and a 
scalar potential. We analyze the conditions on $\xi_{MN}$ that yield partial supersymmetry breaking 
to an $\cN=2$ Minkowski vacuum. The massless and massive $\cN=2$ spectrum comprising 
fluctuations around this vacuum are then determined systematically. We particularly stress the
appearance of massive tensor fields and massive spin-$\sfrac 1 2$ and spin-$\sfrac 3 2$ fermions.
This allows us to derive the key features of the effective $\cN=2$ supergravity theory
arising for the massless fluctuations around the ground state. 
The $\cN=2$ effective action for the massless fields comprises two parts. Firstly, 
there are the classical couplings inherited from the underlying $\cN=4$ theory. 
They are determined by truncating the original theory to the appropriately combined 
massless modes. At energy scales far below the supersymmetry breaking scale 
one might have expected that this determined already the complete $\cN=2$ theory.  
However, as we show in detail in our work, the massive tensor, spin-$\sfrac 1 2$ and spin-$\sfrac 3 2$ 
modes have to actually be integrated out and generically induce non-trivial corrections. In 
fact, using the results of \autoref{sec:5dcherns} one infers that if these massive fields are 
charged under some vector field they generically induce non-trivial one-loop corrections to the 
Chern-Simons terms for the vector. One-loop corrections to the Chern-Simons terms due to massive 
charged spin-$\sfrac 1 2$ fermions have been considered in \cite{Witten:1996qb,Morrison:1996xf}, but 
we stress here that in the $\cN=4$ to $\cN=2$ breaking both massive tensors and gravitini alter 
the result crucially. These kind of one-loop corrections are independent of the mass 
scale of the fields and therefore have to be taken into account in a consistent effective 
theory at scales well below the supersymmetry breaking scale.
 
After this analysis we then extend and use our results on
partial supersymmetry breaking and one-loop Chern-Simons terms
as a tool to investigate consistent truncations of supergravity and string theory.
In principle for a general compactification of some higher-dimensional theory on a compact manifold one has 
to include all massive and massless modes in the derivation of the effective action. 
In contrast, consistent truncations describe the dynamics only for a subset of all these modes.
By definition these modes are chosen such that solutions of the lower-dimensional equations of motion 
lift to solutions of the higher-dimensional equations of motion. 
It is this property that allows one to use the truncated theories as tools for constructing higher-dimensional solutions.
However, recently consistent truncations have also been used for phenomenology in non-Calabi-Yau compactifications.
Consequently, the effective action derived from a consistent truncation should better 
match the genuine effective action with the whole tower of massive modes integrated out.
Setups with partial supergravity breaking will allow us to 
derive necessary conditions for this agreement in theories
where we already know parts of the effective action,
like \textit{e.g.}~Calabi-Yau compactifications. We investigate this issue in the context of one-loop corrections 
to the Chern-Simons terms.

As an application we then make contact with M-theory compactifications on $SU(2)$-structure manifolds.
First we study general consistent truncations of M-theory on $SU(2)$-structure manifolds to
$\cN=4$ gauged supergravity before we restrict to the special case of
Calabi-Yau manifolds with vanishing Euler number, which have $SU(2)$-structure as well, as can be seen by the Poincar\'e-Hopf theorem.
These spaces constitute $\cN=2$ Minkowski vacua of general $\cN =4$ gauged supergravity including massive modes.
The same analysis has been carried out for the type IIA case in \cite{Danckaert:2011ju,KashaniPoor:2013en}.
Since the Chern-Simons terms in the genuine effective action of M-theory on a smooth Calabi-Yau threefold
are not corrected by integrating out massive modes \cite{Cadavid:1995bk,Papadopoulos:1995da,Ferrara:1996hh}, 
we demand that one-loop Chern-Simons terms should also be absent in the effective action 
of a consistent truncation.
For the analyzed example of the Enriques Calabi-Yau
it turns out that the massive modes are not charged under any massless vector,
and one-loop corrections therefore trivially cancel. 
This is one possible way to ensure that consistent truncations on $SU(2)$-structure threefolds 
that are also Calabi-Yau can be compatible with the genuine effective action.  
However, already in the considered consistent truncation for the Enriques Calabi-Yau we miss at the 
massless level a vector multiplet and a hypermultiplet which are not captured by our particular $SU(2)$-structure 
ansatz. Nevertheless, we argue that one can consistently complete the 
Chern-Simons terms including an additional massless vector.

As a second example we consider
a particular consistent truncation of type IIB supergravity on a squashed Sasaki-Einstein
manifold with RR-flux. 
This is again described by five-dimensional $\cN =4$ gauged supergravity, and indeed there are $\cN=2$
vacua that are now AdS \cite{Cassani:2010uw,Liu:2010sa,Gauntlett:2010vu}.\footnote{See also \cite{Tsikas:1986rx,Lu:1999bw,Ceresole:1999zs,Cvetic:2000nc,Hoxha:2000jf,Buchel:2006gb,Gauntlett:2007ma,Skenderis:2010vz,Cassani:2010na,Bena:2010pr,
Bah:2010cu,Liu:2010pq,Liu:2011dw,Halmagyi:2011yd}
for related works on this subject.}
The most prominent example is certainly the five-sphere although our results hold for any squashed Sasaki-Einstein manifold.
In the theory around the vacuum there are massive states that are charged under the gauged $U(1)$ R-symmetry.
Remarkably their one-loop corrections to the gauge and gravitational Chern-Simons terms
cancel in a very non-trivial way. While we are not able to give a precise interpretation of this fact, it is 
an intriguing observation that such cancelations take place. 
We suspect that there could exist an underlying principle which ensures the vanishing of such scale-invariant corrections
in consistent truncations.
Let us however also stress that in 
the AdS case the existence of an effective theory can be generally questioned, since 
the AdS radius is linked to the size of the compactification space. 
It is not hard to see that the squashed Sasaki-Einstein reductions of type IIB are reminiscent of the general $SU(2)$-structure
reductions of M-theory considered before. It was indeed argued that 
there is a relation between these two settings when using T-duality \cite{Gauntlett:2004zh,Gauntlett:2006ai,Gauntlett:2007sm,OColgain:2011ng,Sfetsos:2014tza}
if one includes warping in the $SU(2)$-structure ansatz, which is in general quite difficult and beyond the scope of this thesis.

This part is organized as follows.
In \autoref{sec:susy_break} we review $\cN=4$ gauged supergravity in five dimensions using the embedding tensor formalism
and evaluate the spectrum as well as the relevant parts of the Lagrangian around the vacuum in terms of contracted embedding tensors.
We fully analyze supersymmetry breaking from $\cN=4$ to $\cN=2$ in an Abelian 
configuration with a non-trivial embedding tensor $\xi_{MN}$ in \autoref{sec:ab_mag_gaug}.
We in detail account for 
the tensorial Higgs mechanism as well as the super-Higgs mechanism, and we
determine the $\cN=2$ spectrum and one-loop effective action. We argue that 
all massive multiplets generically induce one-loop corrections to the 
vector couplings of the theory that are independent of the 
supersymmetry breaking scale.
We proceed in \autoref{ch:m_su2}
with the general description of M-theory compactifications on $SU(2)$-structure manifolds. In \autoref{ch:cons_effth}, after
stating some general remarks about the quantum effective action of consistent truncations, we analyze M-theory on the Enriques Calabi-Yau
and type IIB supergravity consistent truncation on a squashed Sasaki-Einstein
manifold.

\chapter{Gauged \texorpdfstring{$\cN =4$}{N=4} Supergravity in Five Dimensions and its Vacua}\chaptermark{Gauged Supergravity in Five Dimensions}
\label{sec:susy_break}

We start this chapter with a short review of some important facts about five-dimensional $\cN=4$ gauged supergravity theories in \autoref{N=4Gen}. 
In \autoref{sec:iso} we provide a tool to extract the propagating degrees of freedom out of the theory since the standard formulation
in \cite{Schon:2006kz} uses vectors and dual tensors on equal footing.
We study the vacua of this setup in \autoref{sec:action}
by deriving the mass terms and charges of the scalar and tensor fields, and also by giving expressions for the
vector masses, field strengths and Chern-Simons terms. The results depend on the precise form of the embedding tensors
contracted with the scalar field VEVs.
Since we are in particular interested in the amount of preserved supersymmetry in the vacuum, we also compute
the mass terms of the gravitini in
terms of the contracted embedding tensors.
Finally, we derive some properties of the subclass of Minkowski vacua in \autoref{sec:mink}.

\section{Generalities} \label{N=4Gen}

Let us at the beginning state the general properties of $\cN =4$ gauged supergravity in five dimensions along the lines of
\cite{Dall'Agata:2001vb,Schon:2006kz}.\footnote{We stress that in our conventions
five-dimensional $\cN =4$ supergravity theories have 16 supercharges and thus are half-maximal supergravities.} First consider 
ungauged Maxwell-Einstein supergravity which couples 
$n$ vector multiplets to a single gravity multiplet. Note that as long as the theory is not gauged, one can equally well replace the
vector multiplets by dual tensor multiplets. 
The gravity multiplet has the field content
\begin{align} \label{grav_multiplet}
 (g_{\mu\nu}, \psi^i_\mu , A^{ij}_\mu , A^0_\mu , \chi^i , \sigma )
\end{align}
with the metric $g_{\mu\nu}$, four spin-$\sfrac 3 2$ gravitini $\psi^i_\mu$, six vectors $(A^{ij}_\mu , A^0_\mu)$, four 
spin-$\sfrac 1 2$ fermions $\chi^i$
and one real scalar $\sigma$. The indices of the fundamental representation of
the R-symmetry group $USp(4)$ are written as $i,j = 1,\dots ,4$. The symplectic form of $USp(4)$, denoted $\Omega$, enjoys the following properties
\begin{align}\label{e:properties_omega}
 \Omega_{ij} = - \Omega_{ji} \, , \qquad \Omega_{ij} = \Omega^{ij} \, , \qquad \Omega_{ij}\Omega^{jk} = - \delta^k_i \, .
\end{align}
Raising and lowering of $USp(4)$ indices is carried out according to the rule
\begin{align}\label{e:raising_lowering}
 V^i = \Omega^{ij} V_j \, , \qquad V_i = V^j \Omega_{ji} \, .
\end{align}
The double index $ij$ labels the \textbf{5} representation of $USp(4)$ defined by the following properties
\begin{align}
 A^{ij}_\mu = - A^{ji}_\mu \, , \qquad A^{ij}_\mu \, \Omega_{ij} = 0 \, , \qquad (A^{ij}_\mu)^* = A_{\mu \, ij} \, .
\end{align}
Since $USp(4)$ is the spin group of $SO(5)$, we will often use the isomorphism
$\mathfrak{so}(5) \cong \mathfrak{usp}(4)$ to switch between representations of both groups. The indices of
the fundamental representation of $SO(5)$ are denoted by $m,n = 1, \dots , 5$, and the Kronecker delta $\delta_{mn}$ is used to raise and lower them.
Moreover all \textit{massless} fermions in this part of the thesis are supposed to be symplectic Majorana
spinors. For further conventions and useful identities consult \autoref{app:spacetime}.
Finally we will often use the definition
\begin{align}
 \Sigma := e^{\sigma / \sqrt 3}\, ,
\end{align}
where $\sigma$ is the real scalar of the gravity multiplet \eqref{grav_multiplet}.

Having introduced the gravity multiplet we can now further couple $n$ vector multiplets labeled by $a,b = 6, \dots , 5+n$. The
indices are again raised
and lowered using the Kronecker delta $\delta_{ab}$. The multiplets have the structure
\begin{align}\label{e:content_vector}
 (A_\mu^a , \lambda^{ia} , \phi^{ija})\, ,
\end{align}
where $A_\mu^a$ denote the vectors, $\lambda^{ia}$ spin-$\sfrac 1 2$ fermions, and the $\phi^{ija}$ scalars in the \textbf{5} of $USp(4)$.

The set of all scalars in the theory span the manifold 
\begin{align} \label{coset_def}
 \cM = \cM_{5,n} \times SO(1,1) \, , \qquad \quad \cM_{5,n} = \frac{SO(5,n)}{SO(5)\times SO(n)}\, ,
\end{align}
where we parametrize the coset $\cM_{5,n}$ by the scalar fields $\phi^{ija}$ in the vector multiplets whereas
the $SO(1,1)$ part is captured by the scalar $\sigma$ in the gravity multiplet.
Hence the global symmetry group of the theory is found to be
$SO(5,n)\times SO(1,1)$.
Note that
\begin{align}
 \textrm{dim}\big(\cM_{5,n}\big) = \textrm{dim}\big(SO(5,n)\big) -  \textrm{dim}\big(SO(5)\big) - \textrm{dim}\big(SO(n)\big) = 5n \, .
\end{align}
We now define $SO(5,n)$ indices $M,N= 1,\dots , 5+n$, which we can
raise and lower with the $SO(5,n)$ metric $(\eta_{MN}) = \textrm{diag}(-1,-1,-1,-1,-1,+1,\dots ,+1)$.
The coupling of the vector multiplets to the gravity multiplet is realized by noting that all vectors in the theory transform 
as a singlet $A^0$ and the fundamental
representation $A^M$ of $SO(5,n)$:
\begin{align}
 (A^0 , A^{ij} , A^n ) \rightarrow (A^0 , A^{M}) \, ,
\end{align}
and they carry $SO(1,1)$ charges $-1$ and $1/2$ for $A^0$ and $A^M$, respectively.
In terms of these representations the generators $t_{MN}$ of $SO(5,n)$ and $t_0$ of $SO(1,1)$ read\footnote{All
antisymmetrizations in this thesis include a factor of $1/n!\,$.}
\begin{align}
 \tensor{t}{_M_N_\, _P^Q} = 2\delta_{[M}^Q \, \eta_{N]P} \, , \qquad \tensor{t}{_0_\, _M^N} = - \frac{1}{2} \,\delta_M^N \, ,
 \qquad \tensor{t}{_M_N_\, _0^0} = 0 \, , \qquad \tensor{t}{_0_\, _0^0} = 1 \, .
\end{align}

The most convenient way to describe the coset space $\cM_{5,n}$ is via the coset representatives
$\cV=(\tensor{\cV}{_M^m} , \tensor{\cV}{_M^a})$, here $m = 1, \dots , 5$ and $a = 6, \dots n+5$
are the indices of the fundamental representations of $SO(5)$ and $SO(n)$, respectively.
The definition is such that local $SO(5)\times SO(n)$ transformations act from the right while global $SO(5,n)$ transformations on $\cV$ act from the left.
It is important to notice that
\begin{align}\label{e:SO(5,N)_index_raising}
 \tensor{\cV}{_M^a} = \eta_{MN}\,\cV^{Na} \, , \qquad \tensor{\cV}{_M^m} = - \eta_{MN}\, \cV^{Nm} \, ,
\end{align}
and also, since
$(\tensor{\cV}{_M^m} , \tensor{\cV}{_M^a})\in SO(5,n)$, we have
\begin{align} \label{eta_viaV}
 \eta_{MN} = - \tensor{\cV}{_M^m} \cV_{N m} + \tensor{\cV}{_M^a} \cV_{N a} \, .
\end{align}
Furthermore we define a non-constant positive definite metric on the coset
\begin{align} \label{M_viaV}
 M_{MN} := \tensor{\cV}{_M^m} \cV_{N m} + \tensor{\cV}{_M^a} \cV_{N a} \, 
\end{align}
with inverse given by $M^{MN}$, which is easy to check.
Lastly we introduce
\begin{align}
 M_{MNPQR} := \varepsilon_{mnpqr}\tensor{\cV}{_M^m}\tensor{\cV}{_N^n}\tensor{\cV}{_P^p}\tensor{\cV}{_Q^q}\tensor{\cV}{_R^r}\, ,
\end{align}
where $\varepsilon_{mnpqr}$ is the (flat) five-dimensional Levi-Civita tensor.

We proceed with the gauging of global symmetries. The different possible gaugings are most conveniently described using the
embedding tensors $f_{MNP}$, $\xi_{MN}$ and $\xi_M$, which are totally antisymmetric
in all indices. They determine the covariant derivative\footnote{Note that 
a gauge coupling constant $g$ can explicitly be included whenever an embedding tensor appears. However for simplicity we take $g=1$ in the following.}
\begin{align} \label{gen_cov_der}
 D_\mu = \nabla_\mu -  A_\mu^M \tensor{f}{_M^N^P} t_{NP} -  A_\mu^0 \, \xi^{MN} t_{MN} -  A^M_\mu \xi^N t_{MN} -  A_\mu^M \xi_M t_{0}\, .
\end{align}
Note that in the ungauged theory the embedding 
tensors are supposed to transform under the global symmetry group.
Fixing a value for the tensor components the global symmetry group is then broken down to a subgroup.
In this thesis we will mostly set $\xi_{M}=0$ since the calculations simplify considerably and several interesting cases are already covered.
However a non-vanishing $\xi_{M}$ might also be included straightforwardly.
Accordingly the covariant derivative \eqref{gen_cov_der} simplifies to
\begin{align}
 D_\mu = \nabla_\mu -  A_\mu^M \tensor{f}{_M^N^P} t_{NP} - A_\mu^0 \, \xi^{MN} t_{MN} \, .
\end{align}
The embedding tensors are further subject to quadratic constraints
which read in the case of $\xi_M = 0$
\begin{align}\label{e:quadr_constr}
 &f_{R[MN}\tensor{f}{_P_Q_]^R}=0\, , & \tensor{\xi}{_M^Q}f_{QNP} = 0 \, .
\end{align}
For vanishing $\xi_M$ the linear constraints on the embedding tensors \cite{Schon:2006kz} are trivially satisfied.
There is an important issue with this kind of nontrivial gaugings which forces us to dualize
some of the vector fields $A_\mu^M$ into two-forms $B_{\mu\nu\, M}$. Therefore we consider 
an action where both $A_\mu^M$ and $B_{\mu\nu\, M}$ are present in order
to write down a general gauged supergravity with $\xi_M = 0$.\footnote{As long as $\xi_M$ vanishes, we do not have to introduce a tensorial counterpart
$B^0_{\mu\nu}$ for $A^0_\mu$.}
Using this approach the tensor fields $B_{\mu\nu\, M}$ carry no on-shell degrees of freedom.
However, they can \textit{eat up} a dynamical vector with three degrees of 
freedom and become massive. This will be treated in \autoref{sec:iso}.

The bosonic Lagrangian of this $\cN=4$ gauged supergravity theory is given by \cite{Dall'Agata:2001vb,Schon:2006kz}
\begin{align} \label{bos_N=4action}
 e^{-1}\cL_{\textrm{bos}}=&
 -\frac{1}{2}R - \frac{1}{4}\Sigma^2 M_{MN}\, \cH^M_{\mu\nu} \cH^{\mu\nu\, N} - \frac{1}{4}\Sigma^{-4}F_{\mu\nu}^0 F^{\mu\nu\, 0}\nn \\
 &-\frac{3}{2}\Sigma^{-2}(\nabla_\mu \Sigma)^2 
  + \frac{1}{16}(D_\mu M_{MN})(D^\mu M^{MN})\nn \\
 &+\frac{1}{16\sqrt 2}\epsilon^{\mu\nu\rho\lambda\sigma} \xi^{MN}B_{\mu\nu\, M} \big( D_\rho B_{\lambda\sigma\, N} 
 +4 \eta_{NP}A^0_\rho \partial_\lambda A^P_\sigma 
 + 4 \eta_{NP}A^P_\rho \partial_\lambda A^0_\sigma \big) \nn \\
 & - \frac{1}{\sqrt 2}\epsilon^{\mu\nu\rho\lambda\sigma}  A_\mu^0 \, \Big( \partial_\nu A^M_\rho \partial_\lambda A_{\sigma\, M}
 + \frac{1}{4}\xi_{MN} A^M_\nu  A^N_\rho \partial_\lambda A^0_\sigma
 - f_{MNP}  A^M_\nu A^N_\rho \partial_\lambda A^P_\sigma \Big ) \nn \\
 & -\frac{1}{4} f_{MNP}\,f_{QRS} \,\Sigma^{-2}\,\Big( \frac{1}{12}M^{MQ}M^{NR}M^{PS} - \frac{1}{4}M^{MQ}\eta^{NR}\eta^{PS} 
 + \frac{1}{6}\eta^{MQ}\eta^{NR}\eta^{PS}\Big)\nn\\
 &- \frac{1}{16} \xi_{MN}\,\xi_{PQ} \,\Sigma^4 \, \Big( M^{MP}M^{NQ}- \eta^{MP}\eta^{NQ} \Big) - \frac{1}{6\sqrt 2} f_{MNP}\, \xi_{QR}\,\Sigma\, M^{MNPQR} \, ,
\end{align}
where $R$ denotes the Ricci scalar, and we define
\begin{align} \label{def-cH}
 \cH^M_{\mu\nu} := 2\, \partial_{[\mu} A^M_{\nu]} -  \tensor{\xi}{_N^M}A_\mu^0 A_\nu^N - \tensor{f}{_P_N^M}A_\mu^P A_\nu^N +
 \frac{1}{2}\, \xi^{MN} B_{\mu\nu\, N}\, ,
\end{align}
as well as
\begin{align}
 F^0_{\mu\nu} := \partial_\mu A^0_\nu - \partial_\nu A^0_\mu \, .
\end{align}
The vectors and dual tensors in this Lagrangian are subject to
vector gauge transformations with scalar parameters $(\Lambda^0 , \Lambda^M)$ as well as
standard two-form gauge transformations with one-form parameters $\Xi_{\mu\, M}$. This
property will be of importance later since it allows us to remove some of the vectors from the action by
gauge transformations. For our choice of gaugings, \textit{i.e.~}$\xi_M =0$, the variation of the vectors reads
\begin{align} \label{general_gaugetransform}
 \delta A^0_\mu = \nabla_\mu \Lambda^0 \, , \qquad \delta A^M_\mu = D_\mu \Lambda^M -\frac{1}{2} \xi^{MN} \Xi_{\mu\, N}\, .
\end{align}

We now continue with the Lagrangian of the gravitino fields. 
To simplify our notation we introduce contractions of the embedding tensors
with the coset representatives
\begin{align}\label{e:dressed_gaugings}
 &\xi^{mn} := \tensor{\cV}{_M^m} \tensor{\cV}{_N^n}\, \xi^{MN} \, , \qquad \xi^{ab} := \tensor{\cV}{_M^a} \tensor{\cV}{_N^b} \,\xi^{MN}  \, , \qquad
 \xi^{am} := \tensor{\cV}{_M^a} \tensor{\cV}{_N^m} \,\xi^{MN}  \, , \nn \\
 &f^{mnp} := \tensor{\cV}{_M^m}  \tensor{\cV}{_N^n} \,\tensor{\cV}{_P^p} f^{MNP} \, , \qquad
 f^{mna} := \tensor{\cV}{_M^m}  \tensor{\cV}{_N^n} \,\tensor{\cV}{_P^a} f^{MNP} \, , \qquad \dots \, .
\end{align}
Note that these objects are field-dependent and acquire a VEV in the vacuum. It is important to realize that
the position of the $SO(5,n)$-indices $M,N$ in \eqref{e:dressed_gaugings} is essential because of \eqref{e:SO(5,N)_index_raising}.
Using this notation we define what will be identified with the gravitino mass matrix
\begin{align}\label{e:shift_matrices}
\textbf{M}_\psi^{ij} := \textbf{M}_\psi^{mn} \,\tensor{\Gamma}{_m_n^i^j}\,  
\end{align}
with
\begin{align}
 & \textbf{M}_\psi^{mn} := - \frac{1}{4\sqrt 2}\Sigma^2 \,\xi^{mn} + \frac{1}{24}\epsilon^{mnpqr}\,f_{pqr} \, , &\Gamma_{mn} := \Gamma_{[m}\Gamma_{n]} \, ,
\end{align}
where $\Gamma_m$ are the $SO(5)$ gamma matrices.
We are now in the position to write down the relevant fermionic terms in the Lagrangian.
For the purpose of this part
we will find it sufficient to only recall the kinetic terms and the mass terms of the gravitini.
The remaining quadratic terms of the fermions can be found in \cite{Dall'Agata:2001vb,Schon:2006kz}.
The relevant part of the Lagrangian reads
\begin{align} \label{fermionic-quadratic}
 e^{-1}\cL_{\textrm{grav}}=&
 -\frac{1}{2} \bar \psi^i_\mu \,\gamma^{\mu\nu\rho}\,\cD_\nu\, \psi_{\rho\, i} 
 +\frac{1 }{2}i\, \textbf{M}_{\psi\, ij}\, \bar \psi^i_\mu \,\gamma^{\mu\nu} \,\psi^j_\nu \, .
\end{align}
The precise form of the covariant derivative is of no importance for the moment, since we are only dealing with the gravitino mass
in this part.
This concludes our discussion of the general properties of $\cN =4$ gauged supergravity in five dimensions.

 \section{Isolation of the Propagating Degrees of Freedom}\label{sec:iso}
The formulation of $\cN=4$ gauged supergravity in terms of embedding tensors, as presented in \cite{Schon:2006kz},
is a very powerful way to implement general gaugings
of global symmetries.
However, in order to study vacua and the resulting 
effective field theories we need to eliminate non-propagating degrees of freedom used in the democratic formulation of 
 \cite{Schon:2006kz}. In particular, we have written down the $\cN=4$ gauged supergravities  
 in terms of vectors and dual tensors.
 We eliminate redundant vectors in the action by tensor gauge transformations
rendering the corresponding dual tensors
the (massive) propagating degrees of freedom. All remaining tensors that are not involved 
in this gauging procedure turn out to decouple in the action and can therefore be consistently set to zero.
In these cases the corresponding vectors constitute the appropriate formulation.
In the following we carry out the necessary redefinition of vectors and tensors explicitly.

The isolation of the appropriate propagating degrees of freedom in $\cN=4$ gauged 
supergravity depends on the form of the embedding tensor
$\xi^{MN}$.\footnote{We again stress that we set $\xi_M =0$ unless stated differently.}
This can easily be seen as follows. Consider the gauge transformations of the vectors $A^M$ \eqref{general_gaugetransform}
as well as the variation of the action with respect to the tensors $B_{\mu\nu\,M}$
\begin{align}
 &\delta A^M_\mu = D_\mu \Lambda^M -\frac{1}{2} \xi^{MN} \Xi_{\mu\, N} \, , 
 & \frac{\delta S}{\delta B_{\mu\nu\,M}} \sim \xi^{MN} (\dots)_N \, .
\end{align}
Note that one can always find orthogonal transformations such that
\begin{align}\label{e:full_rank_rot}
 &(\xi^{MN}) \mapsto \left( \begin{array}{c|c}
 \xi^{\hat M \hat N} &  \textbf{0}^{\hat M \bar N} \Bstrut\\  \hline 
  \textbf{0}^{\bar M \hat N} & \textbf{0}^{\bar M \bar N} \Tstrut
                           \end{array}\right) \, , \\
 & \hat M , \hat N =1,\dots , \rk (\xi^{MN}) \, , \qquad \bar M , \bar N =\rk (\xi^{MN}) + 1,\dots , 5+n \, , \nn
\end{align}
with $(\xi^{\hat M \hat N})$ a full-rank matrix. It is now easy to see that after appropriate partial gauge fixing
one can invert $(\xi^{\hat M \hat N})$ to obtain
\begin{align}
 \delta A^{\hat M}_\mu = - A^{\hat M}_\mu 
\end{align}
using tensor gauge transformations $\Xi_{\mu\, \hat M}$.
The $A^{\hat M}_\mu$ are therefore pure gauge and can be removed from the action. The corresponding tensors $B_{\mu\nu\,\hat M}$
constitute the appropriate formulation. In contrast, we find for the remaining vectors and tensors
\begin{align}
 &\delta A^{\bar M}_\mu = D_\mu \Lambda^{\bar M} \, , 
 & \frac{\delta S}{\delta B_{\mu\nu\,\bar M}} = 0 \, .
\end{align}
The Lagrangian is therefore independent of the $B_{\mu\nu\,\bar M}$, which is why we can set them to zero. We are left with
propagating vectors $A^{\bar M}$ subject to standard vector gauge transformations. To put it in a nutshell, one can see that the propagating
degrees of freedom are captured by $A^{\bar M}_\mu$, $B_{\mu\nu\,\hat M}$.
Moreover, for the pair $B^0_{\mu\nu}$, $A^0_\mu$ it turns out that the tensor $B^0_{\mu\nu}$ does not appear in the action
and $A^0_\mu$ constitutes the field carrying the propagating degrees of freedom.

Note that this procedure easily generalizes if one allows for a non-vanishing $\xi_M$. In this case one just has to replace
$\xi^{MN} \rightarrow 2 Z^{M^0 N^0}$ in the previous calculations, where $M^0 = (0,M)$ and
\begin{align}
 &Z^{MN} = \frac{1}{2} \xi^{MN} \, , & Z^{0M} = - Z^{M0} = \frac{1}{2} \xi^M \, .
\end{align}
One can then rotate $Z^{M^0 N^0}$ into a full-rank part and zero-matrices as in \eqref{e:full_rank_rot}.
The fields $A^0_\mu$ and $B_{0\, \mu\nu}$ then also take part in the procedure of extracting the propagating degrees of freedom.
As already mentioned several times, we nevertheless set $\xi_M = 0$ in the following.

In this thesis we are interested in deriving the
Lagrangian around a vacuum of the $\cN =4$ theory. In order to extract the propagating fields
we therefore slightly modify the approach which we have just described since this proves convenient for our purposes.
We start with the democratic formulation
of $\cN =4$ gauged supergravity reviewed in \autoref{N=4Gen} including the mentioned redundancies. 
Let us then assume that we have found a vacuum in which  all scalars,
\textit{i.e.}~$\langle \tensor{\cV}{_M^m} \rangle$, $\langle \tensor{\cV}{_M^a} \rangle$, $\langle \Sigma \rangle$, 
acquire a VEV. 
In analogy to \eqref{e:dressed_gaugings} we define
\begin{subequations} \label{rotate_AB}
\begin{align}
&B_{\mu \nu}^m  := \tensor{\langle \cV\rangle}{_M^m} B_{\mu \nu}^M \, ,  
&& B_{\mu \nu}^a := \tensor{\langle\cV\rangle}{_M^a} B_{\mu \nu}^M \, , & \\
  &A_\mu^m := \tensor{\langle \cV\rangle}{_M^m} A_\mu^M \, ,  && A_\mu^a :=  \tensor{\langle\cV\rangle}{_M^a} A_\mu^M \, . & 
\end{align}
\end{subequations}
Similarly we can introduce the gauge parameters $(\Lambda^m,\Lambda^a)$ and
$(\Xi_{\mu}^m,\Xi_{\mu}^a )$ by setting 
\begin{subequations}
\begin{align}
   &\Lambda^m:=  \tensor{\langle \cV\rangle}{_M^m} \Lambda^M  \,  , & & \Lambda^a  :=  \tensor{\langle  \cV\rangle}{_M^a}  \Lambda^M \, ,& \\
   &\Xi_{\mu}^m :=   \tensor{\langle \cV\rangle}{_M^m} \Xi_{\mu}^M  \, , &&  \Xi_{\mu}^a :=  \tensor{\langle\cV\rangle}{_M^a}  \Xi_{\mu}^M \, .& 
\end{align}
\end{subequations}
In this rotated basis
the gauge transformations \eqref{general_gaugetransform} read 
\begin{subequations}
\begin{align}
 \label{e:gauge_transf_1} 
 \delta A_\mu^m = & D_\mu \Lambda^m + \frac{1}{2}  \xi^{mn}\, \Xi_{\mu \, n} - \frac{1}{2}  \xi^{ma}\, \Xi_{\mu \, a}\, ,\\
 \label{e:gauge_transf_2} \delta A_\mu^a = & D_\mu \Lambda^a +\frac{1}{2}  \xi^{am}\, \Xi_{\mu \, m} -\frac{1}{2}  \xi^{ab}\, \Xi_{\mu \, b}\, . 
\end{align}
\end{subequations}
The elimination of redundant vectors and tensors is now carried out for the fluctuations around 
the vacuum rather than at a general point in the unbroken theory.

Note that there exist orthogonal
matrices $S$ such that the contracted embedding tensors \eqref{e:dressed_gaugings} transform as
\begin{align}\label{e:xi_trafo}
 &S^T \left( \begin{array}{c|c}
 \xi^{mn} &  \xi^{mb} \Bstrut\\ \hline
  \xi^{an} & \xi^{ab} \Tstrut
                           \end{array}\right) S =
  \left( \begin{array}{c|c}
\xi^{\hat\cM \hat\cN} &  \textbf{0}^{\hat\cM \bar\cN} \Bstrut \\ \hline
  \textbf{0}^{\bar\cM \hat\cN} & \textbf{0}^{\bar\cM \bar\cN} \Tstrut
                           \end{array}\right) \, \\
 &\hat \cM , \hat \cN =1,\dots , \rk (\xi^{MN}) \, , \qquad \bar \cM , \bar \cN =\rk (\xi^{MN}) +1,\dots , 5+n \, , \nn
\end{align}
where $(\xi^{\hat\cM \hat\cN})$ is a full-rank matrix.
In particular one can even choose an orthogonal matrix $S$ such that $\langle \xi^{\hat\cM \hat\cN} \rangle$ is block diagonal
\begin{align}\label{e:fullrank}
 \langle\xi^{\hat\cM\hat\cN} \rangle  
 = \begin{pmatrix}
\gamma_1 \varepsilon &  \cdots & 0 \\ 
\vdots &  \ddots & \vdots \\
 0 & \cdots & \gamma_{n_T} \varepsilon 
                         \end{pmatrix} \, ,
\end{align}
where $n_T = \frac{1}{2} \rk (\xi_{MN})$, which turns out to be the number of complex tensors. Furthermore the
$\gamma_1 , \dots , \gamma_{n_T}$ are constants, and $\varepsilon$ is the two-dimensional epsilon tensor.
The indices $\cM$, $\hat\cM$, $\bar \cM$ are raised and lowered
with the Kronecker delta.
Along the same lines as before, by inverting $\langle\xi^{\hat\cM \hat\cN}\rangle$ and partial gauge fixing,
we find that the propagating degrees of freedom in the vacuum
are captured by $A^{\bar \cM}_\mu$ and $B_{\mu\nu\,\hat \cM}$, where
\begin{align}
&(A^{\cM}_\mu) =
\left( \begin{array}{c}
 A^{\hat \cM}_\mu \Bstrut\\
  A^{\bar \cM}_\mu 
                           \end{array}\right)
 := S^T \left( \begin{array}{c}
 A^{m}_\mu\\
  A^{a}_\mu 
                           \end{array}\right)\, , 
 & (B_{\mu\nu\, \cM}) =
 \left( \begin{array}{c}
 B_{\mu\nu\, \hat\cM} \Bstrut\\
  B_{\mu\nu\, \bar\cM}
                           \end{array}\right)
 := S^T \left( \begin{array}{c}
 B_{\mu\nu\, m} \Bstrut\\
  B_{\mu\nu\, a} 
                           \end{array}\right)\, .
\end{align}
The gauge transformations are defined similarly, and one easily checks that the complement fields $A^{\hat \cM}_\mu$ and $B_{\mu\nu\,\bar \cM}$
can be eliminated from the action.
For later convenience let us also define the dual elements
\begin{subequations}
\begin{align}
 (A^{*\,\cM}_\mu) & =&&
\left( \begin{array}{c}
 A^{*\,\hat \cM}_\mu \Bstrut\\
  A^{*\,\bar \cM}_\mu 
                           \end{array}\right)
&&:= &&S^T \eta S 
 \left( \begin{array}{c}
 \textbf{0}^{\hat \cM} \Bstrut\\
  A^{\bar \cM}_\mu 
                           \end{array}\right) 
                           &&=&& S^T \eta 
 \left( \begin{array}{c}
 A^{m}_\mu \Bstrut\\
  A^{a}_\mu 
                           \end{array}\right)\bigg\vert_{A^{\hat\cM}_\mu \equiv 0}\, , \\
 (B^*_{\mu\nu\, \cM})&=&&
 \left( \begin{array}{c}
 B^*_{\mu\nu\, \hat\cM}\Bstrut\\
  B^*_{\mu\nu\, \bar\cM}
                           \end{array}\right)
 && := &&S^T \eta S 
 \left( \begin{array}{c}
 B_{\mu\nu\, \hat\cM}\Bstrut\\
  \textbf{0}_{\bar\cM} 
                           \end{array}\right)
                           &&=&& S^T \eta  
 \left( \begin{array}{c}
 B_{\mu\nu\, m}\Bstrut\\
  B_{\mu\nu\, a} 
                           \end{array}\right) \bigg\vert_{B_{\mu\nu\, \bar\cM} \equiv 0} \, ,
\end{align}
\end{subequations}
where $\eta = \textrm{diag}(-1,-1,-1,-1,-1,+1,\dots ,+1)$.
Already at this stage it becomes obvious that the number of complex massive tensors is always given by $\frac{1}{2}\rk ( \xi_{MN} )$.
Moreover a closer look at the Lagrangian \eqref{bos_N=4action} shows that the charge of the tensors is
independent of the vacuum. This will become important in \autoref{sec:Higgs}.
Unfortunately for the vectors such simple statements are not possible since most of the properties depend
crucially on the precise form of the vacuum.

\section{The Theory Around the Vacuum}\label{sec:action}

Having studied the redefinition of vectors and tensors
in order to isolate the propagating degrees of freedom
we are now in a position to derive crucial parts of the action around a general vacuum.
In particular, we display
the mass terms and charges of the scalars and tensors as well as the field strengths,
Chern-Simons terms and mass terms of the vectors in a general form which
depends on the (field-dependent) contracted embedding tensors \eqref{e:dressed_gaugings}.
Inserting the expressions for the latter
for a certain example one is interested in then easily yields the precise spectrum and the action.
Furthermore we derive the formulae for the cosmological constant as well as the gravitino masses.

Before writing down the Lagrangian, let us define the fluctuations of the scalars $\sigma$ and $\cV$ around their VEVs
\begin{subequations}
\begin{align}
 &\sigma = \langle \sigma \rangle + \tilde \sigma \, ,  \\
 &\cV = \langle \cV \rangle \exp \big(\,\phi^{ma}[t_{ma}]\,\big) \label{e:vexp} \, ,
\end{align}
\end{subequations}
where $[t_{ma}]_M^{\ N} = 2 \delta_{[m}^{\ \ N} \eta_{a]M}$. The $\phi^{ma}$ capture the unconstrained fluctuations around the VEVs of the coset representatives.
We also define indices $\cM\, , \cN , \dots$ in expressions like $f_{\cM m a}$ using the same transformation
as in \eqref{e:xi_trafo}. 
Furthermore we set
\begin{align}
 \eta_{\cM\cN} := ( S^T \eta S )_{\cM\cN} \, ,
\end{align}
where $S$ is the matrix of \eqref{e:xi_trafo} and $\eta = \textrm{diag}(-1,-1,-1,-1,-1,+1,\dots , +1)$.

The relevant part of the Lagrangian of $\cN = 4$ gauged supergravity around the vacuum then reads
\begin{align}\label{e:vac_lagr}
 e^{-1}\cL = &  \frac{1}{16 \sqrt 2}\epsilon^{\mu\nu\rho\lambda\sigma}  
  \,\xi^{\hat \cM \hat \cN}\, B^*_{\mu\nu\, \hat\cM} \cD_\rho B^*_{\lambda\sigma\, \hat\cN} 
 - \frac{1}{16} \Sigma^2 \, \xi^{\hat\cM\hat\cN} \tensor{\xi}{_{\hat\cM}^{\hat\cP}}\, B^*_{\mu\nu \, \hat\cN} B^{* \,\mu\nu}_{\hat\cP} \nn \\
 & -\frac{1}{4} \Sigma^2 F^{\bar \cM}_{\mu\nu} F_{\bar \cM}^{\mu\nu}
 - \frac{1}{4} \Sigma^{-4} F^0_{\mu\nu} F^{0\, \mu\nu}\nn \\
 & - \frac{\epsilon^{\mu\nu\rho\lambda\sigma}}{\sqrt 2} A_\mu^0 \,\Big ( \partial_\nu A_\rho^{*\, \bar \cM} \partial_\lambda A_{\sigma \, \bar \cM} 
  - f_{\cM\cN\cP}\, A_\nu^{*\, \cM} A_\rho^{*\, \cN}  \partial_\lambda A_\sigma^{*\, \cP} 
   - \frac{1}{4} \xi_{\hat\cN\hat\cP}\, A_\nu^{*\, \hat\cN} A_\rho^{*\, \hat\cP}  \partial_\lambda A_\sigma^{*\, 0} \Big ) \nn \\
 & - \frac{1}{2} \Big ( \cD_\mu \phi^{ma} - \xi^{ma} A_{\mu}^0 - \tensor{f}{_{\cM}^m^a} A_{\mu}^{*\, \cM} \Big ) 
 \Big ( \cD^\mu \phi_{ma} - \xi_{ma} A^{0 \, \mu} - \tensor{f}{^\cN_m_a} A^{*\, \mu}_\cN \Big )\nn \\
& -\frac{1}{2} \partial_\mu \tilde \sigma \, \partial^\mu \tilde \sigma 
  -\frac{1}{2} \textbf{M}^2_{ma\, nb}\, \phi^{ma}\phi^{nb} 
  - \frac{1}{2} \textbf{M}^2 \, \tilde\sigma^2 
  - \textbf{M}^2_{ma} \, \phi^{ma} \tilde \sigma \, ,
\end{align}  
with
\begin{subequations}
\begin{align}
 \cD_\mu \phi^{ma} &:= \partial_\mu \phi^{ma} - A_\mu^0 \, \phi^{nb} \, \big( \tensor{\xi}{_b^a} \delta^m_n -  \tensor{\xi}{_n^m} \delta^a_b \big)
 - A_\mu^{*\, \cM} \, \phi^{nb} \, \big( \tensor{f}{_\cM_b^a} \delta^m_n -  \tensor{f}{_\cM_n^m} \delta^a_b  \big) \, ,\\
\cD_\rho B^*_{\lambda\sigma\, \hat\cN} & :=  \partial_\rho B^*_{\lambda\sigma\, \hat\cN}
 -  \xi^{\hat\cP\hat\cQ}\,\eta_{\hat\cN\hat\cQ}\, A^0_\rho \, B^*_{\lambda\sigma\, \hat\cP} \, , \\
 F^{\bar \cM}_{\mu\nu} &:= 2\, \partial_{[\mu} A_{\nu]}^{\bar \cM} - \tensor{f}{_\cN_\cP^{\bar\cM}}A^{*\, \cN}_\mu A^{*\, \cP}_\nu \, , \\
 F^0_{\mu\nu} &:= 2\,\partial_{[\mu} A_{\nu]}^{0} \, ,
\end{align}
\end{subequations}
and
\begin{subequations}
\begin{align}  
\textbf{M}^2_{ma\, nb}:= &  \Sigma^{-2} \Big ( f_{abp}\tensor{f}{_m_n^p} + f_{abc}\tensor{f}{_m_n^c} + f_{anp}\tensor{f}{_m_b^p} + f_{anc}\tensor{f}{_m_b^c}
 + \delta_{mn} f_{acp}\tensor{f}{_b^c^p} + \delta_{ab} f_{mcp}\tensor{f}{_n^c^p} \Big ) \nn \\
& + \frac{1}{3\sqrt 2} \Sigma \Big ( 3\, \varepsilon_{mnpqr}  \tensor{f}{_a_b^p} \xi^{qr}
+ 6\, \varepsilon_{mnpqr}  \tensor{f}{_a^p^q} \tensor{\xi}{_b^r}
 + \varepsilon_{mnpqr} f^{pqr} \xi_{ab} \nn \\
 & +\frac{3}{2} \delta_{ab} \varepsilon_{mspqr}  \tensor{f}{_n^s^p} \xi^{qr}
 - \delta_{ab} \varepsilon_{mspqr} f^{spq} \tensor{\xi}{_n^r} \Big )  \\
& + \frac{1}{2}\Sigma^{4} \Big ( 2\, \xi_{mn} \xi_{ab} + 2\, \xi_{mb} \xi_{an}
 + \delta_{mn} \xi_{ac} \tensor{\xi}{_b^c} + \delta_{mn} \xi_{ap} \tensor{\xi}{_b^p}
   + \delta_{ab} \xi_{mp} \tensor{\xi}{_n^p} + \delta_{ab} \xi_{mc} \tensor{\xi}{_n^c} \Big ), \nn \\
 \textbf{M}^2 :=  & \Sigma^{-2} \Big (- \frac{1}{9} f_{mnp} f^{mnp} + \frac{1}{3} f_{mna} f^{mna} \Big )
 + \frac{4}{3} \,\Sigma^{4}\, \xi^{ma} \xi_{ma} + \frac{1}{18 \sqrt 2} \,\Sigma\, \varepsilon_{mnpqr} f^{mnp} \xi^{qr}\, , \\
\textbf{M}^2_{ma} := &  - \frac{2}{\sqrt 3} \Sigma^{-2} \tensor{f}{_a^b^n} f_{mbn} 
 + \frac{2}{\sqrt 3} \Sigma^4 \Big ( \xi_{ab} \tensor{\xi}{_m^b} + \xi_{an} \tensor{\xi}{_m^n} \Big ) \nn \\
 & + \frac{1}{6 \sqrt 6} \varepsilon_{mnpqr} \Sigma \Big ( 3\, \tensor{f}{_a^n^p} \xi^{qr} -2\, f^{npq} \tensor{\xi}{_a^r} \Big )  \, .
\end{align}
\end{subequations}
We stress that \eqref{e:vac_lagr} is not the full bosonic Lagrangian around the vacuum since there are additional couplings
which are not displayed. 
However, around an $\cN=2$ vacuum, which is the kind of vacuum we are most interested in,
the included terms together with the residual supersymmetry
turn out to be sufficient to determine the full effective action apart from the metric on the quaternionic manifold. In fact, 
as we discuss in more detail along our analysis in \autoref{Classicaltruncation},   
the effective theory is inferred by knowing the gauge symmetry,
Chern-Simons terms as well as the masses and charges of the fields.
This data is indeed captured by \eqref{e:vac_lagr}, at least for the bosonic sector.
It is also important to keep in mind that all contracted embedding tensors are meant to be evaluated in the vacuum.

Let us comment on some of the properties of the action  \eqref{e:vac_lagr}. Closer inspection of \eqref{e:vac_lagr} shows that 
the scalars $\phi^{ma}$ are coupled to the vectors
with standard minimal couplings as well as with St\"uckelberg couplings. This implies that some of the scalars $\phi^{ma}$
constitute the longitudinal degrees of freedom of massive vectors. We also see that it is in general possible to preserve a non-Abelian
gauge group in the vacuum corresponding to a subset of the $A_\mu^{\bar\cM}$. For this non-Abelian subgroup 
the corresponding Chern-Simons terms can
in general appear. The tensors are in general charged only under a $U(1)$ gauge symmetry.
As already mentioned, the number of massive tensors is given by $\frac{1}{2}\rk ( \xi_{MN} )$ which is obvious
in \eqref{e:vac_lagr} since their mass matrix, determined by $\xi^{\hat \cM \hat \cN}$, is full-rank.
 In contrast, note that the mass matrices of vectors and scalars 
are in general not full-rank.

To proceed further one has to specify the precise form of the contracted 
embedding tensors to study the spectrum and the action case by case.
In particular, one has to diagonalize the mass matrices or gauge-interaction matrices
of all fields, normalize the kinetic terms, and possibly complexify the fields.
We explicitly carry out this procedure for the special case of vanishing $f_{MNP}$ in \autoref{N=2spectrum}
and for examples of consistent truncations in \autoref{sec:enriq} and \autoref{sec:sasa} although not presenting all the details 
of the computations.
The standard form for the Lagrangians of the massive fields are displayed in \eqref{e:lagr_5d}.

To close this general discussion, let us comment on the cosmological constant in the vacuum. 
It can be extracted from the value of the scalar potential, which reads in terms of contracted embedding tensors
\begin{align}\label{e:pot_vac}
 V= - \frac{1}{12}\Sigma^{-2} f^{mnp}f_{mnp} + \frac{1}{4} \Sigma^{-2} f^{mna}f_{mna} + \frac{1}{4}\Sigma^4\, \xi^{am}\xi_{am}
 +\frac{1}{6\sqrt 2}\Sigma\, \varepsilon_{mnpqr} f^{mnp} \xi^{qr}\, .
\end{align}
Furthermore, since we are in particular interested in vacua preserving $\cN=2$ supersymmetry,
it is desirable to formulate a general condition
for a certain set of contracted embedding tensors.
Since massless gravitini are in one-to-one correspondence with preserved supersymmetries,
the remaining amount of supersymmetry in the vacuum can be determined from the mass terms of the gravitini \eqref{fermionic-quadratic}.
The four eigenvalues of the mass matrix $(\tensor{\textbf{M}}{_\psi_\,_i^j})$ denoted by $\pm m_{\psi\pm}$ are given by \cite{Cassani:2012wc}
\begin{align}\label{e:grav_mass}
 m_{\psi\pm} = \sqrt{2\, \textbf{M}_\psi^{mn} \, \textbf{M}_{\psi\, mn} \mp \sqrt{8\, \big(\textbf{M}_\psi^{mn}\, \textbf{M}_{\psi\, mn}\big)^2
 - 16\, \textbf{M}_\psi^{mn}\, \textbf{M}_{\psi\, np}\, \textbf{M}_\psi^{pq}\, \textbf{M}_{\psi\, qm} }}\, .
\end{align}
Additionally the masses of the gravitini receive contributions from a possibly non-trivial cosmological constant $\Lambda = \langle V \rangle$
\begin{align}\label{e:grav_mass_correction}
 \delta m_\psi = \frac{\sqrt 6}{4} \sqrt{-\langle V \rangle}\, .
\end{align}
The condition for preserved $\cN=2$ supersymmetry can then be formulated as
\begin{align}\label{e:n2cond}
 m_{\psi+} - \delta m_\psi \overset{!}{=} 0 \, .
\end{align}

We have now provided all formulae to check, given a set of contracted embedding tensors, if the associated vacuum preserves supersymmetry and has
a non-trivial cosmological constant. The spectrum and the most relevant terms of the Lagrangian are calculated easily using \eqref{e:vac_lagr}.
In the next section we characterize the subclass of Minkowski vacua according to their amount of preserved supersymmetry.

\section{General Properties of Minkowski Vacua}\label{sec:mink}
Let us for this section assume that we have found a vacuum of the original $\cN=4$ theory with
\begin{align}
 \langle V \rangle = 0 
\end{align}
which is by definition of Minkowski type.\footnote{We provide explicit examples of Minkowski vacua in the upcoming chapters.}
Therefore $\delta m_\psi = 0$ in \eqref{e:grav_mass_correction} and the gravitino masses are simply given by \eqref{e:grav_mass}.
Noting that the sum $\textbf{M}_\psi^{mn} \, \textbf{M}_{\psi\, mn}$ is quadratic in each summand there are only three qualitatively different
possibilities for gravitino masses and thus for the amount of preserved supersymmetry in the vacuum which we list in
\autoref{tab:mink_susy}.
\begin{table}[h]
\begin{center}
\begin{tabular}{c|c}

Supersymmetry & Condition \\
\hline
\rule[-.4cm]{0mm}{1.2cm} $\cN =4$ & $\textbf{M}_\psi^{mn} = 0 \quad \forall m,n$ \\
\rule[-.4cm]{0mm}{.9cm}  $\cN=2$ & $\textbf{M}_\psi^{mn}\textbf{M}_{\psi\,np}\textbf{M}_\psi^{pq}\textbf{M}_{\psi\, qm} =
\frac{1}{4} (\textbf{M}_\psi^{mn}\textbf{M}_{\psi\,mn})(\textbf{M}_\psi^{pq}\textbf{M}_{\psi\,pq}) \neq 0$\\
\rule[-.4cm]{0mm}{.9cm}  $\cN=0$ & all others
\end{tabular}
\end{center}
\caption{We collect the characterization for the amount of supersymmetry in Minkowski vacua.}
\label{tab:mink_susy}
\end{table}
We stress that we do not aim for a classification of possible vacua but rather investigate
properties of certain classes of vacua. Since in the following we are mainly interested in
partial supersymmetry breaking vacua from $\cN = 4 \rightarrow \cN =2 $, let us focus on the
solutions to the second
condition in \autoref{tab:mink_susy}
\begin{align}\label{e:n2_cond}
 \textbf{M}_\psi^{mn}\textbf{M}_{\psi\,np}\textbf{M}_\psi^{pq}\textbf{M}_{\psi\, qm} =
\frac{1}{4} (\textbf{M}_\psi^{mn}\textbf{M}_{\psi\,mn})(\textbf{M}_\psi^{pq}\textbf{M}_{\psi\,pq}) \neq 0 \, .
\end{align}
First we bring the antisymmetric matrix $\textbf{M}_\psi^{mn}$ into block diagonal form by orthogonal transformations
\begin{align}
 \textbf{M}_\psi \mapsto \begin{pmatrix}
 m_1\varepsilon & 0 & 0 \\
0 & m_2\varepsilon & 0  \\
0&0&0
                             \end{pmatrix} \, ,
\end{align}
where $\varepsilon$ is the two-dimensional epsilon tensor and $m_1, m_2 \in \mathbb R$.
Then the expression for the gravitino masses \eqref{e:grav_mass} becomes
\begin{align}\label{e:grav_mass_m}
 m_{\psi\pm} = 2 \vert m_1 \mp m_2 \vert \, .
\end{align}
The condition \eqref{e:n2_cond} then just states that
\begin{align}
 m_1 = \pm m_2 \neq 0 
\end{align}
in order to preserve $\cN =2$ supersymmetry.

Finally let us provide the group-theoretical interpretation for \eqref{e:n2_cond}. When we go from $\cN =4$ to $\cN =2$
supersymmetry,
the R-symmetry $USp(4)$ is broken as follows
\begin{align}
 USp(4) \rightarrow SU (2)_R \times SU (2)_F
\end{align}
with $SU (2)_R$ the $\cN =2$ R-symmetry and $SU (2)_F$ the residual flavor symmetry.
Not that $\textbf{M}_\psi^{mn}$ are Lie algebra elements of $\mathfrak{so} (5)\cong\mathfrak{usp} (4)$ 
acting in the fundamental representation,
which can be derived from \eqref{e:shift_matrices}.
The group-theoretical decomposition of traces in the fundamental representation into Casimirs then yields the following constraint
\begin{align}\label{e:group_th_susy}
 \tr_f^{\mathfrak{so} (5)} \textbf{M}_\psi^4 &\overset{!}{=} \tr_f^{\mathfrak{su} (2)_R \oplus \mathfrak{su} (2)_F}
 \textbf{M}_\psi^4 \\
 &= B_f^{\mathfrak{su} (2)_R \oplus \mathfrak{su} (2)_F}
 \ \tr_f^{\mathfrak{su} (2)_R  (\oplus \mathfrak{su} (2)_F}
 \, \textbf{M}_\psi^4  + C_f^{\mathfrak{su} (2)_R \oplus \mathfrak{su} (2)_F} \
\Big ( \tr_f^{\mathfrak{su} (2)_R \oplus \mathfrak{su} (2)_F} \,
 \textbf{M}_\psi^2 \Big)^2 \nn \, .
 \end{align}
One can then look up the values for the Casimirs (\textit{e.g.~}by using the results of \autoref{app:quart_tr})
\begin{align}
 B_f^{\mathfrak{su} (2)_R \oplus \mathfrak{su} (2)_F} &= 0 \,  ,
 &C_f^{\mathfrak{su} (2)_R \oplus \mathfrak{su} (2)_F} &= \frac{1}{4} \, .
\end{align}
Inserting these quantities into \eqref{e:group_th_susy} we then obtain precisely the condition
\eqref{e:n2_cond}.

\chapter{General Solution for Abelian Magnetic Gaugings}\chaptermark{Abelian Magnetic Gaugings}
\label{sec:ab_mag_gaug}

In this chapter we classify all possible vacua of $\cN =4$ gauged supergravity in five dimensions with
\begin{align}
 f_{MNP} = \xi_M = 0 \, .
\end{align}
In particular, for the general category of $\cN = 2$ vacua we derive the spectrum of massive tensors and massive gravitini which
enter via a tensorial Higgs mechanism and a super-Higgs mechanism. Finally we give the complete spectrum and parts of the effective action around
the vacuum including corrections to Chern-Simons terms which are independent of the supersymmetry-breaking scale.

\section{Vacuum Conditions}

In order to find the vacua of the theory with $f_{MNP} = \xi_M = 0$ we consider the scalar potential in \eqref{e:pot_vac} which
now takes the form
\begin{align}
\label{e:potential_1}
 V = \frac{1}{4}\Sigma^4 \xi^{am}  \xi_{am}  \, ,
\end{align}
where we have used \eqref{e:dressed_gaugings}.
The fact that the indices $a$ and $m$ are raised by the 
Kronecker delta implies that the scalar potential is a sum of positive semi-definite terms.

Determining the minima of this potential is trivial. The derivative with respect to $\Sigma$ yields
\begin{align}
 \Big \langle \frac{\partial V}{\partial \Sigma} \Big \rangle = \langle \Sigma^3 \xi^{am}  \xi_{am} \rangle \overset{!}{=} 0\, .
\end{align}
Since the left-hand-side of this equation is a sum of non-negative terms ($\Sigma$ is always positive), the solution simply reads\footnote{Let
us stress once more that $\xi^{am}$ is a field-dependent quantity.}
\begin{align}
\label{e:vacuum_condition}
  \langle\xi^{am}\rangle \overset{!}{=} 0 \qquad\forall a,m \, .
\end{align}
The potential at this point takes the value
\begin{align}\label{e:pot_at_minimum}
 \langle V  \rangle\big\vert_{\langle\xi^{am} \rangle = 0 } = 0 \, .
\end{align}
The remaining derivatives with respect to the scalars in the vector multiplets are trivially 
vanishing since the potential is positive semi-definite. 

In summary, for vanishing embedding tensors $f_{MNP}$, $\xi_M$ the vacua are characterized by the condition
$\langle\xi^{am}\rangle = 0$ for all $a,m$. Due to \eqref{e:pot_at_minimum} all such vacua are necessarily Minkowskian. 
The amount of preserved supersymmetry can therefore be inferred
from \autoref{tab:mink_susy} or \eqref{e:grav_mass_m} using
\begin{align}
 \textbf{M}_\psi^{mn} = - \frac{1}{4\sqrt 2}\Sigma^2 \xi^{mn} \, ,
\end{align}
\textit{i.e.}~by determining the eigenvalues of $\langle\xi^{mn}\rangle$. In particular, we obtain an $\cN =2$ vacuum if
$\langle\xi^{mn}\rangle$ can be brought into the following form using orthogonal transformations
\begin{align}\label{e:orth_xi}
 \langle\xi^{mn}\rangle \mapsto \begin{pmatrix}
 \gamma\varepsilon & 0 & 0 \\
0 & \gamma\varepsilon & 0 \\
0 & 0 & 0
                             \end{pmatrix} 
\end{align}
with $\gamma > 0$.

\section{Tensorial Higgs and Super-Higgs Mechanism}\label{sec:Higgs}
In the rest of this chapter we assume that the vacuum preserves $\cN =2$ supersymmetry. As just mentioned,
this means that $\langle\xi^{mn}\rangle$ can be brought into the form \eqref{e:orth_xi}.
Furthermore we note that because of the vacuum condition
\eqref{e:vacuum_condition} the matrix of contracted
embedding tensors becomes blockdiagonal
\begin{align}
 \left( \begin{array}{c|c}
 \langle\xi^{mn}\rangle &  \langle\xi^{mb}\rangle \Bstrut\\ \hline
  \langle\xi^{an}\rangle & \langle\xi^{ab}\rangle \Tstrut
                           \end{array}\right)  =
  \left( \begin{array}{c|c}
 \langle\xi^{mn}\rangle &  \textbf{0}^{mb} \Bstrut\\ \hline
  \textbf{0}^{an} & \langle\xi^{ab}\rangle \Tstrut
                           \end{array}\right) \, .
\end{align}
In order to single out the propagating degrees of freedom, as explained in \autoref{sec:iso}, we have to determine the 
full-rank part $\langle\xi^{\hat \cM \hat \cN}\rangle$ which we introduced in \eqref{e:xi_trafo} and \eqref{e:fullrank}.
It is easy to see that the constants $\gamma_1, \dots , \gamma_{n_T}$ defined in \eqref{e:fullrank} are obtained
from $\langle\xi^{mn}\rangle$ and $\langle\xi^{ab}\rangle$ by decoupled orthogonal transformations
 \begin{align}\label{e:eigen}
  \langle\xi^{mn}\rangle &\mapsto \begin{pmatrix}
 \gamma_1\varepsilon & 0 & 0 \\
0 & \gamma_2\varepsilon & 0 \\
0 & 0 & 0
                             \end{pmatrix} \, , &
 \langle\xi^{ab}\rangle &\mapsto \begin{pmatrix}
\gamma_3 \varepsilon & \cdots &  &  & \cdots & 0  \\ 
\vdots &  \ddots & & & & \vdots \\
 &  & \gamma_{n_T} \varepsilon & & &  \\
 &  & & 0 & &  \\
\vdots &  & & &  \ddots & \vdots \\
0 & \cdots  &   &   & \cdots  & 0
                         \end{pmatrix} \, .
 \end{align}
Note again that we demand
\begin{align}
 \gamma_1 = \gamma_2 \equiv \gamma > 0 \, ,
\end{align}
which is the $\cN =2$ condition.
We conclude that
\begin{align}
 n_T = \frac{1}{2} \Big (\rk \langle\xi^{mn}\rangle + \rk \langle\xi^{ab}\rangle \Big ) \, .
\end{align}
The index split introduced in \eqref{e:xi_trafo}
\begin{align}
 \cM \rightarrow (\hat\cM,\bar\cM)
\end{align}
into a full-rank part and a null-part induces the corresponding split in \eqref{e:eigen}
\begin{align}
 m &\rightarrow (\hat m , \bar m ) \, , & a &\rightarrow (\hat a , \bar a ) \, ,
\end{align}
again into full-rank parts labeled by $\hat m$, $\hat a$ and null-parts labeled by $\bar m$, $\bar a$.
According to the analysis in \autoref{sec:iso} the propagating tensors are given by $B_{\hat m}$, $B_{\hat a}$ and the propagating
vectors by $A^{\bar m}$, $A^{\bar a}$.

The part of the tensor fields in the Lagrangian \eqref{e:vac_lagr} can now be simplified by using 
the quantities $\gamma,\gamma_{3}, \dots, \gamma_{n_T}$ defined in \eqref{e:eigen}. 
The appearance of the two-dimensional epsilon tensor in these expressions 
makes it natural to define the \emph{complex} tensors out of the $B_{\hat \cM}$
\begin{subequations} \label{complex-tensors}
\begin{align}
 \BB_{\alpha} &:= B_{2\alpha-1} + i B_{2\alpha}\ , &\alpha &= 1,2 \, ,  \\
 \BB_{\check a} &:= B_{2\check a-1} + i B_{2\check a } \, , & \check a &= 3,\ldots, n_T \, . &  
\end{align}
\end{subequations}
One can show that the $\alpha$ index corresponds to the fundamental representation of the $\cN=2$ R-symmetry group $SU(2)_R$. 
Here and in the following we will use boldface symbols to denote complex fields.
Inserting these definitions 
together with  \eqref{e:eigen} into \eqref{e:vac_lagr}
we find for the tensor fields\footnote{In analogy
to the fermions we define $\bar\BB^\alpha := (\BB_\alpha)^*\,$.}
\begin{align}
\label{e:tensor_act_2}
 e^{-1} \cL_{B} = 
 &- \frac{1}{16} \Big[  i  \frac{1}{ \sqrt 2}\epsilon^{\mu\nu\rho\lambda\sigma}  \gamma 
 \bar \BB_{\mu\nu}^\alpha  ( \partial_\rho   \tensor{\BB}{_\lambda_\sigma_\, _{\alpha}} 
 + i  \gamma \tensor{\BB}{_\lambda_\sigma_\,_{\alpha}} A^0_\rho ) + \Sigma^2 \gamma^2 \bar \BB_{\mu\nu}^\alpha \BB^{\mu\nu}_{\alpha} \Big] \nn \\
 &-\frac{1}{16}  \sum_{\check a} \Big[  i   \frac{1}{\sqrt 2}\epsilon^{\mu\nu\rho\lambda\sigma} \gamma_{\check a} 
 \tensor{\bar \BB}{_\mu_\nu_\, _{\check a}}  ( \partial_\rho   \tensor{\BB}{_\lambda_\sigma_\, _{\check a}} 
 - i  \gamma_{\check a} \tensor{\BB}{_\lambda_\sigma_\,_{\check a}} A^0_\rho )  
 + \Sigma^2   \gamma_{\check a}^2 \bar \BB_{\mu\nu\, \check a} \BB^{\mu\nu}_{\check a} \Big]\, .
\end{align}

In the last step we rescale the complex tensors in order to 
bring the action into the standard form \eqref{e:lagr_5d_2}
\begin{align}
 \big( \BB_\alpha  , \BB_{\check a} \big )
 \mapsto \frac{1}{2^{5/4}}
 \bigg(\sqrt \gamma  \BB_\alpha  , \sqrt{\vert\gamma_{\check a}\vert} \BB_{\check a}\bigg )\, . 
\end{align}
We
can now determine the characteristic quantities $\sign (m_\BB),\vert m_\BB\vert ,q_\BB$, \textit{i.e.~}the sign of the mass,
its absolute value
and the charge under the $U(1)$ vector $A^0$
\begin{subequations}\label{tensor_data}
\begin{align} 
 &\sign (m_{\BB_\alpha}) = 1 \, , && \vert m_{\BB_\alpha}\vert  =\frac{1}{\sqrt 2}\Sigma^2 \gamma \, ,&& q_{\BB_{ \alpha}} =  -\gamma \, , \\
 & \sign (m_{\BB_{\check a}})= \sign ( \gamma_{\check a})\, ,&&\vert m_{\BB_{\check a}}\vert =  \frac{1}{\sqrt 2}\Sigma^2  \vert \gamma_{\check a} \vert \, ,
 && q_{\BB_{\check a}} =  \gamma_{\check a}\, . &
\end{align}
\end{subequations}
This concludes our discussion of the massive tensors. We have found that 
evaluated around the $\cN=2$ vacuum there are $n_T$ complex 
massive tensors $(\BB_\alpha,\BB_{\check a})$ with 
standard action \eqref{e:lagr_5d_2} and characteristic 
data \eqref{tensor_data}.

The propagating $U(1)$ vector fields $A^0,A^{\bar m},A^{\bar a}$ stay massless in the vacuum
since the St\"uckelberg couplings to scalars in the Lagrangian \eqref{e:vac_lagr} vanish due to our choice of gaugings
and the vacuum condition $\langle \xi^{am} \rangle =0$.
For convenience we summarize the 
split of the fields induced by $\xi^{MN}$ in \autoref{index_split}.
\begin{table}[h!]
 \centering
\begin{tabular}{c|ccc}
  \rule[-.2cm]{0cm}{.7cm} & Rotation with $\langle \cV \rangle$ &  $\xi^{MN}$-split & Physical degrees\\
\hline
\multirow{4}{*}{$(A^M,B_M)$} & \multirow{2}{*}{$(A^m,B_m)$} & \rule[-.2cm]{0cm}{.7cm}$(A^{\bar m},B_{\bar m})$ &  $A^{\bar m}$  massless \\ 
   & & \rule[-.2cm]{0cm}{.7cm}$(A^{\hat m},B_{\hat m})$ &  $\BB_{\alpha}$ complex, massive\\
   \cline{2-4}
   & \multirow{2}{*}{$(A^a,B_a)$} & \rule[-.2cm]{0cm}{.7cm}$(A^{\bar a},B_{\bar a})$  & $A^{\bar a}$ massless\\  
    & & \rule[-.2cm]{0cm}{.7cm}$(A^{\hat a},B_{\hat a})$ & $\BB_{\check a}$ complex, massive\\
\end{tabular}
\caption{We summarize the natural split of $A^M$ and $B_M$ induced by $\xi^{MN}$.}
\label{index_split}
\end{table}

As we have already mentioned, in the $\cN =2$ broken phase of an $\cN=4$ theory a 
gravitino mass term has to be generated for half of the 
gravitino degrees of freedom. 
This mass arises in the sector of the flavor $SU (2)_F$ subgroup of the $\cN=4$ R-symmetry group $USp(4)$. In fact, 
two gravitini eat up two spin-$\sfrac 1 2$ goldstini from the gravity multiplet and become massive.
In this super-Higgs mechanism the massive gravitini acquire four extra degrees of freedom.
The appropriate description of the massive fields is in
terms of a single Dirac spin-$\sfrac 3 2$ fermion $\Bpsi_\mu$ without a symplectic Majorana condition.
The massive gravitino combines with the two massive complex tensors $\BB_\alpha$ from the former gravity multiplet into
a massive $\cN = 2$ gravitino multiplet $(\Bpsi_\mu,\BB_\alpha)$. The construction of such a half-BPS multiplet 
has been investigated in \cite{Hull:2000cf}.
In the following 
we will briefly discuss the super-Higgs mechanism and determine the mass and $U(1)$ charge of the 
gravitino multiplet. 

Let us first consider the four $\cN =4$ symplectic Majorana gravitini $\psi_\mu^i$
and the spin-$\sfrac 1 2$ fermions in the gravity multiplet $\chi^i$. These split under the breaking
\begin{align}\label{e:usp_break}
 USp(4) \rightarrow SU(2)_R \times SU(2)_F 
\end{align}
into $\psi_\mu^\alpha$, $\psi_\mu^{\dot\alpha}$ and $\chi^\alpha$, $\chi^{\dot\alpha}$, respectively. 
The index $\alpha = 1,2$ refers to the fundamental representation of the $\cN = 2$ R-symmetry
group $SU(2)_R$, while $\dot \alpha = 1,2$ corresponds to the flavor $SU(2)_F$ part. Both indices are raised and lowered
with the epsilon tensor analogous to \eqref{e:properties_omega} and \eqref{e:raising_lowering}.
From the fermionic part of the Lagrangian \cite{Dall'Agata:2001vb,Schon:2006kz} it turns out that
all fermion bilinears involving
$\psi_\mu^\alpha$ and $\chi^\alpha$ vanish in the vacuum
leaving only the kinetic terms for these fields
when one uses the $\cN = 2$ vacuum conditions \eqref{e:n2cond} and \eqref{e:vacuum_condition}.
Thus we find two massless spin-$\sfrac 3 2$ symplectic Majorana fermions $\psi_\mu^\alpha$ and two
massless spin-$\sfrac 1 2$ symplectic Majorana fermions $\chi^\alpha$.
We note that throughout this part
all massless fermionic $\cN = 2$ fields are taken to be symplectic Majorana.

We proceed with the investigation of the remaining fields $\psi_\mu^{\dot\alpha}$ and
$\chi^{\dot\alpha}$. 
The $\chi^{\dot\alpha}$ actually are the goldstini that render the
$\psi^{\dot\alpha}_\mu$ massive and can be removed from the
action by a shift of the gravitini analogous to the one performed in \cite{Hohm:2004rc,Horst:2012ub}.
It is furthermore convenient to merge the two symplectic Majorana fermions $\psi^{\dot\alpha}_\mu$ into a single
unconstrained Dirac
spinor\footnote{We could also choose $\Bpsi_\mu := \psi_\mu^{\dot\alpha = 2}$
which flips the representation and the charge under $A^0_\mu$ since both descriptions
are equivalent.}
\begin{align} \label{def-Bpsi}
 \Bpsi_\mu := \psi_\mu^{\dot\alpha = 1} \, ,
\end{align}
and $\psi_\mu^{\dot\alpha = 2}$ is also replaced appropriately by $\Bpsi_\mu$ using the symplectic Majorana condition.
The Lagrangian then reads
\begin{align} \label{mass_gravitino}
 e^{-1}\cL_{\textrm{mass grav}}=&
 -\bar \Bpsi_\mu \gamma^{\mu\nu\rho}\cD_\nu \Bpsi_{\rho}
 + \frac{1}{\sqrt 2}\Sigma^2 \, \gamma \,
 \bar \Bpsi_\mu \gamma^{\mu\nu} 
 \Bpsi_{\nu}\, ,
\end{align}
with $\cD_\mu \Bpsi_\nu = \partial_\mu \Bpsi_\nu 
 + i  \gamma A^0_\mu  \Bpsi_\nu$, and $\gamma$ is defined in \eqref{e:orth_xi}.

To conclude this section we compare the action \eqref{mass_gravitino} with 
the standard form \eqref{e:lagr_5d_3}
We 
find that $\Bpsi$ is in the $(1,\frac{1}{2})$ representation of
the little group and carries mass and $A^0_\mu$-charge
\begin{align}
 \sign (m_{\Bpsi}) =1 \, ,\qquad \vert m_\Bpsi \vert = \frac{1}{\sqrt 2}\Sigma^2  \gamma \, , \qquad 
 q_\Bpsi  =  -\gamma \, .
\end{align}
These data will be crucial in evaluating the one-loop corrections induced 
by the massive gravitino multiplet in the next section. 
Note that the massive Dirac gravitino $\Bpsi$ indeed combines with the massive tensors $\BB_\alpha$
into a massive gravitino multiplet.

\section{\texorpdfstring{$\cN =2$}{N=2} Mass Spectrum and Effective Action}

In this section we determine the complete spectrum parameterizing the fluctuations around 
the $\cN=2$ vacuum. We determine the masses and $U(1)$-charges of all fields
and show how they reassemble into $\cN = 2$ multiplets in \autoref{N=2spectrum}.
Furthermore, we derive the low-energy effective action of the massless modes 
with particular focus on the data determining the $\cN=2$ vector sector. 
The classical truncation from $\cN=4$ to $\cN=2$ is discussed in \autoref{Classicaltruncation}. 
The crucial inclusion of one-loop quantum corrections due to integrating out 
massive fermions and tensors is discussed in \autoref{oneloopeffects}. 
These induce extra contributions to the metric and Chern-Simons terms that are independent of 
the scale of supersymmetry breaking.

\subsection{The \texorpdfstring{$\cN = 2$}{N=2} Spectrum} \label{N=2spectrum}

The $\cN = 2$ spectrum and its properties can be determined by evaluating 
the $\cN=4$ action in the vicinity of the $\cN=2$ vacuum. 
To read off the masses and charges all kinetic terms and mass terms have 
to be brought into canonical form after spontaneous symmetry breaking. 
This diagonalization procedure is rather lengthy and therefore partially deferred
to \autoref{mass_appendix}. In the following we highlight 
some of the basic steps and summarize the results.

The key ingredients in the mass generation are the gaugings $\xi^{MN}$.
Recall that in the scalar background we rotated $\xi^{MN}$ to $\xi^{mn},\xi^{ab}$
and found the components
\begin{subequations}
\begin{align}
   \xi^{mn} & \rightarrow  \xi^{\hat m \hat n}\, ,& \xi^{\bar m \hat n}& =  \xi^{\bar m \bar m} = 0 \, ,\\ 
   \xi^{ab} &  \rightarrow  \xi^{\hat a \hat b}\, ,& \xi^{\bar a \hat b}& =  \xi^{\bar a \bar b} = 0 \, ,
\end{align}
\end{subequations}
where $\xi^{\hat m \hat n}$ and $ \xi^{\hat a \hat b}$ have maximal rank.
This yielded the natural index split
\begin{align}\label{e:split}
\begin{array}{ccccc}
 m &\rightarrow & (\bar m , \hat m) &\rightarrow & (\bar m , [\alpha 1] , [\alpha 2])\, ,\\ 
 a &\rightarrow & (\bar a , \hat a ) &\rightarrow & (\bar a , [\check a 1],[\check a 2] )\, . 
\end{array}
\end{align}
Here the splitting of $\hat m $ into $[\alpha 1],[\alpha 2]$ and the splitting of $\hat a$ into $[\check a 1],[\check a 2]$
arise due to the block diagonalization in \eqref{e:eigen}
with the first index $\alpha$, $\check a$ labeling the blocks and the second index labeling the two entries of each block.
In order to extract the massless and massive scalar spectrum
recall that in \eqref{e:vexp} we introduced $\phi^{ma}$ as 
the unconstrained fluctuations around the vacuum value 
$ \langle \cV \rangle$. They constitute the scalar degrees of freedom in the $\cN = 2$ effective theory. 
Due to the index split \eqref{e:split} we need to apply the split also to the scalars 
\beq \label{splittingphi}
  \phi^{ma}\ \rightarrow \ \Big (\phi^{\bar m \bar a}, \phi^{\bar m [\check a 1]},\phi^{\bar m [\check a 2]},\phi^{[\alpha 1] \bar a},\phi^{[\alpha 1] [\check a 1]},\phi^{[\alpha 1] [\check a 2]}, 
  \phi^{[\alpha 2] \bar a}, \phi^{[\alpha 2] [\check a 1]},\phi^{[\alpha 2] [\check a 2]}\Big)\ .
\eeq
To treat these more compactly we introduce, just as for the tensors in \eqref{complex-tensors}, the complex scalars
\begin{subequations}\label{compl_scalars}
\bea 
    \Bphi^{\alpha \bar a} &:=& \tfrac{1}{\sqrt{2}} (\phi^{[\alpha1] \bar a} + i \phi^{[\alpha 2] \bar a} )  \ ,\qquad 
    \Bphi^{\bar m \check a} := \tfrac{1}{\sqrt{2}} (\phi^{\bar m [\check a 1]}+i\phi^{\bar m [\check a 2]})\, , \label{compl_scalars_1} \\
   \Bphi^{\alpha \check a}_1 &:=& \tfrac{1}{2} (\phi^{[\alpha 1] [\check a 1]} - \phi^{[\alpha 2] [\check a 2]}
   + i \phi^{[\alpha 2] [\check a 1]} + i \phi^{[\alpha 1] [\check a 2]})\, , \label{compl_scalars_2}\\
   \Bphi^{\alpha \check a}_2 &:=& \tfrac{1}{2} (\phi^{[\alpha 1] [\check a 2]} - \phi^{[\alpha 2] [\check a 1]}
   + i \phi^{[\alpha 2] [\check a 2]} + i \phi^{[\alpha 1] [\check a 1]}) \, .\label{compl_scalars_3}
\eea
\end{subequations}
Note that in this way all $\phi^{ma}$ of the split \eqref{splittingphi} except $\phi^{\bar m \bar a}$
are combined into complex scalars. 

Similarly we proceed for the split of the $\cN=4$ fermions $\lambda^{a}_i$. Note 
that as for the gravitino around \eqref{e:usp_break} one splits $i \rightarrow (\alpha, \dot \alpha)$.
Together with the index split of $a$ given in \eqref{e:split} one has 
\beq \label{split_lambdas}
   \lambda^{a}_i \ \rightarrow \ \Big(\lambda^{\bar a}_\alpha,\lambda_\alpha^{[\check a 1]},\lambda_\alpha^{[\check a 2]},\lambda_{\dot \alpha}^{ \bar a},\lambda_{\dot \alpha}^{[\check a 1]},\lambda_{\dot \alpha}^{[\check a 2]}\Big)\ .
\eeq
It turns out to be convenient to combine all components of $\lambda^a_i$ except of $\lambda^{\bar a}_\alpha$ 
into complex Dirac fermions
\begin{subequations}\label{compl_fermions}
\bea \label{compl_fermions_1}
   \Blambda^{\check a}_\alpha &:=& \tfrac{1}{\sqrt{2}} (\lambda^{[\check a1]}_\alpha+ i \lambda^{[\check a2]}_\alpha)\ ,\\
   \label{compl_fermions_2}
  \Blambda^{\bar a} &:=& \lambda^{\bar a}_{\dot \alpha =1} \ , \quad \Blambda^{\check a}_1 := \frac{1}{\sqrt 2}(\lambda_{\dot \alpha =1}^{[\check a 1]} + i \lambda_{\dot \alpha =1}^{[\check a 2]})\, , \quad 
 \Blambda^{\check a}_2 := \frac{1}{\sqrt 2}(\lambda_{\dot \alpha =1}^{[\check a 1]} - i \lambda^{[\check a 2]}_{\dot \alpha=1})\, . 
\eea
\end{subequations}
To justify the use of \eqref{compl_fermions} we stress that the appearance of all spin-$\sfrac 1 2$ fermions
can be expressed in terms of the unconstrained Dirac spinors $\Blambda^{\check a}_\alpha$, $\Blambda^{\bar a}$, 
$\Blambda^{\check a}_{1,2}$.
Concerning \eqref{compl_fermions_1} the other linear combination
$\tfrac{1}{\sqrt{2}} (\lambda^{[\check a1]}_\alpha - i \lambda^{[\check a2]}_\alpha)$
is related to $\Blambda^{\check a}_\alpha$ by the symplectic Majorana condition.
In \eqref{compl_fermions_2}, by the same reasoning, the linear combinations with $\lambda^a_{\dot \alpha = 2}$ are related to those
involving $\lambda^a_{\dot \alpha = 1}$.
All degrees of freedom of the massive spin-$\sfrac 1 2$ fermions are therefore captured
by the spinors \eqref{compl_fermions} dropping the symplectic Majorana condition.

We are now in the position to summarize the spectrum. 
From the $\cN=4$ gravity multiplet the metric $g_{\mu \nu}$, two 
gravitini $\psi^\alpha_\mu$, two spin-$\sfrac 1 2$ fermions $\chi_\alpha$, two
vectors $A^0,A^{\bar m}$, and one scalar $\Sigma$ remain massless.
These fields group into the $\cN=2$ gravity multiplet $(g_{\mu \nu}, A^{\bar m},\psi^\alpha_\mu)$
and one $\cN=2$ vector multiplet $(A^0,\Sigma,\chi_\alpha)$. Note that the vector multiplet $(A^0,\Sigma,\chi_\alpha)$ is
special since the massive states, such as the tensors and gravitini discussed in \autoref{sec:Higgs}, carry $A^0$-charges.
In order to later derive the quantum effective action for the $A^0$ vector multiplet 
we need to determine these $A^0$-charges. 
$n-2(n_T-2)$ vector multiplets $(A^{\bar a},\phi^{\bar m \bar a},\lambda_\alpha^{\bar a})$ remain massless. 
We have already discussed the massless vectors $A^{\bar a}$ in \autoref{sec:Higgs}. 
We 
check in \autoref{mass_appendix} that the $ \phi^{\bar m \bar a}$ and $\lambda_\alpha^{\bar a}$ are indeed 
massless. 

Recall that an $\cN=2$ hypermultiplet has four real scalars and one Dirac spin-$\sfrac 1 2$ fermion. Using the 
definitions above \eqref{compl_scalars} and \eqref{compl_fermions} one can form the hypermultiplets
\beq 
   (\Bphi^{\alpha \bar a},\Blambda^{\bar a})\, , \quad (\Bphi^{\alpha \check a}_1, \Blambda^{\check a}_1)\, , \quad  (\Bphi^{\alpha \check a}_2, \Blambda^{\check a}_2)\, .
\eeq
The $n-2(n_T-2)$ hypermultiplets $(\Bphi^{\alpha \bar a},\Blambda^{\bar a})$ are always massive since they receive masses 
$\vert m_{\bar a}\vert = \frac{1}{\sqrt 2} \Sigma^2 \gamma$
from a non-trivial $\xi^{\hat m \hat n}$. The hypermultiplets $ (\Bphi^{\alpha \check a}_{1,2}, \Blambda^{\check a}_{1,2})$
can be either massless or massive since their masses have two contributions
from a non-trivial $\xi^{\hat m \hat n}$ and $\xi^{\hat a \hat b}$, respectively.
As we show in \autoref{mass_appendix}, the $\xi^{MN}$-splits \eqref{e:eigen} yield
masses given by
\begin{align} \label{mass_general}
\vert m_{\check a}^1 \vert = \frac{1}{\sqrt 2}\Sigma^2  \vert  \gamma  - \gamma_{\check a}\vert  \, ,\qquad \vert m_{\check a}^2 \vert = \frac{1}{\sqrt 2}\Sigma^2  \vert \gamma + \gamma_{\check a}  \vert \, ,
\end{align}
for the fields $(\Bphi^{\alpha \check a}_{1}, \Blambda^{\check a}_{1})$ and $(\Bphi^{\alpha \check a}_{2}, \Blambda^{\check a}_{2})$, respectively.
This implies that one hypermultiplet is massless whenever 
the condition
\beq  \label{massless_condition}
   \gamma_{\check a} = \pm \gamma 
\eeq
is satisfied. We 
denote the number of such massless hypermultiplets by $n_H$, and name their (pseudo-)real components
$(h^\Lambda_{1,2,3,4}, \lambda^\Lambda_{1,2})$, with $\Lambda = 1,\ldots, n_H$. 
Due to the fact that the hypermultiplets appear in pairs the existence 
of a massless hypermultiplet implies the existence of a massive hypermultiplet with 
mass $\sqrt 2\Sigma^2  \gamma$. Furthermore, one can check that 
one can consistently choose all $\gamma_{\check a}>0$ without 
changing the effective theory.
In summary, one has  $2(n_T -2) - n_H$ massive hypermultiplets with mass \eqref{mass_general}
out of the set $(\Bphi^{\alpha \check a}_{1,2}, \Blambda^{\check a}_{1,2})$.
Together with the $(\Bphi^{\alpha \bar a},\Blambda^{\bar a})$ one finds in total $n-n_H$ massive hypermultiplets.

To complete the summary of the spectrum 
recall that in \autoref{sec:Higgs} we have already identified and analyzed the
$\cN = 2$ massive gravitino multiplet comprising a massive Dirac gravitino $\Bpsi_\mu$ and two
complex massive tensors $\BB_\alpha$. 
Furthermore, we found $n_T-2$ complex massive tensors $\BB_{\check a}$. 
They combine with the Dirac fermions $\Blambda^{\check a}_\alpha$ and the complex scalars $\Bphi^{\bar m \check a}$ into $n_T-2$ complex massive tensor multiplets.

To conclude we list in \autoref{field_decomposition} the decompositions 
of the $\cN = 4$ fields in terms of $\cN =2$ fields along with their masses and charges. 
The reorganization into $\cN=2$ multiplets can be found in \autoref{vacuum_multiplets}.

\setlength\extrarowheight{5pt}
\begin{table}[]
\centering
\begin{tabular}{c|cccc}
$\cN = 4$ fields  & $\cN = 2$ fields & Mass & $\sign (m) $ & $A^0$-charge \\
\hline\hline
\rule[-.3cm]{0cm}{.8cm}  $g_{\mu\nu}$ & $g_{\mu\nu}$ &0& - & 0\\
\hline
\rule[-.3cm]{0cm}{.8cm}   $A^0_\mu$ & $A^0_\mu$ &0& - & 0\\
\hline
\rule[-.3cm]{0cm}{.8cm}   $A_\mu^{ij}$ & $A^{\bar m}_\mu$ &0 &- & 0\\
 \rule[-.3cm]{0cm}{.8cm}  & $\BB_{\mu\nu\,\alpha}$ & $\frac{1}{\sqrt 2} \Sigma^2  \gamma$ & 1 & $ -\gamma$ \\
\hline
 \rule[-.3cm]{0cm}{.8cm}  $\psi_\mu^i$ & $\psi_\mu^\alpha$ & 0 &-& 0\\
 \rule[-.3cm]{0cm}{.8cm} & $\Bpsi_\mu$ & $\frac{1}{\sqrt 2} \Sigma^2  \gamma$ & 1 & $ -\gamma$ \\
\hline
 \rule[-.3cm]{0cm}{.8cm} $\chi_i$ & $\chi_\alpha$ & 0 & -& 0\\
 \rule[-.3cm]{0cm}{.8cm} & $\chi_{\dot\alpha}$ &-& goldstino & -\\
\hline
\rule[-.3cm]{0cm}{.8cm}  $\Sigma$ & $\Sigma$ & 0 &-& 0\\
\hline
\rule[-.3cm]{0cm}{.8cm}  $A_\mu^a$ & $A_\mu^{\bar a}$ &0 & -& 0\\
\rule[-.3cm]{0cm}{.8cm} & $\BB_{\mu\nu\, \check a}$ & $\frac{1}{\sqrt 2}\Sigma^2  \vert \gamma_{\check a} \vert$
& $\text{sign}(\gamma_{\check a})$  & $\gamma_{\check a}$ \\
\hline
\rule[-.3cm]{0cm}{.8cm} $\lambda^a_i$ & $\lambda^{\bar a}_\alpha$ & 0 & -& 0\\
\rule[-.3cm]{0cm}{.8cm}  & $\Blambda^{\check a}_\alpha$ & $\frac{1}{\sqrt 2}\Sigma^2  \vert\gamma_{\check a}\vert$ 
& $\text{sign}(\gamma_{\check a})$& $\gamma_{\check a}$ \\
\rule[-.3cm]{0cm}{.8cm}  & $\Blambda^{\bar a}$ & $\frac{1}{\sqrt 2} \Sigma^2  \gamma$ & -1& $ \gamma$ \\
\rule[-.3cm]{0cm}{.8cm}  & $\Blambda^{\check a}_{1,2}$ & $\frac{1}{\sqrt 2}\Sigma^2  \vert \gamma\mp \gamma_{\check a}     \vert$ 
& $\text{sign}( \pm \gamma_{\check a} - \gamma )$  
& $\pm \gamma_{\check a} - \gamma   $\\
\hline
\rule[-.3cm]{0cm}{.8cm}  $\phi^{m a}$ & $\phi^{\bar m \bar a}$ & 0  & -& 0\\
\rule[-.3cm]{0cm}{.8cm}  & $\Bphi^{\alpha\bar a}$ & $\frac{1}{\sqrt 2} \Sigma^2  \gamma$ & singlet & $\gamma$ \\
\rule[-.3cm]{0cm}{.8cm}   & $\Bphi^{\bar m \check a}$ & $\frac{1}{\sqrt 2}\Sigma^2 \vert\gamma_{\check a}\vert$ & singlet 
& $\gamma_{\check a}$ \\
\rule[-.3cm]{0cm}{.8cm}  & $\Bphi^{\alpha \check a}_{1,2}$ & $\frac{1}{\sqrt 2}\Sigma^2  \vert \gamma \mp \gamma_{\check a}     \vert$ &
singlet& $ \pm \gamma_{\check a} - \gamma$\\
\end{tabular}
\caption{We show the decomposition of the $\cN = 4$ fields. The quantity $\sign (m)$ determines the representation of the little group for 
the massive fields, see \eqref{def_sign_5d}.}
\label{field_decomposition}
\end{table}

\setlength\extrarowheight{5pt}
\begin{table}
 \centering
\begin{tabular}{c|ccc}

\rule[-.3cm]{0cm}{.8cm}  Multiplets & Fields &  Mass & Charge\\
\hline\hline
\rule[-.3cm]{0cm}{.8cm}  1 gravity & $g_{\mu\nu},A^{\bar m}_\mu,\psi_\mu^\alpha$ 
 & 0 & 0\\
\hline
\rule[-.3cm]{0cm}{.8cm}  1 gravitino & $\Bpsi_\mu, \BB_{\mu\nu\,\alpha}$  & $\frac{1}{\sqrt 2} \Sigma^2  \gamma$ & $-\gamma$\\
\hline
\multirow{2}{*}{$(n+5 -2n_T )$ vector} & \rule[-.2cm]{0cm}{.7cm}  $A^0_\mu,\chi_\alpha,\Sigma$  & 0 & 0\\
& \rule[-.3cm]{0cm}{.8cm}  $A_\mu^{\bar a},\lambda^{\bar a}_\alpha,\phi^{\bar m \bar a}$ 
&0 &0\\
\hline
\rule[-.3cm]{0cm}{.8cm}  $n_T -2$ tensor & $\BB_{\mu\nu}^{\check a},\Blambda^{\check a}_\alpha,\Bphi^{\bar m \check a}$ 
 & $\frac{1}{\sqrt 2}\Sigma^2 \vert \gamma_{\check a}\vert$ & $ \gamma_{\check a}$\\
\hline
\multirow{2}{*}{$n$ hyper} & \rule[-.3cm]{0cm}{.8cm} $\Blambda^{\bar a},\Bphi^{\alpha\bar a}$  &$\frac{1}{\sqrt 2} \Sigma^2  \gamma$ & $\gamma$ \\
& \rule[-.3cm]{0cm}{.8cm}  $\Blambda^{\check a}_{1,2},\Bphi^{\alpha \check a}_{1,2}$ 
 & $\frac{1}{\sqrt 2}\Sigma^2  \vert \gamma \mp \gamma_{\check a}    \vert$ & $  \pm \gamma_{\check a} - \gamma  $\\
\end{tabular}
\caption{We depict the $\cN = 2$ multiplets of the vacuum.}
\label{vacuum_multiplets}
\end{table}

\subsection{General \texorpdfstring{$\cN =2$}{N=2} Action and Classical Matching} \label{Classicaltruncation}

We are now in the position to derive the classical $\cN= 2$ effective action 
for the massless modes. In order to do that we at first simply truncate the $\cN=4$ 
action to the massless sector. The discussion of the quantum corrections can then be found in 
the next subsection.

To begin with we recall the canonical form of 
a general $\cN=2$ ungauged supergravity theory.
The dynamics of the gravity-vector sector is entirely specified 
in terms of a cubic potential
\beq \label{canonical-cN}
\mathcal{N} = \tfrac{1}{3!} k_{I J K} M^{I} M^{J} M^{K}\, ,
\eeq
where $M^I, I = 1,\dots, n+6-2 n_T $ are very special real coordinates and $k_{IJK}$ is a 
symmetric tensor. The $M^I$ naturally combine with the vectors $A^I$ of the 
theory. However, since the vector in the gravity multiplet is not accompanied 
by a scalar degree of freedom, the $M^I$ have to satisfy one constraint.  
In fact, the $\cN=2$ scalar field space is identified with the hypersurface 
\begin{equation} \label{very-special-geometry-constraint}
\mathcal{N} \overset{!}{=} 1 \;.
\end{equation}
The gauge coupling function and the metric are obtained as
\begin{equation} \label{gauge-coupling-function}
 G_{I J} = \left[ - \frac{1}{2}\partial_{M^I} \partial_{M^J} \log \mathcal{N} \right]_{\mathcal{N}=1} \ .
\end{equation}
The bosonic two-derivative Lagrangian is then given by
\begin{align} \label{5d_action_canonical}
\cL_{\rm can} =  &-\tfrac{1}{2} R 
 - \tfrac{1}{2} G_{ I J} \partial_\mu M^{I} \partial^\mu M^{J} -\tfrac{1}{4} G_{I J} F^{I}_{\mu \nu} F^{\mu \nu \, J}
\nn \\
&
 + \tfrac{1}{48} \epsilon^{\mu \nu \rho \sigma\lambda} k_{I J K} A^{I}_\mu  F^{J}_{\nu \rho} F^{K}_{\sigma \lambda} 
 - H_{\Lambda\Sigma}^{uv}\, \partial_\mu h^\Lambda_u \, \partial^\mu h^\Sigma_v \;.
\end{align}
Here we included the kinetic term for the hypermultiplet scalars $h^\Lambda_u$ with metric $H_{\Lambda\Sigma}^{uv}$.

The canonical Lagrangian \eqref{5d_action_canonical} has to be compared 
with the truncated $\cN=4$ theory. In our setup we found the vectors $A^I = (A^0,A^{\bar m},A^{\bar a})$, which 
sets the index range for $I$.
The \emph{massless} scalars in the effective theory (except for $\Sigma$) are most conveniently described by 
$SO(5,n)$-rotated elements of the coset space
\begin{align}\label{normalized_Vs}
 \hat\cV := \langle \cV \rangle^{-1} \cV = \text{exp} \big( \phi^{ma} [t_{ma}]\big) \, .
\end{align}
This is in contrast to the analysis of  the \emph{massive} scalar spectrum, for which 
it is efficient to consider the fluctuations $\phi^{ma}$ as it was done in the last section.
Restricting to the $\cN =2$ vector multiplets and
truncating the massive modes
$\phi^{\bar m \hat a}$ and $\phi^{\hat m \bar a}$, the only
remaining elements of the coset space are 
\begin{align} \label{Vhats}
 \tensor{\hat\cV}{_{\bar m}^{\bar m}}\, , \quad \tensor{\hat\cV}{_{\bar m}^{\bar a}}\, , \quad 
 \tensor{\hat\cV}{_{\bar a}^{\bar m}}\, , \quad \tensor{\hat\cV}{_{\bar a}^{\bar b}}\, .
\end{align}
In fact, it turns out that all couplings involving the elements \eqref{Vhats} can be 
expressed as functions of $\tensor{\hat\cV}{_{\bar m}^{\bar a}}$ alone. 
In order to do that one uses the relations
\begin{align}
\tensor{\hat\cV}{_{\bar m}^{\bar m}} &= 
\sqrt{1+ \tensor{\hat\cV}{_{\bar m}^{\bar a}}\tensor{\hat\cV}{_{\bar m}_\, _{\bar a}}}\, , &
 \tensor{\hat\cV}{_{\bar a}^{\bar m}}&=\tensor{\hat\cV}{_{\bar m}^{\bar a}}\, , \\
 \tensor{\hat\cV}{_{\bar a}^{\bar c}}\tensor{\hat\cV}{_{\bar b}_\, _{\bar c}} &= \delta_{\bar a \bar b}
 + \tensor{\hat\cV}{_{\bar a}^{\bar m}} \tensor{\hat\cV}{_{\bar b}^{\bar m}}\, ,  &
 \tensor{\hat\cV}{_{\bar a}^{\bar b}}\tensor{\hat\cV}{_{\bar m}_\, _{\bar b}} &=
 \tensor{\hat\cV}{_{\bar m}^{\bar m}} \tensor{\hat\cV}{_{\bar a}^{\bar m}} \, .  \nn
\end{align}
The element  $\tensor{\hat\cV}{_{\bar m}^{\bar a}}$ itself can be expanded as 
\beq
   \tensor{\hat\cV}{_{\bar m}^{\bar a}}= \text{exp} \big( \phi^{\bar m \bar a} [t_{\bar m \bar a}]\big){_{\bar m}^{\ \, \bar a}} \, ,
\eeq 
after truncating all massive modes.
This implies in particular that $\tensor{\hat\cV}{_{\bar m}^{\bar a}}$ has no dependence on $\phi^{\hat m \hat a}$. 
Therefore, the effective action of the scalars in the $\cN=2$ vector multiplets decouples from the potentially massless
scalars in the hypermultiplets as expected from $\cN=2$ supersymmetry.

The reduced action then takes the simple form
\begin{align}\label{eff_action}
 e^{-1} \cL_{\textrm{class}}=&
 - \frac{1}{2} R - H_{\Lambda\Sigma}^{uv}\, \partial_\mu h^\Lambda_u \, \partial^\mu h^\Sigma_v  
 - \frac{3}{2} \Sigma^{-2}\, \partial_\mu \Sigma \, \partial^\mu \Sigma \nn \\
& -\frac{1}{2} \Big ( \delta_{\bar a \bar b} - 
 \frac{1}{1 + \tensor{\hat\cV}{_{\bar m}^{\bar c}}\, \tensor{\hat\cV}{_{\bar m}_\, _{\bar c}}}
 \tensor{\hat\cV}{_{\bar m}_\, _{\bar a}}\tensor{\hat\cV}{_{\bar m}_\, _{\bar b}}\Big )\,
 \partial_\mu \tensor{\hat\cV}{_{\bar m}^{\bar a}} \, \partial^\mu \tensor{\hat\cV}{_{\bar m}^{\bar b}}
  \nn \\
  &- \frac{1}{4} \Sigma^{-4}\, F^{0}_{\mu\nu} F^{\mu\nu\,0} + \Sigma^2 
   \sqrt{1+\tensor{\hat\cV}{_{\bar m}^{\bar b}}\tensor{\hat\cV}{_{\bar m}_\, _{\bar b}}}\,\tensor{\hat\cV}{_{\bar m}_\, _{\bar a}}\,
   F^{\bar m}_{\mu\nu} F^{\mu\nu\,\bar a} \nn \\ 
 &- \frac{1}{4} \Sigma^2 \Big ( 3 + 2\, \tensor{\hat\cV}{_{\bar m}^{\bar a}}\tensor{\hat\cV}{_{\bar m}_\, _{\bar a}}  \Big )\,
 F^{\bar m}_{\mu\nu} F^{\mu\nu\,\bar m}
 - \frac{1}{4} \Sigma^2 \Big ( \delta_{\bar a \bar b } + 
 2\, \tensor{\hat\cV}{_{\bar m}_\, _{\bar a}} \tensor{\hat\cV}{_{\bar m}_\, _{\bar b}} \Big )\,
 F^{\bar a}_{\mu\nu} F^{\mu\nu\,\bar b} \nn \\
 & +\frac{1}{4\sqrt 2} \epsilon^{\mu\nu\rho\sigma\tau} A_\mu^0 F^{\bar m}_{\nu\rho} F^{\bar m}_{\sigma\tau}  
 - \frac{1}{4\sqrt 2} \epsilon^{\mu\nu\rho\sigma\tau} A_\mu^0 F^{\bar a}_{\nu\rho} F^{\bar a}_{\sigma\tau} \, ,
\end{align}
where $H_{\Lambda\Sigma}^{uv}$ is the metric of the quaternionic manifold parametrized by the scalars in the massless hypermultiplets which
we however do not discuss any further in the work of this thesis.
Therefore by comparison of \eqref{eff_action} with \eqref{5d_action_canonical} we find the identifications
\begin{align} \label{def-Ms}
 M^0 = \frac{1}{\sqrt 2} \Sigma^2\, , \qquad M^{\bar m} = \Sigma^{-1} \tensor{\hat\cV}{_{\bar m}^{\bar m}}\, , \qquad
 M^{\bar a} = \Sigma^{-1} \tensor{\hat\cV}{_{\bar m}^{\bar a}} \, ,
\end{align}
and the real prepotential
\begin{align} \label{Nclass}
 \cN = \frac{1}{2} k_{0 \bar m \bar m} M^0 M^{\bar m} M^{\bar m}
 + \frac{1}{2} k_{0 \bar a \bar a} M^0 M^{\bar a} M^{\bar a}
 = \sqrt 2 M^0 M^{\bar m} M^{\bar m} - \sqrt 2 M^0 M^{\bar a} M^{\bar a} \, .
\end{align}
This result specifies the constant tensors $k_{IJK}$ at the classical level. 
It is interesting to realize that the constraint $\cN \overset{!}{=}1$ translates with the identifications \eqref{def-Ms}
into the condition \eqref{eta_viaV} for the elements of the coset space.
We conclude that the very special real manifold is the 
coset space 
\beq
    SO(1,1) \times \frac{SO(1,n+4-2n_T)}{SO(n+4-2n_T)}\ ,
\eeq
which is the subspace of \eqref{coset_def} spanned by the 
massless scalars in the vector multiplets. 

\subsection{One-loop Effects and Chern-Simons Terms} \label{oneloopeffects}

Now we determine the one-loop corrections to the 
gravity-vector sector of the $\cN=2$ theory specified in \autoref{Classicaltruncation}.
We focus on this sector since the corrections due to integrating 
out massive fields are independent of the supersymmetry breaking 
scale and the masses of the fields running in the loop. Let us stress 
that due to the preserved $\cN=2$ supersymmetry and the fact that the Chern-Simons 
terms can only receive constant corrections the integrating
out process can only perturbatively correct the gravity-vector sector at 
the one-loop level. 

To obtain the one-loop corrected $\cN$ an analysis 
of the Chern-Simons terms is sufficient. The expressions for the latter were stated in \autoref{sec:5dcherns},
and we can simply apply these results to our setup. 
The one-loop corrections arise from integrating out massive 
fields that are charged under some gauge fields $A^I$. 
Since all massive fields are only charged under $A^0$, the classical terms in 
\eqref{Nclass} are unmodified. The fully quantum corrected result for these terms therefore reads (including combinatorial factors)
\begin{align}
k_{0\bar m \bar m} =
 -k_{0\bar a \bar a}= 2 \sqrt 2 \, .
\end{align}
We find that the massive states summarized in \autoref{field_decomposition} induce the 
one-loop couplings
\begin{align} \label{one-loop_k000}
 k_{000} = \frac{1}{2}\Big [ (-1- n + 2n_T) \gamma^3 -2 \sum_{\check a} \vert  \gamma_{\check a} \vert^3
 + \sum_{\check a}  \vert  \gamma -  \gamma_{\check a} \vert^3 + \sum_{\check a}  \vert  \gamma  + \gamma_{\check a} \vert^3 \Big ] \, .
\end{align}
Furthermore, we find also the gravitational one-loop Chern-Simons coupling
\begin{align}\label{one-loop_k0}
 k_0 = -  \Big [ (-1 - n + 2n_T ) \gamma + 10 \sum_{\check a}  \vert \gamma_{\check a} \vert 
 + \sum_{\check a}  \vert \gamma - \gamma_{\check a} \vert + \sum_{\check a}  \vert \gamma + \gamma_{\check a} \vert \Big ] \, , 
\end{align}
which we included for completeness although we did not discuss these higher-curvature terms at the classical level.

The existence of new Chern-Simons couplings implies that the effective theory still sees remnants of the underlying $\cN=4$ theory at 
arbitrarily low energy scales. In fact, the mass of the fields listed in \autoref{vacuum_multiplets} 
can be made arbitrarily large by choosing the VEV of the modulus $\Sigma$. The 
 constants $\gamma$ and $\gamma_{\check a}$ appearing in \eqref{one-loop_k000}  and \eqref{one-loop_k0} are the imaginary parts 
 of the eigenvalues of $\xi^{MN}$ and therefore independent of the VEVs of the fields. 

An interesting case is the one with
\begin{align}
 n &= 3 \, ,& \rk (\xi^{MN}) &=4 \, . 
\end{align}
Remarkably the quantum corrections $k_{000}$ and $k_0$ both vanish for this special choice
since $n_T = \frac{1}{2}\rk (\xi^{MN})$ and the range of the index $\check a$ is zero. It would be worthwhile to understand
the underlying principle which enforces the vanishing of both quantum corrections at the same time.
This is in the very same spirit to our upcoming discussion of effective actions from consistent truncations in \autoref{sec:eff_cons}.

\chapter{M-Theory on \texorpdfstring{$SU(2)$}{SU(2)}-Structure Manifolds}\label{ch:m_su2}

In this chapter we introduce one example for a gauged $\cN = 4$ 
supergravity theory in five dimensions by reducing eleven-dimensional 
supergravity on six-dimensional manifolds $\cM_6$ with $SU(2)$-structure.
In \autoref{basicsofSU(2)} we first recall some basic properties 
of $SU(2)$-structure manifolds. The introduced definitions are then used in 
\autoref{sec:reductionansatz} to formulate the reduction ansatz
specifying a consistent truncation of the full compactification on $\cM_6$ to 
five dimensions. The five-dimensional action is derived in \autoref{sec:dimredaction}
and brought into standard $\cN=4$ supergravity form in \autoref{sec:standardN=4form}. 
This allows us to determine the embedding tensors induced by the 
$SU(2)$-structure and a non-trivial flux background.

\section{Some Basics on \texorpdfstring{$SU(2)$}{SU(2)}-Structure Manifolds}\label{basicsofSU(2)}

Let us begin by recalling some basics on six-dimensional $SU(2)$-structure manifolds $\cM_6$.
See \textit{e.g.~}\cite{Hitchin:2000sk, hitchin2001stable, chiossi2002intrinsic, Gauntlett:2002sc, Gauntlett:2003cy} for properties of general \(G\)-structure manifolds and \cite{Dall'Agata:2004dk, Behrndt:2004mj, Bovy:2005qq, ReidEdwards:2008rd, Lust:2009zb,
Triendl:2009ap, Louis:2009dq, Danckaert:2011ju, KashaniPoor:2013en} for \(SU(2)\)-structure manifolds.
If the structure group of a manifold $\cM_6$ can be reduced to \(SU(2)\), it admits two globally defined, 
nowhere vanishing spinors \(\eta^1\), \(\eta^2\). This can be seen from the fact that two singlets appear in the
the decomposition of the spinor representation ${\bf 4} $ of $Spin(6) \cong SU(4)$ into \(SU(2)\) representations \({\bf 4} \rightarrow {\bf 1} \oplus {\bf 1} \oplus {\bf 2}\).
The existence of these two spinors gives rise to four supersymmetry generators \(\xi^{1,2}_i\) (\(i = 1,2\)) in five dimensions
since we can expand 
the eleven-dimensional supersymmetry generator \(\epsilon\) as 
\begin{equation}
\epsilon = \xi^1_i \otimes \eta^i + \xi^2_i \otimes \eta^{c\, i} \,,
\end{equation}
where $\eta^{c\, i}$ is the charge conjugate spinor to $\eta^i$ and the five-dimensional spinors \(\xi^{1,2}_i\) are symplectic Majorana, see
\autoref{app:spacetime}.
This implies that 
an appropriately chosen reduction
admits $\cN=4$ supersymmetry. 

By contraction with appropriate products of $SO(6)$ $\gamma$-matrices
the globally defined spinors $ \eta^i$ allow to define  three real two-forms \(J^a\), \(a = 1, 2, 3\) forming an \(SU(2)\) triplet,
and a complex one-form $K$. These fulfill the conditions
\begin{equation}\begin{gathered}\label{eq:su2forms}
J^a \wedge J^b = \delta^{ab} \mathrm{vol}_4 \,, \\
K_m K^m = 0 \,,\quad
\bar K_m K^m = 2\ ,\quad
K^m J^a_{mn} = 0 \,,
\end{gathered}\end{equation}
where \(m,n = 1, \dots, 6\) label the local coordinates on $\cM_6$, and $\mathrm{vol}_4$ is a nowhere vanishing four-form on $\cM_6$.
All contractions are performed with the $SU(2)$-structure metric on $\cM_6$. 

These forms define an almost product structure
\begin{equation}
{P_m}^n = K_m \bar K^n + \bar K_m K^n - {\delta_m}^n \, .
\end{equation}
Indeed, it is straightforward to verify that
\begin{align}
 {P_m}^n {P_n}^q = \delta_m^q \, , 
\end{align}
which means that $P$ is a projector, and the 
manifold's tangent bundle can be split into the $+1$ and $-1$ eigenspaces of \(P\)
\begin{equation} \label{tangetsplit}
T\cM_6 = T_2 \cM_6 \oplus T_4 \cM_6 \,,
\end{equation}
where the part $ T_2 \cM_6$ is spanned by $K^1 = \R\, K$ and $K^2 = \I\, K$ each carrying eigenvalue $+1$.
Note that we implicitly identify the tangent and cotangent bundle.

\section{The Reduction Ansatz}\label{sec:reductionansatz}

An appropriate ansatz for the dimensional reduction on manifolds with structure group \(SU(2)\) 
has been worked out in \cite{Danckaert:2011ju,KashaniPoor:2013en}.
The full spectrum of the compactified theory consists of infinitely many modes from which the choice of a particular ansatz keeps only a finite subset.
Such a truncation is called consistent if any of the modes that we keep cannot excite one of the modes we exclude.
This means that there are no source terms for the discarded fields in the reduced action.
In this case any solution of the truncated theory can be uplifted to a solution of the full eleven-dimensional equations of motion.
As explained in \cite{KashaniPoor:2013en}, this can be achieved by choosing the reduction ansatz to be a set of forms on \(\cM_6\) that it is closed under the action of the wedge product \(\wedge\), exterior differentiation \(\dd\) and the Hodge star \(\ast\).

In \cite{Triendl:2009ap} it has been demonstrated how to decompose the field content of type IIA supergravity into representations with respect to the \(SU(2)\) structure group of \(\cM_6\) and arrange it into four-dimensional \(\cN = 4\) multiplets.
The same analysis can be performed for the case of eleven-dimensional supergravity reduced to \(\cN = 4\) supergravity in five dimensions.
The modes transforming as singlets under \(SU(2)\) constitute the five-dimensional gravity multiplet and a pair of vector multiplets,
and every \(SU(2)\)-triplet corresponds to one triplet of vector multiplets.
On the other hand the components of the fields that are doublets under \(SU(2)\) form gravitino multiplets in the \(\cN = 4\) theory.
Since it is not known how to consistently couple gravitino multiplets to gauged \(\cN = 4\) supergravity, these multiplets will be neglected.
This is equivalent to excluding all \(SU(2)\) doublets from the reduction ansatz.
We will further comment on this point in \autoref{sec:enriq}.

Following up these considerations the reduction ansatz now consists of a basis of real one-forms \(v^i\) (\(i = 1,2\)) acting on \(T_2\cM_6\),
and real two-forms \(\omega^I\) (\(I = 1, \dots, \tilde n\)) acting on \(T_4\cM_6\). Forms of odd rank on \(T_4\cM_6\) correspond to doublets of \(SU(2)\) and are thus not included in the ansatz.
These forms are normalized via
\begin{equation}\label{eq:su2ansatzintegral}
\int_{\cM_6} v^1 \wedge v^2 \wedge \omega^I \wedge \omega^J = -\eta^{IJ} \,,
\end{equation}
where \(\eta^{IJ}\) is an \(SO(3,\tilde n-3)\) metric that will be used to raise and lower indices.
For convenience we can also introduce
\begin{align}
 \mathrm{vol}_2^{(0)} &= v^1 \wedge v^2 \, , & -\eta^{IJ} \mathrm{vol}_4^{(0)} &= \omega^I \wedge \omega^J \, ,
\end{align}
which take the role of normalized volume forms on \(T_2\cM_6\) and \(T_4\cM_6\), respectively.

The ansatz has to be chosen in such a way that it is consistent with 
exterior differentiation. Therefore we demand that the differentials of \(v^i\) and \(\omega^I\) obey 
\begin{equation}\begin{aligned}\label{eq:torsion}
\dd v^i &= t^i \,v^1 \wedge v^2 + t^i_I \,\omega^I\,, \\
\dd \omega^I &= T^I_{iJ} \,v^i \wedge \omega^J\,,
\end{aligned}\end{equation}
where the coefficients \(t^i\), \(t^i_I\) and \(T^I_{iJ}\) are related to the torsion classes of $\cM_6$
and have to fulfill the consistency conditions \cite{KashaniPoor:2013en}
\begin{equation}\begin{aligned}\label{eq:torsionconditions}
t^i t^k_I \epsilon_{kj} + t^i_J T^J_{jI} = 0 \,&,\quad
T^I_{iJ} \eta^{JK} t^i_K = 0 \,,\\
T^I_{iJ} t^i - T^I_{iK} \epsilon_{ij} T^K_{jJ} = 0 \,&,\quad
t^i \eta^{IJ} - \epsilon_{ij} T^I_{jK} \eta^{KJ} - \epsilon_{ij} T^J_{jK} \eta^{KI} = 0 \,.
\end{aligned}\end{equation}
Using this basis of forms one now has to expand 
all fields of eleven-dimensional supergravity. In order to 
discuss the reduction of the eleven-dimensional action 
we first expand the $J^a$ and $K$ introduced in \eqref{eq:su2forms}
as
\begin{equation}\label{eq:su2formsexpansion}
J^a = e^{\rho_4/2} \zeta^a_I \omega^I\ ,\qquad 
K = e^{\rho_2/2} (\I\,\tau)^{-1/2} (v^1 + \tau v^2) \,,
\end{equation}
where now the real $\rho_4$, $\rho_2$, $\zeta^a_I$, and complex $\tau$ 
are promoted to five-dimensional space-time scalars.
Together with \eqref{eq:su2forms} we find \(\zeta^a_I \eta^{IJ} \zeta^b_J = -\delta^{ab}\) as well as \(\mathrm{vol}_4 = e^{\rho_4} \mathrm{vol}_4^{(0)}\) and \(K^1 \wedge K^2 = e^{\rho_2} \mathrm{vol}_2^{(0)}\).

The action of the Hodge star on the ansatz is given by
\begin{equation}\begin{aligned}\label{eq:su2ansatzhodge}
\ast\, v^i &= e^{\rho_4}\, \epsilon_{ij} v^j \wedge \mathrm{vol}_4^{(0)} \,, \\
\ast\, \mathrm{vol}_2^{(0)} &= e^{\rho_4}\, \mathrm{vol}_4^{(0)} \,, \\
\ast\, \omega^I &= - e^{\rho_2} {H^I}_J \omega^J \wedge \mathrm{vol}_2^{(0)} \,, \\
\ast \left(v^i \wedge \omega^I\right) &= - \epsilon_{ij} {H^I}_J v^j \wedge \omega^J \,.
\end{aligned}\end{equation}
From the requirement that \(\ast J^a = J^a \wedge K^1 \wedge K^2\) the matrix \({H^I}_J\) can be determined to be \(H_{IJ} = 2\zeta^a_I\zeta^a_J + \eta_{IJ}\).
See \autoref{app:cosetrepr} for a further discussion of its properties.

After this preliminary discussion we are now in a position to give the ansatz for the eleven-dimensional metric. 
More precisely, reflecting the split of the tangent space \eqref{tangetsplit} the metric takes the 
form 
\begin{equation}\label{e:su(2)_metric}
\dd s^2_{11} = g_{\mu\nu} \dd x^\mu \dd x^\nu + e^{\rho_2}g_{ij} (v^i + G^i)(v^j + G^j) + e^{\rho_4} g_{st} \dd x^s \dd x^t \,,
\end{equation}
with \(s,t=1,\dots,4\).
 The \(G^i\) are space-time gauge fields parameterizing the variation of $T_2 \cM_6$.
The metric \(g_{ij}\) can be expressed in terms of \(\tau\)
\begin{equation}
g = \frac{1}{\I\,\tau} \begin{pmatrix}1 & \R\,\tau \\ \R\,\tau & |\tau|^2 \end{pmatrix} 
\end{equation}
such that $e^{\rho_2} g_{ij} v^i v^j = K \bar K$.
Notice that we excluded possible off-diagonal terms of the form \(g_{\mu s}\) and \(g_{is}\) from the ansatz for the metric since they would precisely correspond to \(SU(2)\) doublets.
These terms would give rise to two doublets of additional space-time vectors and four doublets of space-time scalars.

In the following it will be useful to introduce the 
gauge invariant combination
\begin{equation}
\tiv^i = v^i + G^i \, ,
\end{equation}
whose derivative can be calculated using \eqref{eq:torsion}
\begin{equation}
\dd \tiv^i = \dd \!\left(v^i + G^i\right) = \D G^i + t^i \tiv^1 \wedge \tiv^2 - t^i \epsilon_{jk} \tiv^j \wedge G^k + t^i_I \omega^I \,.
\end{equation}
The definition of the covariant derivative \(\D G^i\) can be found in \eqref{eq:covderivvectors}.

Let us next turn to the ansatz for the three form field \(C_3\). 
Using the basis $\tilde v^i,\omega^I$ introduced above, we expand
\begin{equation}\label{eq:c3expansion}
C_3 = C + C_i \wedge \tiv^i + C_I \wedge \omega^I + C_{12} \wedge \tiv^1 \wedge \tiv^2 + c_{iI} \, \tiv^i \wedge \omega^I\,.
\end{equation}
If we had included \(SU(2)\) doublets in the reduction ansatz, we also would have had to expand \(C_3\) in terms of odd forms on \(T_4\cM_6\),
which would give rise to additional fields in five dimensions.\footnote{To make the ansatz 
closed under wedge product it might be necessary in this case to include also additional two-forms on \(T_4\cM_6\) and hence additional \(SU(2)\) triplets.}
For each $SU(2)$ doublet these would be one doublet of two-forms and two doublets of vectors and scalars.
Together with the contributions from the metric we see that for every excluded \(SU(2)\) doublet this resembles precisely a doublet of \(\cN = 4\) gravitino multiplets.

Furthermore, we also consider a possible internal four-form flux for which the most general ansatz is given by\footnote{Notice that
here $n$ counts flux quanta. It is not related to the number of $\cN=4$ vector multiplets, also denoted by $n$ in the previous chapters.}
\begin{equation}
F_4^\mathrm{flux} = n\, \mathrm{vol}_4^{(0)} + n_I\, v^1 \wedge v^2 \wedge \omega^I \,.
\end{equation}
Notice that this is written only in terms of \(v^i\) and not in terms of the gauge invariant quantities \(\tiv^i\) because this would
introduce an unwanted space-time dependency. Moreover \(n\) and \(n_I\) are not completely independent
since it follows from \(\dd F_4^\mathrm{flux} = 0\) that
\begin{equation}\label{eq:fluxcondition}
n \,t^i - n^I t^i_I = 0 \,.
\end{equation}
We finally have to expand the field strength \(F_4 = F_4^\mathrm{flux} + \dd C_3\) 
\begin{equation}\begin{aligned}\label{eq:f4expansion}
F_4 &=  F + F_i \wedge \tiv^i + F_I \wedge \omega^I + F_{12} \wedge \tiv^1 \wedge \tiv^2 + F_{iI} \, \tiv^i \wedge \omega^I \\
&\quad + f_I \, \tiv^1 \wedge \tiv^2 \wedge \omega^I + f \vol_4^{(0)} \,,
\end{aligned}\end{equation}
and obtain the expansion coefficients after evaluating the exterior derivative of \(C_3\) using \eqref{eq:c3expansion}
\begin{equation}\begin{aligned}\label{eq:fieldstrengths}
F &= \dd C + C_i \wedge \D G^i \,, \\
F_i &= \D C_i + \epsilon_{ij} C_{12} \wedge \D G^j \,, \\
F_I &= \D C_I + c_{iI} \D G^i \,, \\
F_{12} &= \D C_{12} \,, \qquad F_{iI} = \D c_{iI} \,, \\
f_I &= n_I + t^i c_{iI} + \epsilon_{ij} T^J_{iI} c_{jJ} \,, \\
f &= n-c_{iI}t^i_J \eta^{IJ} \,. 
\end{aligned}\end{equation}
The four-form flux  and the fact that \(\omega^I\) and \(\tiv^i\) are in general non-closed forms 
induce different non-trivial gaugings. These are encoded by the various appearing covariant derivatives which 
are listed in the next section.

\section{Dimensional Reduction of the Action} \label{sec:dimredaction}

Starting from the bosonic action of eleven-dimensional supergravity
\begin{equation}\label{eq:mtheoryaction}
S = \int_{11} \tfrac{1}{2} (\ast 1) R - \tfrac{1}{4} F_4 \wedge \ast F_4 - \tfrac{1}{12}C_3 \wedge F_4 \wedge F_4\,,
\end{equation}
we will compute a five-dimensional action by compactifying it on \(\cM_6\). We can compare the result with the general description of \(\cN = 4\)
gauged supergravity given in \autoref{N=4Gen} and determine the embedding tensors in terms of geometrical properties of \(\cM_6\).

To compute the reduced five-dimensional action we insert the expansions \eqref{eq:c3expansion} and \eqref{eq:f4expansion} into the eleven-dimensional action \eqref{eq:mtheoryaction} and integrate over the internal manifold using \eqref{eq:su2ansatzintegral}.
The reduction of the Einstein-Hilbert term has been carried out in \cite{KashaniPoor:2013en} and can be adopted without further modifications.
After performing an appropriate Weyl rescaling \(g_{\mu\nu} \rightarrow e^{-\frac{2}{3}(\rho_2 + \rho_4)} g_{\mu\nu}\) to bring the action into the Einstein frame the final result reads 
\begin{equation}\begin{aligned}\label{eq:su2action}
S_{SU(2)} &= \int_5 \biggl\{\tfrac{1}{2} (\ast 1) R_5
- e^{\frac{5}{3}\rho_2 + \frac{2}{3}\rho_4}g_{ij} \D G^i \wedge \ast \D G^j 
-\tfrac{1}{2} (\eta^{IJ} + \zeta^{bI}\zeta^{bJ})\D\zeta^a_I \wedge \ast \D\zeta^a_J\\
& -\tfrac{1}{4}(\I\,\tau)^{-2}\,\D\tau\wedge\ast\D\bar\tau
- \tfrac{5}{12} \D\rho_2 \wedge \ast \D\rho_2
- \tfrac{1}{3}\D\rho_2 \wedge \ast \D\rho_4
-\tfrac{7}{24}\D\rho_4\wedge\ast\D\rho_4 \\
& -\tfrac{1}{4} e^{2(\rho_2+\rho_4)} \left(\dd C + C_i \wedge \D G^i\right) \wedge \ast \left(\dd C + C_j \wedge \D G^j \right) \\
& -\tfrac{1}{4} e^{\frac{1}{3}\rho_2+\frac{4}{3}\rho_4} \left(g^{-1}\right)^{ij} \left(\D C_i + \epsilon_{ik} C_{12} \wedge \D G^k\right) \wedge \ast \left(\D C_j + \epsilon_{jl} C_{12} \wedge \D G^l\right) \\
& -\tfrac{1}{4} e^{\frac{2}{3}\rho_2-\frac{1}{3}\rho_4} H^{IJ} \left(\D C_I + c_{iI} \D G^i\right) \wedge \ast \left(\D C_J + c_{jJ} \D G^j\right) \\
& -\tfrac{1}{4} e^{-\frac{4}{3}\rho_2+\frac{2}{3}\rho_4} \D C_{12} \wedge \ast \D C_{12} - \tfrac{1}{4} e^{-\rho_2-\rho_4} H^{IJ} \left(g^{-1}\right)^{ij} \D c_{iI} \wedge \ast \D c_{jJ} \\
& + \left(\tfrac{1}{4} \dd C + \tfrac{1}{6} C_k \wedge \D G^k\right)\wedge c_{iI}\left(\epsilon^{ij}T^K_{jJ} C_K  + C_{12}t^i_J + \D c_{jJ}\epsilon^{ij}\right) \eta^{IJ} \\
& - \tfrac{1}{6}C_i \wedge \epsilon^{ij}\Bigl( \left(\D C_j + \epsilon_{jk} C_{12} \wedge \D G^k\right) c_{kI}t^k_J + \left(\D C_I + c_{kI} \D G^k\right) \wedge \D c_{jJ}\Bigr)\eta^{IJ} \\
& + \tfrac{1}{6}C_I \wedge \Bigl(\left(\D C_i + \epsilon_{ik} C_{12} \wedge \D G^k\right) \wedge \D c_{jJ}\epsilon^{ij} + \left(\D C_J + c_{lJ} \D G^l\right) \wedge \D C_{12}\Bigr)\eta^{IJ} \\
&+ \tfrac{1}{12} C_{12} \wedge \left(\D C_I + c_{iI} \D G^i\right) \wedge \left(\D C_J + c_{jJ} \D G^j\right)\eta^{IJ} \\
& - \tfrac{1}{6}c_{iI}  \left(\D C_j + \epsilon_{jk} C_{12} \wedge \D G^k\right) \wedge \left(\D C_J + c_{lJ} \D G^l\right) \epsilon^{ij} \eta^{IJ} \\
& - \tfrac{1}{4} n\,\epsilon^{ij} C_i \wedge \left(\D C_i + \epsilon_{ik} C_{12} \wedge \D G^k\right)
 - \left(\tfrac{1}{2} \dd C + \tfrac{1}{4} C_i \wedge \D G^i\right) \wedge \left(n \, C_{12} - n^I C_I\right) \\
& + (\ast 1) \,V \biggl\}\,.
\end{aligned}\end{equation}
The potential term \(V\) is given by
\begin{equation}\begin{split}
V = -\tfrac{5}{8} e^{-\frac{5}{3}\rho_2 - \frac{2}{3}\rho_4} g_{ij} t^i t^j
+ 2 e^{\frac{1}{3}\rho_2 - \frac{5}{3}\rho_4}  g_{ij} t^i_I t^j_J \eta^{IJ} \\
- \tfrac{1}{2} e^{-\frac{5}{3}\rho_2 - \frac{2}{3}\rho_4} (\eta^{IJ} + \zeta^{bI}\zeta^{bJ})\zeta^a_K\zeta^a_L g^{ij} \tilde{T}^K_{iI} \tilde{T}^L_{jJ} \\
+ \tfrac{1}{4}e^{-\frac{8}{3}\rho_2-\frac{5}{3}\rho_4} H^{IJ} f_I f_J
+ \tfrac{1}{4}e^{-\frac{2}{3}\rho_2-\frac{8}{3}\rho_4} f^2 \,.
\end{split}\end{equation}
As mentioned above, we have defined several covariant derivatives.
For the scalars they are given by
\begin{equation}\begin{aligned}\label{eq:covderivscalars}
\D\rho_2 &= \dd\rho_2 - \epsilon_{ij} G^i t^j \,, \\
\D\rho_2 &= \dd\rho_4 + \epsilon_{ij} G^i t^j \,, \\
\D\tau &= \dd\tau - \big((1,\tau)\cdot G \big) \big((1, \tau) \cdot t \big) \,, \\
\D\zeta^a_I &= \dd\zeta^a_I - G^i\tilde{T}^J_{iI}\zeta^a_J \,, \\
\D c_{iI} &= \dd c_{iI} + \epsilon_{ij} t^j_I C_{12} - T^J_{iI} C_J + \epsilon_{ij} G^j t^k c_{kI} - G^j T^J_{jI} c_{iJ} + n_I \epsilon_{ij} G^j \,,
\end{aligned}\end{equation}
whereas those of the vectors read
\begin{equation}\begin{aligned}\label{eq:covderivvectors}
\D G^i &= \dd G^i - t^i G^1 \wedge G^2\,, \\
\D C_I &= \dd C_I + t^i_I C_i + T^J_{iI} C_J \wedge G^i - n_I G^1 \wedge G^2 \,, \\
\D C_{12} &= \dd C_{12} + t^i C_i  - \epsilon_{ij}C_{12} \wedge t^i G^j \,. \\
\end{aligned}\end{equation}
There is also a pair of two-forms \(C_i\) with
\begin{equation}
\D C_i = \dd C_i + \epsilon_{ij} G^j \wedge t^k C_k \,.
\end{equation}

In the next section we compare \eqref{eq:su2action} with the general form of gauged \(\cN = 4\) supergravity.
For this purpose it is necessary to dualize the three-form field \(C\) into a scalar \(\gamma\).\footnote{We stress that the scalar field
$\gamma$ should not be confused with the constant eigenvalues $\gamma$ defined in \eqref{e:orth_xi}.}
Let us therefore collect all terms from the action containing it,
\begin{equation}\begin{aligned}\label{eq:caction}
S_C = &\int -\frac{1}{4}e^{2(\rho_2+\rho_4)} F \wedge \ast F + \frac{1}{2} F \wedge L 
\end{aligned}\end{equation}
with 
\begin{equation}
L = \tfrac{1}{2}c_{iI}\left(\epsilon^{ij}T^K_{jJ} C_K  + C_{12}t^i_J + \D c_{jJ}\epsilon^{ij}\right)\eta^{IJ} - n\, C_{12} + n^I C_I \,.
\end{equation}
The field strength \(F = \dd C + C_i \wedge \D G^i\) fulfills the Bianchi identity
\begin{equation}
\dd F = \D C_i \wedge \D G^i \,,
\end{equation}
which we will impose by introducing a Lagrange multiplier \(\gamma\). Accordingly we add the following term to the action
\begin{equation}
\delta S = -\frac{1}{2} \int \gamma \left(\dd F - \D C_i \wedge \D G^i\right) \,.
\end{equation}
We can now use the equation of motion for \(F\)
\begin{equation}
- e^{2(\rho_2+\rho_4)} \ast F + L + \dd\gamma = 0
\end{equation}
in order to eliminate it from \eqref{eq:caction} and obtain
\begin{equation}\begin{aligned}
S_{\gamma} = &- \frac{1}{4} \int e^{-2(\rho_2+\rho_4)} (\D\gamma + \tfrac{1}{2}c_{iI} \D c_{jJ}\epsilon^{ij} \eta^{IJ}) \wedge \ast (\D\gamma + \tfrac{1}{2}c_{iI} \D c_{jJ}\epsilon^{ij} \eta^{IJ}) \\
&+ \frac{1}{2} \int \gamma \D C_i \wedge \D G^i \,,
\end{aligned}\end{equation}
where the covariant derivative of \(\gamma\) is defined as
\begin{equation}\label{eq:covderivgamma}
\D\gamma = \dd\gamma + \tfrac{1}{2}c_{iI}(\epsilon_{ij}T^K_{jJ} C_K + t^i_J C_{12})\eta^{IJ} - n\, C_{12} + n^I C_I \,.
\end{equation}

Moreover in the general \(\cN = 4\) theory there are no tensors with second order kinetic term. Therefore it is necessary to trade the two-form \(C_i\) for its dual vector \(\tilde C^{\ib}\).
But since \(C_i\) appears additionally in the covariant derivatives of the vectors \(C_I\) and \(C_{12}\), it will be necessary to introduce their duals \(\tilde{C}_I\) and \(\tilde{C}_{12}\) as well.
These dualizations are described for the analog case of the type IIA supergravity reduction 
in \cite{Danckaert:2011ju} and \cite{KashaniPoor:2013en}, thus we will not
perform the explicit calculations again.

\section{Comparison with \texorpdfstring{$\cN = 4$}{N=4} Supergravity} \label{sec:standardN=4form}

As we have described above, the reduced action possesses \(\cN = 4\) supersymmetry. For this reason we will work out how to match it with the general description of gauged \(\cN = 4\) supergravity from \autoref{N=4Gen}.

The arrangement of the vectors into \(SO(5,n)\) representations \(A^M\), \(A^0\) and the form of the scalar metric \(M_{MN}\) can be worked out easiest by switching off all gaugings, \textit{i.e.}~by setting \(t^i = t^i_I = T^I_{iJ} = 0\) and \(n = n_I = 0\).
Since in this way all covariant derivatives become trivial and some of the terms in \eqref{eq:su2action} vanish, it is now very easy to carry out
the dualization of \(C_i\) explicitly.
Afterwards the theory will contain \(5 + \tilde n\) vectors in total, which means that there are \(\tilde n - 1\) vector multiplets and the global symmetry group is given by \(SO(1,1) \times SO(5,\tilde n-1)\).
It is natural to identify \(C_{12}\), which does not carry any indices, with the \(SO(5,\tilde n-1)\) singlet \(A^0\) and the other vectors with \(A^M\), so in summary we have
\begin{equation}\begin{aligned}\label{eq:su2vectors}
A^M &= \left(G^i, \tilde{C}^{\bar\imath}, C^J \right)\,, \\ 
A^0 &= C_{12}\,.
\end{aligned}\end{equation}
The corresponding \(SO(5,\tilde n-1)\) metric is defined as\footnote{Note that in the standard form of gauged supergravity $\eta$ is taken
to be diagonal. Therefore,
in order to compare fields and embedding tensors in this reduction to their standard form, one has to diagonalize $\eta$, which is easily done.}
\begin{equation}\label{eq:su2metric}
\eta_{MN} =
\begin{pmatrix}
0 & \delta_{i\bar\jmath} & 0 \\
\delta_{\bar\imath j} & 0 & 0 \\
0 & 0 & \eta_{IJ}
\end{pmatrix}\,.
\end{equation}
By comparing the kinetic terms of the vectors (in the ungauged theory) with \eqref{bos_N=4action} one obtains the scalar matching
\begin{equation}\label{eq:sigma}
\Sigma = e^{\frac{1}{3}\rho_2-\frac{1}{6}\rho_4}
\end{equation}
and the non-constant coset metric
\begin{equation}\begin{aligned}\label{eq:scalarmetric}
M_{ij} &= e^{\rho_2 + \rho_4} g_{ij} + H_{IJ}\, c^I_i c^J_j + e^{-\rho_2-\rho_4}g^{kl}(\epsilon_{ki}\gamma + \tfrac{1}{2} c_{kI} c^I_i)(\epsilon_{lj}\gamma + \tfrac{1}{2} c_{lI} c^I_j)\,, \\
M_{i\jb} &= e^{-\rho_2-\rho_4} g^{jk}\delta_{j\jb} (\epsilon_{ki}\gamma + \tfrac{1}{2} c_{kI} c^I_i)\,, \\
M_{iI} &= -H_{IJ} c^J_i + e^{-\rho_2-\rho_4} g^{jk} c_{jI} (\epsilon_{ki}\gamma + \tfrac{1}{2} c_{kI} c^I_i)\,, \\
M_{\ib\jb} &= e^{-\rho_2-\rho_4}g^{ij}\delta_{i\ib}\delta_{j\jb}\,, \\
M_{\ib I} &= e^{-\rho_2-\rho_4}g^{ij}\delta_{i\ib}c_{jI}\,, \\
M_{IJ} &= H_{IJ} + e^{-\rho_2-\rho_4}g^{ij}c_{iI}c_{jJ}\,.  
\end{aligned}\end{equation}
From this metric one can also determine the coset representatives \(\cV = ({\cV_M}^m, {\cV_N}^a)\), where \(m\) and \(a\) are \(SO(5)\) or \(SO(\tilde n-1)\) indices, respectively. \(\cV\) is related to the scalar metric via \(M = \cV\cV^T\) and carries the same amount of information. The result can be found in \autoref{app:cosetrepr}.

From \eqref{eq:sigma} and \eqref{eq:scalarmetric} we can determine the general covariant derivatives of the scalars using \eqref{gen_cov_der} and compare them with the results from \eqref{eq:covderivscalars} and \eqref{eq:covderivgamma} in order to derive the embedding tensors
\begin{equation}\begin{aligned}\label{eq:su2embeddingtensors}
\xi_i &= - \epsilon_{ij}t^j\,, \\
\xi_{iI} &= \epsilon_{ij}t^j_I\,, \\
f_{ij\bar\imath} &= \delta_{\bar\imath[i}\epsilon_{j]k}t^k\,, \\
f_{iIJ} &= - T^K_{iI} \eta_{KJ} - \tfrac{1}{2}\epsilon_{ij}t^j \eta_{IJ} \,,
\end{aligned}\end{equation}
and
\begin{equation}\begin{aligned}\label{eq:su2embeddingtensorsflux}
\xi_{ij} &= \epsilon_{ij} n \,, \\
f_{ijI} &= - \epsilon_{ij} n_I \,. \\
\end{aligned}\end{equation}
All other components are either determined by antisymmetry or vanish. One can now use these expressions to calculate the covariant derivatives of the vectors from \eqref{gen_cov_der} and check that they indeed agree with \eqref{eq:covderivvectors}.

For consistency the embedding tensors \eqref{eq:su2embeddingtensors}, \eqref{eq:su2embeddingtensorsflux}
should fulfill the quadratic constraints which
are listed for $\xi_M = 0$ in \eqref{e:quadr_constr} and can be found in full generality
in \cite{Schon:2006kz}.
In order to show that the latter hold it is necessary to use the consistency relations \eqref{eq:torsionconditions} on the matrices \(t^i\), \(t^i_I\) and \(T^I_{iJ}\) as well as the constraint on the flux \eqref{eq:fluxcondition}.

If we neglect the contributions coming from the four-form flux, it is possible to check that \eqref{eq:su2embeddingtensors} is consistent with the results from the type IIA reduction in \cite{KashaniPoor:2013en}.
This is described in \autoref{IIAreduction}.

\chapter{Partial Supergravity Breaking Applied to Consistent Truncations}\chaptermark{Application to Consistent Truncations}
\label{ch:cons_effth}

In this chapter we elaborate on the general discussion 
of supersymmetry breaking in \autoref{sec:susy_break} by investigating 
concrete examples given by consistent truncations of higher-dimensional theories.
In particular we analyze their quantum 'effective action'. In \autoref{sec:eff_cons}
we start with general considerations on the effective action of consistent truncations.
The analysis of one-loop Chern-Simons terms allows us to formulate necessary conditions such that
a consistent truncation gives rise to a physical sensible effective theory.
One class of examples, worked out in \autoref{sec:enriq}, will be provided by the $SU(2)$-structure 
reductions of \autoref{ch:m_su2} with Calabi-Yau vacuum.  Closely related to these kind of reductions is a second class of examples,
consistent truncations of type IIB supergravity on squashed Sasaki-Einstein manifolds which we investigate in \autoref{sec:sasa}.

\section{Quantum Effective Action of Consistent Truncations}\label{sec:eff_cons}

We start by studying the quantum effective action 
which we obtain after $\cN=4 \rightarrow \cN=2$ spontaneous supersymmetry 
breaking. An effective action is obtained by fixing a certain energy scale
and integrating out all modes that are heavier than this scale. In 
five dimensions this is particularly interesting since massive charged modes
induce Chern-Simons terms at one-loop. Importantly, these corrections
do not dependent on the masses of the modes in the loop and are therefore never suppressed.
We are interested in evaluating these terms for
the supersymmetry breaking mechanism in \autoref{sec:susy_break}.
This has already been investigated for purely Abelian magnetic gaugings in \autoref{oneloopeffects}.
A prominent class of more advanced examples for such a breaking pattern
is given by consistent truncations of supergravity.
For instance, if a Calabi-Yau manifold has $SU(2)$-structure, the $\cN=4$ gauged supergravity from the M-theory reduction in the previous
chapter is broken to $\cN=2$ in the vacuum.
It is an interesting question when a consistent truncation also gives rise to a 
proper effective theory. For example, in order to phenomenologically analyze 
non-Calabi-Yau reductions of string theory or M-theory one needs to deal with 
effective theories. 
As we saw in \autoref{sec:5dcherns}, there are one-loop corrections to the Chern-Simons terms.
The fact that these are independent of the mass scale will thus allow
us in the following to investigate the question:
\begin{itemize}
 \item  What are the necessary conditions for a consistent truncation to yield 
 the physical effective theory of the setup below a cut-off scale where all massive 
 modes are integrated out?
\end{itemize}
Clearly a first step is to analyze compactifications of which we know the relevant parts of the effective theory,
like Calabi-Yau compactifications.

In particular a necessary condition for a consistent truncation
to make sense as an effective field theory after integrating out massive modes is that
the one-loop Chern-Simons terms should coincide with the ones in the genuine effective action. Stated differently, the corrections to the Chern-Simons terms
induced by the truncated modes must coincide with the ones which are obtained by taking the full infinite tower of massive modes into account.
For the special case that the relevant parts of the effective theory are already exact at the classical level, as it is the case
for the $\cN=2$ prepotential in Calabi-Yau threefold compactifications of M-theory, the following four possibilities can in principle occur,
such that the fields in the consistent truncation do not contribute at one-loop:
The massive modes
\begin{itemize}
 \item are uncharged.
 \item arrange in long multiplets if the R-symmetry is not gauged.
 \item come in real representations.
 \item cancel non-trivially between different multiplets.
\end{itemize}
The contributions of long multiplets indeed cancel as one can explicitly check
by using \autoref{t:CS_correct} and \autoref{tab:long_mult} for the Minkowski case. This is related to the fact that they have the structure of
special $\cN=4$ multiplets,
which induce no corrections to the Chern-Simons terms.
For Minkowski space we display the two existing long multiplets in \autoref{tab:long_mult}.
\begin{table}
\begin{center}
\begin{tabular}{lrclr}
\multicolumn{2}{l}{\textbf{Long Gravitino Multiplet}}&&\multicolumn{2}{l}{\textbf{Long Vector Multiplet}}\TTstrut\BBstrut\\
\hhline{==~==}
Field Type & $(s_1,s_2)$ && Field Type & $(s_1,s_2)$ \TTstrut\BBstrut \\
\cline{1-2}\cline{1-2}\cline{4-5}\cline{4-5}
1  gravitino & $(1,\frac{1}{2})$ && 1 vector & $(\frac{1}{2},\frac{1}{2})$ \TTstrut\BBstrut\\
2 tensors & $2 \times (1,0)$ && \multirow{2}{*}{4 fermions} & $2 \times (\frac{1}{2},0)$  \TTstrut\BBstrut\\
2 vectors & $2 \times (\frac{1}{2},\frac{1}{2})$ && & $ 2 \times (0,\frac{1}{2})$ \TTstrut\BBstrut\\
\multirow{2}{*}{5 fermions} & $4 \times (\frac{1}{2},0)$ && 4 scalars & $4 \times (0,0)$ \TTstrut\BBstrut \\
\cline{4-5}
& $(0,\frac{1}{2})$ &&& \TTstrut\BBstrut\\
2 scalars & $2 \times (0,0)$ &&& \TTstrut\BBstrut  \\
\cline{1-2}
\end{tabular}
\end{center}
\caption{We display the long 
multiplets of $\cN=2$ supersymmetry in five-dimensional Minkowski space. The fields are labeled by their spins under $SU(2)\times SU(2)$.}
\label{tab:long_mult}
\end{table}
Also real multiplets do not contribute since they are parity-invariant
in contrast to the Chern-Simons terms.

After these general considerations let us now turn to some examples. 
Consider the M-theory reduction on $SU(2)$-structure manifolds of \autoref{ch:m_su2}.
If the compactification space is also Calabi-Yau, the five-dimensional $\cN=4$ gauged supergravity develops an $\cN=2$ vacuum.
This nicely fits into the general pattern of \autoref{sec:susy_break}. Indeed a Calabi-Yau threefold has $SU(2)$-structure 
if and only if its Euler number vanishes.
This can be seen as follows:
A Calabi-Yau threefold has \(SU(3)\) holonomy and thus allows for the existence of one covariantly constant spinor \(\eta^1\).
If the manifold has in addition vanishing Euler number, it follows from the Poincar\'e-Hopf theorem that there exists a nowhere-vanishing vector field \(K^1\).
With this ingredients it is possible to construct a second nowhere vanishing spinor \(\eta^2 = (K^1)^m \gamma_m \eta^1\),
such that the structure group is reduced to $SU(2)$.
This can also be seen without reference to spinors \cite{KashaniPoor:2013en}.
By acting with the complex structure \(J\) on \(K^1\) one obtains a second vector field \(K^2 = J K^1\) and after writing \(J\) and the holomorphic three-form \(\Omega\) as
\begin{equation}
J = J^3 + \tfrac{i}{2} K \wedge \bar K \,,\quad \Omega = K \wedge (J^1 + i J^2) \,,
\end{equation}
it is easy to check that \(K = K^1 + i K^2\) and \(J^a\) fulfill the relations \eqref{eq:su2forms}. 
We could now revert the argument and conclude that a \(SU(2)\) structure manifold with
\begin{equation}\label{eq:CYconditions}
\dd J = \dd \Omega = 0 
\end{equation}
is Calabi-Yau and therefore develops vacua with \(\cN = 2\) supersymmetry.
Using the expansions \eqref{eq:su2formsexpansion} of \(K\) and \(J^a\) we can translate \eqref{eq:CYconditions} into conditions on the five-dimensional fields
\begin{equation}\begin{aligned}\label{eq:5dCYconditions}
(t^1_I + \tau t^2_I)(\zeta^1_J + i \zeta^2_J) \eta^{IJ} &= 0 \,, \\
(T^K_{1I} + \tau T^K_{2I}) (\zeta^1_J + i \zeta^2_J) \eta^{IJ} &= 0 \,, \\
e^{\rho_4/2} T^J_{iI} \zeta^3_J &= \epsilon_{ij} t^j_I e^{\rho_2} \,.
\end{aligned}\end{equation}
These relations have to be used in the analysis of the spontaneous supersymmetry breaking  to \(\cN = 2\) vacua.
In \autoref{app:contracted_gaugings} we use these conditions in order to derive the contracted embedding tensors \eqref{e:dressed_gaugings}
for Calabi-Yau manifolds with vanishing Euler number. Note that the expressions in \autoref{app:contracted_gaugings} still
suffer from scalar redundancies, and it is hard
to eliminate the latter in general using the Calabi-Yau conditions. However, for the special example of the Enriques Calabi-Yau we were able to do so.
Thus we can derive the full spectrum by inserting the contracted embedding tensors into the results of \autoref{sec:action},
and we will actually do so in the next section.
What we will find is that the one-loop Chern-Simons terms do indeed cancel (as in the genuine effective theory)
although very trivially since there are simply no modes in the theory that are charged under a massless vector.
In fact we think that this might be the generic case for Calabi-Yau manifolds because of the following two
heuristic arguments:
\begin{itemize}
 \item Since a Calabi-Yau manifold does not have isometries if the holonomy is strictly $SU(3)$, one would think that the
`KK-vectors' become massive and the massive modes are not charged under massless gauge symmetries.
In particular, the vectors $G^i$ in the ansatz for the metric \eqref{e:su(2)_metric}
\begin{equation}
\dd s^2_{11} = g_{\mu\nu} \dd x^\mu \dd x^\nu + g_{ij} (v^i + G^i)(v^j + G^j) + g_{mn} \dd x^m \dd x^n
\end{equation}
should acquire masses.
\item For Calabi-Yau manifolds with $\chi =0$ and vanishing gaugings $\xi_M$ there are no charged tensors. In fact,
using the Calabi-Yau relations from \eqref{eq:5dCYconditions} it is easy to show that for such manifolds
we have $\xi^{MN}\tensor{\xi}{_N^P}=0$.
Applying also the quadratic constraints to \eqref{bos_N=4action} the vanishing of tensor charges is immediate.
Note that the contributions of tensors was a crucial ingredient in \autoref{sec:ab_mag_gaug} where non-vanishing one-loop Chern-Simons terms
appeared in $\cN=4\rightarrow\cN=2$ supergravity breaking to Minkowski vacua.
\end{itemize}
If massive modes carry no charges under massless vectors in general,
our approach via one-loop Chern-Simons terms imposes no restrictions on the consistent truncation to yield also a proper effective theory.

Let us now turn to the second example of partial supergravity breaking in the context of consistent truncations, type IIB supergravity on a
squashed Sasaki-Einstein manifold which is discussed in \autoref{sec:sasa} in greater detail.
The geometrical reduction to $\cN=4$ gauged supergravity in five dimensions was carried out in \cite{Cassani:2010uw,Liu:2010sa,Gauntlett:2010vu}
and proceeds similarly 
to the M-theory $SU(2)$-structure reduction of \autoref{ch:m_su2}. Again the theory admits $\cN=2$ vacua which however now constitute
AdS backgrounds with gauged R-symmetry. Although it is not really clear if the concept of effective field theory makes sense on such backgrounds,
we nevertheless integrate out massive modes because of the topological origin of the relevant corrections. 
Surprisingly the contributions to the gauge and gravitational one-loop Chern-Simons terms cancel in a  
non-trivial way between
different multiplets.
It would be extremely interesting to find an interpretation for this result.

\section[First Example: M-Theory on the Enriques Calabi-Yau]{First Example: M-Theory on the Enriques Ca-labi-Yau
\sectionmark{M-Theory on the Enriques Calabi-Yau}}
\sectionmark{M-Theory on the Enriques Calabi-Yau}
\label{sec:enriq}

In this section we analyze in detail the spectrum of M-theory on the Enriques Calabi-Yau around the $\cN=2$ vacuum of the $\cN=4$
gauged supergravity using the results of \autoref{sec:action}.
The precise expressions for the embedding tensors in the standard form of $\cN =4$ gauged supergravity
and their contractions with the coset representatives for Calabi-Yau manifolds with $SU(2)$-structure are given
in \autoref{app:contracted_gaugings}.
However, as already mentioned, these quantities still suffer from redundancies of scalar fields which should be eliminated
by using the Calabi-Yau conditions \eqref{eq:5dCYconditions} in order to analyze the setup with the tools of \autoref{sec:action}.
Consequently we focus on the special case of the
Enriques Calabi-Yau where we were able to remove the redundancies.
In the following we derive the spectrum and gauge symmetry in the vacuum of the $SU(2)$-structure reduction
and compare the results to the known Calabi-Yau effective theory.
Besides the fact that the former yields massive states which are absent in the latter, 
the consistent truncation turns out to lack one vector multiplet and one hypermultiplet at the massless level 
compared to the effective theory of the Enriques, analogous to the results in \cite{KashaniPoor:2013en}.
Taking the missing massless vector into account the classical Chern-Simons terms of both
theories may coincide in principle.
Corrections at one-loop to the Chern-Simons terms vanish trivially since there are no modes charged under the massless vectors.

The gauged supergravity embedding tensors $f_{MNP}$, $\xi_{MN}$ of M-theory on the Enriques Calabi-Yau
are evaluated by inserting the expressions \eqref{e:torsion_classes} into \eqref{eq:su2embeddingtensors}. In the standard basis, where
$\eta$ takes the form $\eta = (-1,-1,-1,-1,-1,+1,\dots , +1)$, they read
\begin{align}
 &f_{135} = f_{245} = f_{815} = f_{925} = -f_{13\, 10} = -f_{24\, 10} = -f_{81\, 10} = -f_{92\, 10}=\frac{1}{\sqrt 2} \, ,\nn \\
 &f_{635} = f_{745} = f_{865} = f_{975} = -f_{63\, 10} = -f_{74\, 10} = -f_{86\, 10} = -f_{97\, 10}= -\frac{1}{\sqrt 2} \, , \nn \\
 &\xi_{13} = \xi_{24} = \xi_{81} = \xi_{92} = 
 -\xi_{63} = -\xi_{74} = -\xi_{86} = -\xi_{97} = \frac{1}{\sqrt 2} \, .
\end{align}
As can be inferred form the covariant derivative \eqref{gen_cov_der}, the gauged $SO(5,n)$ symmetry generators $t_{MN}$ are given by
(modulo normalization of the generators)
\begin{align}
 &t_1 := t_{15} + t_{1\, 10} + t_{65} + t_{6\, 10}\, , & t_2 := t_{25} + t_{2\, 10} + t_{75} + t_{7\, 10} \, , \nn \\
 & t_3 := t_{35} + t_{3\, 10} + t_{85} + t_{8\, 10} \, , 
 &t_4 := t_{45} + t_{4\, 10} + t_{95} + t_{9\, 10} \, , \nn \\
 & t_5 := t_{13} + t_{24} + t_{18} + t_{29} + t_{63} + t_{74} + t_{68} + t_{79} \, .
\end{align}
Since all commutators vanish, as one can check easily, the gauge group in the $\cN =4$ theory is $U(1)^5$.

Let us now move to the vacuum. The structure of the embedding tensors contracted with the
coset representatives is derived in \autoref{app:contracted_gaugings}. They read
\begin{align}
 &f_{1,6\,\,\, 3,8\,\,\, 5,10} = f_{2,7\,\,\, 4,9\,\,\, 5,10} = \frac{1}{\sqrt 2} \Sigma^3 \lambda_\xi \nn \\
 &\xi_{1,6\,\,\, 3,8} = \xi_{2,7\,\,\, 4,9} = \lambda_\xi \, ,
\end{align}
where for each index position of the tensors there are two options to choose from.
For convenience we define
\begin{align}
 \lambda_\xi := \frac{1}{\sqrt 2}\, e^{-\frac{1}{2}(\rho_2 + \rho_4)}\, \I \, \tau \, .
\end{align}
The rotation to $\xi_{\cM\cN}$, $f_{\cM\cN\cP}$ \eqref{e:xi_trafo}, which is the appropriate basis
to split off the propagating degrees of freedom, gives the non-vanishing components
\begin{align}
 & \xi_{12} = \xi_{34} = 2\,\lambda_\xi \, , & f_{125} = f_{345} = f_{12\, 10} = f_{34\, 10} = \sqrt 2 \,\Sigma^3 \,\lambda_\xi \, .
\end{align}

The spectrum is calculated by inserting the contracted embedding tensors into \eqref{e:vac_lagr} and bringing the terms in the Lagrangian into standard form.
The fields together with their masses and charges are listed in \autoref{t:Enriques}.
\begin{table}
\begin{center}
\begin{tabular}{lrrr}
\textbf{Multiplet} & \textbf{Mass} & \textbf{Charge} \TTstrut\BBstrut\\
\hline\hline
\rule[-.1cm]{0cm}{.6cm} 1 real graviton multiplet & \multirow{2}{*}{0} & \multirow{2}{*}{\bf0} \TTstrut\BBstrut\\
\rule[-.3cm]{0cm}{.6cm} \((2, 2 \times \tfrac{3}{2}, 1)\) && \TTstrut\BBstrut\\
\hline
\rule[-.1cm]{0cm}{.6cm} 9 real vector multiplets & \multirow{2}{*}{0} & \multirow{2}{*}{\bf0} \TTstrut\BBstrut\\
\rule[-.3cm]{0cm}{.6cm}\((1, 2 \times \tfrac{1}{2}, 0)\) && \TTstrut\BBstrut\\
\hline
\rule[-.1cm]{0cm}{.6cm}  11 real hypermultiplets & \multirow{2}{*}{0} & \multirow{2}{*}{\bf0} \TTstrut\BBstrut\\
\rule[-.3cm]{0cm}{.6cm} \((2 \times \tfrac{1}{2}, 4 \times 0)\) && \TTstrut\BBstrut\\
\hline
\rule[-.1cm]{0cm}{.6cm} 1 complex gravitino multiplet & \multirow{2}{*}{\(m\)} & \multirow{2}{*}{\bf0} \TTstrut\BBstrut\\
\rule[-.3cm]{0cm}{.6cm} \(\left((1,\tfrac{1}{2}), 2 \times (1, 0), 2 \times (\tfrac{1}{2}, \tfrac{1}{2}), 4 \times (\tfrac{1}{2}, 0), (0, \tfrac{1}{2}), 2 \times (0,0)\right)\) && \TTstrut\BBstrut\\
\hline
\rule[-.1cm]{0cm}{.6cm} 1 real vector multiplet & \multirow{2}{*}{\(2mc\)} & \multirow{2}{*}{\bf0} \TTstrut\BBstrut\\
\rule[-.3cm]{0cm}{.6cm} \(\left((\tfrac{1}{2}, \tfrac{1}{2}), 2 \times (\tfrac{1}{2},0)\right)\) &&\TTstrut\BBstrut \\
\hline
\rule[-.1cm]{0cm}{.6cm}  1 complex hypermultiplet & \multirow{2}{*}{\(2m\)} & \multirow{2}{*}{\bf0} \TTstrut\BBstrut\\
\rule[-.3cm]{0cm}{.6cm}  \(\left((\tfrac{1}{2}, 0), 2 \times (0,0)\right)\) && \TTstrut\BBstrut\\
\hline
\end{tabular}
\end{center}
\caption{We depict the spectrum of the $SU(2)$-structure reduction of M-theory on the Enriques Calabi-Yau.}
\label{t:Enriques}
\end{table}
The modes are classified according to their mass, charges under the massless vectors and their spacetime
representations under $\mathfrak{su}(2)$ or $\mathfrak{su}(2) \oplus \mathfrak{su}(2)$, respectively.
Fermions in complex multiplets are Dirac while fermions in real multiplets are taken to be symplectic Majorana.
We set $m=\sqrt 2 \, \Sigma^{2}\lambda_\xi$
and $c=\frac{(1+\Sigma^{-6})^{3/2}}{(1+\Sigma^{-12})^{1/2}}$ .

The massless multiplets are uncharged and consistent with the proper Calabi-Yau effective theory apart from one missing vector multiplet and
one hypermultiplet.
More precisely, the Enriques Calabi-Yau has Hodge numbers $h^{1,1}= h^{2,1} = 11$.
In the effective action of M-theory on Calabi-Yau threefolds one finds $h^{1,1}-1$ vector multiplets and
$h^{2,1}+1$ hypermultiplets while for our consistent truncation on the Enriques Calabi-Yau we find only 9 vector multiplets and 11
hypermultiplets.
This resembles the results in \cite{KashaniPoor:2013en} where the same field content was missing for the analog type IIA setup.
Geometrically the corresponding missing harmonic forms are captured by
$SU(2)$-doublets which we discarded in the reduction of \autoref{ch:m_su2}.
As explained, the doublets correspond to $\cN = 4$ gravitino multiplets, for which
no coupling to standard $\cN =4$ gauged supergravity is known.
Having discussed the massless modes in the vacuum we now turn to the massive spectrum.
We find one long gravitino multiplet, one vector multiplet and one hypermultiplet.
Interestingly no massive field is charged under a massless $U(1)$ gauge symmetry.
For the massive tensors this has already been established on general grounds in the previous section.
Thus we conclude that for the Enriques Calabi-Yau the Chern-Simons terms \eqref{e:def_CS_6d} are trivially not corrected by loops
of fermions or tensors since there are no charged modes in the truncation.

Finally let us also comment on the classical Chern-Simons terms in the reduction. We denote the ten massless vectors in the vacuum of the consistent truncation
by $\tilde A^9_\mu$, $\tilde A^{10}_\mu$, $\tilde A^\mathfrak{a}_\mu$ with $\mathfrak{a} =1,\dots,8$. The $\tilde A^\mathfrak{a}_\mu$ originate from the
$E_8$ nature of the Enriques surface. The classical gauge Chern-Simons couplings are found to be
\begin{align}
& k^{\textrm{trunc}}_{9\, 10\, 9}=2\sqrt 2 \, , & k^{\textrm{trunc}}_{9\mathfrak{a}\mathfrak{a}}=2 \, ,
\end{align}
all others vanish. In the familiar Calabi-Yau effective action the Chern-Simons coefficients reproduce the intersection numbers
of the manifold. For the Enriques Calabi-Yau they read in a suitable basis
\begin{align}\label{e:intersec_enriques}
 & k^{\textrm{eff}}_{9 \, 10 \, 11}=1 \, , & k^{\textrm{eff}}_{9\mathfrak{a}\mathfrak{b}}=A^{E_8}_{\mathfrak{a}\mathfrak{b}} \, ,
\end{align}
where $A^{E_8}$ denotes the Cartan matrix of $E_8$.
If we assume that the missing vector $\tilde A^{11}$ of the consistent truncation
appears together with $\tilde A^{9}$ and $\tilde A^{10}$ in a Chern-Simons term with coefficient
\begin{align}
  k^{\textrm{miss}}_{9 \, 10 \, 11} \neq 0 \, ,
 \end{align}
 we can define
 \begin{align}
  &\hat A^9_\mu := \tilde A^9_\mu \ , &&
  \hat A^{10}_\mu := \tilde A^{10}_\mu \, , &&
  \hat A^{11}_\mu := \sqrt 2\, \tilde A^9_\mu + k^{\textrm{miss}}_{9 \, 10 \, 11} \cdot \tilde A^{11}_\mu \, , 
 \end{align}
 such that in this basis we obtain the Chern-Simons couplings
 \begin{align}
  & k_{9 \, 10 \, 11} =1 \, ,  & k_{9\mathfrak{a}\mathfrak{a}}=2 \, .
 \end{align}
The first one matches with \eqref{e:intersec_enriques}. Concerning the second term we note that
the Cholesky decomposition of $A^{E_8}$ ensures that there exists a field redefinition for the $\hat A^\mathfrak{a}_\mu$
represented by a matrix $T$, which fulfills
\begin{align}
 T^T T = \frac{1}{2} A^{E_8} \, .
\end{align}
It is easy to check that under this redefinition
$k_{9\mathfrak{a}\mathfrak{a}}$ goes to $k^{\textrm{eff}}_{9\mathfrak{a}\mathfrak{b}}$.
These considerations can also be interpreted as a proposition for the Chern-Simons coefficient, 
which involve the missing massless vector $\tilde A^{11}$,
namely $k^{\textrm{miss}}_{9 \, 10 \, 11} \neq 0$. It should be reproduced by the $SU(2)$-doublets.
 
We conclude that for the Enriques Calabi-Yau,
apart from the missing vector multiplet and hypermultiplet, the effective theory of the consistent truncation is consistent
with the genuine Calabi-Yau effective action since it is in principle possible to match the classical Chern-Simons terms of both sides, and more importantly
corrections at one-loop are absent in the consistent truncation because massive modes do not carry any charges.
Since we think that this is the case for generic Calabi-Yau manifolds with vanishing Euler number, 
the analysis of the Chern-Simons terms reveals no restrictions for the consistent truncation to also
yield a proper effective action. This conclusion might change significantly if the 
internal space has isometries and there are massive modes charged under 
massless vectors. We turn to an example that has these features in the next chapter.

\section[Second Example: Type IIB Supergravity on a Squashed Sasaki-Einstein Manifold]{Second Example: 
Type IIB Supergravity on a Squashed Sasaki-Einstein Manifold\sectionmark{Type IIB on a Squashed Sasaki-Einstein Manifold}}
\sectionmark{Type IIB on a Squashed Sasaki-Einstein Manifold}
\label{sec:sasa}

In the following we study a second example of partial supergravity breaking in the context 
of consistent truncations that features a massive spectrum charged under a massless vector.
More precisely, we consider type IIB supergravity
on a squashed Sasaki-Einstein manifold with 5-form flux. This setup admits a consistent truncation to $\cN = 4$ gauged supergravity in five dimensions
which has two vacua, one which breaks supersymmetry completely and one which is $\cN =2$ AdS. We focus on the latter in our analysis.
Since the theory in the broken phase can be described with the results of \autoref{sec:action},
we proceed along the lines of the last section and derive the spectrum and Chern-Simons terms.
The field content turns out to be consistent with \cite{Cassani:2010uw,Liu:2010sa,Gauntlett:2010vu}.
Although there are massive modes charged under the gauged R-symmetry in the vacuum, their corrections to the gauge
and gravitational Chern-Simons terms at one-loop
cancel exactly.

In \cite{Cassani:2010uw} it was shown that in a consistent truncation of type IIB supergravity on a 
squashed Sasaki-Einstein manifold to five-dimensional $\cN=4$ gauged supergravity
the non-vanishing embedding tensors $f_{MNP}$, $\xi_{MN}$ take the form
\begin{align}
 & f_{125}=f_{256}=f_{567}=-f_{157}=-2 \, , \\
& \xi_{12}=\xi_{17}=-\xi_{26}=\xi_{67}=-\sqrt 2 k \, , &\xi_{34}= -3\sqrt 2 \, ,  \nn
\end{align}
where $k$ denotes 5-form flux on the internal manifold.
They encode the gauging of the group $\textrm{Heis}_3 \times U(1)_R$, where a $U(1)_R$ is a subgroup of the R-symmetry group.
The theory admits a vacuum that preserves $\cN =2$ supersymmetry.
If we for simplicity fix the RR-flux to $k=2$, we can use the expressions for the scalar VEVs in \cite{Cassani:2010uw}
to derive the contracted embedding tensors \eqref{e:dressed_gaugings}
\begin{align}
& f_{125}=f_{675}=-f_{175}=-f_{625}=2 \, , \\
& \xi_{12}=\xi_{67}=-\xi_{17}=-\xi_{62}=-2 \sqrt 2  \, , &\xi_{34}= -3\sqrt 2 \, .  \nn
\end{align}
We can now rotate into the basis of \eqref{e:xi_trafo}, in fact we transform $\xi_{\hat\cM\hat\cN}$ already into block-diagonal form.
The non-vanishing gaugings $\xi_{\cM\cN}$, $f_{\cM\cN\cP}$ read
\begin{align}
 & \xi_{12} = 4 \sqrt 2 \, , && \xi_{34} = 3 \sqrt 2 \, ,
 & f_{125} = -4 \, ,
\end{align}
and therefore
\begin{align}
 & \hat \cM = 1,2,3,4 \, , & \bar \cM = 5,6,7 \, .
\end{align}

Carrying out the calculations we find the cosmological constant $\Lambda = -6$, corresponding to an AdS$_5$ background.
Furthermore half of the supersymmetries are broken and the gauge group is reduced
\begin{align}
 \textrm{Heis}_3 \times U(1)_R \rightarrow U(1)_R \, ,
\end{align}
where now the full $U(1)$ R-symmetry of minimal supersymmetry in AdS$_5$ is gauged with gauge coupling $g^2= \sfrac{3}{2}$ .
The complete spectrum of the consistent truncation in the vacuum is depicted in \autoref{t:SE}
where we consulted the categorization of \cite{Ceresole:1999zs}.
The fields are classified
according to their mass, charge under $U(1)_R$ with coupling $g$ and their
representation under the $SU(2)\times SU(2)$ part of the maximal compact subgroup of $SU(2,2|1)$.
\begin{table}
\begin{center}
\begin{tabular}{lcrr}
\textbf{Multiplet} & \textbf{Representation} & \textbf{Mass} & \textbf{Charge}   \\
\hline\hline
\multirow{4}{*}{1 real graviton multiplet}
& $(1,1)$ & 0 & 0\TTstrut\BBstrut\\
& $(1,\frac{1}{2})$ & 0 & -1\TTstrut\BBstrut\\
& $(\frac{1}{2},1)$ & 0 & +1\TTstrut\BBstrut\\
& $(\frac{1}{2},\frac{1}{2})$ & 0 & 0\TTstrut\BBstrut\\
\hline
\multirow{3}{*}{1 complex hypermultiplet}
& $(\frac{1}{2},0)$ & 3/2 & +1\TTstrut\BBstrut\\
& $(0,0)$ & -3 & +2\TTstrut\BBstrut\\
& $(0,0)$ &  0& 0\TTstrut\BBstrut\\
\hline
\multirow{6}{*}{1 complex gravitino multiplet}
& $(\frac{1}{2},1)$ & -5 & +1\TTstrut\BBstrut\\
& $(\frac{1}{2},\frac{1}{2})$ & 8 & 0\TTstrut\BBstrut\\
& $(0,1)$ & 3 & +2\TTstrut\BBstrut\\
& $(0,1)$ & 4 & 0\TTstrut\BBstrut\\
& $(0,\frac{1}{2})$ & -5/2 & +1\TTstrut\BBstrut\\
& $(0,\frac{1}{2})$ & -7/2 & +3\TTstrut\BBstrut\\
\hline
\multirow{9}{*}{1 real vector multiplet}
& $(\frac{1}{2},\frac{1}{2})$ & 24 & 0\TTstrut\BBstrut\\
& $(\frac{1}{2},0)$ & 9/2 & -1\TTstrut\BBstrut\\
& $(0,\frac{1}{2})$ & 9/2 & +1\TTstrut\BBstrut\\
& $(0,\frac{1}{2})$ & 11/2 & -1\TTstrut\BBstrut\\
& $(\frac{1}{2},0)$ & 11/2 & +1\TTstrut\BBstrut\\
& $(0,0)$ & 12 & 0\TTstrut\BBstrut\\
& $(0,0)$ & 21 & -2\TTstrut\BBstrut\\
& $(0,0)$ & 21 & +2\TTstrut\BBstrut\\
& $(0,0)$ & 32 & 0\TTstrut\BBstrut\\
\hline
\end{tabular}
\end{center}
\caption{We depict the spectrum of type IIB supergravity on a 
squashed Sasaki-Einstein manifold in the $\cN=2$ vacuum corresponding to an AdS$_5$ background.}
\label{t:SE}
\end{table}

For our example we find at the classical level\footnote{We do not account for the classical gravitational Chern-Simons term.} 
\begin{align}
 k_{000}^{\textrm{class}}= 4\, \sqrt{\frac{2}{3}} \, .
\end{align}
In order to calculate the quantum corrections, we again use \autoref{t:CS_correct} with the understanding that representations of
$SU(2)\times SU(2) \subset SU(2,2|1)$ in AdS
contribute in the same way as representations of $SU(2)\times SU(2)$ in the Minkowski case.
Although the results of \autoref{t:CS_correct}, derived in \cite{Bonetti:2013ela}, were originally calculated in a Minkowski background, we believe
that they are applicable to AdS as well because of their topological origin.
Remarkably, the one-loop corrections of the massive charged modes to the gauge and gravitational Chern-Simons terms
both cancel in a highly non-trivial way
\begin{align}
 & k_{000}^{\textrm{1-loop}}=0 \, , & k_{0}^{\textrm{1-loop}}=0 \, .
\end{align}
Note that the index zero is now meant to refer to the remaining massless $U(1)_R$ in the vacuum rather than to $A^0$ in the $\cN = 4$ theory.

The interpretation of this result is not as clear as in the last section concerning the Enriques Calabi-Yau. Indeed, 
the naive notion of an effective field theory on AdS backgrounds is not well-defined since the AdS radius is linked to 
the size of the internal space. We nevertheless think that the non-trivial vanishing of the
one-loop Chern-Simons terms is not accidental and should have a clear interpretation. 
One might even suppose that there exists a general principle which ensures the nice behavior of scale-invariant corrections in consistent truncations.
It would be interesting to elaborate more on this.
Related to that, it would also be worthwhile
to find connections to other consistent truncations. The simplest example is certainly the $\cN=8$ consistent truncation to massless modes of
type IIB supergravity on the five-sphere \cite{Cvetic:2000nc}, which is a special Sasaki-Einstein manifold.

\part{Circle-Reduced Gauge Theories, F-Theory and the Arithmetic of Elliptic Fibrations}\label{part:fth}

\chapter{Overview}\label{ch:wotk2_intro}

A study of the consistency of quantum field theories requires to investigate 
their local symmetries both at the classical and at the quantum level. 
In particular, even if such gauge symmetries are manifest in the classical 
theory, they might be broken at the quantum level and induce
a violation of essential current conservation laws. Such 
inconsistencies manifest themselves already
at one-loop level and are known as anomalies which we have already reviewed in \autoref{ch:ano}. 
Four-dimensional quantum field theories with chiral spin-$\sfrac{1}{2}$ fermions for example 
can admit anomalies which signal the breaking of the gauge symmetry. 
Consistency requires the cancelation of these anomalies either by 
restricting the chiral spectrum such that a cancelation among various 
contributions takes place or by implementing a generalized 
Green-Schwarz mechanism \cite{Green:1984sg,Sagnotti:1992qw}. 
The latter mechanism requires the presence of 
$U(1)$ gauged axion-like scalars with tree-level diagrams 
canceling the one-loop anomalies. In six spacetime dimensions 
anomalies pose even stronger constraints since in addition to 
spin-$\sfrac12$ fermions also spin-$\sfrac{3}{2}$ fermions and two-tensors can be 
chiral. Also in this case a generalized Green-Schwarz mechanism can 
be applied to cancel some of these anomalies.

In this part we address the manifestation of anomaly cancelation 
in four-dimensional and six-dimensional field theories from a Kaluza-Klein perspective 
when considering the theories to be compactified on a circle. 
Note that on a circle one can expand all higher-dimensional fields
into Kaluza-Klein modes yielding a massless lowest mode and 
a tower of massive excitations. 
Clearly, keeping track of this infinite set of fields one retains the full information 
about the higher-dimensional theory including its anomalies. 
In a next step one can compute the lower-dimensional effective 
theory for the massless modes only. This requires to integrate out 
all massive states. Of particular interest for the discussion of 
anomalies are the effective lower-dimensional couplings   
that are topological in nature. These do not continuously depend on 
the cutoff scale and might receive relevant quantum corrections 
from integrating out the massive states. Prominent examples are 
three-dimensional gauge Chern-Simons 
terms as well as five-dimensional 
gauge and gravitational Chern-Simons terms which we have introduced in \autoref{ch:CS}.
These couplings are indeed modified at one-loop when integrating out
massive states. In three dimensions only certain massive spin-$\sfrac12$ 
fermions contribute while in five dimensions also massive spin-$\sfrac{3}{2}$ and 
massive self-dual tensors give a non-vanishing shift. In fact, precisely 
those modes contribute that arise from higher-dimensional chiral fields. Therefore
one expects that the Chern-Simons terms of the effective theories encode 
information about the higher-dimensional anomalies. This was recently investigated 
motivated by the study of F-theory effective actions via M-theory in \cite{Grimm:2011fx,Cvetic:2012xn,Bonetti:2011mw,Grimm:2013oga,Cvetic:2013uta,Anderson:2014yva}. With a different motivation similar questions were 
addressed in \cite{Landsteiner:2011iq,Loganayagam:2012pz,Golkar:2012kb,Landsteiner:2012kd,Jensen:2012kj,Jensen:2013kka,Jensen:2013rga,DiPietro:2014bca} in the study of applications of holography. 

The connection between one-loop Chern-Simons terms and 
anomalies in the higher-dimensional theory, while expected to exist, was only shown to be rather indirect. In fact, 
it is not at all obvious how the anomaly cancelation conditions arise for example
from comparing classical and one-loop Chern-Simons terms. While for many concrete 
examples in the framework of F-theory, where these couplings play a prominent role,
it was possible to check anomaly cancelation using the lower-dimensional 
effective theory and Chern-Simons terms arising from M-theory, there was no known systematics behind this 
as of now. In our work we will show that there is an elegant way to actually approach 
this for general quantum field theories by describing symmetry transformations among effective theories that 
exist if and only if higher-dimensional anomalies are canceled. 

Let us consider an effective theory obtained after circle reduction. If the higher-dimensional 
theory admits a gauge group one can use the Wilson-line scalars of the gauge fields 
around the circle to move to the lower-dimensional Coulomb branch. In other words 
one considers situations in which these Wilson line scalars admit a vacuum expectation 
value which we call Coulomb branch parameters in the following. 
The masses of all the massive states are now dependent both on the circle radius if they 
are excited Kaluza-Klein states,  
and on the Coulomb branch parameters if they where charged under the higher-dimensional gauge group. 
With this in mind one can then compute the effective theory 
for the massless modes and focus on the Chern-Simons terms. Since the one-loop
Chern-Simons couplings are not continuous functions of the masses of the integrated-out 
states, they can experience discrete shifts when changing the radius or the 
Coulomb branch parameters.
In particular, the one-loop Chern-Simons terms carry information 
about the representations of the higher-dimensional chiral
spectrum 
supplemented with a table of signs for each state \cite{Grimm:2011fx,Cvetic:2012xn} which coincide with the formal signs of the
Coulomb branch masses.\footnote{Different gauge theory phases of such theories and 
 their relation to geometric resolutions have been recently studied in \cite{Intriligator:1997pq,Grimm:2011fx,Hayashi:2013lra,Hayashi:2014kca,Esole:2014bka,Esole:2014hya,Braun:2014kla}.}
This extra information can be summarized in so-called box graphs introduced 
in \cite{Hayashi:2014kca,Braun:2014kla}, see also \cite{Esole:2014bka,Esole:2014hya}. 
In general however, it is important to also keep track of an 
integer label for each dimensionally reduced state that encodes the 
mass hierarchy between the Kaluza-Klein mass and the Coulomb branch mass.
In other words, depending on the background value 
of the Coulomb branch parameters and the radius, the effective theories can take 
different forms. One thus finds infinitely many values for the 
Chern-Simons coefficients due to the infinite amount of hierarchies of Kaluza-Klein masses
and Coulomb branch masses. However, we show in detail that higher-dimensional large gauge transformations along the circle 
identify different Coulomb branch parameters and effective theories if 
and only if the all gauge anomalies are canceled. 
Our goal is to study this in the context of four- and 
six-dimensional gauge theories with a focus on pure
gauge anomalies and for six-dimensional theories in addition also on mixed gauge-gravitational anomalies.

Furthermore, due to the importance of large gauge transformations along the circle we provide a full classification of such maps
which leave the boundary conditions of
all matter fields invariant. We show that depending on the precise spectrum or global structure of the gauge group
the naive set of what we would call \textit{integer large gauge transformations}
can be enlarged. Indeed for non-simply connected gauge groups the weights of the matter representations can also allow
for \textit{special fractional
large gauge transformations}. It is one of the major goals of this part to identify these group structures of circle-reduced gauge
theories with arithmetic structures on genus-one fibrations via the framework of F-theory and its 
effective physics \cite{Vafa:1996xn,Morrison:1996na,Morrison:1996pp,Ferrara:1996wv,Denef:2008wq,Grimm:2010ks,Bonetti:2011mw,Grimm:2013oga}.
The latter structures can then be invoked to directly show anomaly
cancelation in F-theory compactifications on Calabi-Yau manifolds using one-loop Chern-Simons terms.

In recent years the connection of gauge theories in various dimensions to the geometry of elliptic curves
has been explored intensively 
by using F-theory. In F-theory two auxiliary dimensions
need to be placed on a two-torus whose complex structure 
is identified with the Type IIB axio-dilaton.  Its variations are then encoded by 
the geometry of a two-torus fibration in F-theory. Magnetic sources for the 
axio-dilaton are 7-branes that support gauge theories. Several features 
of these gauge theories can thus be studied using two-torus fibrations.
At first, the F-theory approach seems to suggest 
that the connection between geometry and gauge theories is rather direct. 
However, it turns out that the geometry of elliptic fibrations should 
rather be related to gauge theories compactified on a circle. 
This can be understood by realizing that the volume of the two-torus 
is unphysical in F-theory and that there is no notion of an actual 
twelve-dimensional background geometry. The geometry of the 
elliptic fibration in F-theory is only fully probed in the dual M-theory compactification.
M-theory compactified on an elliptic fibration yields the effective theory of F-theory
compactified on an additional circle. In particular, one is therefore forced to 
relate the geometry of elliptic fibrations with gauge theories on a circle. 
Our focus in this work will be 
on revealing geometric symmetry transformations that correspond to the
large gauge transformations along the circle. 

As a first example of such a relation we will study Abelian gauge theories 
on a circle. In an F-theory compactification the number of massless 
$U(1)$ fields can be related to the number of rational sections (or multi-sections)
minus one \cite{Morrison:1996pp}. 
The  section that is not counted here has to be identified with 
the Kaluza-Klein vector  
obtained from the higher-dimensional metric when placing the gauge theory 
on the additional circle. Recent 
progress in understanding $U(1)$ gauge groups in F-theory can be found for 
example in the references \cite{Grimm:2010ez,Park:2011ji,Morrison:2012ei,Braun:2013yti,Borchmann:2013jwa,Cvetic:2013nia,Grimm:2013oga,Braun:2013nqa,Cvetic:2013uta,Borchmann:2013hta,Cvetic:2013jta,Cvetic:2013qsa,Klevers:2014bqa,Lawrie:2015hia,Cvetic:2015ioa}. The fact that smooth geometries carry information 
about a circle-reduced theory becomes particularly apparent 
in models with rational sections in which the mentioned mass hierarchy between 
Kaluza-Klein masses and lower-dimensional Coulomb branch masses 
is non-trivial \cite{Grimm:2013oga}. 
In other words, it was key in \cite{Grimm:2013oga} that despite 
the fact that massive states have to be integrated out in the circle 
compactified effective theory some cutoff independent information about the massive tower 
has to be kept in order to get the right one-loop Chern-Simons terms. 
For models with only a multi-section, see 
\cite{deBoer:2001px,Braun:2014oya,Morrison:2014era,Klevers:2014bqa,Cvetic:2015moa,Braun:2014qka} for representative works, 
the requirement of a lower-dimensional approach is even more pressing. As discussed in 
\cite{Anderson:2014yva,Garcia-Etxebarria:2014qua,Mayrhofer:2014haa,Mayrhofer:2014laa,Cvetic:2015moa} the multi-section should be understood as a mixing of 
the higher-dimensional $U(1)$s and the Kaluza-Klein vector.  

The first goal of our work is to formalize the relationship between 
geometries with rational sections and circle-reduced gauge theories further.
We carefully identify the Mordell-Weil group acting on rational 
sections as large gauge transformations along 
the circle. The Mordell-Weil group is a discrete 
finitely-generated Abelian group that captures key information about the arithmetic 
of elliptic fibrations. We show that there is indeed a one-to-one correspondence between 
large gauge transformations and basis shifts in the Mordell-Weil group. 
Shifts along the free part of the Mordell-Weil group are identified with integer Abelian large gauge 
transformations while its torsion part is related to special fractional 
non-Abelian large gauge transformations. As a byproduct we explore the 
geometric relationship between the existence of fractional Abelian charges 
of matter states and the presence of a non-Abelian gauge group. 

A second goal is to use our understanding of the arithmetic
for geometries with rational sections to provide evidence for the existence of a natural group law acting on
fibrations with multi-sections. We call this group \textit{extended Mordell-Weil group} despite the fact 
that there is formally no Mordell-Weil group for multi-sections. We also define 
a \textit{generalized Shioda map} that allows to explore the physical implications of the
group action. 
Furthermore, we rigorously establish the correspondence of the proposed group 
action on the divisor level with large gauge transformations around the circle. In many 
examples it is known that there exist geometric transitions from a model 
with several sections to a model with only multi-sections \cite{Morrison:2014era,Anderson:2014yva,Klevers:2014bqa,Garcia-Etxebarria:2014qua,Mayrhofer:2014haa,Mayrhofer:2014laa,Cvetic:2015moa}. Physically 
this corresponds to a Higgsing of charged matter states. By construction, 
the group law of the extended Mordell-Weil group should be inherited from the 
setup with multiple sections. Accordingly, it 
trivially reduces to the standard Mordell-Weil group law in the presence 
of genuine sections. 

The third goal is to extend the discussion to fully include 
matter-coupled non-Abelian gauge theories with gauge group $G$.
Placing these theories on a circle we perform 
large gauge transformations along the circle and explore 
the associated arithmetic structure in the geometry. More precisely, we 
are interested in examining the impact of non-Abelian 
large gauge transformation on the $U(1)^{r+1}$ gauge theory 
obtained in the lower-dimensional Coulomb branch. Here $r$ is 
the rank of $G$ and the additional $U(1)$ is the Kaluza-Klein vector 
of the higher-dimensional metric. 
We show that there indeed is a natural group structure on the set of exceptional 
divisors and rational sections corresponding to these large gauge transformations. 
We also make progress in identifying the geometric symmetry 
corresponding to such transformations. 
First, we show that the transformed exceptional divisors and rational sections can have a standard geometry
interpretation. Second, we employ that  a non-Abelian gauge theory with adjoints can be related
to an Abelian theory with  $r$  $U(1)$s by Higgsing/unHiggsing corresponding to complex structure deformations in
the geometry, at least for the case that there exists a geometric realization of this field theory transition.
Using recent results in \cite{Morrison:2014era,Cvetic:2015ioa} we show
that the postulated group structure on the exceptional divisors gets mapped precisely to 
the  usual Mordell-Weil group law of the geometry corresponding to the Abelian theory. 
All this hints at the existence of a symmetry directly in the geometry associated to the non-Abelian theory that is yet to be formulated explicitly. In fact, on the field theory level this is clear since the circle-compactified theories only differ by a 
non-Abelian large gauge transformation. 
This is obviously a symmetry of an anomaly free gauge theory which indicates that 
the corresponding geometry should be considered physically equivalent to the 
original elliptic fibration. 

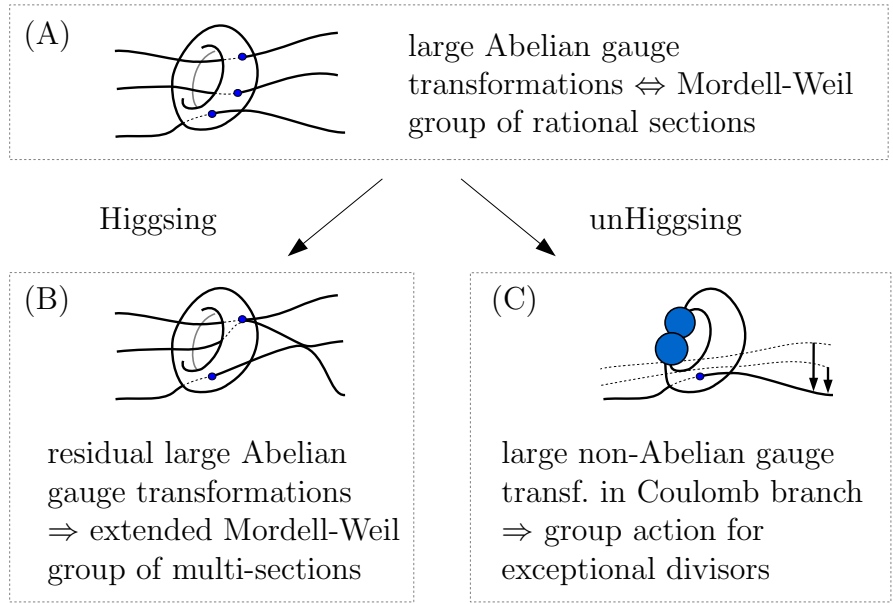
\begin{figure}[t]
\begin{center}
\includegraphics[scale=0.6]{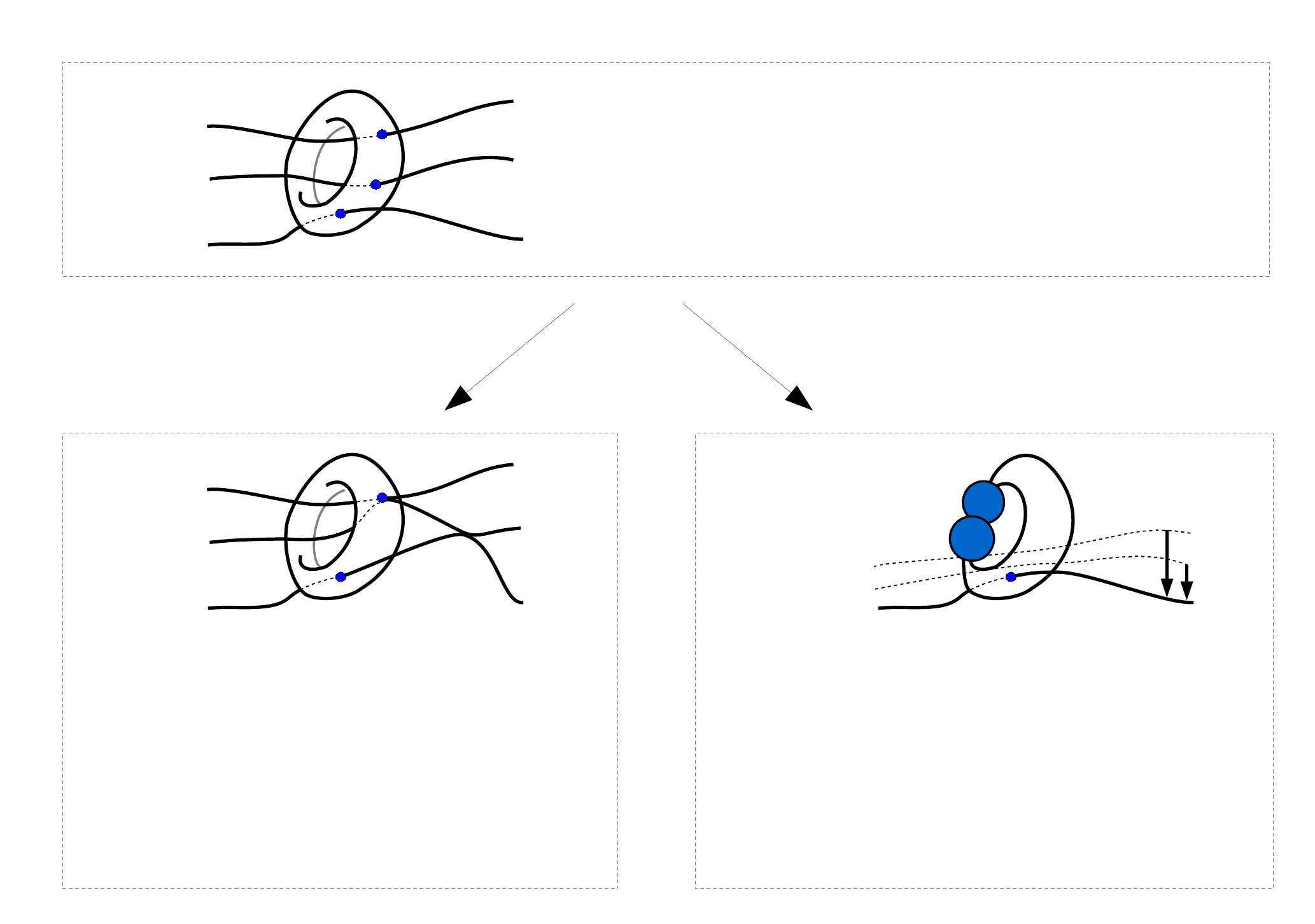}
\begin{picture}(0,0)
\put(-200,195){
\begin{minipage}{6cm} large Abelian gauge \\
transformations  $\Leftrightarrow$ Mordell-Weil   \\
group of rational sections   \\
\end{minipage}}
\put(-335,40){
\begin{minipage}{6cm} residual large Abelian\\
gauge transformations\\
$\Rightarrow$ extended Mordell-Weil\\
group of multi-sections 
\end{minipage}}
\put(-165,40){
\begin{minipage}{6cm} large non-Abelian gauge\\
 transf.~in Coulomb branch\\
$\Rightarrow$ group action for\\
exceptional divisors
\end{minipage}}
\put(-340,220){(A)}
\put(-340,120){(B)}
\put(-165,120){(C)}
\put(-375,150){\begin{minipage}{6cm} \begin{center} 
Higgsing \end{center} \end{minipage}}
\put(-185,150){\begin{minipage}{6cm} \begin{center} 
unHiggsing \end{center} \end{minipage}}
\end{picture}
\vspace*{-.2cm}

\begin{minipage}{12cm}
\caption{Schematic depiction of the various geometric configurations considered in 
this work. The geometries are related by geometric transitions describing 
Higgsing and unHiggsing processes.    \label{HuH-systematics}}
\end{minipage}
\vspace{-1cm}
\end{center}
\end{figure}

Let us 
note that our intuition can be summarized by using \autoref{HuH-systematics}
as follows. First, we are able 
to establish the relation of the Mordell-Weil group to large Abelian gauge transformations
in elliptic fibrations with rational sections depicted in (A).
Using a geometric transition which corresponds to Higgsing in field theory
the resulting fibrations might only admit multi-sections, see (B). 
Therefore, one expects an extended Mordell-Weil group structure for such geometries.
Furthermore, a geometry with rational sections might arise via a geometric 
transition describing a non-Abelian gauge theory. Again this is described by 
a Higgsing in field theory. Such transitions motivate us to transfer the 
Mordell-Weil group structure to a geometry with exceptional divisors. The 
resulting group structure corresponds to large non-Abelian gauge transformation for an 
F-theory effective field theory compactified on a circle and pushed to the Coulomb branch.

It is essential to notice that the investigation of arithmetic structures on the geometry which descend to large gauge transformations
along the circle
allows us to explicitly show anomaly cancelation in F-theory compactifications on genus-one fibered Calabi-Yau fourfolds and 
threefolds. This is obvious since we show at the beginning of this part that the invariance under large gauge transformations
is equivalent to the cancelation of all gauge anomalies in the higher-dimensional theories. Note that the group structure for exceptional
divisors which we propose is not fully established yet. In order words, at the level of homology and also by the intuition of
complex structure deformations we have found convincing
evidence that it exists, but up to now there is no rigorous proof that it is actually geometrically
realized. The same is also true for the extended Mordell-Weil group of multi-sections. In contrast, since the genuine Mordell-Weil group of
rational sections is mathematically established, the corresponding large gauge transformations are symmetries of the theory on the circle,
and the associated anomalies are therefore canceled.

Finally, the discussion of arithmetic structures on elliptic fibrations and their relation to large gauge transformations is also relevant
for understanding the freedom of choice for the zero-section in F-theory. Consider an F-theory compactification on an elliptic fibration
which comes with multiple rational sections generating a non-trivial 
Mordell-Weil group. Then one of these sections has to be chosen as the zero-section
and is matched to the Kaluza-Klein vector in the gauge theory on the circle. Although there should be no preferred choice
for what one calls the zero-section, the effective theories on the circle seem to
differ since the calculation of one-loop Chern-Simons terms yields
different results. However we are able to show that different choices for picking the zero-section are again related by a
large gauge transformation along the circle supplemented by redefining the higher-dimensional $U(1)$ gauge fields.

This part is organized as follows.
We review
the relevant parts for circle compactifications of general four- and six-dimensional gauge theories
to three and five dimensions, respectively,
in \autoref{ch:circle_theories}
and fix our notation. In \autoref{ch:lgts} we describe the action of large gauge transformations along the circle in these theories.
In particular, we show that all four- and six-dimensional gauge anomaly cancelation conditions can be derived from the perspective of the
circle-compactified theories
by imposing that the large gauge transformations act consistently on one-loop Chern-Simons terms in three and respectively five dimensions.
Before we relate these kind of settings to F-theory compactifications on genus-one fibered Calabi-Yau manifolds, 
we give a short introduction into the basic concepts of F-theory in \autoref{ch:f_basics}. Finally, in \autoref{ch:arith}
we match the field-theoretic large gauge transformations along the F-theory circle to arithmetic structures on genus-one fibrations
thereby concretely conjecturing
group structures for exceptional divisors and multi-sections which have not been considered before.
We conclude in \autoref{ch:zero_sec} with the closely related work on
the choice of picking the zero-section in F-theory compactifications on elliptic fibrations.

\chapter{Circle Compactification of Gauge Theories}\label{ch:circle_theories}

\section{General Setup}\label{sec:gauge_gen}

Let us start by introducing some general notions about Abelian and 
non-Abelian gauge theories in four and six spacetime dimensions. Unless stated differently our notation applies
to both kinds of settings. Differences between four and six dimensions will then be highlighted at prominent positions.
We denote by $G$ a simple non-Abelian gauge group\footnote{The generalization to semi-simple gauge groups is straightforward but however omitted
for convenience of presentation.} with
gauge bosons $\hat A$ and Lie algebra $\mathfrak g$. Introducing 
the Lie algebra generators $T_\cI$, $\cI = 1,\ldots,\text{dim}\,\mathfrak g$ we expand 
\beq \label{hatA_expand}
   \hat A = \hat A^\cI T_\cI  = \hat A^I T_I + \hat A^{\boldsymbol{\alpha}} T_{\boldsymbol{\alpha} }
\eeq
where $T_I$, $I = 1 ,\dots ,\rk \mathfrak g$ are the generators of the Cartan subalgebra and $T_{\boldsymbol \alpha}$
are the remaining generators labeled by the roots ${\boldsymbol \alpha}$. 
In addition, we will allow for a number $n_{U(1)}$ of Abelian gauge bosons which are 
denoted by $\hat A^m$ with $m =1,\dots ,n_{U(1)}$.

Since we are in particular interested in anomalies of four- and six-dimensional theories,
let us introduce the relevant fields which induce anomalies at one-loop in the following.
Furthermore, it will also become important that a classical 
four-dimensional or six-dimensional gauge theory does not 
necessarily have to be gauge invariant in order to lead to a consistent quantum theory. In fact, it is well-known 
that often classical gauge non-invariance is required to 
cancel one-loop gauge anomalies induced by chiral fields. Famously, this 
is done by the Green-Schwarz mechanism \cite{Green:1984sg,Sagnotti:1992qw,Sadov:1996zm}. Therefore we also introduce
the fields which participate in the latter.

\vspace{0.7cm}

In \textit{four dimensions} the following modes contribute to anomalies:
\begin{itemize}
 \item Spin-$\sfrac{1}{2}$ Weyl fermions in a representation $R$ of the non-Abelian gauge group and with $U(1)$ charges
 $q_m$ are denoted by $\hat \Bpsi^{\sfrac12} (R,q)$. The covariant derivative for left-handed $\hat \Bpsi^{\sfrac12} (R,q)$
takes the form 
\beq \label{def-Dpsi4}
  \hat \cD_\mu \hat\Bpsi^{\sfrac12} (R,q) = \big(\hat\nabla_\mu 
  - i \hat A^\cI_\mu  T^R_\cI   - i q_m \hat A^m_\mu \big) \hat\Bpsi^{\sfrac12} (R,q)  \, .
\eeq
We expand $\hat \Bpsi^{\sfrac12} (R,q)$ in an eigenbasis $\hat \Bpsi^{\sfrac12} (w,q)$ associated to the weights $w$ of $R$.
They enjoy the property
\begin{align}
 T^R_I \ \hat\Bpsi^{\sfrac12}(w,q) = w_I \ \hat \Bpsi^{\sfrac12}(w,q)
\end{align}
for the Cartan directions, where
$w_I := \langle \boldsymbol{\alpha}^\vee_I , w \rangle $ are the Dynkin labels  and
$\boldsymbol{\alpha}^\vee_I$ is the simple coroot associated to $T_I$. We refer to \autoref{s:lie_conv}
for our conventions in the theory of Lie algebras.
 
 Finally we denote the chiral index of the fields $\hat \Bpsi^{\sfrac12} (R,q)$ by $F_{\sfrac12} (R,q)$
 \begin{align}
  F_{\sfrac12} (R,q) = F_{\sfrac12}^{\textrm{left}} (R,q) - F_{\sfrac12}^{\textrm{right}} (R,q)
 \end{align}
 such that the effective number of chiral modes in the theory at hand is given by $\dim (R)\cdot F_{\sfrac12}(R,q)$.
 
\item The Green-Schwarz mechanism is mediated by axions\footnote{Note
that this index $\alpha$ counts axions and should not be confused with $\boldsymbol{\alpha}$
labeling the roots of $\mathfrak g$.} $\hat \rho_\alpha$, $\alpha = 1,\ldots ,n_{\rm ax}$ with a gauged
shift-symmetry under the $U(1)$ vectors $\hat A^m$. More precisely, their covariant derivative reads
\begin{align}\label{ax_gauging}
 \hat\cD \hat \rho_\alpha = d \hat \rho_\alpha + \theta_{\alpha m} \hat A^m 
\end{align}
with $\theta_{\alpha m}$ constant.
The classical non-gauge-invariant counter-terms are given by
\beq \label{GS4}
   \hat S^{(4)}_{\rm GS} =  - \frac{1}{4}\int   \tensor{\eta}{_\alpha^\beta}\hat\rho_\beta \Big( 
   -\frac{1}{4} a^\alpha \text{tr} \hat \cR \wedge \hat \cR +
   b^\alpha \lambda_{\mathfrak{g}}^{-1} \text{tr}_{f} (\hat F \wedge \hat F)  + b^\alpha_{mn}  \hat F^m \wedge \hat F^n  \Big)\ ,
\eeq
where $ \tensor{\eta}{_\alpha^\beta}$ is a constant square matrix, and
$a^\alpha$, $b^\alpha$ and $b^\alpha_{mn}$ are the Green-Schwarz anomaly coefficients. We denote by $\text{tr}_{f}$ the trace in
the fundamental representation of the gauge algebra.
The expressions $\hat F$ and $\hat F^m$ denote the field strengths of $\hat A$ and $\hat A^m$, respectively, and $\hat \cR$ is the curvature two-form.
The algebra-specific coefficient is given by
\begin{align}
 \lambda_{\mathfrak{g}}^{-1}= \frac{1}{2} \langle {\boldsymbol \alpha}_{\rm max}, {\boldsymbol \alpha}_{\rm max} \rangle  \, ,
\end{align}
 where ${\boldsymbol \alpha}_{\rm max}$
is the root of maximal length (see \autoref{s:lie_conv}).
\end{itemize}

\vspace{0.7cm}

In \textit{six dimensions} we introduce the following types of fields:
\begin{itemize}
 \item Spin-$\sfrac{1}{2}$ Weyl fermions in a representation $R$ of the non-Abelian gauge group and with $U(1)$ charges
 $q_m$ are written as $\hat \Bpsi^{\sfrac 1 2} (R,q)$.
 Note that in the following we do note impose an additional symplectic Majorana condition which is in principle possible in six dimensions.
 We indicated this by using bold symbols.
For left-handed $\hat \Bpsi^{\sfrac 1 2} (R,q)$, \textit{i.e.}~they transform as $\big(\frac{1}{2},0\big)$ of the massless little group
 $SO(4) \underset{\rm locally}{\cong} SU(2) \times SU(2)$,
the covariant derivative reads
\beq \label{def-Dpsi6}
  \hat \cD_\mu \hat\Bpsi^{\sfrac 1 2} (R,q) = \big(\hat \nabla_\mu 
  - i \hat A^\cI_\mu  T^R_\cI   - i q_m \hat A^m_\mu \big) \hat\Bpsi^{\sfrac 1 2} (R,q)  \, .
\eeq
Again, we can expand $\hat \Bpsi^{\sfrac 1 2} (R,q)$ in an eigenbasis $\hat \Bpsi^{\sfrac 1 2} (w,q)$ associated to the weights $w$ of $R$ 
with
\begin{align}
 T^R_I \ \hat\Bpsi^{\sfrac 1 2}(w,q) = w_I \ \hat \Bpsi^{\sfrac 1 2}(w,q)\, .
\end{align}

 Finally we write $F_{\sfrac 1 2} (R,q)$ for the chiral index of the fields $\hat \Bpsi^{\sfrac 1 2} (R,q)$
 \begin{align}
  F_{\sfrac 1 2} (R,q) = F^{\textrm{left}}_{\sfrac 1 2} (R,q) - F^{\textrm{right}}_{\sfrac 1 2} (R,q)
 \end{align}
 such that the effective number of chiral $\big(\frac{1}{2},0\big)$-modes in the theory at hand is given by $\dim (R)\cdot F_{\sfrac 1 2}(R,q)$.
 We stress once more, that in our conventions the $\hat \Bpsi^{\sfrac 1 2} (R,q)$ are not subject to a symplectic Majorana condition.
 
 \item By $T_{\rm sd}$ and 
$T_{\rm asd}$ we denote the number of self-dual and 
anti-self-dual tensors respectively. In the following we will not treat non-Abelian tensor fields, for which no Lagrangian description is known,
but only restrict to Abelian ones. These fields are indeed chiral since self-dual tensors transform as
$(1,0)$ under $SU(2) \times SU(2)$, anti-self-dual tensors as $(0,1)$.
We write $\hat B^\alpha$, $\alpha = 1,\ldots, T_{\rm sd}+T_{\rm asd}$ for the two-form fields, and introduce the chiral index
\begin{align}
 \mathfrak T = T_{\rm sd} - T_{\rm asd} \, .
\end{align}

Furthermore, the $\hat B^\alpha$ can mediate a Green-Schwarz mechanism since
on the one hand one can assign to them
modified field strengths and therefore a non-trivial transformation under six-dimensional gauge transformations
(see \textit{e.g.}~\cite{Bonetti:2011mw,Grimm:2013oga} for a more complete discussion)
\begin{align}
 \delta \hat B^\alpha = d \hat \Lambda^\alpha - \frac{1}{2}a^\alpha \tr \, \hat l d \hat \omega -2 b^\alpha \tr \, \hat \Lambda d \hat A
 -2 b^\alpha_{mn}  \hat \Lambda^m d \hat A^n \ , 
\end{align}
where $\hat l$, $\hat \Lambda$, $\hat \Lambda^m$, $\hat \Lambda^\alpha$
are the parameters of local Lorentz, gauge and two-form transformations, respectively,
and $\hat \omega$ is the spin connection.
On the other hand the $\hat B^\alpha$ can appear with topological couplings
\beq \label{GS6}
    \hat S^{(6)}_{\rm GS} = - \int  \eta_{\alpha \beta } \hat B^\beta\wedge 
    \Big(\frac{1}{4} a^\alpha \text{tr} \hat \cR \wedge \hat \cR +
    b^\alpha \lambda_{\mathfrak{g}}^{-1} \text{tr}_{f} (\hat F \wedge \hat F)  + b^\alpha_{mn}  \hat F^m \wedge \hat F^n  \Big)\ ,
\eeq
where $a^\alpha$, $b^\alpha$ and $b^\alpha_{mn}$ denote the Green-Schwarz coefficients.
The matrix $\eta_{\alpha \beta}$ is constant, symmetric in its indices and its signature consists of $T_{\rm sd}$
positive signs and $T_{\rm asd}$ negative ones.
The $\hat F$ and $\hat F^m$ denote the field strengths of $\hat A$ and $\hat A^m$, respectively,
and $\hat \cR$ is the curvature two-form. The trace in the fundamental representation of $\mathfrak g$ is written as $\text{tr}_{f}$.
We also used
\begin{align}
 \lambda_{\mathfrak{g}}^{-1}= \frac{1}{2}\langle {\boldsymbol \alpha}_{\rm max}, {\boldsymbol \alpha}_{\rm max} \rangle 
\end{align}
with $ {\boldsymbol \alpha}_{\rm max}$
the root of maximal length.

\item We write $\hat \Bpsi^{\sfrac 32}_{\mu}$ for spin-${\sfrac 32}$ fermions. Left-handed $\hat \Bpsi^{\sfrac 32}_{\mu}$ transform as
$\big(1,\frac{1}{2}\big)$ under $SU(2) \times SU(2)$, right-handed ones as $\big(\frac{1}{2},1\big)$. The chiral index is denoted by
\begin{align}
 F_{\sfrac 32} = F_{\sfrac 32}^{\textrm{left}} - F_{\sfrac 32}^{\textrm{right}} \, .
\end{align}
\end{itemize}

In the following \autoref{sec:4d_6d_anomalies} we account for the anomalies which are induced by these fields in full detail. 
For the moment it is worthwhile to realize that both the four-dimensional and six-dimensional settings 
are characterized by Green-Schwarz coefficients $a^\alpha$, $b^\alpha$, $b^\alpha_{mn}$. 
Indeed, in the following we will see that the fields $\hat \rho_\alpha$ and $\hat B^\alpha$ appearing 
in \eqref{GS4} and \eqref{GS6} are both captured by vectors $A^\alpha$ after a circle-compactification and dualization.
This slight abuse of notation will allow us to 
investigate the four-dimensional and six-dimensional case simultaneously at once.

\section{Anomaly Cancelation}\label{sec:4d_6d_anomalies}

In \autoref{sec:anom_lgt} we will show in detail that if one considers the action of large gauge transformations along the circle on
one-loop Chern-Simons, one can recover all gauge anomaly cancelation conditions of the uncompactified theory in a neat way.
Therefore let us shortly collect the anomaly equations in four and six dimensions. For a more general recap
of anomalies in quantum field theory see \autoref{ch:ano} at the beginning of this thesis and the references therein.

\subsection{Four Dimensions}

\begin{figure}[!h]
\centering
 \includegraphics{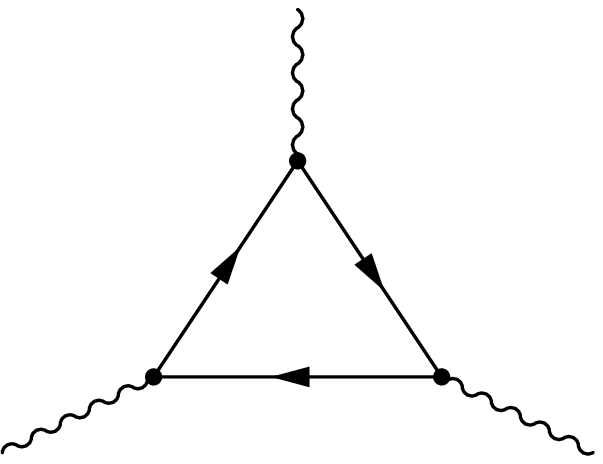}
 \caption{The four-dimensional one-loop anomaly has gravitons or
 gauge bosons as external legs. The modes running in the loop are chiral fermions.}
\label{4dloop}
 \end{figure}

Potential anomalies in four dimensions stem from loops of chiral spin-$\sfrac{1}{2}$ fermions as depicted in \autoref{4dloop}.
As mentioned before, a classical Green-Schwarz mechanism mediated by axions can be exploited in order to cancel the latter.
\begin{figure}[!h]
\centering
 \includegraphics{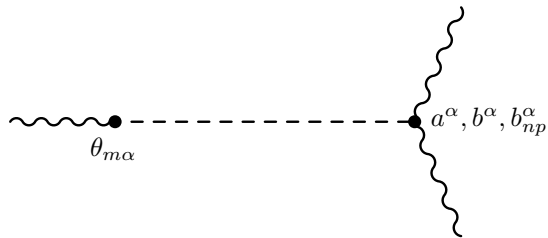}
 \caption{The four-dimensional Green-Schwarz mechanism is mediated by axions. The left external leg is always an Abelian gauge boson $\hat A^m$ 
 while the
 external legs on the right-hand side are either gauge bosons or gravitons.}
 \label{fig:GS4d}
 \end{figure}
In our conventions the one-loop anomaly polynomial for left-handed Weyl fermions which transform in
some representation $R$ (which might also possibly be the singlet representation)
of the non-Abelian gauge group and with $U(1)$ charges $q_m$ takes the form \cite{AlvarezGaume:1983ig}
\begin{align}
 I_{\sfrac{1}{2}}^{\textrm{1-loop}}(R,q) = \, & -\frac{1}{24}\,\textrm{dim}(R) \, q_m  \hat F^m p_1 ( \hat \cR )  + 
 \frac{1}{6} \tr_R \hat F^3 + \frac{1}{2} q_m \hat F^m \tr_R \hat F^2  
 \nn \\ & + \frac{1}{6}\, \textrm{dim}(R) \,
 q_m q_n q_p \hat F^m \hat F^n \hat F^p    \nn \\
 = \, & \frac{1}{48}\,\textrm{dim}(R) \, q_m  \hat F^m \tr \hat \cR^2  + 
 \frac{1}{6} \tr_R \hat F^3 + \frac{1}{2} q_m \hat F^m \tr_R \hat F^2  
 \\ & + \frac{1}{6}\, \textrm{dim}(R) \,
 q_m q_n q_p \hat F^m \hat F^n \hat F^p   \, , \nn
\end{align}
with $\tr_R$ the trace taken in the representation $R$ and the first Pontryagin class given by
\begin{align}
 p_1 ( \hat \cR ) = -\frac{1}{2} \tr \hat \cR^2 \, .
\end{align}

The anomaly polynomial derived from the Green-Schwarz counterterms depicted in \autoref{fig:GS4d} takes the following factorized form
\begin{align}
 I^{\textrm{GS}} &= - \frac{1}{8}\theta_{m\alpha}\Big ( a^\alpha \hat F^m p_1 ( \hat \cR )
 +2 \frac{b^\alpha}{\lambda_{\mathfrak g}} \hat F^m \tr_f \hat F^2
 + 2 b^\alpha_{np} \hat F^m \hat F^n \hat F^p \Big ) \,  \nn \\
 & = \frac{1}{16}a^\alpha \theta_{m\alpha} \hat F^m \tr \hat \cR^2
 -\frac{1}{4} \frac{b^\alpha}{\lambda_{\mathfrak g}} \theta_{m\alpha} \hat F^m \tr_f \hat F^2
 -\frac{1}{4} b^\alpha_{np} \theta_{m\alpha} \hat F^m \hat F^n \hat F^p \, ,
\end{align}
which can be derived straightforwardly by using the anomalous variation of the classical action in the descent equations.

The vanishing condition of the full six-form anomaly polynomial $I_6$ then comprises the one-loop part and the Green-Schwarz contributions
\begin{align}
 I_6 := \sum_{R,q} F_{\sfrac 1 2}(R,q) \ I_{\sfrac{1}{2}}^{\textrm{1-loop}} + I^{\textrm{GS}} \overset{!}{=} 0
\end{align}
This leads to the cancelation conditions\footnote{All
symmetrizations over $n$ indices include a factor of $\frac{1}{n!}$.}
{\allowdisplaybreaks\begin{subequations}
\label{4danomalies}
\begin{empheq}[box=\widefbox]{align}
\label{4d_mG_anomaly} -3  a^\alpha \theta_{m\alpha} &=  \sum_{R,q} \dim(R) \ F_{\sfrac 1 2}(R,q) \, q_m  \ ,  \\ 
 \label{4d_nA_anomaly} 0 &= \sum_{R,q} F_{\sfrac 1 2}(R,q) \, E_R \, , \\
\label{4d_nA_A_anomaly} \frac{1}{2} \frac{b^\alpha}{\lambda_{\mathfrak{g}}} \theta_{m\alpha} &= \sum_{R,q}
F_{\sfrac 1 2}(R,q) \, q_m A_R  \ ,  \\ 
\label{4d_A_anomaly} \frac{3}{2} b^\alpha_{(mn}\theta_{p)\alpha} &=  \sum_{R,q} \dim(R) \  F_{\sfrac 1 2}(R,q) \, q_m q_n q_p \, , 
\end{empheq}
\end{subequations}}
where we employed the definitions
\begin{subequations}\label{e:Casimirs}
\begin{align}
\tr_R \hat F^2 &= A_R \ \tr_f \hat F^2 \, ,  \\
 \tr_R \hat F^3 &= E_R\, \tr_f \hat F^3 \, , \nn \\
 \tr_R \hat F^4 &= B_R \ \tr_f \hat F^4  + C_R \ (\tr_f \hat F^2 )^2 \, . \nn
\end{align}
\end{subequations}
The last definition is only introduced for convenience since it will be used in the six-dimensional setup later.

It will be essential in \autoref{sec:anom_lgt}
that we have managed to express the anomaly cancelation conditions which involve non-Abelian gauge factors
by an alternative representation. 
This is done via replacing the Casimirs $A_R$ and $E_R$ by certain sums over all weights of the given representation $R$.
As we show in \autoref{app:traces}, the equations \eqref{4danomalies} are equivalent to
{\allowdisplaybreaks\begin{subequations}
\label{4danomalies_alternative}
\begin{empheq}[box=\widefbox]{align}
\label{4d_anomaly_alt1} -\frac{1}{4}  a^\alpha \theta_{m\alpha} &= \frac{1}{12} \sum_{R,q}  F_{\sfrac 1 2}(R,q) \sum_{w \in R} \, q_m   \ ,  \\ 
\label{4d_anomaly_alt2}   0 &= \sum_{R,q} F_{\sfrac 1 2}(R,q)\sum_{w \in R}\,w_{I} w_{J} w_{K}  \, , \\
\label{4d_anomaly_alt3} \frac{1}{2} b^\alpha 
\theta_{m\alpha}\, \cC_{IJ} &= \sum_{R,q} F_{\sfrac 1 2}(R,q) \sum_{w \in R}\,q_m w_{I} w_{J}    \ ,  \\ 
\label{4d_anomaly_alt4} \frac{3}{2} b^\alpha_{(mn}\theta_{p)\alpha} &=  \sum_{R,q}   F_{\sfrac 1 2}(R,q) \sum_{w \in R}\, q_m q_n q_p \, .
\end{empheq}
\end{subequations}}
Although we have in principle only rewritten those anomaly cancelation conditions which involve the non-Abelian gauge symmetry,
we once more collect the full set of conditions in this box for later convenience.
Note also that for the same reason all factors of $\dim (R)$ have been rewritten as a sum over weights
of the non-Abelian gauge group. Furthermore, we stress that this way of writing of course involves a lot of redundancy
since there are many more equations due to the appearance of new indices $I,J,K$.
Indeed, as shown in \autoref{app:traces} some equations are trivially fulfilled as a group-theoretical identity,
others are equivalent to each other.

\subsection{Six Dimensions}
\begin{figure}[!h]
\centering
 \includegraphics{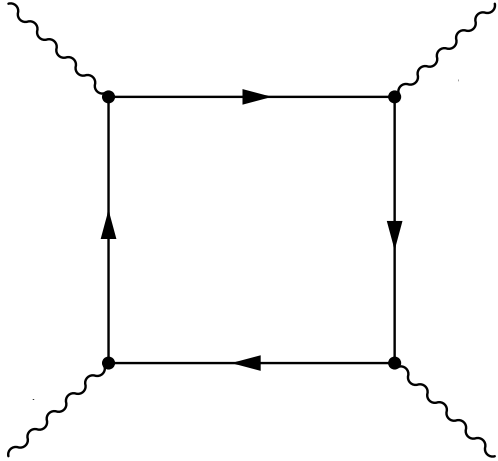}
 \caption{The six-dimensional one-loop anomaly has gravitons or
 gauge bosons as external legs. The modes running in the loop can be chiral spin-$\sfrac{1}{2}$ fermions, spin-$\sfrac{3}{2}$ fermions
or (anti-)self-dual tensors.}
 \label{box6d}
\end{figure}

\begin{figure}[!h]
\centering
 \includegraphics{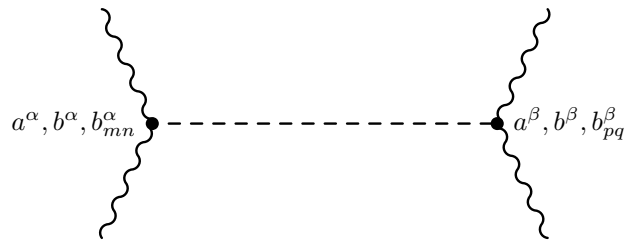}
 \caption{The six-dimensional Green-Schwarz mechanism is mediated by (anti-)self-dual tensors. The external legs can be gauge bosons or
 gravitons.}
 \label{GSpic6d}
 \end{figure}

 The chiral modes in six dimensions which induce anomalies at one-loop are spin-$\sfrac{1}{2}$ and spin-$\sfrac{3}{2}$ fermions
 as well
 as (anti-)self-dual tensors. The corresponding anomalous box diagrams are depicted in
 \autoref{box6d}.
The (anti-)self-dual tensors can additionally participate in a Green-Schwarz mechanism at tree-level as
 illustrated in \autoref{GSpic6d}.
 
 Let us now write down the one-loop anomaly polynomials for the different types of fields in our conventions.
 For a left-handed spin-$\sfrac{1}{2}$ fermion which transforms in a representation $R$ and has $U(1)$ charge $q$, it reads
{\allowdisplaybreaks \begin{align}
  I_{\sfrac{1}{2}}^{\textrm{1-loop}} (R,q) = \, & \frac{1}{360} \dim(R) \  \Big(-4 p_2 (\hat \cR) + 7 p^2_1 (\hat \cR) \Big)
  + \frac{1}{3} \tr_R \hat F^2 p_1(\hat \cR ) \nn \\
  \, & + \frac{1}{3} \dim(R) \  q_m q_n \hat F^m \hat F^n p_1 ( \hat \cR )
   + \frac{2}{3} \tr_R \hat F^4  + \frac{8}{3} q_m \hat F^m \tr_R \hat F^3 \nn \\
 \, & + 4 q_m q_n  \hat F^m \hat F^n  \tr_R \hat F^2 + \frac{2}{3} \dim(R) \  q_m q_n q_p q_q \hat F^m \hat F^n  \hat F^p \hat F^q   \nn \\
  = \, & \frac{1}{360} \dim(R) \  \Big(\tr \hat \cR^4 + \frac{5}{4} (\tr \hat \cR^2)^2 \Big)
  - \frac{1}{6} \tr_R \hat F^2 \tr \hat \cR^2 \\
  \, & - \frac{1}{6} \dim(R) \  q_m q_n \hat F^m \hat F^n\tr \hat \cR^2 
   + \frac{2}{3} \tr_R \hat F^4  + \frac{8}{3} q_m \hat F^m \tr_R \hat F^3 \nn \\
 \, & + 4 q_m q_n  \hat F^m \hat F^n  \tr_R \hat F^2 + \frac{2}{3} \dim(R) \  q_m q_n q_p q_q \hat F^m \hat F^n  \hat F^p \hat F^q   \, , \nn
 \end{align}}
for a self-dual tensor we have
\begin{align}
  I_{\textrm{sd}}^{\textrm{1-loop}} &= \frac{2}{45}  \Big( -7 p_2( \hat \cR )  + p^2_1( \hat \cR ) \Big) \nn \\
  &= \frac{28}{360}  \Big(\tr \hat \cR^4 + \frac{5}{4} (\tr \hat \cR^2)^2 \Big)
  - \frac{1}{8}(\tr \hat \cR^2)^2  \, ,
 \end{align}
and finally for a left-handed spin-$\sfrac{3}{2}$ fermion
\begin{align}
  I_{\sfrac{3}{2}}^{\textrm{1-loop}} &= \frac{1}{72}  \Big( - 196 p_2( \hat \cR )  + 55 p^2_1( \hat \cR ) \Big)  \nn \\
  &= \frac{245}{360}  \Big(\tr \hat \cR^4 + \frac{5}{4} (\tr \hat \cR^2)^2 \Big)
  - (\tr \hat \cR^2)^2 \, .
 \end{align}
 Note that for right-handed fermions and anti-self-dual tensors one picks up a minus sign, respectively. Again
 for the gravitational anomalies we employed the definitions
 of the Pontryagin classes
 \begin{subequations}
 \begin{align}
  p_1 ( \hat \cR ) &= -\frac{1}{2} \tr \hat \cR^2 \, , \\
  p_2 ( \hat \cR ) &= \frac{1}{8} \big ( -2\ \tr \hat \cR^4 + (\tr \hat \cR^2)^2 \big )  \, .
 \end{align}
 \end{subequations}
 
 The classical Green-Schwarz counterterms contribute to the anomaly polynomial in the factorized form
 \begin{align}
  I^{\textrm{GS}} = \, &  \frac{1}{2} \eta_{\alpha\beta} \Big( a^\alpha p_1 ( \hat \cR )
  - 2 \frac{b^\alpha}{\lambda_{\mathfrak g}} \tr_f \hat F^2 - 2 b^\alpha_{mn} \hat F^m \hat F^n \Big ) 
  \Big(  a^\beta p_1 ( \hat \cR )
  - 2 \frac{b^\beta}{\lambda_{\mathfrak g}} \tr_f \hat F^2 - 2 b^\beta_{pq} \hat F^p \hat F^q \Big ) \nn \\
   = \, &  \frac{1}{8} a^\alpha a^\beta \eta_{\alpha\beta} (\tr \hat \cR^2 )^2
  + a^\alpha \frac{b^\beta}{\lambda_{\mathfrak g}} \eta_{\alpha\beta} \ \tr_f \hat F^2 \tr \hat \cR^2 
  + a^\alpha b^\beta_{mn} \eta_{\alpha\beta} \ \hat F^m \hat F^n \tr \hat \cR^2 \\
  \, &+ 2 \frac{b^\alpha}{\lambda_{\mathfrak g}} \frac{b^\beta}{\lambda_{\mathfrak g}} \eta_{\alpha\beta} \ (\tr_f \hat F^2)^2 
  + 4 b^\alpha_{mn} \frac{b^\beta}{\lambda_{\mathfrak g}} \eta_{\alpha\beta} \ \hat F^m \hat F^n \tr_f \hat F^2
  + 2 b^\alpha_{mn} b^\beta_{pq}  \eta_{\alpha\beta} \ \hat F^m \hat F^n \hat F^p \hat F^q \, . \nn
 \end{align}

The cancelation of all anomalies then requires the vanishing of the full eight-form anomaly polynomial $I_8$, consisting of the one-loop
contributions from chiral fermions and (anti-)self-dual tensors as well as the classical Green-Schwarz part
\begin{align}
 I_8 := \sum_{R,q} F_{\fe}(R,q) \ I_{\sfrac{1}{2}}^{\textrm{1-loop}} (R,q) + \mathfrak T \
 I_{\textrm{sd}}^{\textrm{1-loop}}+
 F_{\gr} \ I_{\sfrac{3}{2}}^{\textrm{1-loop}}
 +  I^{\textrm{GS}} \overset{!}{=} 0 \, .
\end{align}
This condition yields the following set of equations
{\allowdisplaybreaks\begin{subequations}\label{e:6d_anom}
\begin{empheq}[box=\widefbox]{align}
\label{e:6d_grav_1}0 &= \sum_{R,q} \dim(R) \ F_{\fe}(R,q) + 28 \mathfrak T  + 245 F_{\gr} \, , \\
\label{e:6d_grav_2}a^\alpha a^\beta \eta_{\alpha\beta} &= \mathfrak T  + 8 F_{\gr} \, , \\
 \label{e:6d_anom_1} 6  a^\alpha \frac{b^\beta}{\lambda_{\mathfrak{g}}} \eta_{\alpha\beta} &= \sum_{R,q} F_{\fe}(R,q) \ A_R  \, , \\
 \label{e:6d_anom_2}  6  a^\alpha b^\beta_{mn} \eta_{\alpha\beta} &= \sum_{R,q} \dim(R) \  F_{\fe}(R,q) \  q_m q_n \, , \\
 \label{e:6d_anom_3}  0 &= \sum_{R,q} F_{\fe}(R,q) \  B_R \, , \\
 \label{e:6d_anom_4}  -3 \frac{b^\alpha}{\lambda_{\mathfrak{g}}} \frac{b^\beta}{\lambda_{\mathfrak{g}}} \eta_{\alpha\beta} &=
 \sum_{R,q} F_{\fe}(R,q) \  C_R \, , \\
 \label{e:6d_anom_5}  0 &= \sum_{R,q} F_{\fe}(R,q) \  q_m E_R \, , \\
 \label{e:6d_anom_6}  - b^\alpha_{mn}\frac{b^\beta}{\lambda_{\mathfrak{g}}} \eta_{\alpha\beta} &= \sum_{R,q} F_{\fe}(R,q) \ q_m q_n A_R \, , \\
 \label{e:6d_anom_7}  -3 b^\alpha_{(mn} b^\beta_{pq)} \eta_{\alpha\beta}
 &= \sum_{R,q} \dim(R) \  F_{\fe}(R,q) \  q_m q_n q_p q_q \, ,
\end{empheq}
\end{subequations}}
with the group theoretical constants $A_R$, $B_R$, $C_R$, $E_R$ defined in \eqref{e:Casimirs}.

Again like in \eqref{4danomalies} and \eqref{4danomalies_alternative} there is an alternative
representation of those anomaly equations which involve the non-Abelian gauge symmetry
as shown in \autoref{app:traces}.
Note that we also display the pure gravitational anomalies in a different manner such that the connection to Chern-Simons terms becomes more
apparent. In particular the new representations \eqref{e:6d_grav_alter_1}, \eqref{e:6d_grav_alter_2} are formed by taking
appropriate linear combinations of the original conditions 
\eqref{e:6d_grav_1}, \eqref{e:6d_grav_2}.
{\allowdisplaybreaks\begin{subequations}\label{e:6d_anom_alernative}
\begin{empheq}[box=\widefbox]{align}
\label{e:6d_grav_alter_1} 4(\mathfrak T + 11 F_{\gr} ) &= \frac{1}{6} \Big (-\sum_{R,q}  F_{\fe}(R,q) \sum_{w \in R}\, 1
- 4 \mathfrak T  + 19 F_{\gr} \Big ) \, , \\ 
\label{e:6d_grav_alter_2} \frac{1}{4} a^\alpha a^\beta \eta_{\alpha\beta} &= \frac{1}{120} \Big (-\sum_{R,q} F_{\fe}(R,q) \sum_{w \in R}\, 1
+ 2 \mathfrak T  - 5 F_{\gr} \Big ) \, , \\
\label{e:6d_alter_3} \frac{1}{2}  a^\alpha b^\beta \eta_{\alpha\beta}\, \cC_{IJ} &= \frac{1}{12} \sum_{R,q} F_{\fe}(R,q) 
  \sum_{w \in R}\, w_{I} w_{J} \, , \\
\label{e:6d_alter_4}  \frac{1}{2} 
a^\alpha b^\beta_{mn} \eta_{\alpha\beta} &= \frac{1}{12} \sum_{R,q}   F_{\fe}(R,q) \sum_{w \in R}\,   q_m q_n \, , \\
\label{e:6d_alter_5} - 3 b^\alpha b^\beta \eta_{\alpha\beta}\, \cC_{(IJ} \cC_{KL)}  
 &= \sum_{R,q} F_{\fe}(R,q) \sum_{w \in R}\, w_{I} w_{J}  w_{K} w_{L} \, , \\
\label{e:6d_alter_6} 0 &= \sum_{R,q} F_{\fe}(R,q) \sum_{w \in R}\, q_m w_{I} w_{J}  w_{K}   \, , \\
\label{e:6d_alter_7} - b^\alpha_{mn} 
b^\beta \eta_{\alpha\beta}\, \cC_{IJ} &= \sum_{R,q} F_{\fe}(R,q) \sum_{w \in R}\, q_m q_n  w_{I} w_{J}\, , \\
\label{e:6d_alter_8} -3 b^\alpha_{(mn} b^\beta_{pq)} \eta_{\alpha\beta}
 &= \sum_{R,q}  F_{\fe}(R,q) \sum_{w \in R}\,   q_m q_n q_p q_q \, .
\end{empheq}
\end{subequations}}
Similar to the situation in four dimensions, demanding that these equations hold true for all index choices $I,J,K,L$ involves of course a lot of
redundancy since some equations are trivially fulfilled as group-theoretical identities and others are equivalent to each other.
However note that the single type of equation \eqref{e:6d_alter_5} comprises both pure non-Abelian gauge anomalies
\eqref{e:6d_anom_3} and \eqref{e:6d_anom_4} as
shown in \autoref{app:traces}.

\section{Circle Compactification}

In the next step we compactify the four- or six-dimensional theory on a circle and push it to the Coulomb branch
by allowing for a non-vanishing Wilson line background for the gauge field component along the 
circle. We stress that we assume generic values for these VEVs throughout this thesis. In particular they should not be
integer multiples of the radius.
Note that we will not account for the full circle reduction but rather review the relevant parts. A complete treatment
(with application to F-theory) can for example
be found in \cite{Grimm:2010ks,Bonetti:2011mw,Grimm:2013oga}.

\subsection{Reduction of the Fields}\label{e:field-red}
First let us fix some notation for convenience.
We stress that all four- or six-dimensional objects carry a 'hat' while the three- or five-dimensional ones
which appear in the circle-reduction do not. 
From the metric in four or six dimensions one finds at lowest level the three-
or five-dimensional metric 
$g_{\mu \nu}$, the Kaluza-Klein vector $A^0$, and the radius modulus $r$ of the circle.
Thus the higher-dimensional line element is expanded according to
\begin{align}\label{e:metric_reduction}
 d \hat s^2 = g_{\mu\nu}dx^\mu dx^\nu + r^2 Dy^2 \, , \qquad Dy := dy - A^0_\mu dx^\mu \, ,
\end{align}
with $x^\mu$ the three- or five-dimensional coordinates, respectively, and $y$ the coordinate along the circle.
Performing the Kaluza-Klein reduction of the vector fields one finds at lowest Kaluza-Klein level  
a number of $\text{dim}\, \mathfrak g$ gauge fields $A^\cI$ and
$\text{dim}\, \mathfrak g$ Wilson line scalars $\zeta^\cI$ from reducing $\hat A$. In addition 
one has  $n_{U(1)}$ $U(1)$ gauge fields $A^m$ and Wilson line scalars $\zeta^m$ 
from reducing $\hat A^m$.
In particular the gauge fields are expanded as
\begin{align} \label{red-AI}
 \hat A^\cI = A^\cI - \zeta^\cI r Dy\  , \qquad \hat A^m = A^m - \zeta^m r Dy \ .
\end{align}
The $A^\cI$ constitute gauge fields of the lower-dimensional version of the gauge group $G$
while the $\zeta^\cI$ transform in the adjoint representation of $G$. In other words, denoting the gauge parameters by $\Lambda^\cI(x)$
and $\Lambda^m(x)$
   one has  
\bea \label{lower-dim-gauge}
   \delta A^\cI &=&  d\Lambda^\cI + f^\cI_{\cJ \cK} \Lambda^\cJ   A^\cK \ , 
   \qquad \  \ \delta \zeta^\cI  =  f^{\cI}_{\cJ \cK} \zeta^\cJ \Lambda^\cK\ ,\\
  \delta A^m &=&  d\Lambda^m  \ , 
   \qquad \qquad \qquad  \qquad \quad \, \delta \zeta^m  =  0\ , \nn
\eea
where $f^\cI_{\cJ \cK} $ are the structure constants of $\mathfrak g$.
It will be crucial to realize later that there is whole class of higher-dimensional gauge
transformations with gauge parameters $\hat \Lambda^\cI(x,y)$
and $\hat \Lambda^m(x,y)$ depending non-trivially on $y$
which are not included in \eqref{lower-dim-gauge}. We will discuss these 
additional transformations in \autoref{ch:lgts} in more detail.

The Coulomb branch of the compactified theory is parametrized by the background values of the scalars $\zeta^\cI$
and $\zeta^m$ by setting 
\beq \label{Coulomb-background}
  \langle \zeta^I \rangle \neq 0 \ , \qquad \langle \zeta^{\boldsymbol \alpha} \rangle = 0 \ , \qquad  \langle \zeta^m \rangle \neq 0 \ ,
\eeq
\textit{i.e.}~giving the Cartan Wilson line scalars a (generic) vacuum expectation value. 
This induces the breaking
\beq
   G \times U(1)^{n_{U(1)}} \rightarrow U(1)^{\rk \mathfrak g} \times U(1)^{n_{U(1)}} \ ,
\eeq
and assigns a mass to the W-bosons $A^{\boldsymbol \alpha}$. Note that 
one has to include the Kaluza-Klein vector $A^0$ in addition, such
that the full three- or five-dimensional massless gauge group is actually $U(1)^{\rk \mathfrak g +n_{U(1)} + 1}$.
We stress that there can be additional massless $U(1)$ vectors which arise from the dualization of former four-dimensional axions
or six-dimensional (anti-)self-dual tensors, respectively, as we will explain in a moment.
However, there are no modes in the theory which carry charges under these vector fields.

The massive fields in the lower-dimensional theory then are precisely the excited Kaluza-Klein modes of 
all higher-dimensional states and the fields that acquire masses on the Coulomb branch.
In particular, also the modes of the higher-dimensional 
charged matter states will gain a mass. 
The three- and five-dimensional spin-$\sfrac{1}{2}$ fermions\footnote{Note
that as for $\hat \Bpsi^{\sfrac 12}$, $\hat \Bpsi^{\sfrac 32}_\mu$
in six dimensions we do not impose a symplectic Majorana condition on the $\Bpsi^{\sfrac 12}$, $\Bpsi^{\sfrac 32}_\mu$
in five dimensions.} $\Bpsi^{\sfrac 12} (w,q)$,
which derive from the former $\hat \Bpsi^{\sfrac 12} (w,q)$
in four and six dimensions
having weight $w$ under the non-Abelian group and $U(1)$ charges $q_m$ as introduced in \autoref{sec:gauge_gen},
obtain a
Coulomb branch mass  $m_{\rm CB}^{w,q}$ in the background \eqref{Coulomb-background}
\beq
   m_{\rm CB}^{w,q} = w_I  \langle \zeta^I \rangle + q_m  \langle \zeta^m \rangle \ .
\eeq
In total the mass of the fields $\Bpsi^{\sfrac 1 2}_{(n)}(w,q)$ at Kaluza-Klein level $n$ in the
lower-dimensional theories reads
\beq \label{e:KK-masses}
 m = m^{w,q}_{\rm CB} + n \,m_{\rm KK} = w_I   \langle\zeta^I \rangle + q_m  \langle\zeta^m \rangle + \frac{n}{\langle r\rangle}\ ,
\eeq
 with $m_{\rm KK} = 1/\langle r \rangle$ being the unit Kaluza-Klein 
mass determined by the background value of the radius. 
Note that a similar analysis can be performed for the Kaluza-Klein modes 
of all other fields, including scalars, W-bosons, and six-dimensional tensor fields.

In particular each six-dimensional (anti-)self-dual tensor $\hat B^\alpha$
yields a whole tower of Kaluza-Klein states after compactification on the circle. 
While the massive modes are genuine tensor fields in five dimensions \cite{Townsend:1983xs,Bonetti:2012fn},
the massless mode can be dualized into a massless five-dimensional vector field $A^\alpha$.
To be more precise, the Kaluza-Klein ansatz for $\hat B^\alpha$ reads  
\beq \label{hatBalpha_red}
  \hat B^\alpha = B^\alpha - \big( A^\alpha - 2 \lambda_{\mathfrak{g}}^{-1} b^\alpha \text{tr}_{f}(\zeta A) - 2 b^\alpha_{mn} \zeta^m A^n \big) \wedge Dy  + \dots \, ,
\eeq
omitting contributions from the spin connection in the expansion. 
Note that the modification of this ansatz with terms proportional to the constant 
Green-Schwarz coefficients $b^\alpha$ and $b^\alpha_{mn}$ defined in \eqref{GS6} is important since 
the six-dimensional tensors have modified field strengths. 
In the classical five-dimensional Coulomb branch parametrized by \eqref{Coulomb-background} the ansatz 
\eqref{hatBalpha_red} including only the massless fields becomes
\beq \label{hatBalpha_red_CB}
   \hat B^\alpha = B^\alpha - \big( A^\alpha - 2   b^\alpha \cC_{IJ} \zeta^I A^J - 2 b^\alpha_{mn} \zeta^m A^n \big) \wedge Dy
   + \dots \ ,
\eeq
where we have introduced the coroot intersection matrix 
$\cC_{IJ} = \lambda_{\mathfrak{g}}^{-1} \text{tr}_{f}(T_I T_J)$ with $T_I$ being 
the Cartan generators in the coroot basis. 
Again, we refer to \autoref{s:lie_conv}, in particular \eqref{e:def_coroot_int_mat} 
and the following paragraph for more details.
In five dimensions the $B^\alpha$ can then be eliminated from the action in favour of the dual vectors
$A^\alpha$ by using the self- or anti-self-duality of $\hat B^\alpha$
(see \cite{Bonetti:2011mw,Grimm:2013oga} for more details). 

The
analog objects to $\hat B^\alpha$ in four dimensions are axions $\hat \rho_\alpha$, $\alpha = 1,\ldots ,n_{\rm ax}$ with a gauged
shift-symmetry under the $U(1)$ vectors $\hat A^m$.
As for the (anti-)-self-dual tensors in six dimensions, after the circle compactification
the Kaluza-Klein zero-modes of $\hat \rho_\alpha$
can be dualized into three-dimensional vectors $A^\alpha$ (see \textit{e.g.}~\cite{Grimm:2010ks,Cvetic:2012xn} for a detailed discussion).

Our main interest in the circle-reduced theories at the massive level
is into parity-violating modes since they derive from chiral higher-dimensional fields and contribute to
one-loop Chern-Simons terms in the effective theory. Therefore let us collect all massive parity violating modes
along with their higher-dimensional origin in \autoref{tab:massive_3d} and \autoref{tab:massive_5d}.
The connection between chirality in four and six dimensions and the choice of a representation
of the Clifford algebra in three and five dimensions is explained in \autoref{ch:CS}.
The Lagrangians for the massive fields in three and five dimensions are given
in \eqref{e:lagr_3d} and \eqref{e:lagr_5d}.

\begin{table}
\begin{center}
\begin{tabular}{c|ccc}
\rule[-5pt]{0pt}{5pt} \bf 4d & \multicolumn{3}{c}{\bf 3d}\\
\hline
\rule{0pt}{15pt}
\multirow{2}{*}{Field}  & \multirow{2}{*}{KK-tower} & 
 \multirow{2}{*}{Mass} & $(A^0, A^I , A^m)$ \\
\rule[-5pt]{0pt}{15pt} &&& Charge \\
\hline
\rule[-10pt]{0pt}{30pt} $\hat \Bpsi^{\sfrac12} (w,q)$  & $\Bpsi^{\sfrac12}_{(n)} (w,q)$ & $m_{\rm CB}^{w,q} +
\frac{n}{\langle r\rangle}$ & $(-n, w_I , q_m)$
\end{tabular}
\end{center}
\caption{Four-dimensional Weyl spinors $\hat \Bpsi^{\sfrac12} (w,q)$ induce a Kaluza-Klein tower of massive three-dimensional Dirac
 spinors $\Bpsi^{\sfrac12}_{(n)} (w,q)$, $n = -\infty , \dots  ,+\infty$ with additional mass contribution on the Coulomb branch.
 The Lagrangian is given by \eqref{e:lagr_3d}.}
\label{tab:massive_3d}
\end{table}
\begin{table}
\begin{center}
\begin{tabular}{cc|cccc}
\multicolumn{2}{c|}{\rule[-5pt]{0pt}{5pt} \bf 6d} & \multicolumn{3}{c}{\bf 5d} \\
\hline
\rule{0pt}{15pt}
\multirow{2}{*}{Field} & \multirow{2}{*}{$\mathfrak{su}(2) \times \mathfrak{su}(2)$} & \multirow{2}{*}{KK-tower} & 
\multirow{2}{*}{$\mathfrak{su}(2) \times \mathfrak{su}(2)$} & \multirow{2}{*}{Mass} & $(A^0, A^I , A^m)$ \\
\rule[-5pt]{0pt}{15pt} &&&&& Charge \\
\hline
\rule[-10pt]{0pt}{30pt} $\hat \Bpsi^{\sfrac12} (w,q)$ & $(\frac{1}{2},0),(0,\frac{1}{2})$ & $\Bpsi^{\sfrac12}_{(n)} (w,q)$ &
$(\frac{1}{2},0),(0,\frac{1}{2})$ & $m_{\rm CB}^{w,q} + \frac{n}{\langle r\rangle}$ & $(-n,w_I,q_m)$ \\
\rule[-10pt]{0pt}{10pt}$\hat B^\alpha$ & $(1,0),(0,1)$ & $\BB^\alpha_{(n>0)}$ & $(1,0),(0,1)$ & $\frac{n}{\langle r\rangle}$ & $(-n,0,0)$ \\
\rule[-10pt]{0pt}{10pt}$\hat \Bpsi^{\sfrac32}_{\mu}$ & $(1,\frac{1}{2}),(\frac{1}{2},1)$ & $\Bpsi^{\sfrac32}_{\mu \, (n)}$
& $(1,\frac{1}{2}),(\frac{1}{2},1)$ & $\frac{n}{\langle r\rangle}$ & $(-n,0,0)$
\end{tabular}
\end{center}
\caption{Six-dimensional spin-$\sfrac 1 2$ Weyl fermions $\hat \Bpsi^{\sfrac12} (w,q)$ induce a Kaluza-Klein tower
of massive five-dimensional spin-$\sfrac 1 2$ Dirac fermions  
$\Bpsi^{\sfrac12}_{(n)} (w,q)$, $n = -\infty , \dots  ,+\infty$ with additional mass contribution on the Coulomb branch.
The Lagrangian is given by \eqref{e:lagr_5d_1}.
Furthermore, (anti-)self-dual tensors in six dimensions $\hat B^\alpha$ yield a tower of massive complex tensors $\BB^\alpha_{(n)}$,
$n = 1 , \dots  ,+\infty$ with Lagrangian \eqref{e:lagr_5d_2}.
Note that $n$ runs only over the positive integers because of the (anti-)self-duality relation. Finally,
six-dimensional spin-$\sfrac 3 2$ Weyl fermions $\hat \Bpsi^{\sfrac32}_{\mu}$ give a Kaluza-Klein tower
of massive five-dimensional spin-$\sfrac 3 2$ Dirac fermions  
$\Bpsi^{\sfrac32}_{\mu \, (n)}$, $n = -\infty , \dots  ,+\infty$ with Lagrangian \eqref{e:lagr_5d_3}. We stress that in our
conventions no additional symplectic Majorana condition is imposed, neither in six nor in five dimensions.}
\label{tab:massive_5d}
\end{table}

Since the Chern-Simons coefficients couple gauge fields, we once more list the vectors which stay massless on the Coulomb branch
along with their higher-dimensional origins in \autoref{tab:vec}. Note that the massless gauge fields are all Abelian
in the effective theory on the Coulomb branch and that there are no modes which
are charged under $A^\alpha$.
\begin{table}
\begin{center}
\begin{tabular}{c|c}
\bf 4d (6d) fields & \bf 3d (5d) massless vectors \\
\hline
$\hat g$ & $A^0$\\
$\hat A$ & $A^I$, $I=1, \dots , \rk \mathfrak g$ \\
$\hat A^m$ & $A^m$, $m=1, \dots , n_{U(1)}$\\
$\hat \rho_\alpha$ ($\hat B^\alpha$) & $A^\alpha$, $\alpha = 1, \dots, n_{\rm{ax}}$ $(T_{\rm{sd}}+T_{\rm{asd}})$
\end{tabular}
\end{center}
\caption{There are in general four different types of massless vectors in the circle-reduced theory: the Kaluza-Klein vector $A^0$,
vectors $A^I$ from higher-dimensional Cartan gauge fields, vectors $A^m$ from higher-dimensional $U(1)$ gauge fields and
dualized vectors $A^\alpha$ which stem from four-dimensional axions or six-dimensional (anti-)self-dual tensors, respectively.}
\label{tab:vec}
\end{table}

\subsection{Chern-Simons Terms}\label{sec:CS_in_circle}
Now that we have discussed the parity violating massive field content as well as the massless vectors
of the theories on the circle, let us investigate in more detail
the special types of topological couplings in
three and five dimensions, namely Chern-Simons terms, which we already introduced in \autoref{ch:CS}.
Recall the general form of these terms in three dimensions
\begin{align}
 S_{\textrm{CS}} = \int \Theta_{\Lambda \Sigma} A^\Lambda \wedge F^\Sigma \, ,
\end{align}
and in five dimensions
\begin{subequations}
\begin{align}
S^{\textrm{gauge}}_{\textrm{CS}} = - \frac{1}{12} \int k_{\Lambda\Sigma\Theta} A^{\Lambda} \wedge F^{\Sigma} \wedge F^{\Theta} \\
 S^{\textrm{grav}}_{\textrm{CS}} = - \frac{1}{4} \int k_{\Lambda} A^{\Lambda} \wedge \tr ( \mathcal{R} \wedge \mathcal{R} ) \, ,
\end{align}
\end{subequations}
where $A^{\Lambda}$ are $U(1)$ vectors with field strengths $F^{\Lambda}$, and $\mathcal{R}$ is the curvature two-form.
The constants $\Theta_{\Lambda \Sigma}$, $k_{\Lambda\Sigma\Theta}$, $k_{\Lambda}$ are called Chern-Simons coefficients,
and in five dimension one can distinguish between gauge and gravitational Chern-Simons terms.
Note that the latter do not exist in three dimensions.

For the special case when the three- and five-dimensional theories arise from a circle compactification pushed to the Coulomb
branch, which we introduced in \autoref{e:field-red},
the index $\Lambda$ labeling the massless vectors splits as $\Lambda = (0, I, m, \alpha)$, as can be inferred from \autoref{tab:vec}.
In these settings it is crucial to distinguish in general between \textit{classical} and \textit{one-loop} Chern-Simons terms. We
define the classical
ones by the property of always being exact at the classical level and never receiving any corrections at one-loop.
From \autoref{tab:massive_3d} and \autoref{tab:massive_5d}
it is clear that they can be characterized by carrying at least one index $\alpha$ since there are no states in the theory which are charged
under $A^\alpha$ and therefore no corrections.\footnote{Actually this is the more accurate definition of \textit{classical}
Chern-Simons terms since in principle it is not required that there is matter which is charged under $A^I$ or $A^m$.}
In general for circle-reduced theories the \textit{classical} Chern-Simons couplings then can be shown to take the special form
\begin{align}\label{e:CS_class_4d}
 \Theta_{\alpha\beta} = 0 \, , \qquad 
 \Theta_{\alpha 0} = 0 \, , \qquad 
 \Theta_{\alpha I} = 0 \, , \qquad 
 \Theta_{\alpha m} = \frac{1}{2}\theta_{\alpha m} \, ,
\end{align}
and respectively
\begin{align}\label{e:CS_cl_6d}
 &k_{\alpha\beta\gamma} =0 \ , &&k_{ 0\alpha\beta} = \eta_{\alpha\beta} \ , &&k_{I\alpha\beta} =0 \ , \nn \\
 &k_{m\alpha\beta} =0 \ , && k_{00\alpha} =0 \ , && k_{IJ\alpha} = -\eta_{\alpha\beta} b^\beta \cC_{IJ} \ , \nn \\
 &k_{mn\alpha} = -\eta_{\alpha\beta} b^\beta_{mn} \ , &&k_{0I\alpha} =0 \ , &&k_{0m\alpha} =0 \ ,\nn \\ 
 &k_{\alpha} = -12\ \eta_{\alpha\beta} a^\beta \ .
\end{align}
All other Chern-Simons coefficients vanish at the classical level and generically receive corrections at one-loop. Thus we call them
\textit{one-loop} Chern-Simons terms. We derive the general results for all different types of these one-loop couplings in
\autoref{sec:one_loop_calc}. For convenience
let us display here only the most important ones which will appear in upcoming calculations
{\allowdisplaybreaks \begin{subequations}\label{e:CS_loop_3d}
\begin{align}
  \Theta_{IJ} &=  \sum_{R,q} F_{\sfrac 1 2}(R,q) \sum_{w \in R}  \ \Big(l_{w,q} +\frac{1}{2} \Big) \ w_I w_J \ 
  \sign  \big(m^{w,q}_{\rm CB}\big) \, , \\
 \Theta_{mn} &=  \sum_{R,q}  F_{\sfrac 1 2}(R,q) \sum_{w \in R} \ \Big(l_{w,q} +\frac{1}{2} \Big) 
 \ q_m q_n \ \sign  \big(m^{w,q}_{\rm CB}\big) \, , \\
 \Theta_{Im} &=  \sum_{R,q}  F_{\sfrac 1 2}(R,q) \sum_{w \in R} \ \Big(l_{w,q} +\frac{1}{2} \Big) 
 \ w_I q_m\ \sign  \big(m^{w,q}_{\rm CB}\big) \, ,
\end{align}
\end{subequations}}
as well as
{\allowdisplaybreaks \begin{subequations}\label{e:CS_loop_5d}
\begin{align}
 k_{IJK} &= \sum_{R,q} F_{\fe}(R,q) \sum_{w \in R} \ \Big(l_{w,q}  +\frac{1}{2} \Big)
 \ w_I w_J w_K \ \sign  \big(m^{w,q}_{\rm CB}\big)\, ,\\
 k_{mnp} &= \sum_{R,q} F_{\fe}(R,q) \sum_{w \in R} \ \Big(l_{w,q}  +\frac{1}{2}\Big) 
 \ q_m q_n q_p \ \sign  \big(m^{w,q}_{\rm CB}\big)\, ,\\
 k_{IJm} &= \sum_{R,q} F_{\fe}(R,q) \sum_{w \in R} \ \Big(l_{w,q}  +\frac{1}{2} \Big)
 \ w_I w_J q_m \ \sign  \big(m^{w,q}_{\rm CB}\big)\, ,\\
 k_{Imn} &= \sum_{R,q} F_{\fe}(R,q) \sum_{w \in R} \ \Big(l_{w,q}  +\frac{1}{2} \Big) 
 \ w_I q_m q_n \ \sign  \big(m^{w,q}_{\rm CB}\big)\, ,\\
 k_{I} &= -2\sum_{R,q} F_{\fe}(R,q) \sum_{w \in R} \ \Big(l_{w,q} +\frac{1}{2} \Big) 
 \ w_I \ \sign  \big(m^{w,q}_{\rm CB}\big) \, , \\
 k_{m} &= -2\sum_{R,q} F_{\fe}(R,q) \sum_{w \in R} \ \Big(l_{w,q} +\frac{1}{2} \Big) 
 \ q_m \ \sign  \big(m^{w,q}_{\rm CB}\big) \, ,
\end{align}
\end{subequations}}
where the non-negative integer $l_{w,q}$ is defined as
\begin{align}\label{e:def_hier}
 l_{w,q} := \Bigg\lfloor \bigg\vert \frac{m^{w,q}_{\rm CB}}{m_{\rm KK}} \bigg\vert \Bigg\rfloor \, .
\end{align}
This quantity was first introduced in \cite{Grimm:2013oga} and captures the hierarchy between the Coulomb branch mass and the Kaluza-Klein mass.
It will become very important later since changes in this hierarchy induce jumps of Chern-Simons terms which have
crucial impact.

Let us conclude by mentioning that for four-dimensional settings on the circle it will become important later that one can
obtain a non-vanishing $\Theta_{\alpha 0}$ and additional classical contributions to $\Theta_{mn}$ and $\Theta_{IJ}$
by switching on circle fluxes
of the axions
\begin{subequations}\label{e:coupling_c_flux}
\begin{align}
\label{e:circle_flux1} \Theta_{\alpha 0} &= \frac{1}{2}\int_{S^1} \langle d \hat \rho_\alpha \rangle  \, , \\
\label{e:circle_flux2} \Theta_{IJ}^{\rm{class}} &= -\frac{1}{2} b^\alpha \cC_{IJ} \int_{S^1} \langle d \hat \rho_\alpha \rangle  \, , \\
\label{e:circle_flux3} \Theta_{mn}^{\rm{class}} &= -\frac{1}{2} b^\alpha_{mn}\int_{S^1} \langle d \hat \rho_\alpha \rangle   \, .
\end{align}
\end{subequations}
In particular we show around \eqref{e:induced_cflux}
that certain large gauge transformations can induce such non-trivial circle fluxes.
Note that also in six-dimensional setups circle fluxes of gauged axions can play an important role
in F-theory compactifications on manifolds without a rational section.
This is however beyond the scope of this thesis and we refer to \cite{Anderson:2014yva} for more details.

\chapter{Symmetries of Gauge Theories on the Circle}\label{ch:lgts}

\section{Classification of Large Gauge Transformations}\label{sec:class_lgt}

In this subsection we discuss in detail the set of gauge transformations of 
an Abelian or non-Abelian theory on a circle that are later translated to a symmetry 
of the geometry of an elliptic fibration in \autoref{ch:arith} using F-theory. Recall that after the compactification 
the effective theory admits \eqref{lower-dim-gauge} as
local symmetries before pushed to the Coulomb branch. In the Coulomb 
branch one simply has a purely Abelian local symmetry.

In addition to the lower-dimensional gauge transformations \eqref{lower-dim-gauge}
we could also have performed a circle-dependent gauge transformation 
and then reduced on the circle $y \sim y+2\pi$. If one preserves the boundary conditions of 
the fields in the compactification ansatz the gauge invariance of the higher-dimensional theory then 
implies that there exists a variety of equivalent  
lower-dimensional effective theories that are obtained after circle reduction of 
the same higher-dimensional theory. 
Gauge transformations that cannot 
be deformed continuously to the identity map are known as \textit{large} gauge transformations.
More concretely, let us consider the effect of a gauge transformation that locally takes the form
\beq \label{LambdaAnsatz}
    \hat \Lambda^\cI(x,y) = \left\{\begin{array}{c}
                             - \fn^I y\\
                             0
                            \end{array} \right. \ , \qquad \hat \Lambda^m(x,y) = - \fn^m y\ ,
\eeq
where $\fn^I$ and $\fn^m$ are constants, and we have included a minus sign for later convenience.  $\fn^I,\fn^m$
will be further restricted below to ensure that \eqref{LambdaAnsatz} is 
in fact a large gauge transformation which preserves the boundary conditions of all fields.
Using the split $\cI = (I ,\boldsymbol{\alpha} )$ as in \eqref{hatA_expand} 
we have set $\hat \Lambda^{\boldsymbol{\alpha}}(x,y)=0$ to ensure that the Coulomb branch values $ \langle \zeta^{\boldsymbol \alpha} \rangle = 0$ in  \eqref{Coulomb-background} are unchanged.\footnote{We stress again that $\boldsymbol{\alpha}$ labels
the roots of the gauge algebra while $\alpha$ counts the axions in four dimensions and the (anti-)self-dual tensors in six dimensions, respectively.} This guarantees that we stay on the 
considered Coulomb branch; all of the following discussions are performed on this background. 
The reduction 
ans\"atze \eqref{red-AI} and \eqref{hatBalpha_red_CB} are also compatible with a gauge transformation \eqref{LambdaAnsatz} if one introduces 
the new quantities 
\beq \label{eq:LGTgeneral_1}
   \tilde r = r\ , \qquad \tilde \zeta^I = \zeta^I + \frac{\fn^I}{r} \ ,  \qquad \tilde \zeta^m = \zeta^m + \frac{\fn^m}{r} \ , 
\eeq
and 
\begin{align}
\label{e:non_abelian_trafo_vectors_4d}
 \begin{pmatrix}
  \tilde A^{0}\\[8pt]
  \tilde A^{I}\\[8pt]  
  \tilde A^{m}\\[8pt]
  \tilde A^{\alpha}
 \end{pmatrix} = 
\begin{pmatrix*}[c]
 1 & 0 & 0  &0 \\[8pt]
 - \fn^I & \delta^{I}_{J} & 0 & 0\\[8pt]
  - \fn^m & 0 & \delta_n^m & 0 \\[8pt]
 \frac{1}{2}\fn^K \fn^L \cC_{KL} b^\alpha +\frac{1}{2}\fn^p \fn^q b^\alpha_{pq} & 
 -\fn^{K} \cC_{KJ} b^\alpha
  &- \fn^p b^\alpha_{pn}& \delta_\beta^\alpha
\end{pmatrix*}\cdot
\begin{pmatrix}
  A^{0} \\[8pt]
  A^{J} \\[8pt]
  A^{n} \\[8pt]
  A^{\beta} \, 
 \end{pmatrix}.
\end{align}
With \eqref{LambdaAnsatz} being compatible with \eqref{red-AI} and  \eqref{hatBalpha_red_CB}
we mean that the form of the reduction ansatz after a gauge transformation 
is unchanged when using the quantities with tildes. 

Some additional remarks are expedient here. First, it is important to stress that the simple shifts in the vector fields 
$\tilde A^I$ only occur for the Cartan direction. 
In the non-Cartan directions, \textit{i.e.}~for the vectors that are massive on the Coulomb 
branch, the non-Abelian structure of $G$ modifies the transformation rule. Second, while 
$ \langle \zeta^{\boldsymbol \alpha} \rangle = 0$ is preserved by \eqref{LambdaAnsatz} the actual 
values for $\langle \zeta^{I} \rangle$ do change in the vacuum. One therefore relates theories 
at different points
on the Coulomb branch. In fact, this is a defining property of a large gauge transformation: they
relate theories at different points in the vacuum manifold of the theory with the same properties, see  \textit{e.g.}~\cite{Harvey:1996ur}.  Third, later on we will consider six-dimensional theories with $(1,0)$ supersymmetry
arising in F-theory. These theories have $T_{\rm sd}=1$ and $T_{\rm asd}\equiv T$. 
Each six-dimensional anti-self-dual tensor is accompanied by a real 
scalar field in the multiplet. After dimensional reduction these scalar fields combine with $T$ of the $A^\alpha$ into 
vector multiplets. Importantly, it was found in \cite{Bonetti:2011mw,Grimm:2013oga} that the redefinition of the five-dimensional scalar fields 
is precisely of the form compatible with \eqref{e:non_abelian_trafo_vectors_4d} (see \textit{e.g.}~(3.30) in \cite{Grimm:2013oga}). In other words, a gauge transformation \eqref{LambdaAnsatz} shifts both the vectors and scalars in a compatible fashion. 
We note that a similar story applies to circle compactifications from four to three space-time dimensions. 
In fact, the transformations \eqref{eq:LGTgeneral_1} and \eqref{e:non_abelian_trafo_vectors_4d} 
are equally valid for this latter case. As noted above the 
vectors $A^\alpha$ are the three-dimensional duals of the former four-dimensional scalars $\hat \rho_\alpha$
appearing in \eqref{GS4}.

Clearly, a gauge transformation \eqref{LambdaAnsatz} also 
requires to transform the Kaluza-Klein modes of all higher-dimensional charged fields. 
Given a general matter state $\Bpsi_{(n)}$ (not necessarily a fermion) at Kaluza-Klein level $n$ in the representation $R$ of $G$
and with charge $q_m$ under $\hat A^m$ we first proceed as described 
after \eqref{def-Dpsi4}, \eqref{def-Dpsi6} and introduce eigenstates $\Bpsi_{(n)}(w,q)$, where $w$ are the weights of $R$.
The transformation \eqref{e:non_abelian_trafo_vectors_4d} mixes these states as
 \begin{align}
\label{e:non_abelian_trafo_charges_4d}
\Bpsi_{(n)}(w,q) \mapsto \Bpsi_{(\tilde n)}(\tilde w,\tilde q) \ ,\qquad \begin{pmatrix}
  \tilde n\\[8pt]
  \tilde w_{I}\\[8pt] 
  \tilde q_{m}\\
 \end{pmatrix} = 
\begin{pmatrix*}[c]
 1 & - \fn^J  & - \fn^n \\[8pt]
 0 & \delta_I^J &0  \\[8pt] 
 0 & 0 & \delta_m^n \\[8pt]
\end{pmatrix*}\cdot
\begin{pmatrix}
  n\\[8pt]
  w_{J}\\[8pt]
  q_n
 \end{pmatrix} \, .
\end{align}
Note that in general this transformation shifts the whole Kaluza-Klein
tower, but there is still no state charged under $\tilde A^{\alpha}$.
Furthermore, imposing that \eqref{e:non_abelian_trafo_charges_4d}
is in fact a consistent reshuffling of the Kaluza-Klein states, which is 
necessary for invariance of the theory, imposes 
conditions on the constants $\fn^I$ and $\fn^m$ that are dependent 
on the spectrum of the theory. 
We will discuss the various choices and conditions 
in the following. 
\vspace{.2cm}

\noindent
\textbf{Integer large gauge transformations}

\noindent
In \eqref{LambdaAnsatz} we have introduced gauge transformations 
that depend on the circle coordinate $y \sim y + 2\pi $. As mentioned before, these 
correspond to  large gauge transformations around the circle if they preserve the 
circle boundary conditions of all fields and wind at least once around the circle. 
Let us now define what we mean by integer large gauge transformations. 
First, if we consider pure gauge theory without charged matter, 
we call a large gauge transformation to be integer if all 
$\fn^I$ and all $\fn^m$ are integers. 
Indeed, the degrees of freedom in \eqref{LambdaAnsatz} are in general characterized by
elements of the homotopy groups
 \beq
    \pi_1(U(1)^{{\rm rk} G}) \cong \mathbb{Z}^{{\rm rk} G}\ , \qquad   
    \pi_1(U(1)^{n_{U(1)}}) \cong \mathbb{Z}^{n_{U(1)}}\,.
 \eeq
Clearly, \eqref{LambdaAnsatz} define maps from $S^1$ into the gauge group (which is purely 
Abelian on the Coulomb branch). These are precisely classified by the first homotopy group of 
the gauge group which in the case at hand consists of tuples of integers.

If one now includes a charged matter spectrum, the invariance of the 
boundary conditions of all these fields dictates the set of large gauge 
transformations. In these cases the space of allowed 
$\fn^I$ and $\fn^m$ has to be quantized. In general the $\fn^I$ 
and $\fn^m$ could still be integer or fractional depending on the 
weights and charges of the matter fields. 
However, for the transformations \eqref{LambdaAnsatz} to be an actual symmetry, 
\textit{i.e.}~a large 
gauge transformation, the following condition for each state $\hat \Bpsi(R,q)$ has to be satisfied:
\beq \label{kk_cond}
  \fn^I w_I + \fn^m q_m \quad \in \ \mathbb{Z}\ , 
\eeq
where $w_I$ are the weights of $R$ and $q_m$ are the $U(1)$ charges.
This condition also arises from the transformation of the Kaluza-Klein level 
in \eqref{e:non_abelian_trafo_charges_4d} and ensures that $\tilde n$ is an integer, which 
implies equivalence of the full Kaluza-Klein towers of the compactified theory by a simple reshuffling.
Now we are in the position to introduce our notion of \textit{integer} large gauge transformations. 
They are spanned by pairs  $(\fn^I , \fn^m)$ satisfying \eqref{kk_cond} and one of the conditions  
\begin{itemize}
 \item[(I)] $\fn^m=0$ and $\fn^I \in \bbZ^*$,
 \item[(II)] $\fn^m \in \bbZ^*$ and  $\fn^I w_I \in \mathbb Q$.
\end{itemize}
It is useful to comment on the class (II) of basis vectors. 
While all $w_I$ reside in an integer lattice 
and therefore do not violate \eqref{kk_cond} for integer $\fn^I$,
the $U(1)$ charges $q_m$ can be fractional. However, we will also consider the set of integer $\fn^m$'s that 
allow a compensation of this fractional contribution to \eqref{kk_cond} by an appropriate 
$\fn^I$-transformation which might be fractional. 
\vspace{.2cm}

\noindent
\textbf{Special fractional large gauge transformations}

\noindent
There is another set of large gauge transformations that will be of importance for us. 
If the arising representations in the spectrum of matter states is special, \textit{e.g.}~if the fundamental representation does not occur, also 
fractional $\fn^I$ might be allowed.
More precisely, we also want to consider pairs $(\fn^I , \fn^m)$ satisfying \eqref{kk_cond}
and
\begin{itemize}
 \item[(III)] $\fn^m=0$ and $\fn^I$ fractional. 
\end{itemize}
We call large gauge transformations satisfying (III) special fractional large gauge transformations.
Note that the conceptual difference between (III) and (I), (II) is that there is always at least one integer
quantity $\fn^I , \fn^m$ in (I) or (II).

It remains to consider the cases where also $\fn^m$ is fractional. As a concrete example this could 
be allowed if the spectrum has special charges such that $\fn^m q_m$ is integer for each state, although there are more general possibilities
involving also the non-Abelian sector.
However, later we find that models which allow for a fractional $\fn^m$ do not appear in our geometric considerations of \autoref{ch:arith}.
For instance in the known 
F-theory examples there are always states that have minimal charge $0<q_m \leq 1$.
Following some folk theorems (see {e.g.}~\cite{Banks:2010zn,Hellerman:2010fv}) this might be true in any theory 
of quantum gravity. In this case the space of all large gauge transformations is spanned 
by $(\fn^I,\fn^m)$ satisfying \eqref{kk_cond} and either (I), (II) or (III). Nevertheless we stress that from a purely field-theoretical
point of view there should be no restriction for also having fractional $\fn^m$ in some settings.

\section{Anomalies from Large Gauge Transformations}\label{sec:anom_lgt}

We now crown our field theory analysis by applying large gauge transformations to Chern-Simons terms.
In particular, demanding that large gauge transformations constitute a symmetry of the theory on the circle (including the
full Kaluza-Klein tower) we are able to derive all gauge anomaly conditions in four and six dimensions.
Note that since there always exist transformations of type (I) and (II) for any kind of field theory spectrum,
our analysis turns out to be totally general.

The guiding principle is that there are two conceptually different ways to evaluate the transformation of the Chern-Simons coefficients
$\Theta_{\Lambda\Sigma}$, $k_{\Lambda\Sigma\Theta}$, $k_{\Lambda}$ under the large gauge transformations
\eqref{e:non_abelian_trafo_vectors_4d},\eqref{e:non_abelian_trafo_charges_4d}.
First rewrite \eqref{e:non_abelian_trafo_vectors_4d} in components as
\begin{align}
 \tilde A^\Lambda = \tensor{L}{^\Lambda_{\Lambda^\prime}}A^{\Lambda^\prime} \, ,
\end{align}
then the Chern-Simons coefficients accordingly have to transform as the dual elements
\begin{subequations}\label{e:dual_trafo}
\begin{align}
 \tilde \Theta_{\Lambda\Sigma} &= \tensor{(L^{-1\,T})}{_\Lambda^{\Lambda^\prime}}
 \tensor{(L^{-1\,T})}{_\Sigma^{\Sigma^\prime}} \, \Theta_{\Lambda^\prime \Sigma^\prime} \, , \\
 \tilde k_{\Lambda\Sigma\Theta} &= \tensor{(L^{-1\,T})}{_\Lambda^{\Lambda^\prime}} \tensor{(L^{-1\,T})}{_\Sigma^{\Sigma^\prime}}
 \tensor{(L^{-1\,T})}{_\Theta^{\Theta^\prime}} \, k_{\Lambda^\prime \Sigma^\prime \Theta^\prime} \, , \\
  \tilde k_{\Lambda} &= \tensor{(L^{-1\,T})}{_\Lambda^{\Lambda^\prime}}
  \, k_{\Lambda^\prime} \, . 
\end{align}
\end{subequations}
On the other hand
the transformed couplings $\tilde \Theta_{\Lambda\Sigma}$, $\tilde k_{\Lambda\Sigma\Theta}$, $\tilde k_{\Lambda}$
can also directly be accessed by evaluating them in the circle-reduced theory characterized by the gauge-transformed parameters.
In the following we will compare both procedures first for the classical terms and then for one-loop terms.

It is obvious from \eqref{e:CS_class_4d}, \eqref{e:CS_cl_6d}
that the classical Chern-Simons couplings (except of $\Theta_{\alpha 0}$ because of \eqref{e:circle_flux1})
only depend on data of the higher-dimensional theory and are
insensitive to the precise form of the circle background. Consistently we find that both procedures of evaluating their transformation
properties yield the same result, namely that they are invariant
\begin{align}
 \tilde\Theta_{\alpha\beta} = \Theta_{\alpha\beta} = 0 \, , \qquad 
 \tilde\Theta_{\alpha I} = \Theta_{\alpha I} = 0 \, , \qquad 
 \tilde\Theta_{\alpha m} = \Theta_{\alpha m} = \frac{1}{2}\theta_{\alpha m} \, ,
\end{align}
and respectively
\begin{align}
 &\tilde k_{\alpha\beta\gamma} = k_{\alpha\beta\gamma} =0 \ , &&\tilde k_{ 0\alpha\beta} = 
 k_{ 0\alpha\beta} = \eta_{\alpha\beta} \ , &&\tilde k_{I\alpha\beta} = k_{I\alpha\beta} =0 \ , \nn \\
 &\tilde k_{m\alpha\beta} = k_{m\alpha\beta} =0 \ , && \tilde k_{00\alpha} = k_{00\alpha} =0 
 \ , && \tilde k_{IJ\alpha} = k_{IJ\alpha} = -\eta_{\alpha\beta} b^\beta \cC_{IJ} \ , \nn \\
 &\tilde k_{mn\alpha} = k_{mn\alpha} = -\eta_{\alpha\beta} b^\beta_{mn} \ , &&\tilde k_{0I\alpha} = 
 k_{0I\alpha} =0 \ , &&\tilde k_{0m\alpha} = k_{0m\alpha} =0 \ ,\nn \\ 
 &\tilde k_{\alpha} = k_{\alpha} = -12\ \eta_{\alpha\beta} a^\beta \ .
\end{align}
However there is one exception to this, namely the classical Chern-Simons coupling $\Theta_{\alpha 0}$. As described in
\eqref{e:circle_flux1} it is sensitive to circle-flux of the axions $\hat\rho_\alpha$. Although we initially started with a setting
without such a background, \textit{i.e.}~with $\Theta_{\alpha 0} = 0$, large gauge transformations can induce a nonzero flux
$\frac{1}{2}\int_{S^1} \langle d \hat \rho_\alpha \rangle$. Indeed, both of our procedures to determine $\tilde\Theta_{\alpha 0}$
yield the same result
\begin{align}\label{e:induced_cflux}
 \tilde\Theta_{\alpha 0} = \frac{1}{2} \fn^m \theta_{\alpha m} \neq \Theta_{\alpha 0}=0 \, .
\end{align}

Let us now determine how the two procedures of evaluating large gauge transformations on Chern-Simons terms are related
for the case of one-loop induced couplings. In contrast to the classical case, first of all the couplings are in general not invariant
under large gauge transformations.
This is somehow expected since the loop-calculations explicitly depend on the details of the circle background, namely the VEVs for the
Wilson lines $\langle \tilde\zeta^I \rangle$ and $\langle \tilde\zeta^m \rangle$ which define the transformed Coulomb branch masses 
\begin{align}
 \tilde m_{\rm{CB}}^{w,q} &= \tilde w_I \langle \tilde \zeta^I \rangle + \tilde q_m \langle \tilde \zeta^m \rangle 
 = w_I \langle \zeta^I \rangle + w_I \frac{\fn^I}{\langle r \rangle} +  q_m \langle \zeta^m \rangle 
 + q_m \frac{\fn^m}{\langle r \rangle} \nn \\
 &= m_{\rm{CB}}^{w,q} + ( \fn^I w_I + \fn^m q_m ) m_{\rm{KK}} \, .
\end{align}
Note that the Coulomb branch mass enters in the formulae for one-loop Chern-Simons terms
\eqref{e:CS_loop_3d}, \eqref{e:CS_loop_5d}
(see also \eqref{e:all_3d_loops}, \eqref{e:all_5d_loops} for the complete list of one-loop Chern-Simons terms)
through $\sign (m_{\rm{CB}})$ and $l_{w,q}$ which is defined in \eqref{e:def_hier}. Thus in general these couplings
indeed do transform which of course poses no problems in principle.
However, if we now compare the results for $\tilde \Theta_{\Lambda\Sigma}$, $\tilde k_{\Lambda\Sigma\Theta}$, $\tilde k_{\Lambda}$
using the transformation rule \eqref{e:dual_trafo}
with the $\tilde \Theta_{\Lambda\Sigma}$, $\tilde k_{\Lambda\Sigma\Theta}$, $\tilde k_{\Lambda}$
which we obtain from directly evaluating the loop-calculations
in the gauge transformed setting, \textit{i.e.}~using $\tilde m_{\rm{CB}}^{w,q}$ in the
formulae for the loops, we seem to get different expressions. More precisely let us define
\begin{subequations}\label{e:deltas}
\begin{align}
 \delta \tilde \Theta_{\Lambda\Sigma} &= \tilde \Theta^{\rm{match}}_{\Lambda\Sigma} - \tilde \Theta^{\rm{dual}}_{\Lambda\Sigma} \, , \\
 \delta \tilde k_{\Lambda\Sigma\Theta} &= \tilde k^{\rm{match}}_{\Lambda\Sigma\Theta} - \tilde k^{\rm{dual}}_{\Lambda\Sigma\Theta} \, , \\
 \delta \tilde k_{\Lambda} &= \tilde k^{\rm{match}}_{\Lambda} - \tilde k^{\rm{dual}}_{\Lambda} \, ,
\end{align}
\end{subequations}
where the labels ``match'' and ``dual'' indicate whether the quantities are evaluated directly by matching
to the loop expressions
using the transformed $\tilde m_{\rm{CB}}^{w,q}$ or by applying the dual transformation \eqref{e:dual_trafo}, respectively.
While we have already mentioned that for all classical Chern-Simons couplings (also $\Theta_{\alpha 0}$)
this difference of calculating the transformation with two types of procedures is always zero, for the one loop expressions
the situation is more subtle.
In particular for \eqref{e:CS_loop_3d} we obtain
{\allowdisplaybreaks \begin{subequations}\label{e:missmatch3d}
\begin{align}
  \delta\tilde \Theta_{IJ} &= \sum_{R,q} F_{\sfrac 1 2}(R,q) \sum_{w \in R} \ w_I w_J  
  \ \bigg[ \Big(\tilde l_{w,q} +\frac{1}{2} \Big) \ \sign \big(\tilde m^{w,q}_{\rm CB}\big) 
  - \Big(l_{w,q} +\frac{1}{2} \Big) \ \sign \big(m^{w,q}_{\rm CB}\big) \bigg ] \nn \\
  &\quad  -  \frac{1}{2} \fn^q b^\alpha \theta_{q\alpha}\, \cC_{IJ} \, , \\
  \delta\tilde \Theta_{mn} &= \sum_{R,q} F_{\sfrac 1 2}(R,q) \sum_{w \in R} \ q_m q_n  
  \ \bigg[ \Big(\tilde l_{w,q} +\frac{1}{2} \Big) \ \sign \big(\tilde m^{w,q}_{\rm CB}\big) 
  - \Big(l_{w,q} +\frac{1}{2} \Big) \ \sign \big(m^{w,q}_{\rm CB}\big) \bigg ] \nn \\
  &\quad  - \frac{3}{2} \fn^q b^\alpha_{(mn}\theta_{q)\alpha} \, ,  \\
  \delta\tilde \Theta_{Im} &= \sum_{R,q} F_{\sfrac 1 2}(R,q) \sum_{w \in R} \ w_I q_m  
  \ \bigg[ \Big(\tilde l_{w,q} +\frac{1}{2} \Big) \ \sign \big(\tilde m^{w,q}_{\rm CB}\big) 
  - \Big(l_{w,q} +\frac{1}{2} \Big) \ \sign \big(m^{w,q}_{\rm CB}\big) \bigg ] \nn \\
    &\quad  -  \frac{1}{2} \fn^L b^\alpha \theta_{m\alpha}\, \cC_{IL}   \, , 
\end{align}
\end{subequations}}
where it is important to notice that for $\tilde\Theta^{\rm{match}}_{IJ}$ and $\tilde\Theta^{\rm{match}}_{mn}$ 
there are besides the standard one-loop contributions \eqref{e:CS_loop_3d}
also the additional classical parts \eqref{e:circle_flux2} and \eqref{e:circle_flux3}
because of the non-zero circle flux of the axions
\begin{align}
 \tilde \Theta^{\rm{class}}_{IJ} &= -  \frac{1}{2} \fn^q b^\alpha \theta_{q\alpha}\, \cC_{IJ}\, , &
 &\tilde \Theta^{\rm{class}}_{mn} = -  \frac{1}{2} \fn^q b^\alpha_{mn} \theta_{q\alpha}\, . \nn
\end{align}

For the six-dimensional theory on the circle we obtain the following relations
{\allowdisplaybreaks \begin{subequations}\label{e:missmatch5d}
\begin{align}
  \delta\tilde k_{IJK} &= \sum_{R,q} F_{\sfrac 1 2}(R,q) \sum_{w \in R}  w_I w_J w_K 
  \ \bigg[ \Big(\tilde l_{w,q} +\frac{1}{2} \Big) \ \sign \big(\tilde m^{w,q}_{\rm CB}\big) 
  - \Big(l_{w,q} +\frac{1}{2} \Big) \ \sign \big(m^{w,q}_{\rm CB}\big) \bigg ] \nn \\
  &\quad + 3 \fn^L b^\alpha b^\beta \eta_{\alpha\beta}\ \cC_{(IJ} \cC_{KL)} \, ,  \\
  \delta\tilde k_{mnp} &= \sum_{R,q} F_{\sfrac 1 2}(R,q) \sum_{w \in R} \ q_m q_n q_p 
  \ \bigg[ \Big(\tilde l_{w,q} +\frac{1}{2} \Big) \ \sign \big(\tilde m^{w,q}_{\rm CB}\big) 
  - \Big(l_{w,q} +\frac{1}{2} \Big) \ \sign \big(m^{w,q}_{\rm CB}\big) \bigg ] \nn \\
  &\quad + 3 \fn^q b^\alpha_{(mn} b^\beta_{pq)} \eta_{\alpha\beta} \, , \\
    \delta\tilde k_{IJm} &= \sum_{R,q} F_{\sfrac 1 2}(R,q) \sum_{w \in R} \ w_I w_J q_m 
  \ \bigg[ \Big(\tilde l_{w,q} +\frac{1}{2} \Big) \ \sign \big(\tilde m^{w,q}_{\rm CB}\big) 
  - \Big(l_{w,q} +\frac{1}{2} \Big) \ \sign \big(m^{w,q}_{\rm CB}\big) \bigg ] \nn \\
  &\quad +  \fn^q b^\alpha b^\beta_{mq} \eta_{\alpha\beta}\ \cC_{IJ}  \, ,  \\
    \delta\tilde k_{Imn} &= \sum_{R,q} F_{\sfrac 1 2}(R,q) \sum_{w \in R} \ w_I q_m q_n
  \ \bigg[ \Big(\tilde l_{w,q} +\frac{1}{2} \Big) \ \sign \big(\tilde m^{w,q}_{\rm CB}\big) 
  - \Big(l_{w,q} +\frac{1}{2} \Big) \ \sign \big(m^{w,q}_{\rm CB}\big) \bigg ] \nn \\
   &\quad +  \fn^L b^\alpha b^\beta_{mn} \eta_{\alpha\beta}\ \cC_{IL}  \, ,  \\
    \delta\tilde k_{I} &= -2\sum_{R,q} F_{\sfrac 1 2}(R,q) \sum_{w \in R} \ w_I
  \ \bigg[ \Big(\tilde l_{w,q} +\frac{1}{2} \Big) \ \sign \big(\tilde m^{w,q}_{\rm CB}\big) 
  - \Big(l_{w,q} +\frac{1}{2} \Big) \ \sign \big(m^{w,q}_{\rm CB}\big) \bigg ] \nn \\
   &\quad + 12 \fn^L a^\alpha b^\beta \eta_{\alpha\beta}\ \cC_{IL}  \, ,  \\
    \delta\tilde k_{m} &= -2\sum_{R,q} F_{\sfrac 1 2}(R,q) \sum_{w \in R} \ q_m
  \ \bigg[ \Big(\tilde l_{w,q} +\frac{1}{2} \Big) \ \sign \big(\tilde m^{w,q}_{\rm CB}\big) 
  - \Big(l_{w,q} +\frac{1}{2} \Big) \ \sign \big(m^{w,q}_{\rm CB}\big) \bigg ] \nn \\
     &\quad + 12 \fn^q a^\alpha b^\beta_{mq} \eta_{\alpha\beta}  \, .  
\end{align}
\end{subequations}}
This mismatch of evaluating the transformation of Chern-Simons couplings via different methods might seem
confusing at first sight. However crucially, it is possible to rewrite \eqref{e:missmatch3d} and \eqref{e:missmatch5d}
using the very important identity
\begin{align}
\Big(\tilde l_{w,q} + \frac{1}{2}\Big) \sign \big(\tilde m^{w,q}_{\rm CB}\big)  - \Big(l_{w,q} + \frac{1}{2}\Big) 
\sign \big(m^{w,q}_{\rm CB}\big) 
=    \fn^L w_{L} + \fn^q q_{q} \, ,
\end{align}
which we prove in \autoref{sec:cb_id}. We obtain for the four-dimensional theory on the circle
{\allowdisplaybreaks \begin{subequations}
\begin{align}
  \delta\tilde \Theta_{IJ} &= \sum_{R,q} F_{\sfrac 1 2}(R,q) \sum_{w \in R} \ w_I w_J  
  \ \big[ \fn^L w_{L} + \fn^q q_{q} \big ]   -  \frac{1}{2} \fn^q b^\alpha \theta_{q\alpha}\, \cC_{IJ}  \, ,  \\
  \delta\tilde \Theta_{mn} &= \sum_{R,q} F_{\sfrac 1 2}(R,q) \sum_{w \in R} \ q_m q_n  
  \ \big[ \fn^L w_{L} + \fn^q q_{q} \big ]   - \frac{3}{2} \fn^q b^\alpha_{(mn}\theta_{q)\alpha} \nn \\
  & = \sum_{R,q} F_{\sfrac 1 2}(R,q) \sum_{w \in R} \ q_m q_n  
     \fn^q q_{q}  - \frac{3}{2} \fn^q b^\alpha_{(mn}\theta_{q)\alpha} \, ,  \\
  \delta\tilde \Theta_{Im} &= \sum_{R,q} F_{\sfrac 1 2}(R,q) \sum_{w \in R} \ w_I q_m  
  \ \big[ \fn^L w_{L} + \fn^q q_{q} \big ]   -  \frac{1}{2} \fn^L b^\alpha \theta_{m\alpha}\, \cC_{IL} \nn \\
  &= \sum_{R,q} F_{\sfrac 1 2}(R,q) \sum_{w \in R} \ w_I q_m  
    \fn^L w_{L}  -  \frac{1}{2} \fn^L b^\alpha \theta_{m\alpha}\, \cC_{IL} \, ,
\end{align}
\end{subequations}}
where we made also use of the algebraic identity
\begin{align}
 \sum_{w \in R} \ w_I = 0 \, ,
\end{align}
which holds for all highest weight representations $R$ and is derived in \cite{Grimm:2013oga}.
For the circle-reduced six-dimensional theory we get
{\allowdisplaybreaks \begin{subequations}
\begin{align}
  \delta\tilde k_{IJK} &= \sum_{R,q} F_{\sfrac 1 2}(R,q) \sum_{w \in R} \ w_I w_J w_K 
  \ \big[ \fn^L w_{L} + \fn^q q_{q} \big ]  + 3 \fn^L b^\alpha b^\beta \eta_{\alpha\beta}\ \cC_{(IJ} \cC_{KL)} \, ,  \\
  \delta\tilde k_{mnp} &= \sum_{R,q} F_{\sfrac 1 2}(R,q) \sum_{w \in R} \ q_m q_n q_p 
  \ \big[ \fn^L w_{L} + \fn^q q_{q} \big ]  + 3 \fn^q b^\alpha_{(mn} b^\beta_{pq)} \eta_{\alpha\beta} \nn \\
  &= \sum_{R,q} F_{\sfrac 1 2}(R,q) \sum_{w \in R} \ q_m q_n q_p 
 \fn^q q_{q}  + 3 \fn^q b^\alpha_{(mn} b^\beta_{pq)} \eta_{\alpha\beta} \, , \\
    \delta\tilde k_{IJm} &= \sum_{R,q} F_{\sfrac 1 2}(R,q) \sum_{w \in R} \ w_I w_J q_m 
  \ \big[ \fn^L w_{L} + \fn^q q_{q} \big ]  +  \fn^q b^\alpha b^\beta_{mq} \eta_{\alpha\beta}\ \cC_{IJ}  \, ,  \\
    \delta\tilde k_{Imn} &= \sum_{R,q} F_{\sfrac 1 2}(R,q) \sum_{w \in R} \ w_I q_m q_n
  \ \big[ \fn^L w_{L} + \fn^q q_{q} \big ]  +  \fn^L b^\alpha b^\beta_{mn} \eta_{\alpha\beta}\ \cC_{IL}  \nn  \\
  &= \sum_{R,q} F_{\sfrac 1 2}(R,q) \sum_{w \in R} \ w_I q_m q_n
   \fn^L w_{L}   +  \fn^L b^\alpha b^\beta_{mn} \eta_{\alpha\beta}\ \cC_{IL}  \, ,  \\
    \delta\tilde k_{I} &= -2\sum_{R,q} F_{\sfrac 1 2}(R,q) \sum_{w \in R} \ w_I
  \ \big[ \fn^L w_{L} + \fn^q q_{q} \big ]  + 12 \fn^L a^\alpha b^\beta \eta_{\alpha\beta}\ \cC_{IL}  \nn \\
  &= -2\sum_{R,q} F_{\sfrac 1 2}(R,q) \sum_{w \in R} \ w_I
  \fn^L w_{L}   + 12 \fn^L a^\alpha b^\beta \eta_{\alpha\beta}\ \cC_{IL}  \, ,  \\
    \delta\tilde k_{m} &= -2\sum_{R,q} F_{\sfrac 1 2}(R,q) \sum_{w \in R} \ q_m
  \ \big[ \fn^L w_{L} + \fn^q q_{q} \big ] + 12 \fn^q a^\alpha b^\beta_{mq} \eta_{\alpha\beta}  \nn \\
  &= -2\sum_{R,q} F_{\sfrac 1 2}(R,q) \sum_{w \in R} \ q_m
   \fn^q q_{q}  + 12 \fn^q a^\alpha b^\beta_{mq} \eta_{\alpha\beta}  \, . 
\end{align}
\end{subequations}}
The interpretation of these potential mismatches is now completely obvious, they are the anomalies in four and six dimensions.
In particular, by taking derivatives with respect to the winding numbers $\fn^L$, $\fn^q$ of the large gauge transformations
we are able to reproduce all anomalies \eqref{4danomalies_alternative} in four dimensions except for the mixed gauge-gravitational anomaly.
Also in six dimensions we get all anomalies \eqref{e:6d_anom_alernative} except for the pure gravitational anomalies.
In detail, the condition $\delta\tilde \Theta_{\Lambda\Sigma}\overset{!}{=}0$
yields
\begin{center}
\begin{tabular}{r|ccc}
\rule[-.3cm]{0cm}{.8cm}  & $\partial_{\fn^L}$ & $\partial_{\fn^L}\partial_{\fn^q}$ & $\partial_{\fn^q}$\\
 \hline
\rule[-.3cm]{0cm}{.8cm} $\delta\tilde \Theta_{IJ}\overset{!}{=}0$ & \eqref{4d_anomaly_alt2} & 0 & \eqref{4d_anomaly_alt3} \\
\rule[-.3cm]{0cm}{.8cm} $\delta\tilde \Theta_{mn}\overset{!}{=}0$ & 0 & 0 &  \eqref{4d_anomaly_alt4} \\
\rule[-.3cm]{0cm}{.8cm} $\delta\tilde \Theta_{Im}\overset{!}{=}0$ & \eqref{4d_anomaly_alt3} & 0 &  0
\end{tabular}
\end{center}
In six dimensions the conditions $\delta\tilde k_{\Lambda\Sigma\Theta}\overset{!}{=}0$ and $\delta\tilde k_{\Lambda}\overset{!}{=}0$
give
\begin{center}
\begin{tabular}{r|ccc}
\rule[-.3cm]{0cm}{.8cm}  & $\partial_{\fn^L}$ & $\partial_{\fn^L}\partial_{\fn^q}$ & $\partial_{\fn^q}$\\
 \hline
\rule[-.3cm]{0cm}{.8cm} $\delta\tilde k_{IJK}\overset{!}{=}0$ & \eqref{e:6d_alter_5} & 0 & \eqref{e:6d_alter_6} \\
\rule[-.3cm]{0cm}{.8cm} $\delta\tilde k_{mnp}\overset{!}{=}0$ & 0 & 0 &\eqref{e:6d_alter_8} \\
\rule[-.3cm]{0cm}{.8cm} $\delta\tilde k_{IJm}\overset{!}{=}0$ & \eqref{e:6d_alter_6} & 0 & \eqref{e:6d_alter_7} \\
\rule[-.3cm]{0cm}{.8cm} $\delta\tilde k_{Imn}\overset{!}{=}0$ & \eqref{e:6d_alter_7} & 0 & 0 \\
\rule[-.3cm]{0cm}{.8cm} $\delta\tilde k_{I}\overset{!}{=}0$ & \eqref{e:6d_alter_3} & 0 & 0 \\
\rule[-.3cm]{0cm}{.8cm} $\delta\tilde k_{m}\overset{!}{=}0$ & 0 & 0 & \eqref{e:6d_alter_4}
\end{tabular}
\end{center}

We have seen that this procedure misses the pure gravitational anomalies in six dimensions. This is of course expected
since in our procedure we probe the spectrum with large \textit{gauge} transformations and not large \textit{Lorentz} transformations.
The fact that we are able to obtain the mixed gauge-gravitational anomalies is due to the appearance of the curvature two-form
in the gravitational Chern-Simons term \eqref{5dgravCS} which is nevertheless probed with large \textit{gauge} transformations.
In contrast, the mixed gauge-gravitational anomaly in four dimensions can not be reproduced with our procedure because
a gravitational Chern-Simons term does not exist in three dimensions.
We are confident that all missing anomalies can be obtained by acting with large Lorentz transformations on Chern-Simons terms,
and we collect evidence for that in \autoref{sec:grav_anomalies}.

Finally let us clarify the intuition behind our procedure and why it has this intriguing connection to anomalies.
By construction, the classical four- and six-dimensional action on the circle is of course invariant
under large gauge transformations (ignoring Green-Schwarz terms). In some sense one can then
interpret our first way for evaluating the transformation of the Chern-Simons couplings, namely by treating them as duals of the vectors
\eqref{e:dual_trafo}, as exploiting the \textit{classical} invariance under large gauge transformation in order to determine them.
The second way for calculating the transformed Chern-Simons coefficients, \textit{i.e.}~by directly evaluating the loops
with the transformed quantity $\tilde m^{w,q}_{\rm CB}$, is only consistent if the invariance under large gauge transformations
is also respected by the quantized theory. For this property to be satisfied it is sufficient that anomalies are canceled.
The remarkable result is that the quantum invariance under large gauge transformations along the circle is actually equivalent
to the cancelation of \textit{all} higher-dimensional gauge anomalies, including also mixed gauge-gravitational anomalies in six dimensions.

The techniques of this chapter were inspired by F-theory compactifications on Calabi-Yau four- and threefolds
which will be the main topic of the upcoming two chapters.
In order to determine the
effective theory in such settings one has to consider a four- or six-dimensional theory, respectively, compactified on a circle.
The Chern-Simons terms then correspond to certain intersections between divisors on the manifold (and homology classes which are dual to flux).
Importantly, what we found is that 
the precise basis of divisors one has to use in order to obtain the correct matching of intersections and Chern-Simons terms
is not uniquely defined. There are rather whole groups of divisors which comprise all possible basis choices. These group structures
precisely correspond to large gauge transformations in the field theory setting.
Abelian large gauge transformations correspond to the well-known Mordell-Weil group of rational sections while for non-Abelian large gauge
transformations we suggest a new group structure on the geometric compactification space.
These groups then establish the cancelation of gauge anomalies in F-theory compactifications on Calabi-Yau manifolds.
We will derive these results in full detail in \autoref{ch:arith}. But before we do so, we have to give a short introduction into
the basic concepts of F-theory in \autoref{ch:f_basics}.

\chapter{The Basic Concepts of F-Theory}\label{ch:f_basics}

In this chapter we provide a very short introduction into the basic ingredients of
F-theory \cite{Vafa:1996xn,Morrison:1996na,Morrison:1996pp}. We stress that our treatment is far from being complete
since we are not doing justice to many in
general important aspects, \textit{e.g.~}the precise derivation of the gauge group and spectrum,
the construction of fluxes, the Sen limit, spectral cover constructions, the duality to
heterotic string theory, the relation to superconformal field theories in six dimension, or T-branes.
It is rather meant as a review of the basic ideas and concepts adjusted in such a way that the special topics
of F-theory effective field theory
which are treated in this thesis can be understood.
Moreover, we require fundamental familiarity with superstring theory. Standard text book references for string theory in general are for
instance \cite{Green:1987sp,Green:1987mn,Polchinski:1998rq,Polchinski:1998rr,Becker:2007zj,Blumenhagen:2013fgp}. For a more
detailed introduction into F-theory we refer to the excellent lecture notes \cite{Denef:2008wq,Weigand:2010wm}.
A nice review of F-theory phenomenology especially for GUT models is given by \cite{Heckman:2010bq}.
Finally, my predecessors included very well-written introductions
to similar topics in F-theory effective field theory
and geometric aspects in their theses \cite{Bonetti:2014,Keitel:2015bma}.

In \autoref{sec:low_energy} we start with a review of the effective physics of type IIB string theory and
M-theory. In particular, we highlight the issue of 7-branes and the need of a non-perturbative formulation
of type IIB theory given by F-theory. We then introduce our working-definition of F-theory in \autoref{sec:duality}
via the duality to M-theory. The basic notion of elliptic fibrations and their implications for F-theory
compactifications are finally treated in \autoref{sec:elliptic}.

\section{Type IIB String Theory and M-Theory}\label{sec:low_energy}
 Up to now there exists no satisfying fundamental description for F-theory, especially not in terms of a twelve-dimensional effective
 action.\footnote{See \cite{Berman:2015rcc} for a recent approach into this direction.} However,
 there are several indirect ways of
 how F-theory can be defined. The most straightforward approach towards this theory is given by considering it as the non-perturbative generalization of
 type IIB string theory with 7-branes and
 varying axio-dilaton background. The latter parametrizes an auxiliary two-torus, which is why F-theory is often considered as
 a twelve-dimensional theory.
 Sometimes the duality between type IIB string theory and M-theory is invoked
 in order to actually define F-theory as M-theory on a torus-fibration with vanishing fiber. As we will see, the vanishing of the
 two-torus grows the
 one additional dimension which is needed for F-theory (via T-duality). It is unknown if the approach via M-theory captures
 all aspects of F-theory, for example it is not clear how a compactification of F-theory on a Calabi-Yau sixfold could be treated, or if
 it is even a consistent background. In this thesis however our approach to F-theory precisely proceeds through this duality to M-theory.
 Note that there is also the important duality of F-theory on an elliptically-fibered K3 to $E_8 \times E_8$ heterotic string theory on a
 two-torus
 (with specification of a vector bundle), which however won't be of importance for the work in this thesis.

\subsection{Low-Energy Description of Type IIB String Theory}
Let us start reviewing the perturbative description of type IIB string theory.
In \autoref{fields_IIB} we list the massless fields in the different sectors of left- and right-movers on the worldsheet. 
\begin{table}
\begin{center}
\begin{tabular}{ccc}
\rule[-.3cm]{0cm}{.8cm} 10d Fields & $SO(8)$ Representations & Sector \\
\hline
\rule[-.3cm]{0cm}{.8cm} $\phi$, $B_2$, $G$ & $\textbf{1}\oplus \textbf{28}_V \oplus \textbf{35}_V$ & NS-NS\\
\rule[-.3cm]{0cm}{.8cm} $\lambda^1$, $\psi^1$ & $\textbf{8}_S\oplus \textbf{56}_S$ & NS-R\\
\rule[-.3cm]{0cm}{.8cm} $\lambda^2$, $\psi^2$ & $\textbf{8}_S\oplus \textbf{56}_S$  & R-NS\\
\rule[-.3cm]{0cm}{.8cm} $C_0$, $C_2$, $C_4$ & $\textbf{1}\oplus \textbf{28}_C \oplus \textbf{35}_C$ & R-R
\end{tabular}
\end{center}
\caption{The massless field content of type IIB string theory in the different sectors of the worldsheet. The index $V$
labels vector-like representations while $S$, $C$ refer to (left- or right-handed) chiral representations of the massless
little group $SO(8)$.}
\label{fields_IIB}
\end{table}
The NS-NS sector consists of the dilaton $\phi$, the Kalb-Ramond field $B_2$ and the metric $G$. The bosonic spectrum is completed by
the differential form fields $C_0$, $C_2$, $C_4$ in the R-R sector accompanied by their magnetic duals.
The fermionic spectrum contains the dilatini $\lambda^1$, $\lambda^2$ and the gravitini $\psi^1$, $\psi^2$. We will omit terms which involve
the fermionic fields since they can be inferred by invoking supersymmetry.
It is convenient to redefine the bosonic fields according to
\begin{align}
H_3 &:= dB_{2} \, , & F_p &:= dC_{p-1} \quad (p=1,3,5) \, ,&  \tau &:= C_0 + i e^{-\phi} \, ,\nn \\
 G_3 &:= F_3 - \tau H_3 \, ,& \tilde F_5 &:= F_5 - \frac{1}{2} C_2 \wedge H_3 + \frac{1}{2} B_2 \wedge F_3 \, .
\end{align}

The low-energy effective action of the massless fields is given by $\cN=(2,0)$ supergravity in ten dimensions, whose pseudo-action
reads in the Einstein frame
\begin{align}\label{e:IIB_action}
  S_{\textrm{IIB}}^{\rm pseudo} = \frac{2\pi}{l_s^8}\int_{M_{10}} R \ast 1 - \frac{1}{2} \frac{d\tau\wedge\ast d\bar{\tau}}{(\textrm{Im}\tau )^2}
  - \frac{1}{2} \frac{G_3 \wedge \ast \bar{G}_3}{\textrm{Im}\tau } 
 - \frac{1}{4}\tilde{F}_5\wedge\ast\tilde{F}_5 - \frac{1}{2} C_4\wedge H_3 \wedge F_3 
 \end{align}
 with $l_s = 2\pi \sqrt{\alpha^\prime}$ the string length and $R$ the ten-dimensional Ricci scalar derived from the metric $G$.
It is crucial to keep in mind that we have written down only a pseudo-action. In fact, one has to additionally impose the self-duality condition
\begin{align}
 \ast \tilde F_5 = \tilde F_5 
\end{align}
\textit{after} deriving the equations of motion from the pseudo-action. Note that no manifestly covariant action for such a self-dual tensor field is available.
It is well-known that the action\footnote{We will omit the prefix \textit{pseudo} in the following.} \eqref{e:IIB_action} is classically invariant under the group $SL(2,\mathbb R)$. Indeed, if we assign
the following transformation properties to the individual fields
\begin{subequations}\label{e:IIB_trafos}
 \begin{align}
  \textrm{for }\begin{pmatrix} a&b\\c&d \end{pmatrix} \in SL(2,\mathbb{R}):\qquad\qquad\begin{pmatrix} C_2 \\ B_2\end{pmatrix} & \mapsto \begin{pmatrix} a&b\\c&d \end{pmatrix} \begin{pmatrix} C_2 \\ B_2 \end{pmatrix}  \ , & \\
  \tau & \mapsto \frac{a \tau + b}{c \tau + d}\ ,&
 \end{align}
  \end{subequations}
and all other fields invariant, it is easy to see that this constitutes a symmetry.
Note that since D(-1) instantons contribute with a factor of $e^{2\pi i \tau}$, the classical $SL(2,\mathbb R)$ is broken at the quantum level
down to $SL(2,\mathbb Z)$. This behavior is very similar to the S-duality of $\cN=4$ super Yang-Mills theory in
four dimensions, in particular it can be invoked to relate regimes of strong and weak string coupling $g_s = e^\phi$.

The supergravity action \eqref{e:IIB_action} also allows for in general
non-perturbative solutions which correspond to (electric or magnetic) sources for the
generalized gauge potentials $B_2$, $C_0$, $C_2$, $C_4$. The R-R fields are sourced by D-branes while
for the Kalb-Ramond field the fundamental string and the NS5-brane constitute the electric and magnetic sources, respectively.
We summarize the fields together with their respective sources in \autoref{IIB_branes}.
\begin{table}
\begin{center}
\begin{tabular}{rll}
\rule[-.3cm]{0cm}{.8cm} Gauge Potential & Electric Source & Magnetic Source\\
\hline
\rule[-.3cm]{0cm}{.8cm} $B_2$ & fundamental string & NS5-brane\\
\rule[-.3cm]{0cm}{.8cm} $C_0$ & D(-1)-brane & D7-brane \\
\rule[-.3cm]{0cm}{.8cm} $C_2$ & D1-brane  & D5-brane \\
\rule[-.3cm]{0cm}{.8cm} $C_4$ & D3-brane  & D3-brane
\end{tabular}
\end{center}
\caption{We display the generalized gauge potentials of type IIB supergravity with the corresponding electric and magnetic sources.}
\label{IIB_branes}
\end{table}

The following discussion now is essential in order to understand the nature and necessity for F-theory. We focus in more detail on the different
types of branes and we elaborate on the fundamental difference between codimension-two branes on the one hand
and lower-dimensional branes on the other hand.
Already at this stage we draw attention to the following fact: Because of the $SL(2, \mathbb Z)$ transformation properties
\eqref{e:IIB_trafos} of the massless
fields the sources for the latter, \textit{i.e.}~the branes and strings, mix in general under
the action of $SL(2, \mathbb Z)$. In the following we will
therefore generally denote the sources which extend over $p+1$ dimensions as $p$-branes.
It is important to recognize the special asymptotic behavior of 7-branes in contrast to $p$-branes with $p < 7$.
More precisely, consider general codimension-$n$ $p$-branes, \textit{i.e.}~$p= 9-n$. In the $n$ directions normal to the brane the latter looks
like a point-like source and we find a Poisson equation for the sourced fields $\Phi$
\begin{align}
 \Delta^{(n)} \Phi (r) \sim \delta^{(n)} (r) \, 
\end{align}
with $r$ the radial coordinate in the $n$ normal directions to the brane.
The solutions to this equation are of course well-known and depend crucially on $n$
\begin{subequations}
\begin{align}
 \Phi (r) &\sim \frac{1}{r^{n-2}} \quad \textrm{for } n>2 \, , \\
 \Phi (r) &\sim \mathrm{log} (r) \quad \textrm{for } n=2 \, .
\end{align}
\end{subequations}
From this heuristic discussion one realizes that for $n>2$ the effect of backreaction of the branes drops off as long as one moves away from
the brane far enough. For
a more detailed discussion see for instance \cite{Weigand:2010wm}.
However, this is not true anymore for $n=2$, \textit{i.e.}~for 7-branes, since the
logarithm has a totally different asymptotic behavior and, what seems even more severe, a branch cut.

Let us investigate in detail how this puzzle concerning 7-branes in type IIB is resolved.
We start with a D7-brane, which constitutes a
magnetic source for $C_0$. In terms of $SL(2,\mathbb Z)$ representations $C_0$ combines together with the string coupling $g_s$
into the complex axio-dilaton $\tau$
\begin{align}
 \tau = C_0 + \frac{i}{g_s} \, .
\end{align}
From supersymmetry considerations it is known that $\tau$ must be a holomorphic function in $z := x^8 + i x^9$ with $x^8, x^9$ parameterizing the
space perpendicular to the brane. It can be shown that in the vicinity of the D7-brane located at $z_0$ the solution for $\tau$ takes the form
\begin{align}\label{e:taud7}
 \tau (z) = \tau_0 + \frac{1}{2\pi i} \ln (z - z_0) + \dots 
\end{align}
omitting regular terms. As already mentioned, the solution exhibits a branch cut which means that it is a multivalued function.
Indeed, when we circle once around this D7-brane at $z_0$, we find the monodromy
\begin{align}\label{tau_mono}
 \tau \mapsto \tau +1  \quad \Rightarrow \quad C_0 \mapsto C_0 +1 \, .
\end{align}
Although this behavior might seem odd, we stress that using the
$SL(2, \mathbb Z)$ symmetry of type IIB \eqref{e:IIB_trafos} resolves the puzzle in a very elegant fashion.
In fact, we define the transformation $T$
\begin{align}
 T := \begin{pmatrix}
 1 & 1 \\
 0 & 1
      \end{pmatrix}
 \in SL(2, \mathbb Z)
\end{align}
such that it precisely acts on $\tau$ as in \eqref{tau_mono}.

This special interplay between the $SL(2,\mathbb Z)$ symmetry and the monodromy behavior of
the D7-brane is actually
only the tip of the iceberg. We have seen that the $SL(2,\mathbb Z)$ symmetry mixes the fields $C_2$ and $B_2$.
Thus we should combine the electric sources of these fields, \textit{i.e.}~the fundamental string and the D1-brane, into an $SL(2, \mathbb Z)$ doublet, the
$(p,q)$-string carrying charge $p$ under $B_2$ and charge $q$ under $C_2$. Thus the fundamental string is represented by $(1,0)$ and the
D1-brane is written as $(0,1)$. A $(p,q)$-7-brane is then defined as the object on which
a $(p,q)$-string can end, generalizing the notion of the D7-brane.
The monodromy around a $(p,q)$-7-brane takes the general form
\begin{align}\label{e:mono}
 M_{p,q} = \begin{pmatrix}
 1-pq & p^2  \\
 - q^2 & 1+ pq
           \end{pmatrix} \in SL(2, \mathbb Z) \, .
\end{align}
Note that for each single $(p,q)$-7-brane in the theory it is possible to transform it into a D7-brane
using $SL(2,\mathbb Z)$ transformations. However, globally this does not work for all
7-branes at the same time, at least in the generic case.

In the following we will show how the standard type IIB setup with conventional D7-branes and O7-planes fits into this pattern.
For type IIB orientifold compactifications one usually places four D7-branes on top of one O7-plane in order to achieve tadpole cancelation
locally (a D7 brane carries one unit of charge while an O7-plane contributes $-4$ units). This special configuration results in a constant $\tau$
over the whole compactification space, and since $\tau_0$ is a modulus, the string coupling can be chosen to be perturbatively
small $g_s = e^\phi \ll 1$ everywhere. In fact, the  O7-plane is described by
the combination of a $(3,-1)$-7-brane and a $(1,-1)$-7-brane
as one can check by evaluating the corresponding monodromy \eqref{e:mono} around the brane system.

Generically however, tadpole cancelation does not necessarily have to be accomplished locally
which then results in a non-constant axio-dilaton. However, it is still possible to find a weak-coupling limit in complex structure moduli space, \textit{i.e.}~by a limit on $\tau_0$.
For this rewrite the solution \eqref{e:taud7} as (ignoring regular terms)
\begin{align}
 \tau (z) = \frac{1}{2\pi i} \ln \frac{z - z_0}{\lambda}\, , \quad \textrm{with }\,\, \ln \lambda := -2\pi i \tau_0 \, . 
\end{align}
For the string coupling we find
\begin{align}
 \frac{1}{g_s} = - \frac{1}{2\pi} \ln \Big \vert \frac{z-z_0}{\lambda} \Big \vert \, . 
\end{align}
In particular, if $\vert z -z_0 \vert \ll \vert \lambda \vert$, then $g_s$ becomes small. So the limit of weak coupling in complex structure
moduli space is given by $\vert \lambda \vert \rightarrow \infty$ since, loosely speaking, the region 
with $\vert z -z_0 \vert \ll \vert \lambda \vert$
becomes large. Note however that the axio-dilaton still has a varying profile generically since the
latter depends on the configuration of branes.
For a more detailed analysis of the weak coupling limit we refer to the original paper of Sen \cite{Sen:1997gv}.

Finally, we are now in the position to formulate the F-theory conjecture:
The framework of F-theory interprets the $SL(2,\mathbb Z)$ as the parametrization symmetry
of an actual two-torus $T^2$ (in the limit of vanishing volume).
The axio-dilaton $\tau$ constitutes the complex-structure modulus of this $T^2$, and a varying background for $\tau$ corresponds to a non-trivial
torus-fibration structure over some
base space. The problem of finding consistent axio-dilaton profiles is then geometrically recast into the task
of constructing genuine $T^2$-fibrations. As we will see later in more detail, the whole information about 7-branes and their backreaction
is encoded in the geometry of the elliptic fibration. In this thesis we are mainly concerned with compactifications down to effective four-
and six-dimensional supergravity theories with minimal supersymmetry. These effective settings can be
obtained by considering F-theory compactifications on $T^2$-fibered, compact
Calabi-Yau four- and threefolds, respectively. They describe (non-perturbative) type IIB compactifications on
the complex-three- or two-dimensional base spaces of
the $T^2$-fibrations.

\begin{figure}
 \centering
 \includegraphics[scale=1]{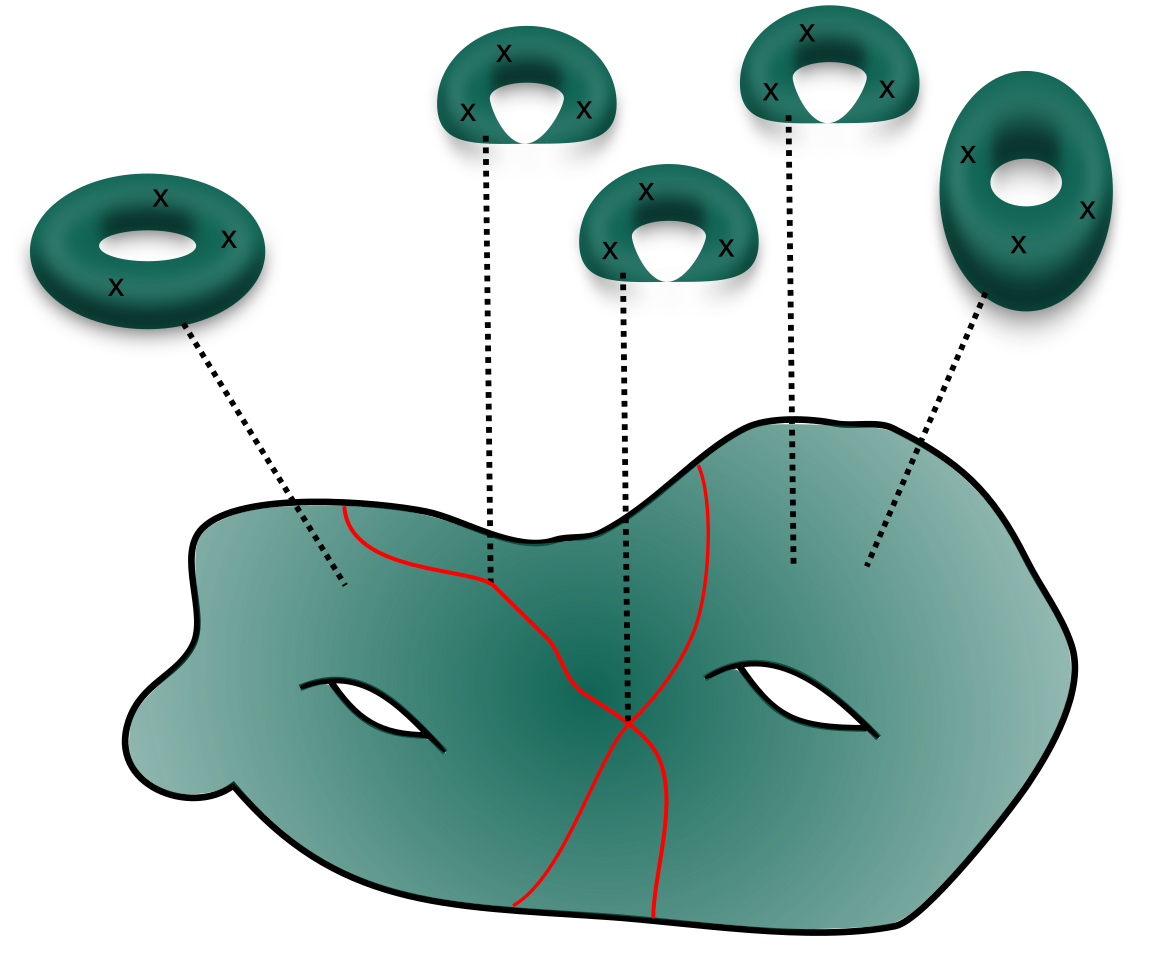}
 \caption{We schematically depict a $T^2$-fibration over some base space.
Degenerations of the fiber indicate the presence of 7-branes. These pinching tori can in principle
 appear over isolated points in the base or over higher-dimensional submanifolds (indicated by the red curves).
 The crosses denote rational sections of the fibration. We will introduce $T^2$-fibrations in \autoref{sec:elliptic} in more detail.}
 \label{fig:fibr}
\end{figure}

From \eqref{e:taud7} it is easy to see how one can
locate the position of 7-branes in the base space since the axio-dilaton, \textit{i.e.}~the complex
structure of the torus, diverges at the 7-brane loci. In order to find the latter we therefore have to look for degenerations of the $T^2$-fibration,
\textit{i.e.}~for loci in the base over which the torus pinches as depicted in \autoref{fig:fibr}.
This is part of \autoref{subsec:matter}
where we sketch how the information of the gauge group, matter and Yukawa couplings is encoded in the geometry.
Before we come to that, we first explain in \autoref{sec:duality} the duality between F-theory and M-theory which is also sometimes referred
to as the definition of F-theory. As a preparation we shortly review the low-energy effective physics of M-theory.

\subsection{Low-Energy Description of M-Theory}\label{sec:low_energy_m}

At low energies M-theory can be approximated by eleven-dimensional supergravity.
The field content of the latter is summarized in \autoref{tab:M_fields}.
\begin{table}
\begin{center}
\begin{tabular}{cc}
\rule[-.3cm]{0cm}{.8cm} 11d Fields & $SO(9)$ Representations \\
\hline
\rule[-.3cm]{0cm}{.8cm} $G$,  $C_3$ & $\textbf{44}\oplus\textbf{84}$ \\
\rule[-.3cm]{0cm}{.8cm} $\psi$ & $\textbf{128}$  \\
\end{tabular}
\end{center}
\caption{We display the field content of eleven-dimensional supergravity in terms of representations of the massless
little group $SO(9)$.}
\label{tab:M_fields}
\end{table}
It consists of the metric $G$, the gravitino $\psi$ and
a three-form field $C_3$ (accompanied by the dual six-form field) with field strength $G_4 = dC_3$.
The bosonic effective action of M-theory is given by
\begin{align}\label{e:eff_action_m}
 S_M = \frac{2\pi}{l_M^9} \int_{M_{11}} R \ast 1 - \frac{1}{2} G_4 \wedge \ast G_4 + \frac{1}{12} C_3 \wedge G_4 \wedge G_4
 + l_M^6 \,  C_3 \wedge I_8 (\cR) + \dots \, ,
\end{align}
with $l_M$ the Planck length in eleven dimensions.
Note that the part of the action which involves the
gravitino $\psi$ can be derived by invoking the power of supersymmetry.
Furthermore, in addition to the standard two-derivative supergravity action
we also included the higher-curvature contribution $I_8 (\cR)$ \cite{Vafa:1995fj,Duff:1995wd} in the effective action.
$I_8 (\cR)$ is a polynomial of degree four in the curvature two-form $\cR$, and it
plays a crucial role in anomaly and tadpole cancelation
In particular,
for compactifications on Calabi-Yau threefolds
$I_8 (\cR)$ induces a gravitational Chern-Simons term \eqref{5dgravCS} which lifts to a Green-Schwarz counterterm in six dimensions
canceling gravitational anomalies.

Finally the electric sources for $C_3$ are given by M2-branes whose world-volume theory can be described by the famous
ABJM theory \cite{Aharony:2008ug}.
The magnetic sources constitute M5-branes whose world-volume theories are the mysterious six-dimensional $\cN = (2,0)$
superconformal field theories
\cite{Strominger:1995ac,Witten:1995em}. For a single M5-brane one faces a free theory of an (Abelian) $(2,0)$ tensor multiplet
which is well-understood. In the case of a stack of M5-branes the situation is much more subtle since one expects an interacting theory of
non-Abelian tensors for which the existence of an action is not expected. Although few is known about such theories,
let us collect at least some
properties that have been derived:
\begin{itemize}
 \item They follow an $ADE$-pattern.
 \item There are some results about the conformal anomalies and its relations to R-symmetry
 anomalies \cite{Henningson:1998gx,Harvey:1998bx,Bastianelli:2000hi,Yi:2001bz,Gaiotto:2009we,Maxfield:2012aw}. In particular
 the conformal anomaly of $N$ M5-branes scales like $N^3$.
 \item The effective theory on a circle is given by $\cN=4$ supersymmetric Yang-Mills theory in five dimensions. The gauge algebra is
 of the $ADE$-type of the higher-dimensional theory. For some first steps towards investigating properties
 of six-dimensional superconformal field theories via a circle-compactification to five dimensions
 exploiting the Kaluza-Klein tower and the power of Chern-Simons terms we refer to \textit{e.g.~}\cite{Bonetti:2012st,Ohmori:2014kda}.
 \item Classical string constructions are type IIB compactifications on $\mathbb C^2 / \Gamma$ with
 $\Gamma$ a discrete subgroup of $SU(2)$\cite{Witten:1995zh}.
 \item The F-theory realization is given by compactifications on non-compact Calabi-Yau threefolds of the form $B\times T^2$ with 
 $B=\mathbb C^2 / \Gamma$. Again $\Gamma$ is a discrete subgroup of $SU(2)$.
 There are also attempts to classify and investigate six-dimensional
 $\cN=(1,0)$ and $\cN=(2,0)$ superconformal field theories by constructing non-compact elliptically-fibered
 Calabi-Yau threefolds for F-theory compactifications, for recent work in this direction see \cite{Heckman:2013pva,DelZotto:2014hpa,Heckman:2014qba,
 Haghighat:2014vxa,DelZotto:2014fia,Heckman:2015bfa,DelZotto:2015isa,Gadde:2015tra,Heckman:2015ola,Bertolini:2015bwa,
 Bhardwaj:2015oru,Heckman:2016ssk,Font:2016odl,Morrison:2016djb}.
\end{itemize}

\section{Defining F-Theory via M-Theory}\label{sec:duality}

In the previous section we approached F-theory via generalizing type IIB with a generic configuration of 7-branes which takes their
backreaction into account. However, 
often the duality between type IIB string theory and M-theory is actually invoked in order to define F-theory compactifications
as M-theory compactified
on torus fibrations whose fiber volume vanishes in a certain limit to be explained later. While it is not clear if all
compactifications
of F-theory can be approached via M-theory,\footnote{For example it is not clear how to treat compactifications on
torus-fibered Calabi-Yau sixfolds
or if they are consistent at all.}
in many instances it serves as a more convenient definition to work with. Also in this thesis
F-theory compactifications on torus-fibered Calabi-Yau manifolds are always understood in the dual M-theory setting. In the following
we will shortly review this important duality.

\subsection{Fibering the Duality}
Obviously, M-theory on a two-torus is dual to type IIB theory on a circle since the compactification of
M-theory on one circle gives type IIA theory and
T-duality along the second circle leads to type IIB theory on the circle with dual radius. We now explain
how this procedure can be applied also fiberwise for non-trivial torus fibrations. We closely follow the excellent treatment
in \cite{Denef:2008wq}
where more details can be found.

Let us consider an M-theory compactification on an elliptic fibration over some nine-dimensional spacetime manifold $M_9$.
Technically
an elliptic fibration is a $T^2$-fibration with a rational section. We will introduce the notions of elliptic fibrations and
rational sections together with their nice
mathematical properties in more detail
in \autoref{sec:elliptic}. However, for the moment
it is only important to know that, if the $T^2$-fibration has a rational section (\textit{i.e.}~constitutes an elliptic fibration),
the metric does not have off-diagonal components between base and fiber, and it thus
takes the form\footnote{This ansatz is only correct in the limit $v \rightarrow 0$ which we are however going to take in the end.}
\begin{align}\label{Mth_metric}
 ds_M^2 = \frac{v}{\tau_2}\Big ( (dx + \tau_1 dy )^2 + \tau_2^2 dy^2 \Big ) + ds_9^2
\end{align}
with $x$ and $y$ parameterizing the torus with complex structure $\tau = \tau_1 + i \tau_2$ and volume $v$. For a non-trivial
fibration structure both moduli $\tau$ and $v$ depend on the coordinates of $M_9$.
We stress that also $T^2$-fibrations without sections are consistent backgrounds. Indeed, in
\cite {Morrison:2014era,Anderson:2014yva,Klevers:2014bqa,Garcia-Etxebarria:2014qua,Mayrhofer:2014haa,Mayrhofer:2014laa,Cvetic:2015moa}
it is shown
that fibrations without sections signal the presence of discrete symmetries in the effective theory. Note that
except for
\autoref{sec:multi_group} we will restrict to only elliptic fibrations in this thesis. 

Let us now proceed by applying the duality between
M-theory and type IIB string theory fiberwise to \eqref{Mth_metric}.
In order to do so we choose the cycle parametrized by $x$ as the circle which mediates the duality between M-theory
and type IIA string theory. Afterwards we then T-dualize to type IIB string theory along the second cycle parametrized by $y$.
Suppose moreover that $M_9$ is a product of a compact complex-$n$-dimensional K\"ahler manifold $B_n$ and $(9-2n)$-dimensional Minkowski space.
We choose the torus fibration over $B_n$ to be a Calabi-Yau $(n+1)$-fold in order to preserve supersymmetry.
A careful analysis gives the following relations between the
different quantities of type IIB string theory and M-theory on the elliptic fibration:
\begin{align}
ds^2_{IIB} = -(dx^0)^2 + (dx^1)^2 + \dots + (dx^{8-2n})^2 +  \frac{l_M^6}{v^2} dy^2 + ds_{B_n}^2 \, , \qquad  C_0 + \frac{i}{g_s} = \tau
\end{align}
with $\tau$ the complex structure of the torus and $ds_{IIB}$ the line element in the Einstein frame.
The matching of the remaining differential form fields can also be derived straightforwardly by expanding the M-theory
three-form in harmonic forms which represent
the cycles of the elliptic fiber and again tracing their fate under the duality chain fiberwise.
Note that different type IIB fluxes enjoy a uniform description in the dual M-theory setting in terms of
$G_4$-flux \cite{Witten:1996md}.\footnote{Since most of the time the
construction of appropriate $G_4$-fluxes won't play any crucial role in this thesis, 
we refrain from treating this vast and poorly understood subject in our introduction. At prominent positions we will only mention
the facts which we need for our work.}
We finally summarize the mapping of sources between type IIB string theory and M-theory in \autoref{tab:source_map}.
\begin{table}
\begin{center}
\begin{tabular}{rll}
\rule[-.3cm]{0cm}{.8cm} \textbf{M-theory Object}& \textbf{Fiber Cycle}  & \textbf{Type IIB  Object}\\
\hline\hline
\rule[-.3cm]{0cm}{.8cm} M2-brane & none & D3-brane \\
\rule[-.3cm]{0cm}{.8cm}  & $(p,q)$ & $(p,q)$-string \\
\hline
\rule[-.3cm]{0cm}{.8cm} M5-brane & none & Kaluza-Klein-monopole\\
\rule[-.3cm]{0cm}{.8cm}  & $(p,q)$ & $(p,q)$-5-brane\\
\rule[-.3cm]{0cm}{.8cm}  & $T^2$ & D3-brane\\
\hline
\rule[-.3cm]{0cm}{.8cm} Kaluza-Klein-monopole & $(p,q)$ & $(p,q)$-7-brane \\
\hline
\end{tabular}
\end{center}
\caption{We depict the mapping between brane wrappings/degenerations of cycles in the fiber of M-theory
compactifications and branes in type IIB string theory.}
\label{tab:source_map}
\end{table}
For the precise correspondence between all fields on both sides of the duality see \textit{e.g.~}\cite{Denef:2008wq}.
In the next step we send $v \rightarrow 0$. This limit somehow mysteriously grows an additional non-compact dimension,
and at the same time it implements full Poincar\'e invariance of the resulting $(10-n)$-dimensional Minkowski spacetime.
Although there has been some progress in recent years, this limit is still not understood in full detail.

\section{F-Theory and Elliptic Fibrations}\label{sec:elliptic}

In this section we introduce the important mathematical notion of elliptic fibrations, which are the most important geometric objects of
interest in F-theory compactifications. It is unavoidable that the following discussion requires some basic knowledge
in algebraic geometry and topology.

\begin{figure}
\centering
\includegraphics[scale=3]{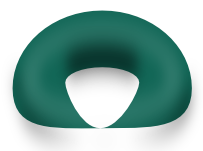}
\caption{We depict a singular elliptic curve where a certain $(p,q)$-cycle has collapsed to zero size.}
\label{fig:sing_tor}
\end{figure}

Elliptic fibrations are defined as fibrations of elliptic curves over a fixed base space, thus we need to define elliptic curves first.
An elliptic curve is a non-singular (in our setups \textit{complex}) algebraic curve of genus one with one marked point, the base point $\mathcal O$.
Loosely
speaking, it is a $T^2$ with a special point singled out.
A convenient way to represent elliptic curves is by embedding them as hypersurfaces into weighted projective space. Since they are
in particular Calabi-Yau onefolds, the degree of the hypersurface equation has to match the sum of all scaling
weights.\footnote{This property implements the vanishing of the first Chern class.
It therefore holds for all Calabi-Yau hypersurfaces in weighted projective space. Compare for instance
the \textit{Quintic} in $\mathbb P^4$.} 
The three most prominent representations are the cubic $\mathbb P_{1,1,1}[3]$, the quartic $\mathbb P_{1,1,2}[4]$ and the sextic
$\mathbb P_{2,3,1}[6]$. We will work with the sextic in the following since it is the most common one in the physics literature.
One can show that after a suitable coordinate redefinition every hypersurface of degree six in $\mathbb P_{2,3,1}$
can be transformed into what is called \textit{Weierstrass form}
\begin{align}\label{Weierstrass}
 y^2 - x^3 - fxz^4 -gz^6 = 0 
\end{align}
with general complex coefficients $f,g$, and $x,y,z$ are the homogenous coordinates of $\mathbb P_{2,3,1}$.
It is easy to see that one can specify a base point $\mathcal O$ with \textit{rational} coordinates independently of $f$ and $g$,
namely $\mathcal{O} := [1,1,0]$. This point will become important when we pass to elliptic fibrations.
Furthermore, we have already mentioned that in the framework of F-theory one also has to consider degenerations of elliptic curves,
since they indicate the presence of 7-branes.
The procedure to figure out from a given Weierstrass equation \eqref{Weierstrass} if the associated elliptic curve degenerates
is in fact not very complicated.
One can show that Weierstrass model is singular if and only if the discriminant $\Delta$, given by
\begin{align}
 \Delta = 27 g^2 + 4f^3 \, ,
\end{align}
vanishes for the given values of $f$ and $g$. In this case the elliptic curve indeed pinches as depicted in \autoref{fig:sing_tor}.

Now we pass to fibrations of elliptic curves over a complex-$n$-dimensional K\"ahler base space $B_n$ with local coordinates $u_i$.
In order to define the fibration structure the former constants $f,g$ in \eqref{Weierstrass}
are promoted to functions in the local coordinates $u_i$, or more accurately formulated,
they become global sections of certain line bundles over the base $B_n$. The line bundles corresponding to the different variables
are fixed by demanding that we want the total space to be Calabi-Yau.
One finds, as for example nicely explained in \cite{Weigand:2010wm,Klevers:2014bqa}, that
\begin{subequations}\label{e:bundle_elements}
\begin{align}
 x &\in H^0 (B_n, K_{B_n}^{-2}) \, ,& y &\in H^0 (B_n, K_{B_n}^{-3}) \, , & z \in H^0 (B_n, \mathcal O_{B_n}) \, , \\
  f &\in H^0 (B_n, K_{B_n}^{-4}) \, ,&  g &\in H^0 (B_n, K_{B_n}^{-6}) \, ,&
\end{align}
\end{subequations}
with $K_{B_n}^{-p}$ the $(-p)$-th power of the canonical bundle of the base space, and $\mathcal O_{B_n}$ is
the trivial bundle over $B_n$. In fact, the former hypersurface in the weighted projective
space $\mathbb P_{2,3,1}$ is promoted to a hypersurface in the weighted projective
bundle $\mathbb P_{2,3,1}(K_{B_n}^{-2}\oplus K_{B_n}^{-3}\oplus \mathcal O_{B_n})$.
The conditions \eqref{e:bundle_elements} are then the global generalizations of the local necessity that the degree of the hypersurface
has to coincide with
the sum over all scaling weights in the ambient projective space.
Finally, we consider the analog object of the base point $\mathcal O$ for elliptic fibrations, the (rational) zero-section.
We have 
mentioned earlier that a genus-one fibration is called an elliptic fibration if it admits at least one rational section, \textit{i.e.}~the section has to cut
out a rational point in each elliptic fiber, and one of the (maybe multiple) rational sections has to be picked as the zero-section. The physical
significance of
this choice and the remaining rational sections will be explained in \autoref{sec:ec_group_structures}
and \autoref{ch:zero_sec}.
Note that fibrations which are defined by a Weierstrass equation \eqref{Weierstrass} always constitute
elliptic fibrations since the $(z=0)$-locus defines a rational section independently of $f$ and $g$. Conversely, one can show
that up to flop transitions any elliptic fibration can be described by \eqref{Weierstrass}. Note that if a ``section'' does not
cut out rational points in each fiber but irrational roots of an algebraic equation, there will be branch cuts. Thus as one moves around the
fibration, monodromies exchange the different roots, and the ``section'' actually cuts out several points in the fiber
whose number is given by the degree of the defining
algebraic equation.
These objects are therefore from now on referred to as multi-sections. They will become important later when we
consider F-theory compactifications on geometries without rational section
in \autoref{sec:multi_group}, but for now we restrict to elliptic fibrations. We stress that nevertheless elliptic fibrations do indeed
possess besides of rational sections also multi-sections, and we will take a first step towards uncovering their significance
in \autoref{sec:graded_MW}.

\subsection{Gauge Symmetry, Matter and Yukawas}\label{subsec:matter}

In F-theory the main interest is usually in locating the degenerations of the elliptic fiber since they indicate the presence of 7-branes.
Therefore we have to look for vanishing loci
of the discriminant $\Delta = 27 g^2 + 4f^3$, which depends on the local coordinates of the base $u_i$.
Suppose that the discriminant vanishes to order $N$
over a complex-codimension-one
locus $S^{\rm b}$ in the base, \textit{i.e.}~a divisor of the base.
This indicates that there are, formulated in the language of type IIB, $N$ coincident 7-branes wrapping
$S^{\rm b}$. For the case of $K3$, \textit{i.e.}~Calabi-Yau twofolds,
Kodaira investigated the different types of fiber degenerations resulting in the famous $ADE$-classification of
singularities.
In the following we will explain how this classification is to be understood by looking at resolutions of the degenerate fibers.
Along the way we will argue that the singularity types over complex-codimension-one in the base precisely constitute the gauge algebras
of the F-theory model.
For higher-dimensional Calabi-Yau manifolds than Calabi-Yau twofolds additional monodromies come into play, and the possible singularity types
are extended to the non-simply laced $B$- and $C$-series as well as $F_4$ and $G_2$, see \textit{e.g.}~\cite{Bershadsky:1996nh}.

\begin{figure}
\begin{center}
   \includegraphics[scale=0.6]{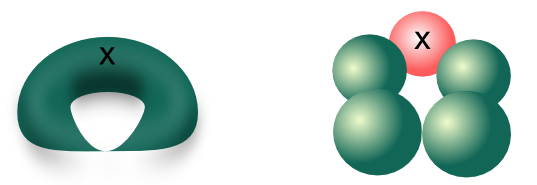}
   \put(-140,35){\huge$\overset{\textrm{\tiny blow-up}}{\rightarrow}$}
\end{center}
\caption{The singular elliptic fiber of type $A_4$ is blown up by introducing a tree of four $\mathbb P^1$s. The original fiber component
is marked in red and constitutes the affine node of the extended Dynkin diagram of $A_4$. The cross marks the zero-section.}
\label{fig:blow_up}
  \end{figure}

To work directly with singular elliptic fibrations in F-theory is quite hard, partly because we do not understand M-theory on singular
spaces very well. Therefore we will in the rest of this thesis always resolve the singularities in the fiber, assuming
implicitly that there
exists a resolution which preserves the Calabi-Yau condition. Following the duality of \autoref{sec:duality}
M-theory compactified on the \textit{resolved} elliptic fibration (with
still finite fiber size)
corresponds to going to the Coulomb branch of the F-theory setting on the circle of finite size.\footnote{Note that
one can also deform the singularities in order to obtain a smooth
geometry. This would correspond to a Higgsing of the gauge group.}
Importantly, the blow-up of the singularities over complex-codimension-one
in the base introduces a tree of $\mathbb P^1$s
in the fiber, which intersect (together with the original fiber component) as the (affine extension of the) Dynkin diagram
of a simple Lie algebra. This explains the classification of singularities started by Kodaira.
Note that there can also be codimension-one singularities which only lead to a degeneration of the fiber
but do not render the total space singular. These so-called $I_1$-singularities are therefore not part of the resolution process and 
do not introduce blow-up $\mathbb P^1$s.
In the framework of F-theory the singularity type directly corresponds to the non-Abelian gauge algebra induced by the
stack of 7-branes wrapping the divisor $S^{\rm b}$.
Expanding the M-theory three-form along the
resolution $\mathbb P^1$s (fibered over $S^{\rm b}$) gives the Cartan fields of the gauge theory while wrapping chains of $\mathbb P^1$s
by M2-branes provides amongst others the massive W-bosons. When the $\mathbb P^1$s collapse to zero size,
the W-bosons actually become massless as it should
be for the singular geometry.
We depict the resolution process for a specific example in \autoref{fig:blow_up}.
In principle it
is possible that there are singular fibers over several different complex-codimension-one loci in the base, which corresponds to a
product of simple non-Abelian gauge algebras. For simplicity we will assume in this thesis that there is only one simple non-Abelian gauge algebra
in the F-theory compactification at hand. The generalization to semi-simple gauge algebras is straightforward but the additional
index would reduce the readability. Note that the global structure of the gauge group is encoded
in torsional rational sections which will be introduced in \autoref{sec:MWtorsion}.
To complete our discussion of codimension-one singular fibers note that at this stage not only non-Abelian gauge multiplets
are induced, but also a certain sets of so-called bulk matter states which are counted by sheaf extension groups can arise,
see \textit{e.g.~}\cite{Weigand:2010wm} for more details.

In a general F-theory compactification one can also find a number of $U(1)$ gauge symmetries. Their nature is rather different from
the one of non-Abelian gauge symmetries since they are not localized on divisors in the base but in contrast depend
explicitly on the global structure of the
geometry. In fact, the number of independent rational sections (minus the zero-section) of the elliptic fibration
gives the number of Abelian gauge factors. We will discuss this in more detail in \autoref{sec:ec_group_structures}
when we introduce the Mordell-Weil group
of rational sections. In particular we will clarify what we mean by ``independent rational sections''.
Moreover, for compactifications of F-theory on Calabi-Yau 
fourfolds there are still additional $U(1)$s which do not correspond to rational sections but are counted
by $h^{2,1}$. These so-called bulk $U(1)$s will be neglected in this thesis since there is no matter
which is charged under them, and they never enter
in the discussions of this work. However, they nevertheless encode interesting physics and have also recently attracted
renewed attention \cite{Greiner:2015mdm,Corvilain:2016kwe}.

So far we have only considered fiber degenerations over a divisor $S^{\rm b}$ in the base, which corresponds to a non-Abelian gauge algebra
in the F-theory effective field theory. Let us now pass to higher codimension degenerations, which also encode interesting physics.
We have to distinguish between two different cases. First, the singularity at complex-codimension-two could arise at the
intersection of two divisors $S^{\rm b}_1$, $S^{\rm b}_2$ in the base on which non-Abelian gauge symmetries are located.
Note that actually $S^{\rm b}_1$ and $S^{\rm b}_2$
can be identical, and for our restriction to one simple non-Abelian algebra they indeed are. 
Second, the singularity could be
isolated at complex-codimension-two in the base. These two cases lead to different physical behaviors:
\begin{itemize}
 \item From the intuition of intersecting branes it seems clear that colliding $S^{\rm b}_1$ and $S^{\rm b}_2$ should correspond to
an intersecting stack of 7-branes inducing matter states. The easiest way to see what exactly happens is to work with the resolved geometry
in the M-theory picture. The $\mathbb P^1$s over $S^{\rm b}_1$ and $S^{\rm b}_2$ intersect as dictated by the the gauge algebras $\mathfrak{g}_1$ and
$\mathfrak{g}_2$, respectively. We denote their adjoint representations by $\textrm{ad}_{\mathfrak{g}_1}$, $\textrm{ad}_{\mathfrak{g}_2}$.
At the intersection of $S^{\rm b}_1$ and $S^{\rm b}_2$ the number of $\mathbb P^1$s enhances, and the latter intersect
according to the (affine extension of the) Dynkin diagram of an enhanced simple Lie algebra $\mathfrak{g}_3$ with the rank
given by the sum of the ranks of $\mathfrak{g}_1$ and $\mathfrak{g}_2$.
In M-Theory M2-branes can wrap the $\mathbb P^1$s at the intersection yielding formally the adjoint representation of the enhanced algebra
$\textrm{ad}_{\mathfrak{g}_3}$. 
The latter however does not correspond to an actual gauge theory, and the states therefore
have to be decomposed into representations of the true gauge algebras $\mathfrak{g}_1 \oplus \mathfrak{g}_2$
\begin{align}
 \mathfrak{g}_3 & \rightarrow (\mathfrak{g}_1 \oplus \mathfrak{g}_2) \\
   \textrm{ad}_{\mathfrak{g}_3} &\rightarrow (\textrm{ad}_{\mathfrak{g}_1},1) \oplus (1,\textrm{ad}_{\mathfrak{g}_2})
   \oplus \sum_i (R^i_1,R^i_2) \, ,   \nn
\end{align}
with $R^i_1$, $R^i_2$ denoting the representations which complete the decomposition. These precisely provide the additional matter states at the
intersection. Recall that the adjoint representation of the gauge theory is located at complex-codimension-one. 

\item Isolated singularities at complex-codimension-two are conifold singularities.
Once they get resolved by a single $\mathbb P^1$, an M2-brane can wrap the latter
and induce matter states in this way.
These however are generically singlets under the non-Abelian gauge group but carry
non-trivial $U(1)$-charge. 
\end{itemize}

Starting with Calabi-Yau fourfolds there can also appear complex-codimension-three singularities.
It his not hard to verify that these induce Yukawa couplings. This can be derived in the M-theory picture by again considering
an enhanced gauge algebra of now three colliding divisors $S^{\rm b}$ in the base.
We will not further comment on this topic since it is not important for this thesis.
Yukawa couplings are nevertheless extremely interesting (in particular for model building purposes),
but unfortunately they are poorly understood
in global models though some results exist for local settings. Again we refer to \cite{Weigand:2010wm} for more details and references.

At this point it is crucial to draw attention to a very important point. While compactifications on Calabi-Yau threefolds to
six-dimensional $\cN = (1,0)$ theories are always necessarily chiral, this is not true for compactifications
on Calabi-Yau fourfolds to four-dimensional $\cN = 1$ theories. In the former the chiral index of a matter state
over codimension-two in the base is simply given by the homology class of the codimension-two
locus in the base, \textit{i.e.~}by counting a number of points. In contrast, in four dimensions chirality is only introduced in the presence of
$G_4$-flux. In particular, for a matter state in a representation $(R,q)$,
which is located on a surface $\cS_{(R,q)}$ at codimension-two in the base, the chiral index
is given by
\begin{align}
 \chi(R,q) = \int_{\cS_{(R,q)}}G_4 \, .
\end{align}

Furthermore, in compactifications on Calabi-Yau threefolds (anti-)self-dual tensors arise from M5-branes wrapping vertical divisors which are 
pullbacks from divisors in the base to the whole fibration. Indeed this leads to string states in six dimensions.
In compactifications on Calabi-Yau fourfolds the same setting yields a number of axions.

Before we conclude let us stress
 that F-theory allows for richer possibilities concerning the gauge
groups and matter states than
the standard type II D-brane settings. Since in the latter only two-index representations can occur (strings begin and end on D-branes),
the only possible gauge algebras are the classical ones of type $A$, $B$, $C$ or $D$. In contrast, the richer monodromy structure of
$(p,q)$-strings ending on $(p,q)$-branes allows one to form so-called string junctions in a well-defined manner. These
then give rise to more general
gauge algebras and representations than in the type II case \cite{Gaberdiel:1997ud,DeWolfe:1998zf}.

\subsection{The Topology of Elliptic Fibrations in F-Theory}\label{sec:top_torus}

We now investigate in detail how topological quantities of Calabi-Yau fourfolds and threefolds enter in the effective
field theory of the corresponding F-theory compactification. We stress that the results which we review in here are crucial for the
original work in this thesis.

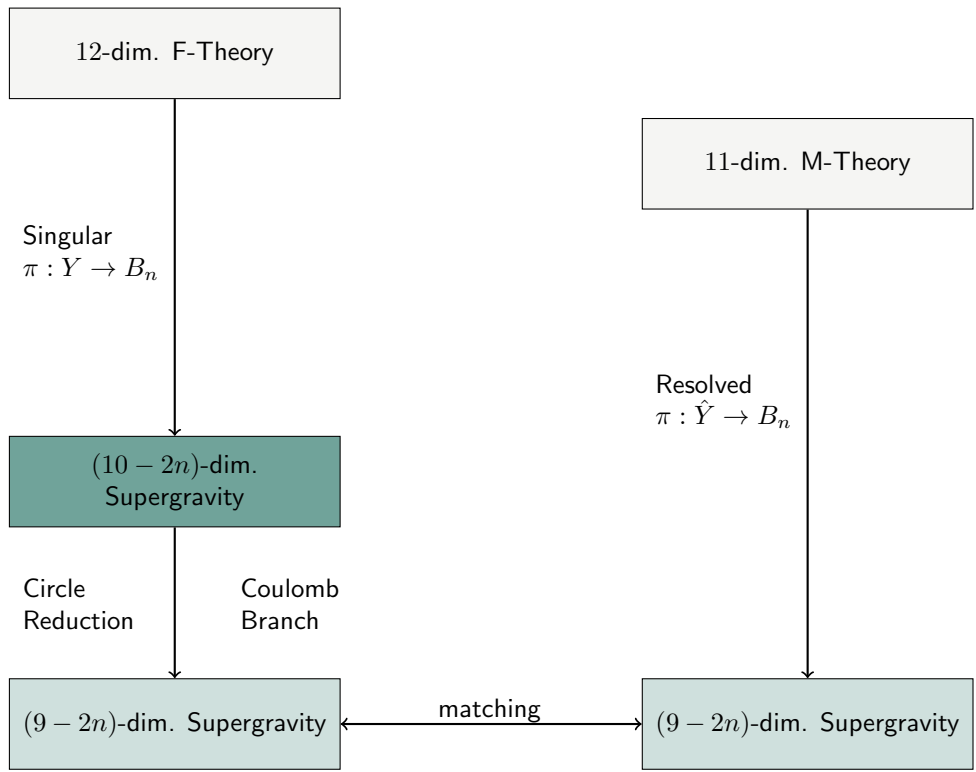
\begin{figure}
\begin{center}
\begin{tikzpicture}
[node distance = 1cm, auto,font=\footnotesize,
every node/.style={node distance=3cm},
comment/.style={rectangle, inner sep= 5pt, text width=4cm, node distance=0.25cm, font=\footnotesize\sffamily},
force1/.style={rectangle, draw, fill=mppgreen!60, inner sep=5pt, text width=4cm, text badly centered, minimum height=1.2cm, font=\footnotesize\sffamily},
force2/.style={rectangle, draw, fill=mppgray!30, inner sep=5pt, text width=4cm, text badly centered, minimum height=1.2cm, font=\footnotesize\sffamily},
phantom/.style={text width=4cm,minimum height=1.2cm},
force3/.style={rectangle, draw, fill=mppgreen!20, inner sep=5pt, text width=4cm, text badly centered, minimum height=1.2cm, font=\footnotesize\sffamily}] 

\node [force2] (fth) {$12$-dim. F-Theory};
\node [phantom,  below=0.25cm of fth] (p1) {};
\node [force2, right=4cm of p1] (mth) {$11$-dim. M-Theory};
\node [force1,  below=3cm of p1] (eff) {$(10-2n)$-dim. Supergravity};
\node [force3,  below=2cm of eff] (efff) {$(9-2n)$-dim. Supergravity};
\node [phantom,  below=2cm of eff] (p2) {};
\node [force3,  right=4cm of p2] (effm) {$(9-2n)$-dim. Supergravity};

\node [comment, below=1.5cm of fth] (singcy) {Singular\\ $\pi: Y \rightarrow B_n$};
\node [comment, below=0.5cm of eff] (circ) {Circle\\ Reduction};
\node [comment, right=-1.5cm of circ] (cb) {Coulomb\\ Branch};
\node [comment, below=2cm of mth] (rescy) {Resolved\\ $\pi: \hat Y \rightarrow B_n$};
\node [comment, right=1.1cm of efff] (dual) {matching\\ $\,$};

\path[->,thick] 
(fth) edge (eff)
(eff) edge (efff)
(mth) edge (effm);

\path[<->,thick] 
(efff) edge (effm);
\end{tikzpicture} 
\end{center}
\caption{We show a schematic diagram of
how the duality between F-theory and M-theory can be used to infer the effective action
of F-theory on elliptically-fibered singular Calabi-Yau manifolds $Y$. The circle compactification of a general
minimal $(10-2n)$-dimensional supergravity theory is pushed to the Coulomb branch and matched to
M-theory compactified on the resolved space $\hat Y$. Sending the volume of the blow-up $\mathbb P^1$s to zero
corresponds to going to the origin of the Coulomb branch in the dual picture. The limit of vanishing elliptic fiber
on the M-theory side corresponds to the decompactification limit of the circle on the F-theory side.}
\label{fig:schematic_picture}
\end{figure}

As already mentioned, in order to derive the effective theory of F-theory compactifications on elliptically-fibered Calabi-Yau manifolds
we first consider
M-theory on the resolved spaces, and obtain three-dimensional or five-dimensional $\cN = 2$ supergravity theories, respectively.
These are matched to general four-dimensional $\cN =1$ or six-dimensional $\cN =(1,0)$ supergravities on a circle of finite radius.
The resolution of the geometry
further forces us to move to Coulomb branch of the circle-compactified
theories \cite{Vafa:1996xn,Morrison:1996na,Morrison:1996pp,Ferrara:1996wv,Denef:2008wq,Grimm:2010ks,Bonetti:2011mw,Grimm:2013oga}.
For convenience we depict the duality in \autoref{fig:schematic_picture}.
Via this matching procedure the parameters of the unknown four-dimensional and six-dimensional effective theories can be obtained in terms
of the data describing the elliptic fibration. In this thesis we only restrict to the topological sector
of the matching procedure, more precisely we consider in detail matchings of Chern-Simons terms in field theory with intersection numbers on
Calabi-Yau manifolds.

In \autoref{ch:circle_theories} we already discussed the compactification of 
general four- and six-dimensional gauge theories on a circle. This of course also includes the minimal supergravity theories in these dimensions
which appear in F-theory compactifications on Calabi-Yau manifolds.
Moreover, we described large gauge transformations along the circle as additional symmetries.
Assuming that the M-theory to F-theory duality provides a correct approach 
to understand the system, we have to suspect that the smooth Calabi-Yau geometry should share the 
symmetries of the gauge theories on a circle as well. 
Thus let us start with describing the M-theory compactifications and specify the matching to the circle-reduced theories.

First we establish some geometric notions of the Calabi-Yau manifold which we compactify on. 
We denote the resolved Calabi-Yau space by $\hat Y$, and assume that it constitutes an elliptic fibration over some base space
$B_n$, with the corresponding projection given by $\pi: \hat Y \rightarrow B_n$. As announced, we will assume that 
the fibration has at least one rational section. 
A set of linearly independent (minimal) rational sections of the elliptic fibration is 
denoted by $s_0$, $s_m$, $t_r$ where one arbitrary section $s_0$
is singled out as the so-called zero-section. The sections $t_r$ will be purely torsional, while the $s_m$
are assumed to be non-torsional. We will have to say more about rational sections and 
this distinction in \autoref{sec:ec_group_structures}.
Furthermore, there might exist a divisor $S^{\rm b}$ in the base $B_n$ of the resolved space $\hat Y$ 
over which the fiber becomes reducible with the individual irreducible components
intersecting as the (affine extension of the) Dynkin diagram of the gauge algebra. Fibering these
over the corresponding codimension-one locus $S^{\rm b}$ in $B_n$ yields the blow-up divisors of $\hat Y$ 
which we denote by $D_I$.

In the following we define a basis of divisors $D_\Lambda = (D_0 , D_I , D_m , D_\alpha)$ on the resolved space $\hat Y$ 
in the correct frame
such that the
corresponding gauge fields obtained from the expansion of the M-theory three-form
\begin{align}\label{e:expansion_threeform}
 C_3 = A^0\wedge [D_0] + A^m\wedge [D_m]+ A^I\wedge [D_I] + A^\alpha\wedge [D_\alpha]
\end{align}
can be matched properly to the gauge fields in the circle-reduced theory in \autoref{tab:vec}.
In this expression $[D]$ denotes the Poincar\'e-dual 
two-form to the divisor $D$ in $\hat Y$.

\begin{itemize}
 \item Divisors $D^{\rm b}_\alpha$ 
 of the base $B_n$ define the vertical divisors $D_\alpha:=\pi^{-1}(D^{\rm b}_\alpha)$ via pullback.
 For each $D^{\rm b}_\alpha$ in $B_n$ there is an axion in the four-dimensional F-theory compactification and an (anti-)self-dual tensor in the
 six-dimensional setting, respectively. Supersymmetry implies $T_{\rm sd} = 1$ (this is the tensor in the gravity multiplet)
 and thus we find 
 \bea
  n_{\rm ax} &=& h^{1,1}(B_3)\ \ \textrm{in four dimensions}\ , \\
     T_{\rm asd} &\equiv & T = h^{1,1}(B_2)-1\ \ \textrm{in six dimensions}\ ,    \nn
 \eea
 with $T_{\rm sd}, T_{\rm asd}, n_{\rm ax}$ as defined in \autoref{sec:gauge_gen}, and $T$ denotes the number of
 six-dimensional tensor multiplets.

 For Calabi-Yau fourfolds it is also necessary to introduce vertical four-cycles $\cC^\alpha:=\pi^{-1}(\cC_{\rm b}^\alpha)$ which
are the pullbacks of curves $\cC_{\rm b}^\alpha$ in the base intersecting the $D^{\rm b}_\alpha$ as
\begin{align}\label{e:def_metric}
 \tensor{\eta}{_\alpha^\beta} = D^{\rm b}_\alpha \cdot \cC^\beta_{\rm b}
\end{align}
with $\tensor{\eta}{_\alpha^\beta}$ a full-rank matrix. 
For Calabi-Yau threefolds the analogous intersection matrix
\begin{align} \label{etaalphabeta}
 \eta_{\alpha\beta} := D^{\rm b}_\alpha \cdot D^{\rm b}_\beta
\end{align}
is used to raise and lower indices $\alpha, \beta$.
The matrices \eqref{e:def_metric} and \eqref{etaalphabeta} are matched to the corresponding expressions in the
four- and six-dimensional Green-Schwarz terms \eqref{GS4} and \eqref{GS6}, respectively. 
 
 For later convenience we also define the projection of two arbitrary divisors $D, D^\prime$ as
\begin{align}\label{def_proj}
 \pi (D\cdot D^\prime) :=
 \begin{cases}
 \big( D \cdot D^\prime \cdot \cC^\beta\big)\ \tensor{\eta}{^{-1}_\beta^\alpha}\ D_\alpha   &\textrm{in three dimensions,} \\
 \big(D \cdot D^\prime  \cdot D_\beta\big) \ \tensor{\eta}{^{-1\,}^\beta^\alpha}\ D_\alpha  & \textrm{in five dimensions.}
 \end{cases}
\end{align}
Furthermore we write $\pi_{\cM I}$ for the intersection number of a section $s_{\cM}$ with a blow-up divisor $D_I$
restricted to the elliptic fiber $\mathcal E$
\begin{align}
 \pi_{\cM I} := \cap (s_{\cM}, D_I)\big\vert_{\mathcal E}\, .
\end{align}  

 \item We denote the divisor associated to the zero-section $s_0$ by $S_0\equiv Div(s_0)$. 
 The divisor $D_0$ is then defined by shifting $S_0$ as
 \begin{align}\label{e:old_base}
 D_0 = S_0 - \frac{1}{2}\pi(S_0 \cdot S_0)\ .
\end{align}
 The corresponding vector $A^0$ in \eqref{e:expansion_threeform} is identified with the Kaluza-Klein vector in the 
 circle-reduced F-theory setting, {i.e.}~with $A^0$ in \eqref{e:metric_reduction}. 
 
 We stress that since all rational section enjoy the property that they always square to the canonical class of the base, we have
 \begin{align}\label{e:sqare_k}
  \pi(S_0 \cdot S_0) = K\, ,
 \end{align}
with $K$ the canonical class of the base.
 
 \item The $D_I$ denote the blow-up divisors and yield the 
 Cartan gauge fields $A^I$ in \eqref{e:expansion_threeform}. This implies that $I=1,\dots , \rk G$.

 \item Given a set of rational sections $s_m$ 
 the $U(1)$ divisors $D_m$ are defined via the so-called Shioda map. Denote by $S_m \equiv {Div}(s_m)$ the 
 divisor associated to $s_m$. The Shioda map ${D}(\cdot)$ reads
 \beq\label{e:old_shioda}
 D(s_m) \equiv D_m = S_m - S_0 - \pi \Big (  (S_m -
 S_0  ) \cdot S_0 \Big )
 + \pi_{mI} \, \mathcal{C}^{-1\,IJ}D_J  \, ,
\eeq
where $\cC_{IJ}$ is the coroot intersection matrix \eqref{e:def_coroot_int_mat} of the gauge algebra $\mathfrak{g}$
derived from the intersection of the blow-up divisors.
 Via \eqref{e:expansion_threeform} the $D_m$ yield the 
 Abelian gauge fields in the F-theory setting such that $m=1,\dots , n_{U(1)}$.
 
 \item The crucial property of the purely torsional sections $t_r$ is that they have 
 no non-trivial image under the Shioda map. Denoting the divisors associated to $t_r$ 
 by $T_r = Div(t_r)$ one has \cite{Mayrhofer:2014opa}
 \beq \label{Shioda_tr}
  D(t_r) = T_r - S_0 - \pi \Big (  (T_r -
 S_0  ) \cdot S_0 \Big )
 + \pi_{rI} \, \mathcal{C}^{-1\,IJ}D_J = 0\ ,
 \eeq
 which, as the other expressions above, should be read in homology.
\end{itemize}
By the Shioda-Tate-Wazir theorem $(D_0 , D_I , D_m , D_\alpha)$ indeed form a basis of the Ner\'on-Severi group of divisors\footnote{For
Calabi-Yau manifolds
the Ner\'on-Severi group coincides with the Picard group which is why we will identify both groups throughout this thesis.} (times $\mathbb Q$).
Let us briefly mention that with this basis of divisors the charges and weights of matter states can be computed in a very simple
fashion geometrically.
Suppose that a holomorphic curve $\cC$ is wrapped by an M2-brane, they are precisely given by the intersections
\begin{align}
 (n, w_I, q_m) = (D_0 \cdot \cC, D_I \cdot \cC, D_m \cdot \cC)
\end{align}
with $n$ the Kaluza-Klein level, $w_I$ the Dynkin labels and $q_m$ the $U(1)$-charges of the state.

It is important to realize that the definition of the base divisor ensures in contrast to \eqref{e:sqare_k} that
\begin{align}\label{e:base_orth}
 \pi(D_0 \cdot D_0) = 0 \, ,
\end{align}
and the Shioda map enjoys the orthogonality properties
\begin{align}\label{e:shioda_orth}
 \pi(D_m \cdot D_\alpha) = \pi(D_m \cdot D_I) = \pi(D_m \cdot D_0) = 0 \, ,
\end{align}
which are essential in order to perform the F-theory limit correctly. The blow-up divisors $D_I$ and the vertical divisors $D_\alpha$
further satisfy the properties 
\beq \label{blowup-cond}
   \pi (D_I \cdot D_\alpha) = \pi(D_I \cdot D_0) = \pi(D_\alpha \cdot D_\beta) = 0\ . 
\eeq

Via the matching of the M-theory compactification to the circle-reduced theory some
intersections of the divisor basis
$\pi (D_\Lambda \cdot D_\Sigma) \equiv  \pi (D_\Lambda \cdot D_\Sigma)^\alpha D_\alpha$ can be nicely 
related to four- and six-dimensional supergravity data
\begin{subequations}\label{evaluate_DD}
\begin{align} 
 \pi(D_I \cdot D_J)^\alpha &=
   -\cC_{IJ}\ b^\alpha \, , \\
 \pi(D_m \cdot D_n)^\alpha &=
    -b^\alpha_{mn} \, , \\
    \pi(D_0 \cdot D_\beta)^\alpha &=
   \delta^\alpha_\beta \, , 
\end{align}
\end{subequations}
where the last equality implicitly encodes the matching of the intersection
matrices $\tensor{\eta}{_\alpha^\beta}$, $\eta_{\alpha\beta}$ in \eqref{e:def_metric}, \eqref{etaalphabeta}
with the corresponding objects in the
Green-Schwarz-terms because of the definitions \eqref{def_proj}.
The $b^\alpha, b^\alpha_{mn}$ are the Green-Schwarz couplings appearing in \eqref{GS4} and \eqref{GS6}.
These relations hold both for Calabi-Yau three- and fourfolds. 
The Green-Schwarz coefficients $b^\alpha$ are equivalently obtained 
as 
\beq
     S^{\rm b} = b^\alpha D_\alpha^{\rm b}\ , 
\eeq
where $S^{\rm b}$ was the divisor in $B_n$ supporting the non-Abelian gauge group.

Since we are in particular interested in the matching of Chern-Simons terms between the circle-reduced theory and the M-theory
compactification, let us approach this topic in the following. The origin of Chern-Simons terms gauge theories on the circle
has already been investigated in
\autoref{sec:CS_in_circle}. In particular recall that there are classical contributions as well as quantum corrections at one-loop
which have to be included for a proper matching procedure \cite{Grimm:2011fx,Bonetti:2011mw,Cvetic:2012xn,Grimm:2013oga,Cvetic:2013uta,
 Anderson:2014yva}. In contrast, in
 the M-theory compactifications all Chern-Simons terms are on equal footing, and the massive modes are
 automatically integrated out since we are already facing an effective theory. In terms of geometrical and flux data
 the Chern-Simons couplings are given by
 \begin{align}
  \Theta_{\Lambda\Sigma} = - \frac{1}{4} D_{\Lambda} \cdot D_{\Sigma} \cdot [ G_4 ] \, ,
  \end{align}
for compactifications on Calabi-Yau fourfolds with $[G_4]$ the Poincar\'e-dual to four-form flux, and
\begin{subequations}
\begin{align}
 k_{\Lambda\Sigma\Theta} &= D_{\Lambda} \cdot D_{\Sigma} \cdot D_{\Theta} \, , \\
 k_{\Lambda} &= D_{\Lambda} \cdot [c_2 ] \, ,
\end{align}
\end{subequations}
for compactifications on Calabi-Yau threefolds with $[c_2 ]$ the Poincar\'e-dual to the second Chern class of the resolved
total space. Note that the special set of Chern-Simons couplings $k_{\alpha\Lambda\Sigma}$ has already been discussed since they
appear in the expressions $\pi(D_\Lambda \cdot D_\Sigma)$.
For convenience we list the one-loop Chern-Simons terms for the circle-reduced theory in \autoref{sec:one_loop_calc}
and the relevant intersection numbers along with their matching to Chern-Simons terms and supergravity data in \autoref{app_inter}.

 We highlight that, since one-loop induced Chern-Simons terms carry information about the number of 
 matter fields, the matching to M-theory allows to translate information about the spectrum into the geometric 
 data of the resolved space. This is the underlying reason why will be able in the following chapter to relate
 field-theoretic large gauge transformations to geometric symmetries on elliptic fibrations, and in particular
 show anomaly cancelation in F-theory compactifications on elliptically-fibered Calabi-Yau manifolds.

\chapter{Arithmetic Structures on Genus-One Fibrations}
\chaptermark{Arithmetic on Genus-One Fibrations}
\label{ch:arith}

In this chapter we finally
connect the results of \autoref{ch:lgts} with the general considerations in \autoref{sec:top_torus}.
In particular we identify arithmetic structures
on genus-one fibrations with large gauge transformations of gauge theories on the circle in F-theory.

In \autoref{sec:ec_group_structures} we focus on geometries with rational 
 sections and the corresponding Abelian parts of the gauge theory. The Mordell-Weil group action is introduced and mapped to large gauge 
 transformations along the circle.
 We also discuss the impact of torsion from this perspective.
 In \autoref{sec:multi_group} we then turn to geometries with multi-sections. 
 Insights obtained by using Higgs transitions allow us to define an extended 
 Mordell-Weil group of multi-sections and a generalized Shioda map.  
 In \autoref{Arithmetics_nonAb} we extend the analysis further 
 to cover non-Abelian gauge groups. We argue for the existence of 
 a group law on the exceptional divisors and rational sections that is
 shown to be induced by large gauge transformations of the Cartan gauge fields. 
 We geometrically motivate its existence 
 by explicitly considering Higgsings to Abelian gauge theories.

\section[Arithmetic Structures on Fibrations with Rational Sections]{Arithmetic Structures on Fibrations with Rational Sections
\sectionmark{Arithmetic on Fibrations with Rational Sections}} 
\sectionmark{Arithmetic on Fibrations with Rational Sections}
\label{sec:ec_group_structures}

In this section we argue that the arithmetic structures of 
elliptic fibrations with multiple rational sections correspond to certain 
large gauge transformations introduced in \autoref{sec:class_lgt}.  
The considered arithmetic is encoded by the so-called \textit{Mordell-Weil 
group of rational sections} which we introduce in more detail in \autoref{MWgeneralities}.
In the same subsection we also discuss how the geometric 
Mordell-Weil group law translates to a general group law for rational sections 
in terms of homological cycles.
The free generators of the Mordell-Weil group correspond 
to Abelian gauge symmetries in the effective F-theory action. 
In \autoref{sec:FreeMWShifts} we show that group actions of the free part of the Mordell-Weil group 
are in one-to-one correspondence to specific integer large gauge transformations along the F-theory circle.
A similar analysis for the torsion subgroup is performed in \autoref{sec:MWtorsion}. 
We find that it precisely captures special fractional non-Abelian large gauge transformations 
introduced in \autoref{sec:class_lgt} due to the presence of a non-simply connected
non-Abelian gauge group.

\subsection{On the Mordell-Weil Group and its Divisor Group Law} \label{MWgeneralities}
 
The most famous arithmetic structure on an elliptic curve is encoded by the Mordell-Weil group.
The Mordell-Weil group is formed by the rational points of an elliptic curve endowed 
with a certain geometric group law (see \textit{e.g.}~\cite{Silverman}).
The rational points on the generic elliptic fiber of an elliptic fibration $Y$ directly extend to rational sections
and form a finitely generated Abelian group which is called the Mordell-Weil group of rational sections
$\textrm{MW}(Y)$. Thus it splits into a free part and a torsion subgroup
\begin{align} \label{splitMW}
 \textrm{MW}(Y) \cong \bbZ^{\rk \textrm{MW}(Y)} \oplus \bbZ_{k_1} \oplus \dots \mathbb 
    \oplus \, \bbZ_{k_{n_{\rm tor}}}  \, .
\end{align}
Having chosen one (arbitrary) zero-section as the neutral element of the Mordell-Weil group, $\rk \textrm{MW}(Y)$ rational sections generate the free part
and $n_{\rm tor}$ rational sections generate the torsion subgroup. The precise group law on the generic fiber (in Weierstrass form) may be looked up for example
in \cite{Silverman}. We denote the addition of sections $s_1,s_2$ 
using the Mordell-Weil group law by `$\oplus$', \textit{i.e.}~we write $s_3= s_1 \oplus s_2$ with $s_3$ being the new
rational section.
Since, as noted before, the rational sections $s_\cM$ of an elliptic fibration define 
divisors $S_\cM \equiv {Div}(s_\cM)$, we will investigate how the group law is translated to divisors.
More precisely, we will derive the divisor class
\begin{align}
 {Div}( s_1 \oplus \fn s_2 )\, , \quad \fn \in \mathbb Z \, ,
\end{align}
where $\fn s_2 = s_2 \oplus \ldots \oplus s_2$ with $\fn$ summands. 
In 
contrast the addition in homology of divisor classes associated to sections
is denoted by `$+$' .
Extending the treatment in \cite{Morrison:2012ei} the group law written in homology is uniquely determined by the three conditions:
\begin{enumerate}
 \item The Shioda map $D(s_{\cM})$ introduced in  \eqref{e:old_shioda} is a homomorphism from the Mordell-Weil group to the Ner\'on-Severi group (times $\mathbb Q$)
 \begin{align}
  D(s_1 \oplus \fn s_2) = D(s_1) + \fn \, D(s_2) \, .
 \end{align}

 \item A section $s_\cM$ intersects the generic fiber $\mathcal E$ exactly once
 \begin{align}
  S_\cM \cdot \mathcal E = 1 \, .
 \end{align}
 \item In the base $B_n$ a divisor $S_\cM$ associated to a section squares to the canonical class of the base $K$, {i.e.}
 \begin{align}
  \pi (S_\cM \cdot S_\cM) = K \, .
 \end{align}
\end{enumerate}
Taking these constraints into account the group law for two sections $s_1, s_2$ on the level of divisors can then be derived to
be of the following form
\begin{align}\label{e:MW_law}
 {Div}(s_1 \oplus \fn s_2 ) &= S_1 + \fn (S_2 - S_0)
 - \fn \pi \Big( ( S_1 - \fn S_0 )\cdot 
  ( S_2 - S_0  ) \Big )  \, ,  
\end{align}
where $S_0$ denotes the divisor associated to the zero-section $s_0$. 
We stress that we assumed that blow-up divisors do not contribute to the group-law. 
This can be derived easily in a general ansatz by enforcing that ${Div}(s_1 \oplus s_0 ) = {Div}(s_0 \oplus s_1 )  = {Div}(s_1)$.
In other words, the appearance of blow-up divisors in the ansatz always violates the Abelian structure of the group.

It is well known that the Shioda map as an injective homomorphism \eqref{e:old_shioda} transfers this group structure
to the Ner\'on-Severi group (times $\mathbb Q$) of divisors modulo algebraic equivalence.
Therefore it is reasonable to ask how a Mordell-Weil
group action on the elliptic fibration effects the circle-reduced supergravity. We will find that the free part of the Mordell-Weil group
corresponds to certain Abelian large gauge transformations while the torsion subgroup manifests in special fractional
non-Abelian large gauge transformations.
As we discussed in \autoref{sec:anom_lgt} these arithmetic structures allow to 
establish the cancelation of all pure Abelian and mixed Abelian-non-Abelian gauge anomalies in 
the effective field theory of F-theory (as well as Abelian gauge-gravitational anomalies in six-dimensional models).

\subsection{The Free Part of the Mordell-Weil Group}
\label{sec:FreeMWShifts}
 
\begin{figure}
\begin{center}
\begin{tikzpicture}[scale = 0.85, axis/.style={very thick, ->, >=stealth'}]
 \foreach \x in {0,...,3} 
  { \foreach \y in {0,...,3} 
      {\fill (\x*1.5cm,\y*1.5cm) circle (0.07cm); }}
 \draw[axis]  (0cm,0cm) -- (1.43cm,0cm) node(xline)[left] {};
 \draw[axis]  (0cm,0cm) -- (0cm,1.43cm) node(yline)[below] {};
 \node at (-0.3cm,-0.3cm) {$s_0$};
 \node at (1.7cm,-0.4cm) {$s_{m_1}$};
 \node at (-0.6cm,1.5cm) {$s_{m_2}$};
 \node at (3.3cm,3.3cm) {$\fn^n s_{n}$};
 \draw[->, dashed, >=latex] (0cm,0cm) -- (2.93cm,2.93cm) ;
 \draw[->, dashed, >=latex] (1.5cm,0cm) -- (4.43cm,2.93cm) ;
 \draw[->, dashed, >=latex] (0cm,1.5cm) -- (2.93cm,4.43cm) ;
  \foreach \x in {8,...,11} 
  { \foreach \y in {0,...,3} 
      {\fill (\x*1.5cm,\y*1.5cm) circle (0.07cm); }}
  \draw[axis]  (15cm,3cm) -- (16.43cm,3cm) node(xline)[left] {};
  \draw[axis]  (15cm,3cm) -- (15cm,4.43cm) node(yline)[below] {};
 \node at (14.7cm,2.7cm) {$\tilde s_0$};
 \node at (16.7cm,2.6cm) {$\tilde s_{m_1}$};
 \node at (14.4cm,4.5cm) {$\tilde s_{m_2}$};
   \draw[->, thick, >=latex] (6.3cm,2cm) -- (10.2cm,2cm) ;
    \node at (8cm,3.5cm) {
\begin{tabular}{l}
$\tilde s_0  \ = s_0 \oplus \fn^n s_{n}$ \\
$ \tilde s_m = s_m \oplus \fn^n s_{n}$ 
                                                                       \end{tabular}
} ;
\end{tikzpicture}
\end{center}
\caption{We depict the process of shifting the basis of the free Mordell-Weil group by a vector $\fn^n s_{n}$. Note that
in addition also possible torsional generators $t_r$ get shifted.}
\label{fig:MWshift}
\end{figure}
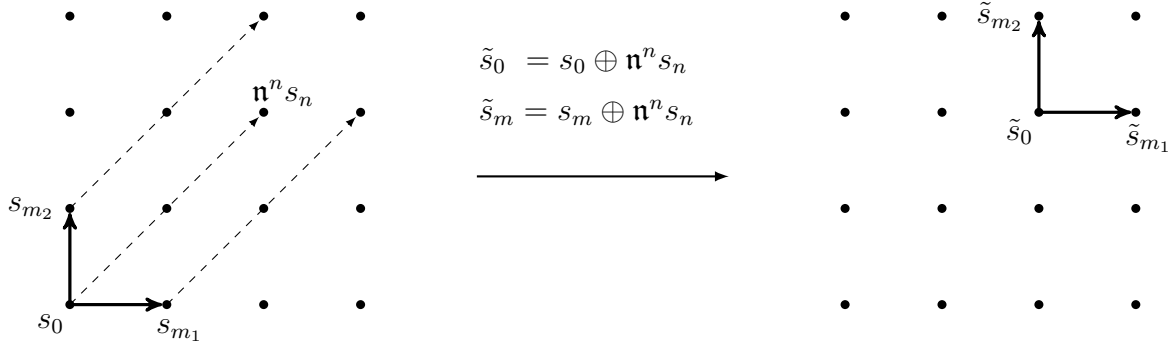

Let us first consider the free part of the Mordell-Weil group.
On the elements of the Mordell-Weil basis, consisting of the 
zero-section $s_0$, the free generators $s_m$, and the torsional
generators $t_r$, we now perform a number of $\fn^m \in \mathbb Z$ shifts into the directions of the free generators $s_{m}$,
\textit{i.e.}~we find a new Mordell-Weil basis given by
\begin{align}
\label{eq:MWshift}
 &\tilde s_0 := s_0 \oplus \fn^n s_{n} \, , &&  \tilde s_m := s_m \oplus \fn^n s_{n} \, , &&
 \tilde t_r := t_r \oplus \fn^n s_{n} \, ,
\end{align}
where $ \fn^n s_{n}=  \fn^1 s_{1} \oplus \ldots  \oplus \fn^{n_{U(1)}} s_{n_{U(1)}}$ and 
each summand $\fn^1 s_1, \fn^2 s_2, \ldots$ is evaluated using the Mordell-Weil group law.
For illustration we depict the transformation in \autoref{fig:MWshift}.
Our goal will be to translate these shifts to the divisor basis $D_\Lambda = (D_0 , D_I , D_m , D_\alpha)$
introduced in \autoref{sec:top_torus}, and then identify the corresponding large 
gauge transformation. 

We use the formula \eqref{e:MW_law} to derive the change in the definition of the $U(1)$ divisors $D_m$, 
the Cartan divisors $D_I$ and the base divisor $D_0$. Note that the divisors $\tilde D_\alpha = D_\alpha$ are unchanged 
under this transformation. Explicitly we find 
\begin{subequations}  \label{e:MW_shift_divisors_1}
\begin{align} 
\label{free_shift_Ab} \tilde D_m &= D_m - \fn^n \, \pi (D_{n}\cdot D_m)   , \\[.1cm]
\label{DIshift} \tilde D_I &= D_I - \fn^K \, \pi (D_K\cdot D_I) 
 \ , \\
 \tilde D_0  
& =  D_0 + \fn^n \,D_{n} + \fn^J\, D_J   -\frac{\fn^n \fn^p}{2} \pi (D_{n}\cdot D_{p}) -\frac{\fn^J \fn^L}{2} \pi (D_J\cdot D_L) \ .
\end{align} 
\end{subequations}
where we have defined 
\beq \label{def-associatedkK}
   \fn^I := - \fn^n \, \pi_{n J} \,\mathcal{C}^{-1\,JI} \ . 
\eeq
Using the expressions \eqref{evaluate_DD} one further evaluates
\begin{align} \label{D-transform1}
 \begin{pmatrix}
  \tilde D_{0}\\[8pt]
  \tilde D_{I}\\[8pt]  
  \tilde D_{m}\\[8pt]
  \tilde D_{\alpha}
 \end{pmatrix} = 
\begin{pmatrix*}[c]
 1 & \fn^J & \fn^n  &  \frac{\fn^p \fn^q}{2} b^\beta_{pq}  +\frac{\fn^K \fn^L}{2} \cC_{KL} b^\beta \\[8pt]
0 & \delta^{J}_I & 0 &   \fn^K \cC_{IK} b^\beta \\[8pt]
  0 & 0 & \delta_m^n & \fn^p b^\beta_{mp} \\[8pt]
 0 & 
 0
  & 0& \delta^\beta_\alpha
\end{pmatrix*}\cdot
\begin{pmatrix}
  D_{0} \\[8pt]
  D_{J} \\[8pt]
  D_{n} \\[8pt]
  D_{\beta} \, 
 \end{pmatrix}.
\end{align}
It is now straightforward to check that, using \eqref{D-transform1} and the large gauge transformations 
\eqref{e:non_abelian_trafo_vectors_4d} with $(\fn^m,\fn^I)$ as in \eqref{eq:MWshift} and \eqref{def-associatedkK}, one finds 
\beq
    C_3 = A^\Lambda \wedge [D_{\Lambda}] = \tilde A^\Lambda \wedge [\tilde D_{\Lambda}] \ . 
\eeq
This implies that the Mordell-Weil shift \eqref{eq:MWshift} indeed induces a large gauge transformation 
 as discussed in \autoref{sec:class_lgt}. More precisely, the 
pair $(\fn^m,\fn^I)$ are in general the basis vectors of type (II) and realize an Abelian 
large gauge transformation combined with a fractional non-Abelian large gauge transformation. 
It is also straightforward to check that the quantization condition 
\eqref{kk_cond} is satisfied for the pair $(\fn^m,\fn^I)$ inferred from the geometry. 
In fact, one finds 
\beq
   \fn^I w_I + \fn^n q_n =  \fn^n ( - \pi_{n J} \,\mathcal{C}^{-1\,JI} w_I +  q_n)  = \fn^n \Big(S_n - S_0 - \pi \big (  (S_n -
 S_0  ) \cdot S_0 \big) \Big)\cdot \cC \ ,
\eeq
where we have used that $w_I = D_I \cdot\cC$ and $q_m = D_m\cdot \cC$ are the charges of a matter state $\hat \Bpsi(w_I , q_m)$
arising from an 
M2-brane wrapped on the curve $\cC$. The condition  \eqref{kk_cond} then follows 
from the fact that $\fn^m \in \bbZ$ and  
the appearing intersections between divisors $S_n,S_0$ and the curve $\cC$ are always integral. From its definition \eqref{def-associatedkK} it 
is also clear that $\fn^I$ are either zero or fractional due to the appearance of the inverse $\cC^{-1\, IJ}$.

Let us make a few comments concerning the derivation and 
interpretation of \eqref{e:MW_shift_divisors_1} and \eqref{D-transform1}.
First, it seems from counting the number of conditions \eqref{eq:MWshift} and \eqref{e:MW_shift_divisors_1}
that the former conditions cannot suffice to fix the complete transformation law. In fact, 
the shift of the non-Abelian Cartan divisors $D_I$ to $\tilde D_I$ is not immediately inferred from \eqref{eq:MWshift}
but appears to be crucial to make the transformation well-defined. To derive \eqref{e:MW_shift_divisors_1} one first 
starts with the transformation to $\tilde D_0$ by evaluating $\tilde S_0 = Div(s_0 \oplus \fn^n s_{n})$ and using 
\eqref{e:old_base} which is 
straightforward. If one tries to proceed in a similar fashion for $\tilde D_m$ one realizes that the evaluation 
of the Shioda map \eqref{e:old_shioda} for $\tilde D_m$ in the transformed divisors formally 
requires also to use  new $\tilde D_I$, which is not fixed by \eqref{eq:MWshift}. However, note that the 
shift \eqref{DIshift} is uniquely fixed by requiring that the $\tilde D_I$ again behave as genuine blow-up divisors.
More precisely, we find that  \eqref{DIshift} is fixed if 
the three conditions
\begin{align}
 \pi (\tilde D_I \cdot \tilde D_\alpha) &\overset{!}{=} 0 \, , &  \pi (\tilde D_I \cdot \tilde D_J) &\overset{!}{=}  \pi (D_I \cdot D_J) \, , &
 \pi (\tilde D_I \cdot \tilde D_0) &\overset{!}{=} 0 \, .
\end{align}
are to be satisfied for the new divisors. These are simply the conditions \eqref{blowup-cond} and \eqref{evaluate_DD} in the $\tilde D_\Lambda$ basis. Having fixed $\tilde D_I$ the transformed $\tilde D_m$ in \eqref{free_shift_Ab} are determined 
uniquely.

Let us again emphasize that the non-Abelian part of this gauge transformation is absolutely essential. 
On the one hand it is non-zero if and only if $\pi_{m I} \neq 0$
for some $D_I$. On the other hand we find fractional $U(1)$-charges if and only if $\pi_{m I} \neq 0$. This can easily
be seen in the Shioda map \eqref{e:old_shioda}. The $U(1)$ charge $q_m$ of an M2-brane state wrapping a holomorphic curve $\cC$ is given by
the intersection of $\cC$ with $D_m$. A fractional contribution to the charge therefore can only arise from the last term in \eqref{e:old_shioda}
since $\mathcal{C}^{-1\,IJ}$ in general has fractional components proportional to $\mathrm{det}(\cC)^{-1}$.
In fact, since a section can only intersect nodes with Coxeter label
equal to one, the last term
in \eqref{e:old_shioda}
always vanishes for the simple Lie algebras $E_8, F_4, G_2$, which do not have nodes with Coxeter label one, and they are precisely the only simple Lie algebras
having integer $\mathcal{C}^{-1\,IJ}$, or equivalently $\mathrm{det}(\cC)=1$.\footnote{This is related
to the fact that the center of the corresponding universal covering group is trivial. We will elaborate more on this
fact in the part about the torsion subgroup of the Mordell-Weil group.} 
The Coxeter labels for the simple Lie algebras can be found in \autoref{t:lie_conventions}.
To put it in a nutshell, if and only if there are fractional $U(1)$ charges, the free Mordell-Weil group action induces Abelian large gauge transformations
supplemented by non-zero fractional non-Abelian large gauge transformations. Effectively, in 
the presence of fractional $U(1)$ charges a pure Abelian large gauge
transformation with integer winding $\fn^m \in \mathbb Z$ is in general ill-defined. 
What makes it well-behaved is precisely the additional contribution
from the fractional non-Abelian large gauge transformation (which is by itself also ill-defined)
which compensates for the fractional part in the Abelian sector. This matches the gauge theory discussion of \autoref{sec:class_lgt}.

It should be stressed that not all redefinitions \eqref{e:MW_shift_divisors_1} have an immediate geometric interpretation. 
Away from the singular loci over $B_n$ that are resolved
the transformation  $D_m \rightarrow \tilde D_m$ is induced by the action of the Mordell-Weil group, which has a 
known geometric origin as addition of points in the fiber (see \textit{e.g.}~\cite{Silverman}). We are not 
familiar of how the latter geometric group law is extended to the non-Abelian singularities or to 
their resolutions. This prevents us from identifying a geometric interpretation of $D_I \rightarrow \tilde D_I$. 
Nevertheless the considered arithmetic operations are well-defined on the level of divisors and consistently 
include also the blow-up divisors for example in the Shioda map. 
This suffices to infer information about the effective theory after compactification on this 
space. We will encounter similar transformations in the general discussion of non-Abelian large gauge 
transformations in \autoref{Arithmetics_nonAb}.
A possible way to resolve this puzzle could be provided by the theory of schemes in connection with the minimal model program.
In particular for the resolved space the Mordell-Weil group could be only well-defined as an arithmetic structure on a whole scheme rather than on
individual algebraic varieties. Or loosely speaking, there exists a branched cover of the elliptic fibration on which the arithmetic structure
is properly defined.
See \textit{e.g.~}\cite{Silverman2} for more details.

Finally, we conclude the discussion about the free part of the Mordell-Weil group with a comment about $G_4$-flux.
As we have seen around
\eqref{e:induced_cflux}, Abelian large gauge transformations of four-dimensional theories generically induce
circle-fluxes $\frac{1}{2}\int_{S^1} \langle d \hat \rho_\alpha \rangle$
resulting in a non-vanishing Chern-Simons coupling $\tilde\Theta_{\alpha 0} \neq 0$.
Note that we started with $\Theta_{\alpha 0} = 0$.
In F-theory compactifications this coefficient is given by
\begin{align}
 \Theta_{\alpha 0} = -\frac{1}{4} D_\alpha \cdot D_0 \cdot [G_4] \, ,
\end{align}
and the vanishing of it constrains the choices for $G_4$-flux. If we now also want
$\tilde\Theta_{\alpha 0}$ to vanish, we should manually switch on in field theory additional compensating
flux $-\frac{1}{2}\int_{S^1} \langle d \hat \rho_\alpha \rangle$ after the large gauge transformation
such that the net flux adds up to zero. In the F-theory picture this corresponds to imposing that also the
$G_4$-flux transforms after a Mordell-Weil shift according to
\begin{align}\label{e:trafo_g_flux}
 \tilde G_4 = G_4 - \fn^n \, \big(D_\alpha \cdot D_n \cdot [G_4]\big)\, \tensor{\eta}{^{-1}_\beta^\alpha}\, \cC^\beta \, .
\end{align}
It is then easy to show that this compensates for the unwanted contributions of \eqref{e:coupling_c_flux},
\textit{i.e.~}setting
them to zero.

\subsection{The Torsion Part of the Mordell-Weil Group} 
\label{sec:MWtorsion}

In a similar spirit we can show that a non-trivial torsion subgroup in \eqref{splitMW} is connected to special 
fractional non-Abelian large gauge transformations, \textit{i.e.}~to the basis vectors $(\fn^I,\fn^m)$ of type (III) introduced
in \autoref{sec:class_lgt}.
For a torsional section $t_r$, $r=1,\ldots,n_{\rm tor}$, we use the key fact that its image under the Shioda map vanishes, {cf.}~\eqref{Shioda_tr}. As discussed in \autoref{sec:top_torus} this implies that they do not define divisors that 
appear in the Kaluza-Klein expansion \eqref{e:expansion_threeform} and thus do not give rise to massless gauge fields 
in the effective theory.

 Despite the fact that torsional sections do not define massless gauge fields in the effective field theory, the 
 Mordell-Weil group action along these sections nevertheless results in a non-trivial 
 transformation  in the circle-reduced theory. In order to show this 
 we perform $\fn^r \in \mathbb Z$ shifts along the torsional generators $t_r$
 using the Mordell-Weil group law \eqref{e:MW_law}. The derivation 
 proceeds in a similar fashion as the one in \autoref{sec:FreeMWShifts}.
 In fact, keeping in mind that $D(t_r)=0$ one can use \eqref{e:MW_shift_divisors_1} to infer 
\begin{align}
\tilde D_m &= D_m \ ,\\
\tilde D_I &= D_I - \fn^K  \pi (D_K\cdot D_I) \ , \\ 
 \tilde D_0  
& =  D_0  + \fn^J D_J   - \frac{\fn^J \fn^L}{2} \, \pi (D_J\cdot D_L) \, ,
\end{align}
where we have defined, similar to \eqref{def-associatedkK}, that 
\beq
  \fn^I := -  \fn^r \ \pi_{rJ} \,\mathcal{C}^{-1\,JI}\ .
\eeq
Just as in \autoref{sec:FreeMWShifts}, in general $\fn^I$ will be fractional due to the appearance of the inverse matrix $\mathcal{C}^{-1\,JI}$. 
In other words the transformations induced by $\fn^r$ correspond to special fractional 
non-Abelian large gauge transformations parametrized by pairs $(\fn^I, \fn^m=0)$, introduced 
as case (III) in \autoref{sec:class_lgt}.

The fact that torsion in the Mordell-Weil group allows for the presence of special fractional 
non-Abelian large gauge transformations is not unexpected. As discussed in \cite{Aspinwall:1998xj,Mayrhofer:2014opa}, torsion in the Mordell-Weil group indicates that the gauge group is 
not simply connected, and therefore certain representations of the Lie algebra do not appear on the level of the group, \textit{i.e.}~the weight 
lattice of the group is coarser. Because of this fact also certain fractional non-Abelian
large gauge transformations are compatible with the circle boundary conditions. Indeed the 
torsional shifts exhaust all possible fractional large gauge
transformations, which is evident from considering the center of the universal covering group as
in section 3.3 of \cite{Mayrhofer:2014opa}.
Note that since the adjoint representation is always present in terms of gaugini,
the possible set of special fractional large gauge transformations, which might be restricted by the global structure of the group,
can be derived by demanding
\begin{align}
 \fn^I w_I^{\rm adj} \in \mathbb Z
\end{align}
with $w^{\rm adj}$ the weights of the adjoint representation.

\section[Arithmetic Structures on Fibrations with Multi-Sections]{Arithmetic Structures on Fibrations with Mul-ti-Sections
\sectionmark{Arithmetic on Fibrations with Multi-Sections}}
\sectionmark{Arithmetic on Fibrations with Multi-Sections}
\label{sec:multi_group}

In this section we aim to generalize the discussion of \autoref{sec:ec_group_structures} to 
Calabi-Yau geometries that admit a genus-one fibration that does not have a rational section.
These setups always come with multi-sections that no longer cut out rational points of the genus-one 
fiber but rather roots which are exchanged over branch cuts in the base $B_n$. On such 
genus-one fibrations with only multi-sections there is no known arithmetic structure analog to
the Mordell-Weil group. However, our understanding of the F-theory effective action 
associated to such geometries, which will have $U(1)$ gauge group factors if there is more than one 
independent multi-section, and the possibility to perform Abelian large gauge transformations in 
these setups suggest that an arithmetic structure should equally exist on genus-one fibrations without section. 
We will collect evidence for the existence of this structure which we name the \textit{extended Mordell-Weil group}
and study its key properties.

Our considerations will be driven by two facts. First, we will make use of the fact that 
a genus-one fibration with only multi-sections can often be related by a geometric transition to 
elliptic fibrations with multiple rational sections. Physically this corresponds to an unHiggsing 
of Abelian gauge fields \cite{Morrison:2014era,Anderson:2014yva,Klevers:2014bqa,Garcia-Etxebarria:2014qua,Mayrhofer:2014haa,Mayrhofer:2014laa,Cvetic:2015moa}. Following the divisors through this transition we are able 
to reverse-engineer on the level of divisor classes the group law on the geometry without rational sections from the 
Mordell-Weil group law 
in the unHiggsed geometry. We note that at this point
we can only determine the extended Mordell-Weil group law up to vertical divisors, which will be the task of \autoref{multi_group}.
This definition however will allow us to uniquely define a 
 \textit{generalized Shioda map} in \autoref{sec:gen_Shioda}. The latter defines divisors associated to 
massless Abelian  gauge symmetries from the generators of the postulated extended Mordell-Weil group, \textit{i.e.}~from the multi-sections. Finally, in \autoref{sec:ExtendedMWLGT}
 we show that translations in the extended Mordell-Weil group correspond to Abelian large gauge transformations.

\subsection{A Group Action for Fibrations with Multi-Sections} \label{multi_group}
  
We now present an extension of the results from the last subsection to
F-theory compactified on genus-one fibrations without section. These geometries come with multi-sections which mark points in the elliptic fiber that are exchanged
over branch cuts in the base $B_n$. If they mark a set of $n$ points in the fiber, we call the multi-section an $n$-section. 
For a genus-one fibration one can always birationally
move to the Jacobian fibration, which replaces each independent $n$-section 
by a rational section and therefore constitutes an elliptic fibration.
Importantly the genus-one fibration and its Jacobian describe the same F-theory effective action 
in four or six dimensions \cite{deBoer:2001px,Braun:2014oya}. It is therefore clear that the presence of at least two 
homologically independent multi-sections indicates the existence of massless 
$U(1)$ gauge fields in the four- or six-dimensional F-theory effective field theory.\footnote{A single 
multi-section gives one massless $U(1)$ in the three- or five-dimensional effective theory, which captures 
the degree of freedom of the circle Kaluza-Klein vector.} In particular, the associated Jacobian fibration 
of a genus-one fibration with more than one multi-section will have a non-trivial Mordell-Weil group.
One can therefore ask how to identify the divisor classes associated to massless $U(1)$ gauge 
symmetries already in the genus-one fibrations. This is relevant \textit{e.g.}~for the computation of 
$U(1)$-charges or the computation of anomaly coefficients. 
Furthermore, we will argue for the existence of a group law for multi-sections.

To address these issues we first have to introduce some additional facts about 
fibrations with multi-sections and state our assumptions. 
First, recall that an $n$-section $s^{(n)}$ with divisor class 
$S^{(n)} = Div(s^{(n)})$ fulfills  
\beq
	S^{(n)}\cdot f=n\,,
\eeq
where $f$ is the class of the genus-one fiber. 
This implies that
\beq
	S^{(n)}\cdot D_\alpha\cdot D_{\beta}=n\, D^b_\alpha\cdot D^b_{\beta}\,,\qquad S^{(n)}\cdot D_\alpha\cdot D_{\beta}\cdot D_{\gamma}=n\, D^b_\alpha\cdot D^b_{\beta}\cdot D_{\gamma}^b\,,
\eeq
where the first equation applies for threefolds and the second for fourfolds. 
Second, note that it is always possible to 
find a basis of multi-sections in homology that are all of the same degree \cite{Braun:2014oya}, 
\textit{i.e.}~they cut out the same number of points in the fiber. We denote the number 
of such multi-sections by $n_{\rm ms}$ and assume $n_{\rm ms}  \geq 2$. 
We denote such a basis of $n$-sections by $s_0^{(n)},s_m^{(n)}$, $m=1,\ldots ,n_{\rm ms}-1$ 
and demand that it is minimal in the sense that there does not exist any multi-section in the geometry
that cuts out $n-1$ or fewer points.\footnote{From
now on we will always require that the considered basis of multi-sections is of this type.}
We have also singled out an arbitrary multi-section which we labeled by $0$.
The divisors associated to these sections are denoted by $S_0^{(n)} = Div(s_0^{(n)})$
and $S_m^{(n)}= Div(s_m^{(n)})$ in accord with our previous notation.

To propose a group law we will work with the following assumption for genus-one fibrations throughout this section:
\begin{itemize}
\item
We assume that there exists a specialization of the complex structure of 
the fibration such that each $n$-section $s_0^{(n)}$ and $s_{m}^{(n)}$ 
splits into $n$ rational sections $s_0^1,\ldots, s_0^n$ and $s_m^1, \ldots , s_m^n$. 
After 
resolving the singularities in the new geometry we will denote the resulting 
space by $\hat Y_{\rm uH}$, where we indicate that this geometry captures 
the unHiggsing from a field-theoretic point of view.
In the following we impose that the rational sections $s_0^1,\ldots, s_0^n$ and
$s_m^1, \ldots , s_m^n$ are the generators of the Mordell-Weil group
supplemented by the zero-section of the elliptic fibration $\hat Y_{\rm uH}$. 
We expect however that the following discussion can be extended to 
the more general situation in which these rational 
sections only generate a sublattice of the Mordell-Weil lattice.
With this simplification the divisor homology groups of $\hat Y$ and $\hat Y_{\rm uH}$ are generated 
as follows:
\beq \label{spanH_ms}
   H_{p} (\hat Y) = \langle S_0^{(n)},  S_m^{(n)}, D_\alpha \rangle\ ,  \qquad 
   H_{p} (\hat Y_{\rm uH}) = \langle S_0^{1},\ldots, S_0^n ,S_m^1,\ldots  , S_m^{n}, D_\alpha' \rangle \ ,
\eeq
where $p=4$ for Calabi-Yau threefolds and $p=6$ for Calabi-Yau fourfolds.
Note that we will in the following assume that the theory has no non-Abelian 
gauge groups. In other words, we do not include exceptional divisors in \eqref{spanH_ms}.

\item We also introduce an \textit{unHiggsing map} $\varphi$ from the 
divisors of $\hat Y$, \textit{i.e.}~the fibrations admitting multi-sections, to the divisors 
of $\hat Y_{\rm uH}$,
\beq
	\varphi:\quad H_p (\hat{Y})\,\,\hookrightarrow \,\, H_p (\hat{Y}_{\text{uH}})\,.
\eeq
Here we have indicated that the map is injective. In addition, we require it to be an injective ring homomorphism
from the full intersection ring on $\hat{Y}$ into that of $\hat{Y}_{\text{uH}}$.
This map is defined to identify the $n$-sections with 
$n$ rational sections on the divisor level:
\beq \label{SplitS}
 \varphi(S_0^{(n)})=  S_0^1 + \dots + S_0^n \, ,\qquad \varphi(S_m^{(n)})= S_m^1 + \dots + S_m^n \, .
\eeq
We do not consider torsional sections in the following 
discussion. We furthermore assume that the map $\varphi$ 
acts trivially on the remaining divisors $ D_\alpha$ and is linear on the 
vector space of divisors, i.e.
\beq \label{prop_map}
    \varphi(\nu^i S_{i}^{(n)} + \nu^\alpha D_\alpha ) = 
    \nu^i S_i^1 + \dots + \nu^i S_{i}^n  + \nu^\alpha D'_\alpha   \, ,
\eeq
for some constants $(\nu^i,\nu^\alpha)$. Note that $D_\alpha$ and $D_\alpha'$ 
actually define the same divisor classes since they both ascend from the same 
divisors in the base $B_n$ common to both $\hat Y$ and $\hat Y_{\rm uH}$.

\end{itemize}

Note that only a single example of a geometry with more than one independent multi-section has been studied in the
literature \cite{Klevers:2014bqa} which is given by
an embedding of the fiber as a hypersurface into $\mathbb P^1 \times \mathbb P^1$. In these setups one finds two independent two-sections
which do indeed
split into four sections in the prescribed way by blowing up the fiber ambient space to $dP_3$.\footnote{It is important to notice that the two toric
two-sections
of $\mathbb P^1 \times \mathbb P^1$ do not exclusively 
split into the four toric sections of $dP_3$. 
One rather has to pick four appropriate elements of the Mordell-Weil lattice 
of the blow-up that are not necessarily torically realized.}

Let us make the following preliminary ansatz for a group structure placed on the set of multi-sections
written down in homology similar to \eqref{e:MW_law}:
Choose one $n$-section $s_0^{(n)}$ as what we call the \textit{zero-$n$-section} or \textit{zero-multi-section}. 
Then two arbitrary
$n$-sections $s_1^{(n)},s_2^{(n)}$ are added according to
\begin{align}\label{e:gen_MW_law_ansatz}
 Div(s_1^{(n)} \oplus \fn s_2^{(n)} ) := S_1^{(n)} + \fn(S_2^{(n)} - S_0^{(n)} )
 + \lambda^\alpha D_\alpha  \, .
\end{align}
Making the definition \eqref{e:gen_MW_law_ansatz} precise would require to determine the 
constants $\lambda^\alpha$. However, we will argue in the following that these are not uniquely 
determined, which can be traced back to the fact that there exist divisor classes corresponding 
to genuine multi-sections that differ only in their vertical parts induced by the base homology. This implies that 
we need to talk about equivalence classes $[\cdot ]$ of divisors associated to multi-sections defined modulo vertical part. 
Furthermore, we will in the following provide evidence  that $Div(s_1^{(n)} \oplus \fn s_2^{(n)})$
defines a divisor class representing an actual $n$-section in the geometry when neglecting the 
vertical part. Let us stress again that our approach
just allows us to investigate how the group law for multi-section is defined in terms of homology classes.

\definecolor{blue_standard}{rgb}{0,0,255}
\begin{figure}
\begin{center}
\includegraphics[scale=0.6]{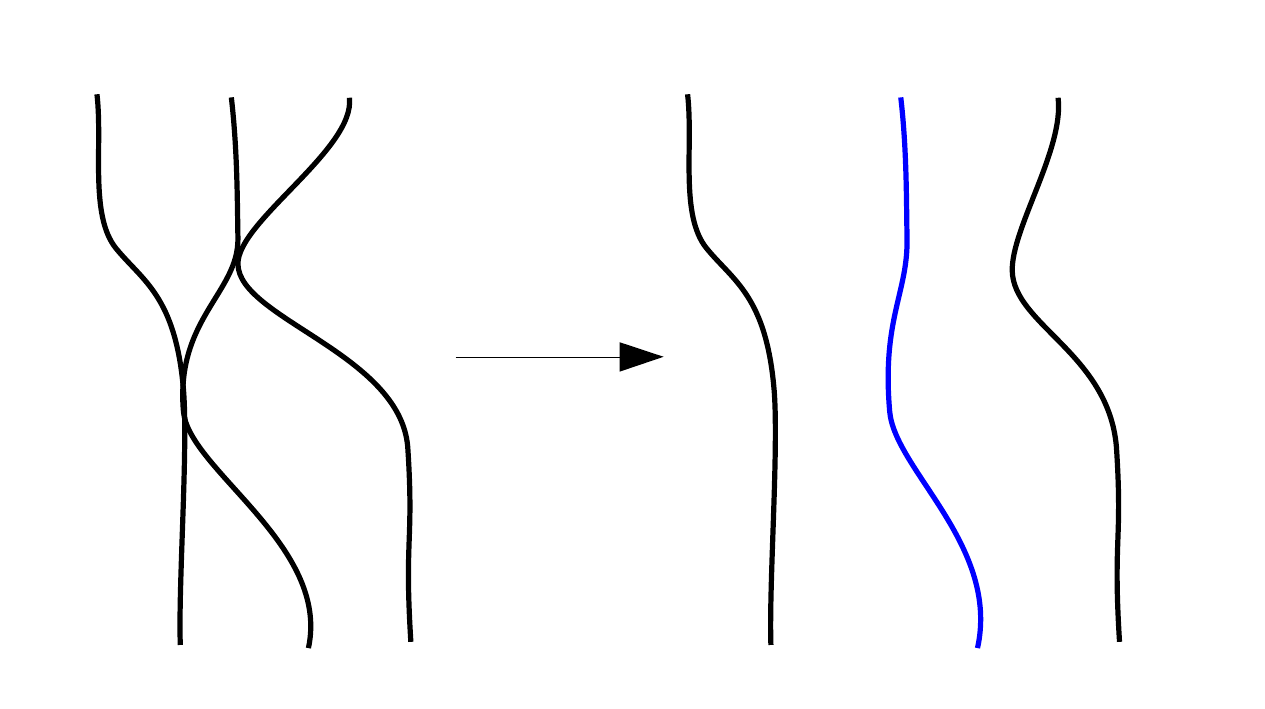}
\begin{picture}(0,0)
\put(-215,15){$s^{(3)}_0$}
\put(-105,15){$s^{1}_0$}
\put(-70,15){\textcolor{blue_standard}{$s^{2}_0$}}
\put(-45,15){$s^{3}_0$}
\put(-220,-10){zero-3-section}
\put(-90,-10){\textcolor{blue_standard}{zero-section}}
\end{picture}
\vspace*{.5cm}

\begin{minipage}{12cm}
\caption{The zero-$n$-section is chosen to contain the zero-section after unHiggsing to a setting 
with rational sections only. \label{pick-zero}}
\end{minipage}
\end{center}
\end{figure}
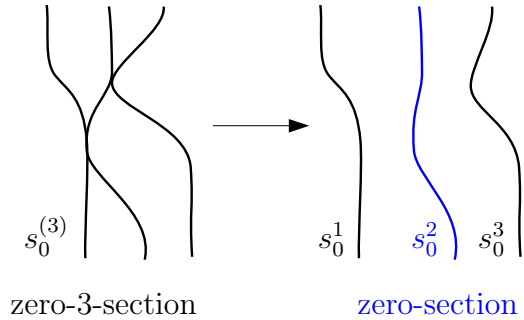

We first want to provide evidence that there is indeed 
a multi-section associated to $\tilde S^{(n)} \equiv Div(s_1^{(n)} \oplus \fn s_2^{(n)})$
as defined in \eqref{e:gen_MW_law_ansatz}. In order 
to do that we will check in which ways $\varphi( \tilde S^{(n)})$ 
can split into a sum of $n$ sections in the homology of $\hat Y_{\rm uH}$.
Let us denote such a set of $n$ linear independent 
sections of $\hat Y_{\rm uH}$ by $\{ \hat s^i \} $, and demand that
\begin{align}\label{e:splitting}
\varphi\big( S_1^{(n)} + \fn(S_2^{(n)} - S_0^{(n)} ) + \lambda^\alpha D_\alpha \big) \overset{!}{=} \sum_{i=1}^{n} \hat S^i \, ,
\end{align}
where $\lbrace \hat S^i = Div(\hat s^i)  \rbrace$ is the associated set of linearly 
independent divisors in $\hat Y_{\rm uH}$. It turns out that there are infinitely many possibilities to define an
appropriate set of sections $\lbrace \hat s^i \rbrace$.
For example, choosing one arbitrary element $s_0^l$ (for fixed $l$) as the zero-section (see \autoref{pick-zero} where \textit{e.g.}~$l=2$),
there is the very simple choice
\begin{align} \label{choice_si}
 \hat s^i := s_1^i \oplus \fn s_2^i \ominus \fn s_0^i \, ,
\end{align}
which gives the right structure \eqref{e:splitting} upon using \eqref{prop_map} and the conventional Mordell-Weil group
law \eqref{e:MW_law}. Clearly, the ansatz \eqref{choice_si}
allows us to fix the $\lambda^\alpha$ specifying the vertical part in \eqref{e:gen_MW_law_ansatz}. 
The existence of an appropriate set of $\hat s^i$ indicates that there is indeed a multi-section 
in the divisor class $Div(s_1^{(n)} \oplus \fn s_2^{(n)} )$ when fixing the $\lambda^\alpha$
via \eqref{e:splitting}, \eqref{choice_si} and \eqref{e:MW_law}.

\definecolor{blue_standard}{rgb}{0,0,255}
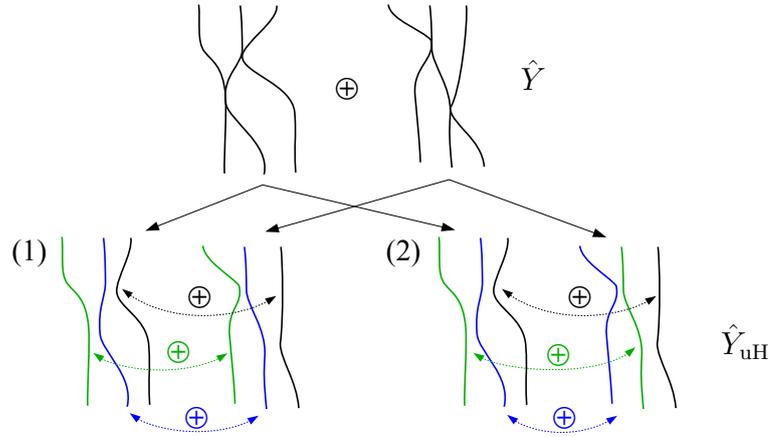
\begin{figure}
\begin{center}
\includegraphics[scale=0.4]{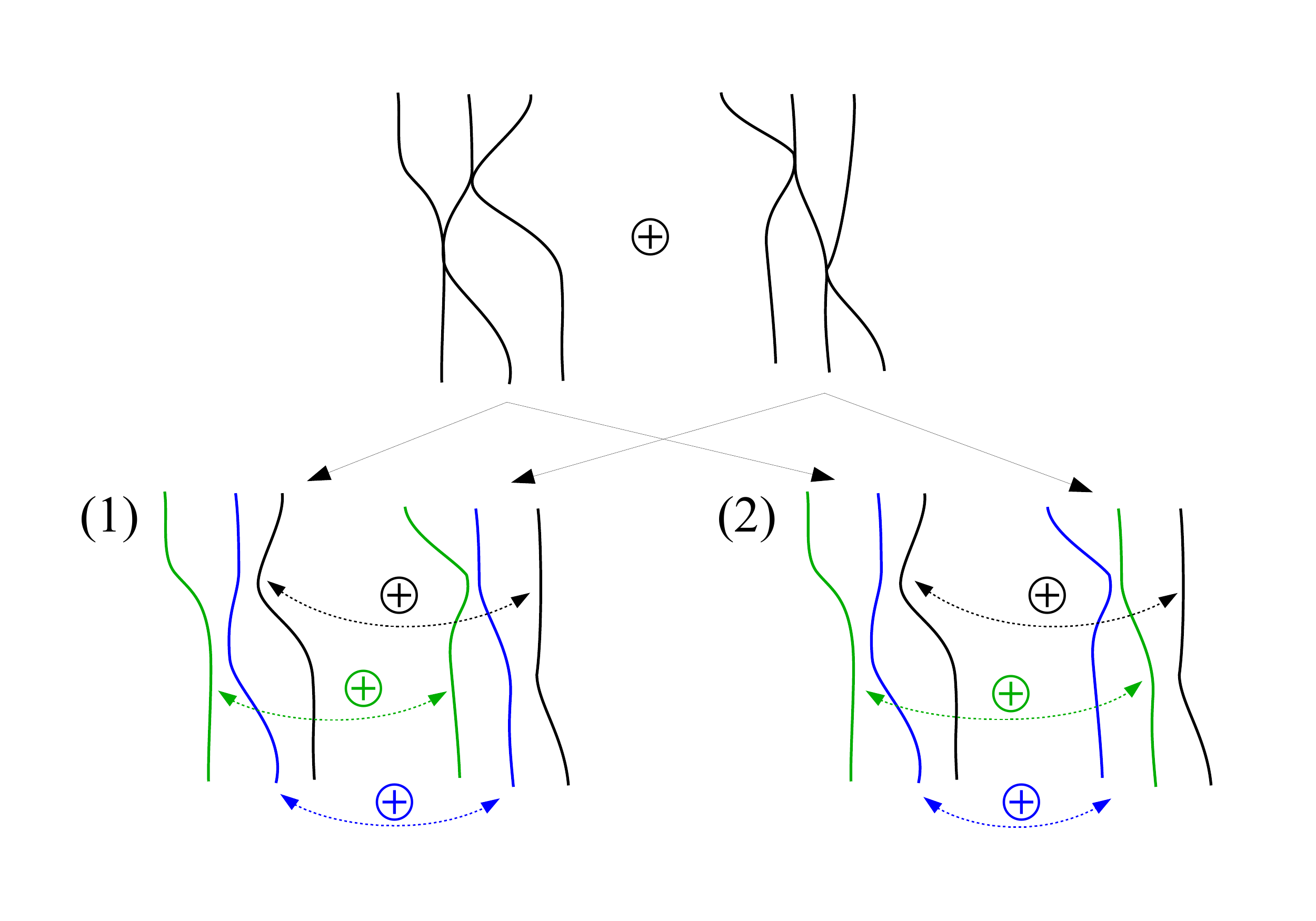}
\begin{picture}(0,0)
\put(-85,150){$\hat Y$}
\put(-10,50){$\hat Y_{\rm uH}$}
\end{picture}
\vspace*{-.5cm}

\begin{minipage}{12cm}
\caption{Moving from $\hat Y$ to the unHiggsed phase $\hat Y_{\rm uH}$ there are in general 
many ways to add the individual sections. We schematically indicate 
two choices in (1) and (2), which in general differ by their vertical parts when 
considering the associated multi-section. \label{add-sections}}
\end{minipage}
\end{center}
\end{figure}

However, using merely the 
existence of sections $\hat s^i $ in $\hat Y_{\rm uH}$ satisfying \eqref{e:splitting} does not seem to fix
the class $Div(s_1^{(n)} \oplus \fn s_2^{(n)} )$  uniquely. In fact, 
other appropriate sets $\lbrace\hat s'^i \rbrace$ can be obtained if one picks two arbitrary sections out of
$\lbrace \hat s^i \rbrace$ and adds a third arbitrary chosen section to one of the latter while subtracting it from the other one by using the Mordell-Weil group law on $\hat Y_{\rm uH}$. Such a freedom of choice
is schematically depicted in \autoref{add-sections}. The set $\lbrace\hat s'^i \rbrace$ 
can be used to satisfy \eqref{e:splitting} but will generally yield a different set of constants $\lambda^\alpha$
compared to the choice \eqref{choice_si}. In other words, 
the contribution from vertical divisors in \eqref{e:gen_MW_law_ansatz} is a priori not
uniquely fixed by the compatibility conditions that we impose. 
This should be contrasted with the situation for genuine sections
(\textit{i.e.}~$n=1$) in which this part is fixed by demanding
that the section squares to the canonical class in the base of the elliptic fibration. 
For multi-sections this relation to the canonical class of the base is in general not valid.\footnote{This 
can also be understood after applying the map $\varphi$ to $\hat Y$, since two of the individual 
sections $s_i^{1}$, $s_i^{2}$ arising from the $i$th $n$-section can have non-trivial intersection.}
In fact, for concrete examples
one can verify that there exist multi-sections that differ only by their vertical parts.\footnote{One can consider for example fibrations where the fiber is embedded
as a hypersurface into $\mathbb P^1 \times \mathbb P^1$. In these setups one can find four toric multi-sections from which two are independent, and the other two do indeed
differ by vertical parts from the latter in certain examples.}
Therefore there is no immediate way that we are aware of to infer the possible vertical parts of a multi-section. 
As stressed above this also reflects our inability to give a unique choice for the splitting \eqref{e:splitting}
and the fixation of the constants $\lambda^\alpha$ in \eqref{e:gen_MW_law_ansatz}.

It is therefore natural to define the group law \eqref{e:gen_MW_law_ansatz} only in terms of
equivalence classes of multi-sections modulo vertical divisors as
\begin{align}\label{e:gen_MW_law}
 \widehat{Div}(  [ s_1^{(n)}  ] \oplus \fn [ s_2^{(n)} ]   ):=   \big[ S_1^{(n)} + \fn(S_2^{(n)} - S_0^{(n)} ) \big]  \, .
\end{align}
In this expression we indicate that the divisors as well as the multi-sections should only 
be considered modulo vertical parts.  $ \widehat{Div}$ maps between equivalence classes 
of multi-sections and equivalence classes of divisors and reduces on representatives 
to $Div$.
Let us stress that this does not imply that one can add arbitrary vertical divisors to the right-hand side
of this equation and find an actual multi-section in the geometry. The formulation in \eqref{e:gen_MW_law}
loses some information about divisor classes supporting multi-sections. Crucially, this information  
turns out to be irrelevant in the discussion of the generalized Shioda map and therefore does not affect the 
considered physical implications of the setup.

\subsection{The Generalized Shioda Map} \label{sec:gen_Shioda}

In order to investigate the physical implications of the group law studied in \autoref{multi_group} 
for the effective field theory we first need to define a generalized Shioda map
for multi-sections. Recall that the considered multi-sections 
were arbitrarily subdivided as $s_0^{(n)}$, $s_{m}^{(n)}$, $m=1,\ldots,n_{\rm ms}-1$, 
where we called $s_0^{(n)}$ the zero-$n$-section.
The generalized Shioda map associates divisors $D^{(n)}_m \equiv D(s_m^{(n)})$ to 
the $n$-sections $s_m^{(n)}$. In addition one has to define a map from $s_0^{(n)}$
to a divisor $D_{0}^{(n)}$ generalizing \eqref{e:old_base}, also to be defined below.
The Poincar\'e-dual two-forms $[D_{0}^{(n)}]$ and $[D^{(n)}_m]$ 
can then appear in the Kaluza-Klein expansion of the M-theory three-form 
$C_3$ generalizing \eqref{e:expansion_threeform} as 
\beq
   C_3 = A^\alpha \wedge [D_\alpha] +  A^0 \wedge [D^{(n)}_0]+   A^m \wedge [D_m^{(n)}]\ ,
\eeq
where we recall that we assume the absence of exceptional divisors $D_I$ associated 
to a non-Abelian gauge group in this section. The vectors $A^m$ 
correspond to massless linear combinations of former $U(1)$ gauge fields in the six- or four-dimensional effective 
theory, while $A^0$ will contain (in a massless linear combination) the degree of freedom arising from the 
Kaluza-Klein vector along the circle when performing the F-theory 
uplift from three or five to four or six dimensions.

If we have $n_{\rm ms} \geq 2$, as we will assume in the following, we have to define generalized Shioda maps yielding the $U(1)$ divisors $D_m^{(n)}$.
To this end, we once again borrow results from the unHiggsed geometry $\hat Y_{\rm uH}$.
As was explained in \cite{Morrison:2014era,Anderson:2014yva,Klevers:2014bqa,Garcia-Etxebarria:2014qua,Mayrhofer:2014haa,Mayrhofer:2014laa,Cvetic:2015moa}  the transition from the unHiggsed geometry $\hat Y_{\rm uH}$ 
to the genus-one fibration is described by a Higgsing in the effective
five- or three-dimensional field theory of certain matter states charged under a 
linear combination of the $n \times n_{\rm ms}$  Abelian gauge fields. 
We aim to find the proper linear combinations of these $U(1)$s 
that constitute the \textit{massless} $U(1)$ vectors after the Higgsing.
To begin with we consider the divisor classes $D'_{m,l}$ on $\hat Y_{\rm uH}$ given
by 
\begin{align}\label{e:multi_div_sum}
  D'_{m,l}  : = \sum_{i=1}^{n} D_m^i - \sum_{\substack{i=1\\i\neq l}}^n D_0^i \, ,
\end{align}
where $D_m^i \equiv D(s_m^i)$, $D_0^i = D(s_0^i)$ denote the Shioda maps of $s_m^i , s_0^i$, with $i\neq l$, and we have chosen an 
element $s_0^l$ (fixed $l$) as the zero-section on $\hat Y_{\rm uH}$.
We want our ansatz to be invariant under exchanging individual sections which come from the same multi-section since
we aim to write all expressions in terms of the map $\varphi$ as defined in \eqref{SplitS}.
Note that the ansatz \eqref{e:multi_div_sum} is invariant under 
the exchange of the components $s_m^i , s_0^{i\neq l}$ of a given multi-section $s_m^{(n)}$ and the zero-multi-section $s_0^{(n)}$, respectively,
and it is therefore almost consistent with the map \eqref{SplitS}.
However, the definition \eqref{e:multi_div_sum} of $D_{m,l}^{\prime}$ still 
depends on the choice of the zero-section $s_0^l$ and is thus not 
invariant under the exchange of \textit{all} the components of the zero-multi-section $s_0^{(n)}$ in this respect.
We therefore included an additional index $l$ in the notation. 
Inserting the explicit expressions for the Shioda maps $D_m^i , D_0^i$ and 
using \eqref{SplitS} we obtain
\begin{align}\label{e:multi_blow_up_shioda}
  D_{m,l}^{\prime}  = \varphi( S_m^{(n)} - S_0^{(n)}) - \pi \big( \varphi  (S_m^{(n)} -
 S_0^{(n)} ) \cdot S_0^l \big) \, .
\end{align}

Thus it is clear that the expression \eqref{e:multi_blow_up_shioda} is still problematic 
if one wants to move solely to the phase of the genus-one fibration since \eqref{e:multi_blow_up_shioda}
manifestly depends on the choice
of the zero-section $s_0^l$ on $\hat Y_{\rm uH}$. However, as it follows from the upcoming discussion in \autoref{ch:zero_sec}, different choices of the zero-section
are just related by large gauge transformations in the effective field theory. Therefore it seems logical to treat all phases
with different zero-section $s_0^l$ on equal footing. We thus average over all these choices and 
use again \eqref{SplitS} to obtain
\beq \label{average_D'}
   \frac{1}{n}\sum_{l} D_{m,l}^{\prime} = \varphi( S_m^{(n)} - S_0^{(n)}) - \frac{1}{n} \pi \big( \varphi  (S_m^{(n)} -
 S_0^{(n)} ) \cdot  \varphi  (S_0^{(n)})\big)\ .
\eeq
Since $\varphi$ is a ring homomorphism and all pieces lie in the image 
of $\varphi$, we can now drop the map $\varphi$ in this expression and consistently define
a generalized Shioda map $D^{(n)}_m$ for the multi-section $s_m^{(n)}$ without reference to an unHiggsed phase
\begin{align}\label{gen_Shioda}
 D^{(n)}_m :=S_m^{(n)} - S_0^{(n)} - \frac{1}{n} \pi \Big ( \big (S_m^{(n)} -
 S_0^{(n)} \big ) \cdot S_0^{(n)} \Big ) \, .
\end{align}
By construction it is evident that $U(1)$ charges $q_m$ of matter in the genus-one fibration without rational sections are calculated by intersecting the associated curves
with $D^{(n)}_m$. Note that this intersection is independent of the vertical contribution in $D_m$.
Furthermore \eqref{gen_Shioda} is a generalization of the map given in \cite{Klevers:2014bqa} (without the factor $\frac{1}{n}$), where the
authors consider fibers embedded into $\mathbb P^1 \times \mathbb P^1$.
We expect that our definition of $D^{(n)}_m$ yields the correct $U(1)$ divisors in 
order to study the effective field theory of F-theory on genus-one fibrations
directly without explicit reference to an unHiggsed geometry $\hat Y_{\rm uH}$ or the Jacobian of $\hat Y$.
Exploring this effective theory in detail is however beyond the scope 
of the work in this thesis.

Further indication that $D^{(n)}_m$ is an important object of the genus-one fibration is provided by the fact that
the definition \eqref{gen_Shioda} only depends on the equivalence classes $\big[S_m^{(n)}\big],\big[S_0^{(n)}\big]$.
Indeed, it is easy to check that \eqref{gen_Shioda} even provides a homomorphism from the generalized Mordell-Weil group \eqref{e:gen_MW_law}
to the Ner\'on-Severi group. Note that both of these conditions are extremely restrictive.

In a similar fashion we can construct the divisor $D_0^{(n)}$ appearing in \eqref{e:expansion_threeform}. It is the cycle that is dual to the massless linear combination
of the Kaluza-Klein
vector and a set of $n-1$ $U(1)$ vectors that are massive in the higher-dimensional theory \cite{Anderson:2014yva}. 
 These correspond to the individual constituents of the zero-multi-section under the splitting \eqref{SplitS}.
In analogy to \eqref{e:multi_div_sum} we first make the ansatz
\begin{align}\label{invariantKKsum}
 D'_{0,l} := n\,   D_0 +  \sum_{\substack{i=1\\i\neq l}}^n D_0^i \, ,
\end{align}
where $D_0^i = D (s_0^i)$ are the Shioda maps with a chosen zero-section $s_0^l$.
This expression is again as before invariant under the exchange of the individual sections $s_0^{i\neq l}$
modulo vertical divisors.
We stress that $D_0$ denotes the divisor yielding the Kaluza-Klein vector in $\hat Y_{\rm uH}$ and is therefore given as in  \eqref{e:old_base} by 
\begin{align}\label{definition_base_div}
  D_0 = S^l_0 - \frac{1}{2} \pi(S^l_0 \cdot S^l_0) \, .
\end{align}
Using in \eqref{invariantKKsum} the explicit expressions for 
the Shioda maps as well as \eqref{definition_base_div} and \eqref{SplitS} we obtain
\begin{align}
 D'_{0,l} = \varphi (S_0^{(n)}) + \frac{n}{2}K - \pi \big (\varphi (S_0^{(n)}) \cdot S_0^l \big ) 
\end{align}
with $K$ the canonical class of the base.
Averaging over all zero-section choices as in \eqref{average_D'} we get
\begin{align}
 \frac{1}{n} \sum_l D'_{0,l} = \varphi (S_0^{(n)}) + \frac{n}{2}K - \frac{1}{n} \pi \big (\varphi (S_0^{(n)}) \cdot \varphi (S_0^{(n)}) \big ) \, .
\end{align}
Now we can drop the map $\varphi$ using the similar arguments as above, and
are therefore able to define $D_0^{(n)}$ as
\begin{align}
 D_0^{(n)} := S_0^{(n)} + \frac{n}{2}K - \frac{1}{n} \pi \big (S_0^{(n)} \cdot S_0^{(n)} \big ) \, ,
\end{align}
which is a generalization of \eqref{e:old_base}.

\subsection{Extended Mordell-Weil Group and Large Gauge Transformations} \label{sec:ExtendedMWLGT}

In this final subsection we show that, similar to the genuine Mordell-Weil group of rational sections, 
translations in the extended Mordell-Weil lattice are in one-to-one correspondence with Abelian 
large gauge 
transformations in the effective field theory. As before, the formulation of this group law on the 
divisor level will be completely sufficient for the question we aim to address due to the uniqueness of the generalized Shioda maps.

We begin by shifting the basis of multi-sections by $\fn^m$-times the $n$-section $s^{(n)}_{m}$ as
 \begin{align}
  \big[ \tilde s^{(n)}_0 \big] &:=  \big[s_0^{(n)} \big] \oplus \fn^m \big[ s_{m}^{(n)}\big]\, , \\
  \big[ \tilde s^{(n)}_m \big] &:= \big[s_m^{(n)}\big] \oplus \fn^m \big[ s_{m}^{(n)}\big]\, \, . \nn
 \end{align}
Using the group law \eqref{e:gen_MW_law} and inserting the resulting divisor classes into 
the generalized Shioda map \eqref{gen_Shioda} we
then find that the transformation of $D^{(n)}_m$ is given by
\begin{align}\label{e:multi_sec_shift}
 D^{(n)}_m \mapsto D^{(n)}_m - \frac{\fn^p}{n} \pi \big ( D^{(n)}_{p} \cdot D^{(n)}_m\big ) \, ,
\end{align}
which differs by a factor of $\frac{1}{n}$ in the vertical part from \eqref{free_shift_Ab}.
We emphasize that in this evaluation the ambiguity in the vertical parts is 
absent after applying the generalized Shioda map.

Let us now analyze the large gauge transformations from a field theory perspective. 
Recall that in general the actual Kaluza-Klein vector mixes in the Higgsed phase with other $U(1)$s 
as dictated by the zero-multi-section $s_0^{(n)}$ \cite{Anderson:2014yva,Garcia-Etxebarria:2014qua,Mayrhofer:2014haa,Mayrhofer:2014laa,Cvetic:2015moa}. While there 
are $n-1$ massive $U(1)$s parametrized by $s_0^{(n)}$ only a single $U(1)$ remains massless.
To simplify the treatment of the large gauge transformations in such a situation, 
we again can consider the unHiggsed phase. This will allow us to show
that \eqref{e:multi_sec_shift} is induced by large gauge transformations.
In particular we consider the different splits corresponding to the divisor $D^{(n)}_{m}$. Note that $D^{(n)}_{m}$
was obtained in \eqref{average_D'} by averaging over all divisors $D_{m,l}^{\prime}$, defined in \eqref{e:multi_div_sum},
which together represent the different choices for the zero-section.
Focusing now on a particular divisor $D_{m,l}^{\prime}$ with zero-section $s_0^l$, we find that the dual gauge field in the unHiggsed phase reads
\begin{align}
 A^{\prime\,m,l} = \frac{1}{n} \Big ( \sum_{i=1}^{n} A_i^{m} - \sum_{\substack{i=1\\i\neq l}}^n A_i^0 \Big ) \, ,
\end{align}
with $A_i^0, A_i^{m}$ dual to $D_0^i, D_{m}^i$.
Our main interest is in the form of the large gauge transformation for this vector field $A^{\prime\,m,l}$.
Therefore let us apply large gauge transformations with winding $\fn^m_i$
of the individual constituents $A_{i}^{m,l}$.
We find 
\begin{align}
  A_{i}^{m} \mapsto  A_{i}^{m} - \fn_i^m A_l^0 \, , && A_{i}^{0} \mapsto  A_{i}^{0} \, ,
\end{align}
where $A_l^0$ denotes the Kaluza-Klein vector.
We conclude that the large gauge transformations act on $A^{\prime\, m,l}$ as
\begin{align}
 A^{\prime\,m,l} \mapsto A^{\prime\,m,l} - \frac{\sum_{i=1}^n \fn^m_i}{n} A_l^{0} \, .
\end{align}
Using the results from \autoref{sec:ec_group_structures} we conclude that the dual divisors
transform as
\begin{align}
 D_{m,l}^{\prime} \mapsto  D_{m,l}^{\prime} - \frac{\sum_{i=1}^n \fn^p_i}{n} \pi \big ( D_{p,l}^{\prime} \cdot D_{m,l}^{\prime} \big ) \, .
\end{align}
Averaging now as in \eqref{average_D'} over the different choices for the zero-section
we can finally infer that the genuine $U(1)$ divisors $D_m^{(n)}$ in the Higgsed phase transform as
\begin{align}
  D^{(n)}_m \mapsto  D^{(n)}_m - \frac{\sum_{i=1}^n \fn^p_i}{n} \pi \big ( D^{(n)}_{p} \cdot D^{(n)}_m \big ) \, .
\end{align}
This is precisely what we get from \eqref{e:multi_sec_shift} for appropriate choices of the $\fn^m_i$.
We finally conclude that shifts in the generalized Mordell-Weil group correspond to Abelian large gauge transformations
in the Higgsed phase.

\section[Arithmetic Structures on Fibrations with Exceptional Divisors]{Arithmetic Structures on Fibrations with Exceptional Divisors
\sectionmark{Arithmetic on Fibrations with Exceptional Divisors}}  
\sectionmark{Arithmetic on Fibrations with Exceptional Divisors}
\label{Arithmetics_nonAb}

In this section, we focus on elliptic fibrations $\hat{Y}$ with codimension-one singularities leading to
non-Abelian gauge groups with matter in F-theory. The resolution of 
singularities of the elliptic fibration at codimension-one in the base $B_n$ requires introducing a set 
of blow-up divisors. In \autoref{sec:defGroupStructureExeptional} we define a novel group action on the set of these divisors  in $\hat{Y}$.
We are guided by two principles in defining this group structure, one geometric and one field-theoretic one. 

First, we employ the geometric
fact that many geometries $\hat{Y}$ with a Higgsable non-Abelian gauge group can be connected by
a number of extremal transitions, corresponding to Higgsing in field theory, 
to a geometry $\hat{Y}_{\rm H}$ with a purely Abelian gauge group,  \textit{i.e.}~a number of rational sections. 
Under this transition, the Cartan $U(1)$s inside the non-Abelian gauge group are mapped to 
$U(1)$s associated to the free generators of the Mordell-Weil group of the Higgsed 
geometry $\hat{Y}_{\rm H}$.  The postulated group structure on the blow-up divisors of the non-Abelian theory is 
then nothing but the translational symmetry in the Mordell-Weil group of the Higgsed theory 
that has been shown to be a geometric symmetry in \autoref{sec:ec_group_structures}. 
In \autoref{sec:defGroupStructureExeptional} we will assume that such a Higgs transition exists 
and exploit it to define the group structure on $\hat{Y}$. 
We show this correspondence  explicitly in the simplest case of  an adjoint Higgsing of $SU(2)$ 
to $U(1)$ in \autoref{sec:unHiggsing} 
and use induction on the number of $U(1)$s to generalize to higher rank groups. 
Thus, we see that the non-Abelian group structure is required by consistency under motion in the
moduli space of F-theory.

Second, we show in \autoref{sec:defGroupStructureExeptional} that in the  effective field theory the postulated group action manifests itself 
simply as non-Abelian large gauge transformations and is therefore trivially a symmetry
in an anomaly-free theory. Thus,
we claim that the non-Abelian group action should have a direct geometric interpretation  on $\hat{Y}$
and does generally exist for any non-Abelian setup, even for those lacking Higgsings to Abelian 
theories.

We note that application of the results from \autoref{sec:anom_lgt}
implies that the geometric symmetries postulated here imply the cancelation of all pure 
and mixed non-Abelian gauge anomalies in the effective action of F-theory compactifications on elliptically
fibered Calabi-Yau three- and fourfolds.

\subsection{A Group Action for Exceptional Divisors}
\label{sec:defGroupStructureExeptional}

As outlined at the beginning of this section, we define in the following a group structure  on the
set of resolution divisors of codimension-one singularities of an elliptic fibration $\hat{Y}$. We first motivate
the group structure geometrically by the connection between Abelian and non-Abelian
gauge groups via (un)Higgsing. Then we show that the postulated group law is identified with
non-Abelian large gauge transformations, which are automatically a symmetry of the effective
theory. Furthermore, we show that the postulated group law 
leaves key classical intersections on $\hat{Y}$ invariant. In particular the intersections of
the transformed exceptional divisors yield the same Cartan matrix as before and the transformed 
rational sections obey again the defining intersection properties of rational sections discussed 
in \autoref{sec:top_torus}.

We will start with a purely Abelian theory specified by an elliptic fibration $\hat{Y}_{\rm H}$ with a 
Mordell-Weil group generated by elements $s'_{m}$, $m=1,\ldots, n_{U(1)}$. 
Although the following arguments hold in general, we will assume that the Mordell-Weil group has 
no torsion elements. 
We consider an unHiggsing
to a geometry $\hat{Y}$ where a subset of the rational sections are 
turned into exceptional divisors $D_I$ corresponding to a non-Abelian gauge group $G$. 
As discussed systematically in \cite{Morrison:2014era,Cvetic:2015ioa}, such an unHiggsing is a 
tuning in the complex structure of $\hat{Y}_{\rm H}$ such that certain
rational sections coincide globally in the tuned geometry. 
Thus $\hat{Y}$ will have a lower-rank Mordell-Weil group with generators denoted by $s_n$, 
$n=1,\ldots,\tilde{n}_{U(1)}$ for $\tilde{n}_{U(1)}< n_{U(1)}$. 

We focus here on the 
simplest situation possible corresponding to a rank preserving unHiggsing,  \textit{i.e.}~a situation 
with $\text{rk}(G)=n_{U(1)}-\tilde{n}_{U(1)}$. Then
the  non-Abelian gauge theory associated to $\hat{Y}$ is Higgsed back to the 
original Abelian gauge theory specified by $\hat{Y}_{\rm H}$ via matter in the adjoint representation. 
Thus the divisor groups of $\hat{Y}_{\rm H}$ and $\hat{Y}$ are of the same dimension and 
generated by the following elements, respectively:
\beq
	H_p (\hat{Y}_{\rm H})=\langle S'_0,S'_n,S'_I,D'_\alpha\rangle\,,\qquad
	H_p(\hat{Y})=\langle S_0,S_n,D_I,D_\alpha\rangle\,,
\eeq
where $p=4$ for Calabi-Yau threefolds and $p=6$ for Calabi-Yau fourfolds.
Here $S_0'$, $S'_n$ and $S'_I$ are divisor classes associated to the rational sections on 
$\hat{Y}_{\rm H}$, $S_0$ and $S_n$ are divisor classes  of the sections on $\hat{Y}$. 
$D_\alpha'$ and $D_\alpha$ are divisors that ascent from divisors in $B_n$ and 
define in fact the same classes in $\hat Y_{\rm H}$ and $\hat Y$. The 
index $I=1,\ldots,n_{U(1)}-\tilde{n}_{U(1)} $ is the same for both geometries and labels the 
sections on $\hat{Y}_{\rm H}$ that are mapped to exceptional divisors $D_I$ associated to the group
$G$ on $\hat{Y}$.

We propose that  the unHiggsing $\hat{Y}_{\rm H}\rightarrow \hat{Y}$ induces a map
$\varphi$ from the divisor group of $\hat{Y}_{\rm H}$  to that of  $\hat{Y}$,
\beq
	\varphi:\quad H_p(\hat{Y}_{\rm H})\,\,\rightarrow\,\,H_p(\hat{Y})\,,
\eeq
with certain properties to be defined next. We we will argue explicitly in \autoref{sec:unHiggsing} that the (un)Higgsing processes 
described in \cite{Morrison:2012ei,Morrison:2014era,Cvetic:2015ioa} implies the existence of a 
map $\varphi$ as described now.

We require $\varphi$ to be a bijective ring homomorphism from the full intersection ring
on $\hat{Y}_{\rm H}$ to that on $\hat{Y}$, \textit{i.e.}~to commute with the intersection pairing of 
divisors and to be linear. The image of $\varphi$ on the generators of $H_p(\hat{Y}_{\rm H})$  with
$p=4$ (for threefolds) or $p=6$ (for fourfolds)  is given by
\beq \label{eq:unHiggsing}
\varphi( S_0')= S_0 \,,\quad
\varphi(S_{n}')= S_{n} \, , \quad \varphi\big(D(s'_{I})\big)=D_{I}\,, \quad \varphi(D'_{\alpha})=D_{\alpha}\,.
\eeq
We emphasize that $\varphi$ maps the  Shioda map $D(s'_{I})$ of the rational section $s_I'$ 
to a Cartan divisor $D_I$ of the unHiggsed gauge group $G$ on 
$\hat{Y}$. Note however that 
\eqref{eq:unHiggsing} implies that $\varphi$ does \textit{not} necessarily map the Shioda map $D(s'_n)$ of 
a section $s'_n$ on $\hat{Y}_{\rm H}$ to the 
Shioda map $D(s_n)$ of $s_n$ on $\hat{Y}$. This is clear as the formula for  $D(s_n)$
according to \eqref{e:old_shioda} involves the Cartan divisors on $\hat{Y}$ that are 
absent on $\hat{Y}_{\rm H}$ and consequently do not appear in the formula for $D(s_n')$.

We are now in the position to investigate the image of a translation in the Mordell-Weil group
of $\hat{Y}_{\rm H}$ under the map $\varphi$ to the unHiggsed geometry $\hat{Y}$.
We are particularly interested in shifts by rational sections $s'_I$, whose associated 
Shioda maps $D(s_I')$ map to Cartan divisors $D_I$ in $\hat{Y}$.
To this end we recall the action of a Mordell-Weil translation on $\hat{Y}_{\rm H}$ on its divisor group.
First, we express the Mordell-Weil 
translations on $\hat{Y}_{\rm H}$  conveniently in terms of the $D(s'_I)$. 
Shifting the Mordell-Weil lattice on $\hat{Y}_{\rm H}$
by a vector $\oplus\, \fn^{ I} s'_{I}$ we rewrite
\eqref{e:MW_law} for all sections $s'_{\cM} :=\{ s'_0, s'_m\}$ as
\begin{align} \label{eq:GroupLawDm}
 Div(s'_{\cM}\oplus \fn^{ I} s'_{I})  &= S'_{\cM} + \sum_{I}\fn^{I}D(s'_{I}) -
 \frac{1}{2}\sum_{I, J}\fn^{I} \fn^{J} \pi\big(D(s'_{I}) \cdot D(s'_{J})\big) \nn \\
 &\,\quad - \sum_{ I}\fn^{ I} \pi\big(S'_\cM \cdot D(s'_{I})\big) \, .
\end{align}
We also recall the general Mordell-Weil group action on a Shioda map $D(s'_m)\equiv D'_m$ of a section 
$s'_m$ as given in 
\eqref{free_shift_Ab}.
We now perform the unHiggsing by applying the ring homomorphism $\varphi$, employing
\eqref{eq:unHiggsing}, to the formulae in
\eqref{eq:GroupLawDm} and \eqref{free_shift_Ab}. We find the 
following transformation of divisor classes of sections and Cartan divisors on 
$\hat{Y}$ by lifting the Mordell-Weil translations on $\hat{Y}_{\rm H}$:
\begin{subequations} \label{eq:DsZeroNodeShift}
\begin{align} 
 \tilde S_{0} &= S_{0} + \sum_{I}\fn^{I}D_I - \frac{1}{2}\sum_{I,J}\fn^I \fn^J \pi(D_I \cdot D_J) \, , \\
 \tilde S_{n} &= S_{n} + \sum_{I}\fn^{I}D_I - \frac{1}{2}\sum_{I,J}\fn^I \fn^J \pi(D_I \cdot D_J)
 - \sum_{I}\fn^{I} \pi( {S}_n \cdot D_I) \, , \\
  \tilde{D}_I &= D_I - \sum_{J}\fn^{J} \pi (D_J \cdot D_I) \, .
\end{align}
\end{subequations}
We note that the map $\varphi$ simply amounts to 
$\sum_{I} \fn^{I}D(s'_{I}) \mapsto \sum_{I}\fn^{I}D_I$, as follows from \eqref{eq:unHiggsing}. 
Then, we have additionally used $\pi(S_0\cdot D_I)=0$  in the first equation since the zero 
section does not pass through the Cartan divisors on $\hat{Y}$ as expected and $\pi(D_n\cdot D_I)=0$ by definition of the Shioda map on the unHiggsed geometry 
$\hat{Y}$.

From a field theory point of view it is clear that the shifted classes \eqref{eq:DsZeroNodeShift} 
correspond to non-Abelian large gauge transformations along the Cartan subalgebra.
Indeed one finds that under \eqref{eq:DsZeroNodeShift} the natural F-theory divisor basis on $\hat{Y}$
transform as 
\begin{align} \label{D-transformZeroNode}
 \begin{pmatrix}
  \tilde D_{0}\\[8pt]
  \tilde D_{I}\\[8pt]  
  \tilde D_{n}\\[8pt]
  \tilde D_{\alpha}
 \end{pmatrix} = 
\begin{pmatrix*}[c]
 1 &  \fn^J & 0  & \frac{\fn^K \fn^L}{2} \cC_{KL} b^\beta \\[8pt]
0 & \delta^{J}_{I} & 0 &   \fn^K \cC_{IK} b^\beta \\[8pt]
  0 & 0 & \delta_n^k & 0 \\[8pt]
 0 & 
 0
  & 0& \delta^\beta_\alpha
\end{pmatrix*}\cdot
\begin{pmatrix}
  D_{0} \\[8pt]
  D_{J} \\[8pt]
  D_{k} \\[8pt]
  D_{\beta} \, 
 \end{pmatrix}\,,
\end{align}
where we have used \eqref{evaluate_DD} and again recall that $\pi(D_n\cdot D_I)=0$.
Indeed, \eqref{D-transformZeroNode} is precisely  the formula for a non-Abelian large 
gauge transformation given in \eqref{eq:LGTgeneral_1} along the non-Abelian Cartan gauge 
fields $A^I$ with windings $\fn^I$, {i.e.}~for $\fn^m=0$, so that the combination 
$C_3=A^\Lambda\wedge [D_\Lambda]$ remains invariant. 

In the following we will impose these shifts in F-theory compactifications with non-Abelian gauge 
symmetry independently of an existing
adjoint Higgsing to the maximal torus of $G$. 
We conclude this by showing that the transformed divisor
classes \eqref{eq:DsZeroNodeShift} on $\hat{Y}$
obey the key properties \eqref{e:shioda_orth} and \eqref{blowup-cond} of new Cartan divisors 
and new rational sections, respectively, so that the gauge algebra and the rank of the 
Mordell-Weil group are invariant. First of all let us note that the $\tilde S_{\cM} = \{\tilde S_{0}, \tilde S_{n}\}$ define good divisor classes for sections. Indeed we find that
\begin{align}
 \tilde S_{0} \cdot \tilde D_I &= 0 \, \\
  \tilde S_{\cM} \cdot \cE &= 1 \,  \\
  \pi (\tilde S_{\cM} \cdot \tilde S_{\cM} ) &= K \, .
\end{align}
Second we check that also the other classical intersection numbers such as \eqref{evaluate_DD} 
for the divisors $\tilde D_I$ are not changed.  This indicates that there might exist a new geometric interpretation
of the transformed divisors $(\tilde S_0 ,\tilde S_n,\tilde D_I)$ 
as sections and exceptional divisors in an associated geometry.
It also hints to the existence of a geometric interpretation for the Mordell-Weil translations
lifted from $\hat{Y}_{\rm H}$ to $\hat{Y}$.

To close this subsection, let us note that we can push the analogy to the elliptic fibration 
with rational sections even further by defining a so-called \textit{zero-node} $\Sigma_0$.
We introduce $\Sigma_0$  as
\begin{align}
 \Sigma_0 := \sum_{I}\fn^{I}D_I\ .
\end{align}
Using this definition the transformations \eqref{eq:DsZeroNodeShift} can be rewritten in a simpler 
form eliminating all explicit $\fn^I$-dependence. The freedom to make a shift by a large 
non-Abelian gauge transformation then translates to `picking a zero-node' in the lattice
which is spanned by the blow-up divisors. 
This is analogous to determining the origin in the Mordell-Weil lattice.
For the latter case we have argued in \autoref{sec:ec_group_structures} that this should constitute 
an actual symmetry of the M-theory to F-theory limit and therefore implies cancelation 
of Abelian anomalies. We will also discuss a closely related topic in \autoref{ch:zero_sec}.
For the non-Abelian large gauge transformations and 
the group action introduced here such a geometric symmetry principle has yet 
to be established but would guarantee the cancelation of all non-Abelian anomalies.

 \subsection{Arithmetic Group Structures from Higgs Transitions}
 \label{sec:unHiggsing}

Given a gauge theory with non-Abelian gauge group $G$ and matter in the adjoint 
representation, we can Higgs to $U(1)^{r}$ with $r=\text{rk}(G)$ by switching on VEVs
along the Cartan generators in the adjoint. The inverse process is called unHiggsing a $U(1)$ 
symmetry. Various examples of unHiggsing $U(1)$ symmetries in F-theory have been considered,
see \textit{e.g.}~the most recent works 
\cite{Morrison:2012ei,Morrison:2014era,Klevers:2014bqa,Cvetic:2015ioa} on 
unHiggsings of up to two $U(1)$s. We will 
employ the unHiggsing of $U(1)$ symmetries to non-Abelian groups in the following in order to 
provide further geometrical evidence for the existence of the group structure on the
exceptional divisors postulated in \autoref{sec:defGroupStructureExeptional}. For simplicity 
we focus here on the simplest case of the unHiggsing of one $U(1)$ to $SU(2)$ as studied in
\cite{Morrison:2012ei,Morrison:2014era}. By induction over the number of $U(1)$s, as 
suggested in \cite{Cvetic:2015ioa}, the obtained results are expected to generalize to 
higher-rank non-Abelian gauge groups. Note that basic knowledge of toric geometry is required to understand the following discussion.

It has been shown by Morrison and Park in \cite{Morrison:2012ei} that the normal form of a 
general elliptic fibration with a Mordell-Weil group of rank one, \textit{i.e.}~a single $U(1)$, is a 
Calabi-Yau hypersurface $\hat{Y}_H$ with elliptic fiber given as  the quartic hypersurface in 
the blow-up of $\mathbb{P}^2(1,1,2)$, denoted $\text{Bl}_1\mathbb{P}^2(1,1,2)$. 
This space has a toric description.
Denoting the projective coordinates  on $\text{Bl}_1\mathbb{P}^2(1,1,2)$ by $[u:v:w:e]$, 
where $e=0$ is the exceptional divisor of the blow-up (with map $[u:v:w:e]\mapsto [ue:v:we]$ to 
$\mathbb{P}^2(1,1,2)$), the hypersurface equation which defines the 
elliptic fibration can be brought into the form\footnote{Note that the coefficient of $ew^2$ is set to one in order to avoid the
$\mathbb{Z}_2$-singularity at $u=v=0$ which  would give rise to a codimension-one singularity of type $I_2$, 
\textit{i.e.}~an SU(2) gauge group in F-theory, see \cite{Anderson:2014yva,Klevers:2014bqa}
for an analysis of the geometry with this additional singularity.}
\beq \label{eq:resolvedQuartic}
	e w^2+bv^2w=u(c_0u^3e^3+c_1u^2e^2v+c_2uev^2+c_3v^3)\,.
\eeq
The coefficients $c_i$, $i=0,1,2,3$, are sections in specific line bundles that are determined by 
the the requirement that \eqref{eq:resolvedQuartic} defines a well-defined section of a line 
bundle on the base $B_n$ and obeys the Calabi-Yau condition:
\beq
\text{
\begin{tabular}{c|c}
\text{Section} & \text{Class}\\
\hline
	$[c_0]$&$-4K-2[b]$\rule{0pt}{13pt} \\
	$[c_1]$&$-3K-[b]$\rule{0pt}{12pt} \\
	$[c_2]$&$-2K$\rule{0pt}{12pt} \\
	$[c_3]$&$-K+[b]$\rule{0pt}{12pt} \\
	$[b]$&$[b]$\rule{0pt}{12pt} \\
\end{tabular}
}
\eeq
Here, we denote the divisor class of a section by $[\cdot]$ and $-K$ is the anti-canonical 
divisor of $B_n$. Note that the class $[b]$ of the divisor  $b=0$ is a free parameter of the 
Calabi-Yau manifold $\hat{Y}$. 

The two rational sections of the elliptic fibration are given by
\beq \label{eq:ratSections}
	s'_0:\quad [0:1:1:-b]\,,\qquad s'_1:\quad  [b:1:c_3:0]\,,
\eeq
where we picked $s'_0$ as the zero-section.\footnote{This convention deviates from the 
one chosen in \cite{Morrison:2012ei} but is physically equivalent as we show in \autoref{ch:zero_sec}.}
The Shioda map \eqref{e:old_shioda} of the section $s'_1$ reads
\beq
\label{eq:ShiodaSigma1}
D(s'_1)=S'_1-S'_0+K-[b]=S'_1-S'_0-[c_3]\,,
\eeq
where we denoted the  homology class of the two sections by $S'_0=Div(s'_0)$ and $S'_1=Div(s'_1)$.

The divisor $D(s'_1)$ supports the $U(1)$ of the F-theory compactification 
on \eqref{eq:resolvedQuartic} as can be seen from the expansion \eqref{e:expansion_threeform}
of the M-theory three-form. Shifting the origin in the Mordell-Weil lattice, as discussed in 
\autoref{sec:FreeMWShifts}, yields the new divisor classes \eqref{free_shift_Ab} that were 
shown to correspond to large Abelian gauge transformations.

The unHiggsing of the $U(1)$ to an $SU(2)$ is performed by tuning 
$b\mapsto 0$ in the elliptic fibration \eqref{eq:resolvedQuartic}, as discussed in \cite{Morrison:2012ei,Morrison:2014era}, so that the rational sections in \eqref{eq:ratSections} coincide 
globally except for the locus $c_3=0$. As the fiber is toric, we blow up 
at $u=e=0$, which amounts to replacing
\beq \label{eq:buQuartic}
	u\mapsto ue_1\,,\qquad e\mapsto ee_1\,,
\eeq
where $e_1=0$ is a new divisor. 
The hypersurface equation for the unHiggsed geometry $\hat{Y}$
after blow-up reads
\beq \label{eq:resolvedQuartic1}
	e w^2=u(c_0u^3e^3e_1^6+c_1u^2e^2e_1^4v+c_2uee_1^2v^2+c_3v^3)\,.
\eeq

The single remaining section on $\hat{Y}$, now denoted by 
$s_0$, is described by $e_1=0$ after blow-up, with 
coordinates $[u:v:w:e:e_1]$ reading
\beq
	s_0:\quad [1:1:1:c_3:0]\,
\eeq
showing that $s_0$ is holomorphic.
We note that the blown-up hypersurface has a Kodaira singularity of type $I_2$ at $c_3=0$ 
corresponding to an $SU(2)$ gauge group in F-theory.  Indeed, by setting $c_3=0$ in \eqref{eq:resolvedQuartic1} 
we obtain 
\beq
	e \big(w^2-u(c_0u^2e^2e_1^6+c_1uee_1^4v+c_2 e_1^2v^2)\big )=0\,,
\eeq
which describes two $\mathbb{P}^1$s intersecting at two points. Thus we identify $S^{SU(2)}=\{c_3=0\}$ as the divisor supporting the $SU(2)$ gauge group.  
As the zero-section $s_0$ passes through the $\mathbb{P}^1$ given by $e=0$, we 
determine the class of the Cartan divisor $D_1$ as 
\beq
	D_1=[c_3]-[e]\,.
\eeq
Furthermore, we see that the divisor $u=0$ does not intersect the hypersurface 
\eqref{eq:resolvedQuartic1}, \textit{i.e.}~$\hat{Y}\cap \{u=0\}=0$, due to the Stanley-Reisner ideal of the blown-up ambient space.
Using these observations, we infer that the pull-back of the Shioda map \eqref{eq:ShiodaSigma1}
of the original rational section $s'_1$ to the  unHiggsed geometry $\hat{Y}$ 
reads
\beq
	D(s_1')\mapsto [e]-[c_3]=-D_1\,.
\eeq
Clearly, we have $S_0'\mapsto S_0$ while vertical divisor $D_\alpha$ map trivially.
These are, up to the irrelevant sign in the map of $D(s_1')$, precisely the properties of the map $\varphi$ defined in
\eqref{eq:unHiggsing}.

In summary, we see that the Shioda map of the rational sections is mapped, up to sign, to the 
Cartan divisor of the unHiggsed $SU(2)$ gauge group on $\hat{Y}_{\text{uH}}$. Consequently, the Mordell-Weil shift of a 
rank one Mordell-Weil group, as introduced in \autoref{sec:FreeMWShifts}, is mapped under 
the transition corresponding to the unHiggsing to $SU(2)$ to a similar shift of
divisors, where the  Shioda map is replaced by the Cartan divisor of the 
$SU(2)$ (the sign can be absorbed by the integer $\fn$ in \eqref{eq:MWshift}). In addition, a 
similar replacement should apply for unHiggsing a higher rank Mordell-Weil group by induction on 
its rank as discussed in \cite{Cvetic:2015ioa}. This is expected to
establish the existence of the group law postulated in \autoref{sec:defGroupStructureExeptional}
on the Cartan divisors of any non-Abelian gauge group 
in F-theory that can be Higgsed in an adjoint Higgsing to a purely Abelian gauge group. We 
propose that this group law exists even for those non-Abelian groups that can not be Higgsed,
such as the non-Higgsable clusters in \cite{Morrison:2012np}.

\chapter{The Freedom of Picking the Zero-Section in F-Theory}
\chaptermark{Picking the Zero-Section in F-Theory}
\label{ch:zero_sec}

Finally, let us conclude this part by presenting the work and results which actually had served as the inspiration for what we have discussed
up to now. The original aim was to clarify the implication of the fact that one is free to choose the zero-section in F-theory in any way.
More precisely, given a full set of independent sections of an elliptic fibration one has to single out one as the zero-section.
F-theory does not seem to impose constraints on which section has to be chosen, they are rather all on equal footing.
On the other hand
it has been known that the intersection numbers which are matched to data in the circle-reduced supergravity
are not invariant when comparing different choices. In order to understand
how F-theory deals with this fact we investigate how the basis of divisors defined in \autoref{sec:top_torus} transforms.

We start with a setting that has $s_0$ as the chosen zero-section and generating sections $s_m$ which
correspond to Abelian gauge symmetries.
Let us now consider the same geometry but with a different choice
for the zero-section $\hat s_0$ and the generating sections $\hat s_m$.
We denote the quantities in the new F-theory setup by a 'hat'.
In order to compare both choices
we first have to split the index labeling the generating sections, \textit{i.e.}~the higher-dimensional $U(1)$s, in the old setting 
\begin{align}
 m \rightarrow m^\circ, m^c \, ,
\end{align}
where $m^\circ$ refers to the single section $s_{m^\circ}$ which we pick as the new zero-section $\hat s_0$, and
$s_{m^c}$ denotes the remaining generating sections, the complement to $s_{m^\circ}$ in $\lbrace s_{m}\rbrace$. We set
\begin{align}
 \hat s_0 &= s_{m^\circ} \, , &
 \hat s_{m^\circ} &= s_0 \, ,& 
 \hat s_{m^c} &= s_{m^c}  \, .
\end{align}

\begin{figure}
\begin{center}
\begin{tikzpicture}[scale = 0.84, axis/.style={very thick, ->, >=stealth'}]
 \foreach \x in {0,...,2} 
  { \foreach \y in {0,...,2} 
      {\fill (\x*2cm,\y*2cm) circle (0.1cm); }}
 \draw[axis]  (0cm,0cm) -- (1.9cm,0cm) node(xline)[left] {};
 \draw[axis]  (0cm,0cm) -- (0cm,1.9cm) node(yline)[below] {};
 \node at (-0.3cm,-0.3cm) {$s_0$};
 \node at (2.2cm,-0.4cm) {$s_{m^\circ}$};
 \node at (-0.6cm,2cm) {$s_{m^c}$};
  \foreach \x in {6,...,8} 
  { \foreach \y in {0,...,2} 
      {\fill (\x*2cm,\y*2cm) circle (0.1cm); }}
  \draw[axis]  (14cm,0cm) -- (12.1cm,0cm) node(xline)[left] {};
  \draw[axis]  (14cm,0cm) -- (12.07cm,1.93cm) node(yline)[below] {};
   \node at (11.7cm,-0.3cm) {$\hat s_{m^\circ}$};
   \node at (14.2cm,-0.4cm) {$\hat s_{0}$};
   \node at (11.4cm,2cm) {$\hat s_{m^c}$};
   \draw[->, thick, >=latex] (6cm,2cm) -- (9cm,2cm) ;
    \node at (7.5cm,3.5cm) {
\begin{tabular}{l}
$\hat s_0 \,\,\,\, =  s_{m^\circ}$ \\
$\hat s_{m^\circ} =  s_0 $ \\
$\hat s_{m^c} = s_{m^c} $ 
                                                                       \end{tabular}
} ;
   \foreach \x in {3,...,5} 
  { \foreach \y in {-4,...,-2} 
      {\fill (\x*2cm,\y*2cm) circle (0.1cm); }}
  \draw[axis]  (8cm,-8cm) -- (9.9cm,-8cm) node(xline)[left] {};
  \draw[axis]  (8cm,-8cm) -- (8cm,-6.1cm) node(yline)[below] {};
  \node at (7.7cm,-8.3cm) {$\tilde s_0$};
  \node at (10.2cm,-8.4cm) {$\tilde s_{m^\circ}$};
  \node at (7.4cm,-6cm) {$\tilde s_{m^c}$};
  \draw[->, thick, >=latex] (1cm,-2cm) -- (4cm,-6cm) ;
  \node at (1cm,-5.1cm) {
\begin{tabular}{l}
$\tilde s_0 \,\,\,\, = s_0 \oplus s_{m^\circ}$ \\
$\tilde s_{m^\circ} = s_{m^\circ} \oplus s_{m^\circ}$ \\
$\tilde s_{m^c} = s_{m^c} \oplus s_{m^\circ}$ 
                                                                       \end{tabular}
} ;
  \draw[->, thick, >=latex] (12cm,-6cm) -- (15cm,-2cm) ;
  \node at (15.2cm,-5.1cm) {
\begin{tabular}{l}
$\hat s_0 \,\,\,\, =\tilde s_0 $ \\
$\hat s_{m^\circ} = \tilde s_0 \ \widetilde\ominus \ \tilde s_{m^\circ}$ \\
$\hat s_{m^c} =  \tilde s_{m^c} \ \widetilde\ominus \ \tilde s_{m^\circ}$ 
                                                                       \end{tabular}
} ;
\end{tikzpicture}
\end{center}
\caption{In the first line we depict the change of the zero-section choice from $s_0$ to $\hat s_0 := s_{m^\circ}$. In the
original setting the generators of the Mordell-Weil lattice are defined by $s_{m^\circ}$ and $s_{m^c}$ while in the new setting
they are given by $\hat s_{m^\circ}$ and $\hat s_{m^c}$. The crucial observation is that this map factorizes in a convenient way.
In particular we first shift the whole basis of the lattice by $s_{m^\circ}$. These shifts were investigated in \autoref{sec:FreeMWShifts}
and were shown to correspond to integer Abelian large gauge transformations in the theory on the circle. The second
map leaves the zero-section invariant and only redefines the generators.
Since the Shioda map is a homomorphism, this corresponds just to a simple redefinition of four-  or six-dimensional $U(1)$ gauge fields.
Finally note that $\widetilde \ominus$ indicates that the
Mordell-Weil group law is defined with respect to the zero-section $\tilde s_0$ while $\oplus$ is defined with respect to the
original zero-section
$s_0$.}
\label{fig:zero_sec}
\end{figure}
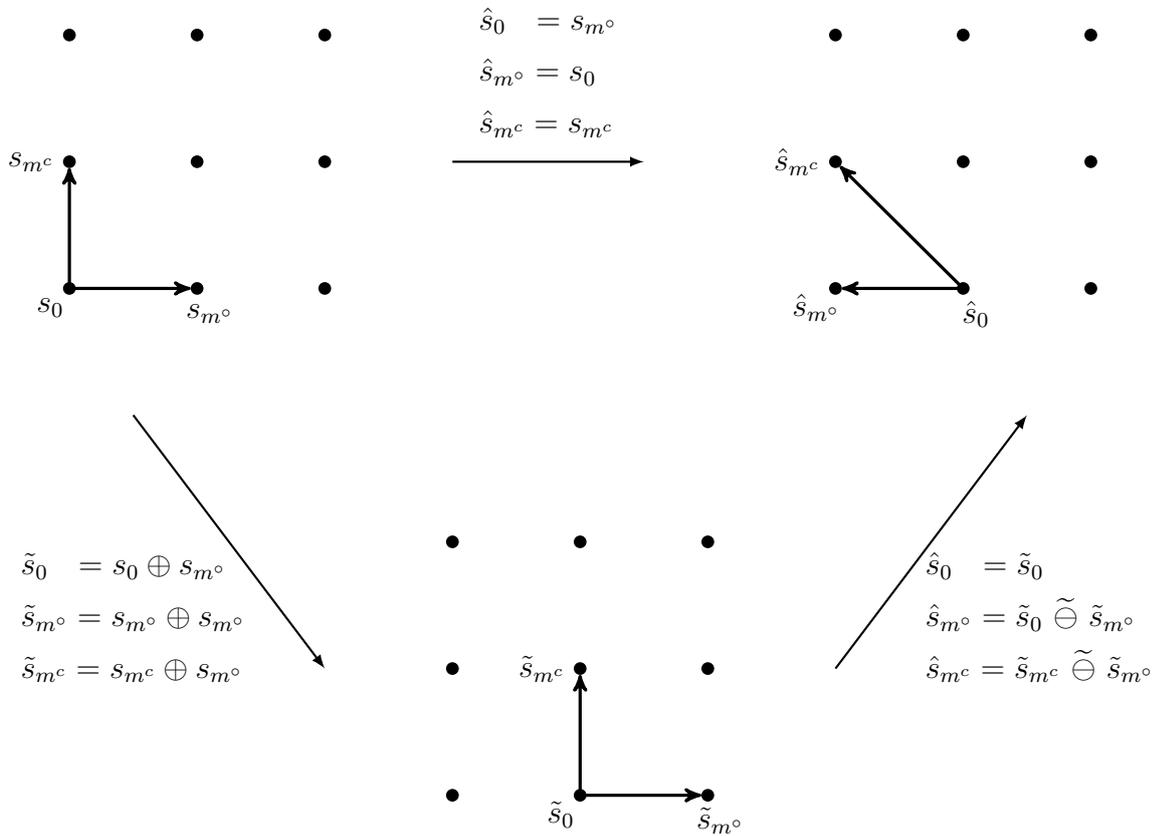

Before we infer the transformation of the basis of divisors which we defined in \autoref{sec:top_torus},
let us introduce the following notation
\begin{subequations}
 \begin{align}
 \tensor{A}{_I^J} &:= \delta_I^J - \pi_{{m^\circ} I} (\delta_I^J + \tensor{a}{_I^J}) \, , \\
 \tensor{a}{_I^J} &:= a^J \quad \forall I \, , \\
 \fn^J &:= - \pi_{{m^\circ} K} \,\mathcal{C}^{-1\,KJ}
\end{align}
\end{subequations}
with $a^J$ the Coxeter labels defined in \eqref{def:cox}.
Using \eqref{e:old_base}, \eqref{e:old_shioda} we then obtain
\begin{align}
 &\begin{pmatrix}
  \hat D_{0}\\[8pt]
  \hat D_{I}\\[8pt] 
  \hat D_{m^\circ}\\[8pt]
  \hat D_{m^c}\\[8pt]
  \hat D_{\alpha}
 \end{pmatrix} = 
\begin{pmatrix*}[c]
 1 & \fn^J & 1  & 0 & \frac{1}{2} b^\beta_{{m^\circ}{m^\circ}}  +\frac{\fn^K \fn^L}{2} \cC_{KL} b^\beta \\[8pt]
0 & \tensor{A}{_I^J} & 0 & 0 &   \pi_{{m^\circ} I} b^\beta \\[8pt]
0 & 0 & -1 & 0 & - b^\beta_{{m^\circ}{m^\circ}} \\[8pt]
  0 & 0 & -1 & \delta_{m^c}^{n^c} & - b^\beta_{{m^\circ}{m^\circ}} + b^\beta_{{m^\circ}{m^c}} \\[8pt]
 0 & 
 0 & 0
  & 0& \delta^\beta_\alpha
\end{pmatrix*}\cdot
\begin{pmatrix}
  D_{0} \\[8pt]
  D_{J} \\[8pt]
  D_{m^\circ} \\[8pt]
  D_{n^c} \\[8pt]
  D_{\beta} \, 
 \end{pmatrix} \\ \nn \\
  &=\underbrace{\begin{pmatrix*}[c]
 1 & 0 & 0  & 0 & 0 \\[8pt]
0 & \tensor{A}{_I^M} & 0 & 0 &   0 \\[8pt]
0 & 0 & -1 & 0 & 0 \\[8pt]
  0 & 0 & -1 & \delta_{m^c}^{p^c} & 0 \\[8pt]
 0 & 
 0 & 0
  & 0& \delta^\gamma_\alpha
\end{pmatrix*}}_{\substack{\textrm{redefintion of lattice generators,} \\
\textrm{redefintion of blow-up divisors } \equiv\\ \textrm{redefinition of 4d/6d gauge fields}}}  \cdot 
\underbrace{\begin{pmatrix*}[c]
 1 & \fn^J & 1  & 0 & \frac{1}{2} b^\beta_{{m^\circ}{m^\circ}}  +\frac{\fn^K \fn^L}{2} \cC_{KL} b^\beta \\[8pt]
0 & \delta^{J}_M & 0 & 0 &   \fn^K \cC_{MK} b^\beta \\[8pt]
0 & 0 & 1 & 0 &  b^\beta_{{m^\circ}{m^\circ}} \\[8pt]
  0 & 0 & 0 & \delta_{p^c}^{n^c} &  b^\beta_{{m^\circ}{p^c}} \\[8pt]
 0 & 
 0 & 0
  & 0& \delta^\beta_\gamma
\end{pmatrix*}}_{\substack{\textrm{basis shift in the Mordell-Weil lattice } \equiv\\ \textrm{large gauge transformation}}} \cdot
\begin{pmatrix}
  D_{0} \\[8pt]
  D_{J} \\[8pt]
  D_{m^\circ} \\[8pt]
  D_{n^c} \\[8pt]
  D_{\beta} 
 \end{pmatrix} \nn \, .
\end{align}
Using the results of \autoref{sec:FreeMWShifts} we see immediately that this map factorizes into the large gauge transformation
\eqref{D-transform1},
corresponding to a shift in the Mordell-Weil lattice with $\fn^n = \delta^n_{m^\circ}$, and a
simple redefinition of $U(1)$ divisors. We explain in \autoref{fig:zero_sec} in detail why this form is indeed to be expected
from the perspective of the Mordell-Weil lattice. We stress that also the definition of the blow-up divisors changes
if $\pi_{{m^\circ} I} = 1$ for some index $I$. In this case the new zero-section $\hat s_0$ intersects the blow-up divisor $D_I$ in the old basis.
However, the zero-section should not intersect a blow-up node but rather the affine node which is why we have to perform the following
redefinition for $\pi_{{m^\circ} I} = 1$
\begin{subequations}
\begin{align}
 \hat D_{\rm affine} & = D_I  \, , \\
\hat D_I &= D_{\rm affine} \equiv S - a^J D_J 
\end{align}
\end{subequations}
with $a^J$ the Coxeter labels and $D_{\rm affine}$ the divisor which corresponds to the affine node of the extended Dynkin diagram.
We note that the components of the tuple of blow-up divisors
in the new basis $( \hat D_1 , \dots , \hat D_{\rk \mathfrak g} )$
still has to be permuted in order to get the standard
intersections in terms of the coroot intersection matrix.

Let us conclude by mentioning that using the results of \autoref{sec:anom_lgt} and \autoref{sec:FreeMWShifts} it
is obvious that the freedom of the choice for the zero-section
is equivalent to the cancelation of pure and mixed Abelian gauge anomalies in F-theory compactifications on Calabi-Yau fourfolds and threefolds.

\part{Unpublished Results and Ideas}\label{part:add}

\chapter{More on the Arithmetic Structure of Genus-One Fibrations}
\chaptermark{Arithmetic of Genus-One Fibrations}

The treatment in this chapter mainly refers to the results of \autoref{ch:arith} concerning arithmetic structures on genus-one
fibrations and their connection to F-theory effective physics.

\section{Constraints on the F-Theory Spectrum }
It was pointed out in \autoref{sec:class_lgt} that there is another possibility how one might fulfill \eqref{kk_cond}. In fact,
the spectrum could be such that some of the $\fn^m$ can be chosen to be fractional. We illustrate this in the following
examples:\footnote{One might include additional singlets and a Green-Schwarz mechanism in order to cancel potential anomalies in our examples,
and obtain a consistent effective theory.
For the considered cases this should always be possible.}
\begin{itemize}
 \item \framebox{$SU(2)\times U(1)$ with matter $\bf{2_1}$}
 
 Choose the windings $(\fn^I) = (\fn^m) = (\frac{1}{2})$.
 
 We can now easily verify that \eqref{kk_cond} is fulfilled for all states:
 
 \textbf{adjoint representation} ($w^\pm = \pm 2$, $w^0 = 0$, $q=0$)
 \begin{align}
  \fn^m q_m + \fn^I w^\pm_I &= \pm 1 \quad \in \mathbb Z \nn \\
  \fn^m q_m + \fn^I w^0_I &= 0 \quad \in \mathbb Z \nn 
 \end{align}
\textbf{fundamental matter} ($w^{\pm}= \pm 1$, $q=1$)
 \begin{align}
  \fn^m q_m + \fn^I w^+_I &= 1 \quad \in \mathbb Z \nn \\
  \fn^m q_m + \fn^I w^-_I &= 0 \quad \in \mathbb Z \nn 
 \end{align}
 \item \framebox{$U(1)^2$ with matter $\bf{1_{(1,1)}}$}
 
 Choose the windings $(\fn^m) = (\frac{1}{2}, \frac{1}{2})$.
 
  Then, \eqref{kk_cond} takes the form:
 
 \textbf{charged matter} $\big( \vec q = (1, 1)\big)$
  \begin{align}
  \fn^m q_m = 1 \quad \in \mathbb Z \nn 
 \end{align}
 \item \framebox{$SU(3)\times U(1)$ with matter $\bf{3_1}$}
 
 Choose the windings $(\fn^m) = (-\frac{1}{3})$, $ (\fn^I) = (\frac{1}{3},\frac{2}{3})$.
 
  Investigating \eqref{kk_cond} we obtain:
 
 \textbf{adjoint representation} 
 
 One can check that $\fn^I w_I \in \mathbb Z$ for all weights of the adjoint representation.
 This is clear since $(\fn^I) = (\frac{1}{3},\frac{2}{3})$
 are precisely the fractional numbers which parametrize the \textit{special fractional large gauge transformations}
 for a non-simply connected gauge group $SU(3)/ \mathbb Z_3$ as introduced in \autoref{sec:class_lgt} and
 along with torsion in the Mordell-Weil group in \autoref{sec:MWtorsion}.
 Both gauge groups of course share the
 same adjoint representation since the algebra is identical.
 
 \textbf{fundamental matter} $\big(w^{+}= (1,0)$, $w^- = (0,-1)$, $w^0 = (-1,1)\big)$
 \begin{align}
  \fn^m q_m + \fn^I w^+_I &= 0 \quad \in \mathbb Z \nn \\
  \fn^m q_m + \fn^I w^-_I &= -1 \quad \in \mathbb Z \nn \\
  \fn^m q_m + \fn^I w^0_I &= 0 \quad \in \mathbb Z \nn 
 \end{align}
\end{itemize}
As already mentioned in \autoref{sec:class_lgt},
we conjecture that the F-theory spectrum is always such that fractional values for $\fn^m$ can never lead to
\begin{align}
 \fn^m q_m + \fn^I w_I \in \mathbb Z \, .
\end{align}
The indication that this might be true is twofold:
\begin{itemize}
 \item By checking the generic spectra of \cite{Klevers:2014bqa}
 we have verified that in all these cases fractional values for $\fn^m$ are not possible.
 \item From \eqref{e:MW_shift_divisors_1} and its connection to \eqref{eq:MWshift}
 it is obvious that a fractional value of $\fn^m$ would correspond in the Mordell-Weil group to the addition of
 a fraction of a generating section. However, the Mordell-Weil generators are minimal, and it seems hard to make sense out of the addition of
 \textit{fractional generators}.
\end{itemize}

Indeed, there are many known compactifications in F-theory which share the same gauge groups with our three examples.
However, in the F-theory setting there seems to be always additional matter which forbids the use of fractional $\fn^m$. These enlarged
settings
generically look like
\begin{itemize}
 \item \framebox{$SU(2)\times U(1)$ with matter $\bf{2_1}, \bf{2_0}, \bf{1_1}$}
 \item \framebox{$U(1)^2$ with matter $\bf{1_{(1,1)}}, \bf{1_{(1,0)}}, \bf{1_{(0,1)}}$}
 \item \framebox{$SU(3)\times U(1)$ with matter $\bf{3_1}, \bf{3_0}, \bf{1_1}$}
\end{itemize}

We have also mentioned in \autoref{sec:class_lgt} that there might exist precise physical reasons for why the spectrum in F-theory
seems to be always such that fractional $\fn$ are not allowed. Indeed, folk theorems about the consistency of quantum gravity
theories constrain especially the $U(1)$-charges of states in the theory. They
have attracted recent attention in terms of the Weak Gravity Conjecture
\cite{ArkaniHamed:2006dz}
and its extensions. Particularly interesting seems to be the (Sub)Lattice Weak Gravity Conjecture
\cite{Heidenreich:2015nta,Heidenreich:2016jrl,Heidenreich:2016aqi,Montero:2016tif}
which might be connected to the seemingly forbidden fractional shifts in the Mordell-Weil lattice.
This could be subject to future research.

\section{Towards a Graded Mordell-Weil Pseudo-Ring}\label{sec:graded_MW}
 In this section we take a first step towards combining the arithmetic structures for rational sections and multi-sections, \textit{i.e.~}the 
 genuine Mordell-Weil group and the conjectured extended Mordell-Weil group, into a single mathematical structure which we call the \textit{graded
 Mordell-Weil pseudo-ring}.
 
 In \autoref{sec:multi_group} we defined the extended Mordell-Weil group of multi-sections only for fibrations
 which lack a rational section. This was inspired geometrically by Higgs transitions to geometries with rational sections, and field-theoretically
 by investigating large gauge transformations. However, typically fibrations which do have rational sections
 also admit multi-sections though their associated homology classes are not linearly independent from the ones of genuine sections.
 For instance, realizing the fiber as the sextic
 $\mathbb P_{2,3,1}[6]$ the fibration
 generically admits one toric rational section, one toric two-section and one toric three-section. They correspond
 to the three edges of the defining reflexive polytope. Another interesting example is given by embedding the fiber as a generic hypersurface
 into $dP_1$. It was found in \cite{Klevers:2014bqa} that this setting has two independent rational sections, one of which is non-torically
 realized. Thus the Mordell-Weil group has rank one. It is easy to verify in this example that already on the toric level
 there are two two-sections and one
 three-section which correspond to the remaining edges of the polytope.
 It seems plausible to suspect that also in such settings there exists for each $n \leq n_{\rm max}$ an
 extended Mordell-Weil group of $n$-sections as defined in \autoref{sec:multi_group}. We will call the latter
 the $n$-extended Mordell-Weil group, and we assume that $n$ is bounded by some natural number $n_{\rm max}$
 which should depend on the precise geometrical setting one is considering.
 The reason why we think that such structures could be present and also useful is twofold.
 First, writing down
 the group law in terms of cycles for the extended Mordell-Weil group in \autoref{sec:multi_group} can be done
 without reference to the non-existence of rational sections. However, this time the arguments of Higgs transitions and large gauge transformations
 don't seem to apply immediately.
 Second, multi-sections in settings without section were shown to capture the information about massive $U(1)$s. Our intuitions of type IIB 
 setups let us suppose that massive Abelian gauge symmetries should be generically present in F-theory compactifications, whether or not 
 the geometry admits a section. We think that this information is 
 precisely encoded in the multi-sections of the fibration even in the presence of 
 genuine sections.
 
 We are now in the position to define
 the graded Mordell-Weil pseudo-ring $\cM\cW$ as the formal direct sum of all $n$-extended Mordell-Weil groups $\mathrm{MW}^{(n)}$
 \begin{align}
  \cM\cW := \bigoplus_{n=1}^{n_{\rm max}} \mathrm{MW}^{(n)} .
 \end{align}
As in \autoref{sec:multi_group} we assume for simplicity that there are now exceptional divisors present in the geometry.
 
Having written down $\cM\cW$ as a set we now define the pseudo-ring operations. We start with the addition '$\oplus$' of two
elements $\mathpzc s_1 ,\mathpzc s_2 \in \cM\cW$. Expanding
\begin{subequations}\label{e:expandMW}
\begin{align}
 \mathpzc s_1 = \bigoplus_{n=1}^{n_{\rm max}} [s_1^{(n)}]\, , \quad \textrm{with } [s_1^{(n)}] \in  \mathrm{MW}^{(n)}\, ,\\
 \mathpzc s_2 = \bigoplus_{n=1}^{n_{\rm max}} [s_2^{(n)}]\, , \quad \textrm{with } [s_2^{(n)}] \in  \mathrm{MW}^{(n)} \, ,
\end{align}
\end{subequations}
we set
\begin{align}
 \mathpzc s_1 \oplus \mathpzc s_2 := \bigoplus_{n=1}^{n_{\rm max}} \Big( [s_1^{(n)}] \oplus [s_2^{(n)}] \Big ) \, . 
\end{align}
Note that  the expression $[s_1^{(n)}] \oplus [s_1^{(n)}]$ denotes the addition in the $n$-extended Mordell-Weil group
which we introduced in \autoref{multi_group}. This operation on elements of the pseudo-ring
inherits the individual Abelian group structures of the $n$-extended Mordell-Weil groups. The zero-element of the addition $\mathpzc s_0$
is obviously given by the sum of all zero-multi-sections $[s_0^{(n)}]$ of the individual $n$-extended Mordell-Weil groups
\begin{align}
 \mathpzc s_0 := \bigoplus_{n=1}^{n_{\rm max}} [s_0^{(n)}] \, .
\end{align}
We now sketch how the ring multiplication, denoted by '$\otimes$', should look like. Let $\mathpzc s_1 ,\mathpzc s_2 \in \cM\cW$
with expansions \eqref{e:expandMW}, then we set
\begin{align}
 \mathpzc s_1 \otimes \mathpzc s_2 := \bigoplus_{n=2}^{n_{\rm max}} \
 \bigoplus_{p+q = n} \Big( [s_1^{(p)}] \otimes [s_2^{(q)}] \Big ) \, . 
\end{align}
Note that without extending this structure there is no multiplicative identity element, \textit{i.e.~}we only have a pseudo-ring,
and we still have to specify how the operation $[s_1^{(p)}] \otimes [s_2^{(q)}]$ is defined on equivalence classes of
$n$-sections.
We conjecture that
\begin{align}
 \otimes : \mathrm{MW}^{(p)} \times \mathrm{MW}^{(q)} \rightarrow \mathrm{MW}^{(p+q)}
\end{align}
is indeed well-defined and fulfills the following condition
\begin{align}
 \widehat{Div} \Big([s_1^{(p)}] \otimes [s_2^{(q)}]\Big ) =
 \begin{cases}
   \ \ \big [ S_1^{(p)} + S_2^{(q)} \big ] \\
   \ \ [0]
 \end{cases}
\end{align}
with $[S_1^{(p)}]$ and $[S_2^{(q)}]$ the divisors classes (modulo vertical divisors) associated to $[s_1^{(p)}]$ and $[s_2^{(q)}]$.
In particular, '$\otimes$' should be commutative. We have
also assumed that it might happen that this product vanishes at some point in order to get
a finite value for $n_{\rm max}$.
Note that it is not clear if there actually exists
a $(p+q)$-section in the geometry with divisor class $[S_1^{(p)} + S_2^{(q)}]$ or how the multiplication looks like algebraically
(not in terms of homology).
One might also again suspect, similar to the discussion at the end of \autoref{sec:FreeMWShifts},
that one has to consider some kind of branched cover
or scheme of the fibration where $\cM\cW$ is well-defined.
However we stress again that, although the $n$-extended Mordell-Weil group
structure was strongly motivated in settings without rational sections,
there is no evidence that it is also realized geometrically (beyond homology level) in setups that do have rational sections.
These questions along with the physical implications of the full suggested pseudo-ring structure have to be investigated in future research.

\chapter{More on Anomalies}

This chapter provides a first step towards generalizing the results of \autoref{ch:lgts}
and especially \autoref{sec:anom_lgt}.

\section{Gravitational Anomalies}\label{sec:grav_anomalies}

We managed to derive all gauge anomaly cancelation conditions in four and six dimensions from large gauge transformations
in \autoref{sec:anom_lgt}.
In particular we derived these constraints from demanding that after an additional circle compactification
gauge transformations act in a consistent way. In six dimensions we were even able
to derive the mixed gauge-gravitational anomaly equations. Thus we are still missing two kind of
anomaly equations:
\begin{itemize}
 \item The mixed gauge-gravitational anomaly in four dimensions \eqref{4d_anomaly_alt1}
 \begin{align}\label{e:4gravan}
   -\frac{1}{4}  a^\alpha \theta_{m\alpha} = \frac{1}{12} \sum_{R,q}  F_{\sfrac 1 2}(R,q) \sum_{w \in R} \, q_m \, .
 \end{align}
 \item The two pure gravitational anomalies in six dimensions \eqref{e:6d_grav_alter_1}, \eqref{e:6d_grav_alter_2}
 \begin{subequations}
 \begin{align}
\label{e:6gravan1} 4(\mathfrak T + 11 F_{\gr} ) &= \frac{1}{6} \Big (-\sum_{R,q}  F_{\fe}(R,q) \sum_{w \in R}\, 1
- 4 \mathfrak T  + 19 F_{\gr} \Big ) \, , \\ 
\label{e:6gravan2} \frac{1}{4} a^\alpha a^\beta \eta_{\alpha\beta} &= \frac{1}{120} \Big (-\sum_{R,q} 
F_{\fe}(R,q) \sum_{w \in R}\, 1
+ 2 \mathfrak T  - 5 F_{\gr} \Big ) \, .
 \end{align}
 \end{subequations}
\end{itemize}
The reason why our procedure so far misses these conditions is obvious since gravitational anomalies should be probed
with large local Lorentz transformations rather than with large gauge transformations.\footnote{Note that it was possible to derive the mixed
gauge-gravitational anomaly in six dimensions only because of the presence of gravitational Chern-Simons terms in five dimensions.
Such terms do not exist for three-dimensional theories.}

Indeed, the form of certain one-loop Chern-Simons couplings already reveals the structure of these anomalies, and suggests that we
might be able to extract the latter by acting with large local Lorentz transformations:
\begin{itemize}
 \item For the mixed gauge-gravitational anomaly in four dimensions we have
 \begin{align}
  \Theta_{0m} = \frac{1}{12}\sum_{R,q}  F_{\sfrac 1 2}(R,q) \sum_{w \in R} \ \Big(1+ 6\ l_{w,q} \ \big(l_{w,q} +1 \big )\Big) \ q_m\, .
 \end{align}
 \item For the two pure gravitational anomalies in six dimensions we have
 \begin{subequations}
  \begin{align}
 k_{0} &= \frac{1}{6} \bigg (- \sum_{R,q} F_{\fe}(R,q) \sum_{w \in R} \Big( 1 + 6 \  l_{w,q} \ \big(l_{w,q} +1 \big) \Big ) 
-4 \mathfrak T + 19 F_{\gr} \bigg ) \, , \\
   k_{000} &= \frac{1}{120} \bigg (- \sum_{R,q} F_{\fe}(R,q)\sum_{w \in R} \Big( 1-30 \ l_{w,q}^2 \ \big (l_{w,q} +1 \big)^2 \Big )  
 +2 \mathfrak T -5 F_{\gr}\bigg ) \, .
  \end{align}
 \end{subequations}
\end{itemize}
To be more precise, using the notation of \eqref{e:deltas} we describe in \autoref{e:CS_grav_anom}
which anomaly
cancelation condition we expect to obtain from which Chern-Simons coupling.
\begin{table}
\begin{center}
\begin{tabular}{cccr|c}
\multicolumn{1}{r|}{\rule[-.3cm]{0cm}{.8cm}}&Large Lorentz Transf. &$\qquad\qquad$&& Large Lorentz Transf. \\
 \cline{1-2}\cline{4-5}
\multicolumn{1}{r|}{\rule[-.3cm]{0cm}{.8cm}$\delta\tilde \Theta_{00}\overset{!}{=}0$}&0&&$\delta\tilde k_{000}\overset{!}{=}0$&\eqref{e:6d_grav_alter_2} \\
\multicolumn{1}{r|}{\rule[-.3cm]{0cm}{.8cm}$\delta\tilde \Theta_{0I}\overset{!}{=}0$}&0&&$\delta\tilde k_{00I}\overset{!}{=}0$ & 0 \\
\multicolumn{1}{r|}{\rule[-.3cm]{0cm}{.8cm}$\delta\tilde \Theta_{0m}\overset{!}{=}0$}&\eqref{4d_anomaly_alt1}&&$\delta\tilde k_{00m}\overset{!}{=}0$ & 0  \\
\multicolumn{1}{r|}{\rule[-.3cm]{0cm}{.8cm}$\delta\tilde \Theta_{IJ}\overset{!}{=}0$}&0&&$\delta\tilde k_{0IJ}\overset{!}{=}0$ & \eqref{e:6d_alter_3} \\
\multicolumn{1}{r|}{\rule[-.3cm]{0cm}{.8cm}$\delta\tilde \Theta_{mn}\overset{!}{=}0$}&0&&$\delta\tilde k_{0mn}\overset{!}{=}0$ & \eqref{e:6d_alter_4} \\
\multicolumn{1}{r|}{\rule[-.3cm]{0cm}{.8cm}$\delta\tilde \Theta_{Im}\overset{!}{=}0$}&0&&$\delta\tilde k_{0Im}\overset{!}{=}0$ & 0 \\
\rule[-.3cm]{0cm}{.8cm}&&&$\delta\tilde k_{IJK}\overset{!}{=}0$ & 0 \\
\rule[-.3cm]{0cm}{.8cm}&&&$\delta\tilde k_{mnp}\overset{!}{=}0$ & 0 \\
\rule[-.3cm]{0cm}{.8cm}&&&$\delta\tilde k_{IJm}\overset{!}{=}0$ & 0 \\
\rule[-.3cm]{0cm}{.8cm}&&&$\delta\tilde k_{Imn}\overset{!}{=}0$ & 0 \\
\rule[-.3cm]{0cm}{.8cm}&&&$\delta\tilde k_{0}\overset{!}{=}0$ & \eqref{e:6d_grav_alter_1} \\
\rule[-.3cm]{0cm}{.8cm}&&&$\delta\tilde k_{I}\overset{!}{=}0$ & 0 \\
\rule[-.3cm]{0cm}{.8cm}&&&$\delta\tilde k_{m}\overset{!}{=}0$ & 0 \\
\end{tabular}
\end{center}
\caption{We depict which anomaly equations we expect to obtain from a consistent action of large
local Lorentz transformations on Chern-Simons couplings in the respective theory on the circle.}
\label{e:CS_grav_anom}
\end{table}

Finally let us note that, although this reasoning seems to proceed in a straightforward manner, there are a lot of subtleties
involved. For instance one has to keep track of how precisely the spacetime representations of the matter fields, the Wilson line
moduli and the radius change under a local Lorentz transformation. Furthermore, comparing the Chern-Simons terms with the anomaly equations,
we are already familiar of how the structure of the left-hand sides of
\eqref{e:4gravan} and \eqref{e:6gravan2} can in principle be obtained from geometry, namely via intersections of divisors $D_\alpha$.
In contrast it is not at all clear how to obtain the left-hand side of
\eqref{e:6gravan1}, and there might be some higher-derivative couplings involved which have not been considered
before.

In the spirit of \autoref{ch:arith} it
would be extremely interesting to uncover the symmetry structure of elliptically-fibered Calabi-Yau manifolds which corresponds
to large local Lorentz transformations in F-theory compactifications, and therefore ensures the expected cancelation of gravitational
anomalies beyond the treatment in \cite{Cvetic:2012xn,Grimm:2013oga} where this is shown using a different argument.

\section{Chern-Simons Terms and Anomalies Revisited}

This section does not provide completely new ideas but makes some of our results in \autoref{sec:anom_lgt} more precise.
We have shown that demanding that large gauge transformations act consistently on Chern-Simons terms
is equivalent to the cancelation of gauge anomalies in the higher-dimensional theory. In fact, it is now
even possible to write down the higher-dimensional anomaly polynomial in terms of the variation of Chern-Simons couplings \eqref{e:deltas}
in the theory
on the circle. We find using
\begin{align}
 \tr_R \hat F^{k} = \sum_{w \in R} w_{I_1}\dots w_{I_k} \, \hat F^{I_1}\dots \hat F^{I_k}
\end{align}
and introducing the generalized index $\hat I = (I,m)$ labeling gauge fields
\begin{subequations}\label{e:anompol}
 \begin{align}
  I_6 &= \frac{1}{6}\,  \partial_{\fn^{\hat I}} \big(\delta\tilde\Theta_{\hat J \hat K} \big)\, \hat F^{\hat I} \hat F^{\hat J} \hat F^{\hat K}
  +\dots \, ,  \\
    I_8 &= \frac{1}{12}\, \partial_{\fn^{\hat I}} \big(\delta\tilde k_{\hat J} \big)\, \hat F^{\hat I} \hat F^{\hat J}  \tr \hat \cR^2 +
    \frac{2}{3}\,  \partial_{\fn^{\hat I}} \big(\delta\tilde k_{\hat J \hat K \hat L} \big)\, \hat F^{\hat I} \hat F^{\hat J} \hat F^{\hat K} \hat F^{\hat L}
    + \dots \, ,
 \end{align}
\end{subequations}
where we are missing the contribution from mixed gauge-gravitational anomalies in four dimensions
and the one from pure gravitational anomalies in six 
dimensions indicated by the 'dots'.

What makes the precise relations \eqref{e:anompol} so powerful is the fact that we now have access to global anomalies. Note that in \autoref{sec:anom_lgt}
we have only shown that the anomaly \textit{cancelation conditions} can be derived from one-loop Chern-Simons terms rather than the precise form
of the anomaly polynomial. Global anomalies however do not necessarily have to be canceled in the effective theory in order to obtain a consistent
quantum theory, and one is often interested in the precise form of the anomaly rather than a cancelation condition.
Using \eqref{e:anompol} this is now possible. Indeed,
in order to compute the anomaly polynomial of the global symmetry group one is interested in, one introduces associated background gauge
fields and evaluates the one-loop Chern-Simons terms in the theory on the circle pushed to the Coulomb branch of the background gauge fields. 

With this procedure it is for instance for some theories straightforward to compute R-symmetry anomalies.
However, we stress that there might appear complications when the structure of the Kaluza-Klein tower is unclear.
This happens for example for interacting six-dimensional (2,0) SCFTs. Although we know that the effective theory on the circle is
a five-dimensional $\cN =4$ supersymmetric Yang-Mills theory. The precise dualization prescription and the charges of the Kaluza-Klein modes
are still speculative. One might indeed go the other way round and use the established conjecture for the global anomalies
\cite{Bershadsky:1997sb,Freed:1998tg,Intriligator:2000eq,
 Yi:2001bz,Maxfield:2012aw,
 Monnier:2013kna,Monnier:2013rpa,Monnier:2014txa,Intriligator:2014eaa,Ohmori:2014pca,Ohmori:2014kda}
in order
to understand the structure of the Kaluza-Klein tower.
For convenience we list the
different six-dimensional supermultiplets along with the
spacetime representation and R-symmetry representation of their component fields in \autoref{app:multi}.

 \part{Conclusions}\label{part:concl}

\chapter{Closing Remarks and Future Directions}

In this thesis we investigated topological aspects of quantum field theory and string theory.
Furthermore a deep connection between gauge theories on the circle and the arithmetic of genus-one fibrations was uncovered using the framework
of F-theory.
Of particular importance were corrections to Chern-Simons couplings at one-loop
in three and five dimensions since these turned out to capture crucial features of the underlying theories.

In the first part we considered supersymmetry breaking of five-dimensional $\cN=4$ gauged supergravity. We derived essential quantities in the theory
around the respective vacuum without aiming at a classification of vacua. Special emphasis was put on $\cN=2$ vacua
of the $\cN=4$ theories, and for the case of solely Abelian
magnetic gaugings we were even able to derive the complete $\cN=2$ effective theory in the gravity-vector sector.
The latter includes one-loop corrections to the Chern-Simons terms, \textit{i.e.~}corrections 
to the $\cN=2$ prepotential. Importantly,
since they are independent of the supersymmetry breaking scale, they have to be included at any energy. We found that for a special choice of the
Abelian magnetic gaugings and the spectrum these corrections can vanish. The same cancelation occurs for more general $\cN=4$ gaugings
with $\cN=2$ vacua. Indeed, such breaking patterns naturally arise in consistent truncations of supergravity and string theory.
Therefore we derived a consistent truncation of M-theory on $SU(2)$-structure manifolds to five-dimensional $\cN=4$ gauged supergravity.
The latter theory (with different gaugings) also arises in a well-studied truncation of type IIB supergravity on squashed Sasaki-Einstein manifolds.
Both examples allow for $\cN=2$ vacua, and we studied them in the context of effective field theory. More precisely,
although consistent truncations originally had been developed to derive particular solutions to the higher-dimensional theories,
recently they
have also been used in the literature as effective theories arguing that quantum corrections can be safely neglected at low energies.
In principle this can be in conflict with the scale-invariant corrections we just described. Therefore, the latter should better coincide for
consistent truncations and their genuine effective actions in order for this procedure to make sense. We formulated necessary conditions
for this non-trivial obstruction. Surprisingly, in our two examples of consistent truncations with breaking pattern $\cN=4 \rightarrow \cN=2$
the scale-invariant corrections vanish. For M-theory on $SU(2)$-structure manifolds this is trivial since there are simply no charged states
in the $\cN=2$ vacuum of the consistent truncation. The latter is a Calabi-Yau manifold of which we know the genuine effective theory. In particular, since
one-loop correction are also absent there, the consistent truncation is in this respect consistent with the genuine effective theory.
In contrast, for type IIB supergravity on a squashed Sasaki-Einstein manifold there are charged states in the $\cN=2$ vacuum. Nevertheless integrating them out
does not induce scale-invariant corrections because a very non-trivial cancelation between all contributions takes place. Note that the notion of an effective
theory is not so clear here since we are in anti-de Sitter space. It would be interesting to find out if it constitutes a general feature of consistent truncations
that scale-invariant corrections are well-behaved. Turning this argument around, one could identify field-theoretically cancelations of scale-invariant
corrections in gauged supergravity and try to find a corresponding consistent truncation. For instance, for purely Abelian magnetic gaugings we found
a non-trivial cancelation of the gauge and the gravitational Chern-Simons term
for the special choice $n=3$ and $\rk (\xi^{MN})=4$. We suspect that there is an underlying principle to be uncovered for this case.

The topological properties of Chern-Simons couplings were
then used in the second part of this thesis to investigate anomaly cancelation in higher-dimensional gauge theories.
More precisely, we considered general matter-coupled four- and six-dimensional theories on a circle
pushed to the Coulomb branch, and classified large gauge transformations which
preserve the boundary conditions of the matter fields along the circle. It was easy to see that there exist in principle two different approaches to evaluate
the mapping of one-loop Chern-Simons couplings under these non-trivial transformations. The \textit{classical} way is to treat these coefficients as the duals
the the vector fields. The \textit{quantum} way to infer their mapping is to directly evaluate the loop-calculation with the gauge-transformed
quantities, in particular the gauge-transformed Wilson lines. Demanding consistency of both approaches we were able to derive all four- and six-dimensional
gauge anomaly cancelation conditions. By probing one-loop corrections to the gravitational Chern-Simons term we even obtained the six-dimensional mixed
gauge-gravitational anomaly constraints. We then applied our findings to the framework of F-theory and genus-one fibrations.
In particular, the fact that the F-theory effective action is determined by matching a circle-reduced gauge-theory with M-theory on a genus-one fibration 
allowed us to derive a detailed dictionary between boundary conditions preserving large gauge transformations along the circle on the one hand and arithmetic structures
of genus-one fibrations on the other hand. Indeed, integer 
Abelian large gauge transformations were identified with the free part of the Mordell-Weil group of rational
sections. For the case that the genus-one fibration does not admit a section there can still be Abelian large gauge transformations in the associated theory
on the circle. Accordingly we conjectured the new arithmetic structure called \textit{extended Mordell-Weil group}, and formulated the group law in terms of homological
cycles. While a general algebraic proof of the existence of this group is still missing, we found further evidence by investigating Higgs transitions of example geometries.
In the non-Abelian sector we were able to match so-called special fractional large gauge transformations to the torsion subgroup of the Mordell-Weil group.
Ordinary integer non-Abelian large gauge transformations again led us to defining a novel associated arithmetic structure on the elliptic fibration. We
again wrote down
the group operation in terms of homological cycles. As for the extended Mordell-Weil group there is yet no proof of this structure beyond homology level,
but we once more gathered further evidence by considering geometric examples of Higgs transitions.
For future research it would be extremely 
interesting to verify this group structures on an algebraic level, \textit{i.e.~}to find a general proof for geometric transitions
associated to the conjectured group operations. This would fully proof the cancelation of the corresponding anomalies. We think that
in general to achieve this one has to investigate branched covers of genus-one fibrations in terms of schemes, and define the group structures on these
more abstract objects. In fact there seems to be a striking relation to the minimal model program in algebraic geometry. 
This goes however beyond the work in this thesis.
Let us mention that in this respect it could be fruitful to extend our results
to F-theory compactifications to two or eight dimensions. In the latter case one has to consider K3 manifolds which are quite well understood,
and one might even be able to prove the existence of some of the conjectured group structures there as a first step.

In \autoref{part:add} we already highlighted some directions into which one could proceed from this. Let us shortly collect them again here.
First of all, it seems that F-theory spectra are always such that fractional Abelian large gauge transformations in the effective theory are not possible.
We argued that this might be due to constraints from quantum gravity, and it is 
also hard to imagine a group structure on genus-one fibrations which
would realize them as a geometric symmetry. One might get some new insights by making these points more precise.
In particular there might even be a connection to the recently investigated (Sub)Lattice Weak Gravity Conjecture.
Second, we made a first step
in unifying the Mordell-Weil group of rational sections with the conjectured extended Mordell-Weil groups of multi-sections into a single framework
called \textit{graded Mordell-Weil pseudo-ring}. We think that this might be key to understanding massive $U(1)$s in F-theory much better.
Moreover, the quintessence of our discussion on geometric symmetries is that arithmetic structures on genus-one fibrations
are mapped to gauge theories on the circle and vice versa. This fact is very constraining since the dictionary has to make use of homomorphisms.
Note that we discussed only Calabi-Yau compactifications of F-theory in this thesis, but enforcing that such a dictionary always exists
might give us a tool to go beyond Calabi-Yau level. In fact, although we had known the effective theory for Calabi-Yau compactifications
before we carried out our analysis,
much of the structure
turned out to be directly dictated by the homomorphisms. 

Finally, field-theoretically we proposed an idea how one could approach gravitational anomalies via Chern-Simons terms, but there are still
a lot of subtleties
which need to be worked out. Much more settled is the application of our results to global anomalies and in particular R-symmetries.
However, in order to treat them for superconformal field theories like the 6d $(2,0)$ non-Abelian tensor theories, there is always the issue
that one has to fully understand the complete Kaluza-Klein tower after circle-compactification.
Another direction to continue could be to investigate other kinds of topological terms like Wess-Zumino terms, and find out if a similar analysis can 
teach us new lessons.

\appendix

\part{Appendices}\label{part:app}

\chapter{Spacetime Conventions and Identities}\label{app:spacetime}

We shortly state the conventions of differential geometry in five dimensions used in this thesis.
Curved five-dimensional spacetime indices are denoted by Greek letters $\mu,\nu,\dots$. Antisymmetrizations
of any kind are always done with weight one, \textit{i.e.}~include a factor of $1/n!\,$.
We use the $(-,+,+,+,+)$ convention for the five-dimensional metric $g_{\mu\nu}$, and we adopt the negative sign in front of the Einstein-Hilbert term.
Moreover we set
\begin{align}
 \kappa^2  = 1 \, .
\end{align}
The Levi-Civita tensor with curved indices $\epsilon_{\mu\nu\rho\lambda\sigma}$ reads
\begin{align}
 \epsilon_{01234} = + e \, , \qquad  \, , \epsilon^{01234} = - e^{-1}\, ,
\end{align}
where $e = \sqrt{-\det g_{\mu\nu} }\,$.

The five-dimensional spacetime gamma matrices are denoted by $\gamma_\mu$ and satisfy
\begin{align}
 \lbrace \gamma_\mu , \gamma_\nu \rbrace = 2 g_{\mu\nu} \, .
\end{align}
Antisymmetrized products of gamma matrices are defined as
\begin{align}
 \gamma_{\mu_1 , \dots , \mu_k } := \gamma_{[ \mu_1}\gamma_{\mu_2}\dots \gamma_{\mu_k ]}\, . 
\end{align}
The convention for the charge conjugation matrix $C$ is such that
\begin{align}
 C^T = -C = C^{-1}
\end{align}
and it fulfills
\begin{align}
 C \gamma_\mu C^{-1} = ( \gamma_\mu )^T \, .
\end{align}
All massless spinors in five dimensions are meant to be symplectic Majorana, that is in the $\cN=4$ theory they are subject to the condition
\begin{align}
 \bar \chi^i := (\chi_i)^\dagger \gamma_0 = \Omega^{ij} \chi_j^T C \, ,
\end{align}
where $i,j =1, \dots , 4$ and $\Omega^{ij}$ is the symplectic form of $USp(4)$ defined in \eqref{e:properties_omega}.
In the $\cN=2$ theory the symplectic Majorana condition reads
\begin{align}
 \bar \chi^\alpha := (\chi_\alpha)^\dagger \gamma_0 = \varepsilon^{\alpha\beta} \chi_\beta^T C \, ,
\end{align}
where $\alpha,\beta = 1,2$, $\varepsilon^{\alpha\beta}$ is the two-dimensional epsilon tensor.

\chapter{Derivation of the \texorpdfstring{$\cN =2$}{N=2} Mass Terms and Couplings}
\chaptermark{\texorpdfstring{$\cN =2$}{N=2} Mass Terms and Couplings}
\label{mass_appendix}

In this chapter we explicitly derive the masses for the spin-$\sfrac 1 2$ fermions and scalars 
induced by the supergravity breaking in \autoref{sec:ab_mag_gaug}.
We also evaluate the charges of the spin-$\sfrac 1 2$ fermions under the Abelian gauge field $A^0$. 
Note that the Lagrangian
of a massive spin-$\sfrac 1 2$ Dirac spinor is given in \eqref{e:lagr_5d_1}.

\section{Fermion Masses}
Let us now investigate the masses which the $\cN = 4$ gaugini acquire from supersymmetry breaking.
The relevant mass terms can be found in \cite{Dall'Agata:2001vb,Schon:2006kz}. In our notation they take the following form
\begin{align}
\label{e:fermion_quadratic}
 e^{-1}\cL_{\lambda , \, \textrm{mass}} &= i \Big ( \frac{1}{2 \sqrt 2} \Sigma^2 \xi_{ab}\delta^j_i
 - \frac{1}{2}\tensor{\textbf{M}}{_\psi_\, _i^j}\delta_{ab}\Big ) \bar \lambda^{ia}\lambda^b_j  \nn \\
 &=  0 \cdot \bar \lambda^{\alpha \bar a}\lambda^{\bar b}_\alpha
  + \frac{1}{2 \sqrt 2} i \Sigma^2 \xi_{\hat a \hat b} 
 \bar \lambda^{\alpha \hat a}\lambda^{\hat b}_\alpha
  - \frac{1}{2}i \tensor{\textbf{M}}{_\psi _\, _{\dot\alpha}^{\dot\beta}}\delta_{\bar a \bar b}\bar 
  \lambda^{\dot\alpha \bar a}\lambda^{\bar b}_{\dot\beta} \nn \\
 &\quad + i \Big ( \frac{1}{2 \sqrt 2} \Sigma^2 \xi_{\hat a \hat b} \delta^{\dot\beta}_{\dot\alpha}
 - \frac{1}{2}\tensor{\textbf{M}}{_\psi _\, _{\dot\alpha}^{\dot\beta}}
 \delta_{\hat a \hat b}\Big ) \bar \lambda^{\dot\alpha \hat a}\lambda^{\hat b}_{\dot\beta}\, .
 \end{align}
 We proceed by diagonalizing these terms. In order to do so we discuss the four different types of fields separately.
 
\begin{itemize} 

\item $\lambda^{\bar a}_\alpha$ \\
We observe that the $\lambda^{\bar a}_\alpha$ stay massless.
Thus all together we find a number of $2(n+4-2n_T)$ massless spin-$\sfrac 1 2$ fermions supplemented by a symplectic Majorana condition.
They constitute the fermionic part of massless vector multiplets in the vacuum.

\item $\lambda^{\hat a}_\alpha$ \\
For the fermions $\lambda^{\hat a}_\alpha $
we write the mass terms using the split \eqref{split_lambdas} as
\begin{align}
 \frac{1}{2 \sqrt 2} i \Sigma^2 \xi_{\hat a \hat b}\bar\lambda^{\alpha \hat a}\lambda^{\hat b}_{\alpha}&=
  \frac{1}{2 \sqrt 2} i \Sigma^2 \sum_{\check a}\gamma_{\check a}\varepsilon_{kl}
 \bar \lambda^{\alpha [\check a k]}\lambda^{[\check a l]}_{\alpha} \nn \\
 & = \frac{1}{\sqrt 2}\Sigma^2 \sum_{\check a}\gamma_{\check a} 
 \bar\Blambda^{\alpha\check a}\Blambda_\alpha^{\check a}\, ,
\end{align}
with $k,l$ both taking values $1,2$.
Here we redefined the fermions by introducing $\Blambda^{\check a}_\alpha$ as in \eqref{compl_fermions_1}
and drop the symplectic Majorana condition such that the mass terms become diagonal.
Let us now have a look at how the corresponding kinetic terms transform under this redefinition
\begin{align}
 -\frac{1}{2}\sum_{\check a}\big( \bar\lambda^{\alpha [\check a 1]}\slashed \partial \tensor{\lambda}{_\alpha^{[\check a1]}}
 + \bar\lambda^{\alpha [\check a 2]}\slashed \partial \tensor{\lambda}{_\alpha^{[\check a 2]}}\big)=
 -\sum_{\check a}\bar{\Blambda}^{\alpha \check a }\slashed \partial \Blambda_\alpha^{\check a}\, .
\end{align}
The computation of the charge under $A^0$ proceeds as for the mass terms. The covariant derivatives can be found in
\cite{Dall'Agata:2001vb,Schon:2006kz}.
We find
\begin{align}
 \vert m_{\Blambda^{\check a}_\alpha} \vert  = \frac{1}{\sqrt 2} \Sigma^2  \vert \gamma_{\check a}\vert\, , \qquad
 \sign (m_{\Blambda^{\check a}_\alpha}) =  \sign \gamma_{\check a} \, , \qquad 
 q_{\Blambda^{\check a}_\alpha} =  \gamma_{\check a}\, .
\end{align}

\item $\lambda^{\bar a}_{\dot \alpha}$\\
The structure of mass terms of the fermions $\lambda^{\bar a}_{\dot \alpha}$
is similar to those of the massive gravitini. In particular, the diagonalization 
procedure of the gravitino mass terms automatically
diagonalizes the mass terms of the $\lambda^{\bar a}_{\dot\alpha }$. Again we move from symplectic Majorana spinors to
Dirac spinors $\Blambda^{\bar a}$ using \eqref{compl_fermions_2}.
We find
\begin{align}
 \vert m_{\Blambda^{\bar a}} \vert = \frac{1}{\sqrt 2}  \Sigma^2 \gamma \, , \qquad
 \sign ( m_{\Blambda^{\bar a}} ) = -1 \, ,\qquad  q_{\Blambda^{\bar a}} =  \gamma \, .
\end{align}

\item $\lambda^{\hat a}_{\dot\alpha }$\\
The mass terms after the split \eqref{split_lambdas} become
\begin{align}
 i \sum_{\check a} \Big ( \frac{1}{2 \sqrt 2} \Sigma^2 \gamma_{\check a} \varepsilon_{kl} \delta^{\dot\beta}_{\dot\alpha}
 - \frac{1}{2}\tensor{\textbf{M}}{_\psi _\, _{\dot\alpha}^{\dot\beta}}
 \delta_{kl}\Big ) \bar \lambda^{\dot\alpha [\check a k]}\lambda^{[\check a l]}_{\dot\beta}\, .
\end{align}
 We redefine and use Dirac spinors $\Blambda^{\check a}_1$ and $\Blambda^{\check a}_2$ given in \eqref{compl_fermions_2}. 
The mass terms then become
\begin{align}
  \sum_{\check a} \Big [ \frac{1}{\sqrt 2} \Sigma^2 \gamma_{\check a} 
 ( \bar{\Blambda}^{\check a}_1 \Blambda_1^{\check a} - 
 \bar{\Blambda}^{\check a}_2 \Blambda_2^{\check a} ) -
 \frac{1}{\sqrt 2} \Sigma^2  \gamma ( \bar{\Blambda}^{\check a}_1 \Blambda_1^{\check a} +  
 \bar{\Blambda}^{\check a}_2 \Blambda_2^{\check a} ) \Big ]\, .
\end{align}
The kinetic terms are unaffected.
We conclude that
\begin{subequations}
\begin{align}
 &\vert m_{\Blambda^{\check a}_1}\vert = \frac{1}{\sqrt 2}\Sigma^2  \vert  \gamma - \gamma_{\check a}   \vert \, ,& \qquad 
 &\vert m_{\Blambda^{\check a}_2}\vert = \frac{1}{\sqrt 2}\Sigma^2  \vert  \gamma + \gamma_{\check a}  \vert \, ,& \\
 &\sign (m_{\Blambda^{\check a}_1})= \sign (\gamma_{\check a} - \gamma  ) \, , & \qquad 
 &\sign(m_{\Blambda^{\check a}_2})= \sign (- \gamma_{\check a} - \gamma ) \, , &\\
 &q_{\Blambda^{\check a}_1}= \gamma_{\check a} - \gamma   \, ,& \qquad 
 &q_{\Blambda^{\check a}_2}= - \gamma_{\check a} - \gamma  \, .&
\end{align}
\end{subequations}

\end{itemize}

\section{Scalar Masses}
Lastly we investigate the scalar degrees of freedom in the vacuum (except of $\Sigma$).
In order to derive the scalar masses we insert the expansion \eqref{e:vexp}
into the scalar potential written down in \eqref{bos_N=4action}
\begin{align}
 e^{-1} \cL_{\textrm{pot}} = - \frac{1}{16} \xi^{MN} \xi^{PQ} \Sigma^4
 &\Big [\langle\cV\rangle\exp \big (  \sum_{m,a} \phi^{ma} [t_{ma}]\Big ) 
 \exp \big (  \sum_{n,b} \phi^{nb} [t_{nb}]\big )^T \langle\cV \rangle^T\Big ]_{MP} \times \nn \\
 &\Big [\langle\cV\rangle \exp \big (  \sum_{p,c} \phi^{pc} [t_{pc}]\big ) 
 \exp \big (  \sum_{q,d} \phi^{qd} [t_{qd}]\big )^T \langle\cV \rangle^T \Big ]_{NQ} \, .
\end{align}
To read off the mass terms of the scalars, we focus on the terms quadratic in $\phi^{ma}$
\begin{align}
 e^{-1} \cL_{\phi, \, \textrm{mass}} = - \frac{1}{16} \Sigma^4 \phi^{ma} \phi^{nb}
 (8 \xi_{mn} \xi_{ab} + 4 \delta_{mn} \xi_{ac} \tensor{\xi}{_b^c} + 4 \delta_{ab} \xi_{mp} \tensor{\xi}{_n^p})\, .
\end{align}

According to the index split \eqref{e:split} the scalar fields arrange in four different groups: 
\begin{itemize}
 \item $\phi^{\bar m \bar a}$ \\
 The mass terms for these fields vanish:
 \begin{align}
  e^{-1} \cL_{\phi, \, \textrm{mass}} = 0 \, .
 \end{align}
 Thus we find $n + 4 - 2 n_T$ massless real scalar fields $\phi^{\bar m \bar a}$.
 
 \item $\phi^{\hat m \bar a}$ \\
 The mass terms of these modes receive one contribution from the gauging $\xi^{mn}$
 \begin{align}
  e^{-1} \cL_{\phi, \, \textrm{mass}} = - \frac{1}{4}\gamma^2 \Sigma^4 \phi^{\hat m \bar a} \phi_{\hat m \bar a} \, .
 \end{align}
 We can now complexify the scalars as in \eqref{compl_scalars_1}  into the $2(n+4 - 2 n_T)$ massive complex scalars $\Bphi^{\alpha \bar a}$
 with mass\footnote{We stress that the kinetic terms
  for the scalars here and in the following are always automatically canonically normalized, even after
  the field redefinitions carried out in this section. This one can check explicitly by inserting the expansion
  \eqref{e:vexp} into the $\cN =4$ scalar kinetic terms.}
 \begin{align}
  m_{\Bphi^{\alpha \bar a}} = \frac{1}{\sqrt 2}  \Sigma^2 \gamma \, .
 \end{align}
 
 \item $\phi^{\bar m \hat a}$ \\
 There is now solely a mass contribution from the gaugings $\xi^{ab}$
 \begin{align}
  e^{-1} \cL_{\phi, \, \textrm{mass}} = -\frac{1}{4}\Sigma^4 \sum_{\hat a} \gamma_{\check a}^2 \phi^{\bar m \hat a}
  \phi^{\bar m \hat a } \, .
 \end{align}
  Using the definition \eqref{compl_scalars_1} 
  one identifies $n_T - 2$ massive complex scalar fields $\Bphi^{\bar m \check a}$ with mass
  \begin{align}
   m_{\Bphi^{\bar m \check a}} = \frac{1}{\sqrt 2} \Sigma^2 \vert \gamma_{\check a} \vert \, .
  \end{align}

  \item $\phi^{\hat m \hat a}$ \\
  We now face mass contributions both from $\xi^{mn}$ and $\xi^{ab}$
  \begin{align}
    e^{-1} \cL_{\phi, \, \textrm{mass}} = - \frac{1}{16}\Sigma^4 \sum_{\check a , \alpha} 
    (8 \gamma \gamma_{\check a} \varepsilon_{\dot\alpha\dot\beta}\varepsilon_{kl}
    + 4 \gamma^2  \delta_{\dot\alpha\dot\beta}\delta_{kl}
    + 4 \gamma_{\check a}^2  \delta_{\dot\alpha\dot\beta}\delta_{kl} )
    \phi^{[\alpha\dot\alpha ][\check a k]} \phi^{[\alpha\dot\beta ][\check a l]} \, .
  \end{align}
  One can check that the mass terms are diagonalized by the redefinitions 
  \eqref{compl_scalars_2} and \eqref{compl_scalars_3} to $4 n_T -8$ complex scalars
  $\Bphi^{\alpha \check a}_1$ and 
   $\Bphi^{\alpha \check a}_2$ with masses
  \begin{align}
   m_{\Bphi^{\alpha \check a}_1} = \frac{1}{\sqrt 2} \Sigma^2 \vert \gamma - \gamma_{\check a} \vert\, ,  \qquad
   m_{\Bphi^{\alpha \check a}_2 } = \frac{1}{\sqrt 2} \Sigma^2 \vert \gamma + \gamma_{\check a} \vert  \, .
  \end{align}
\end{itemize}

\chapter{The Coset Representatives and Contracted Embedding Tensors for \texorpdfstring{$SU(2)$}{SU(2)}-Structure Manifolds}
\chaptermark{\texorpdfstring{$SU(2)$}{SU(2)}-Structure Manifolds}
\label{app:cosetrepr}
\section{The Coset Representatives \texorpdfstring{$\cV$}{V}}
From the expressions for $M_{MN}$ in \eqref{eq:scalarmetric} we can 
extract representatives \(\cV = ({\cV_M}^m, {\cV_M}^a)\) of the coset space
\begin{equation}
\frac{SO(5,\tilde n-1)}{SO(5) \times SO(\tilde n-1)} \,,
\end{equation}
where \(m = 1,\dots,5\) and \(a = 6,\dots,5+n\) denote \(SO(5)\) and \(SO(n)\) indices.\footnote{Note that the indices $n$, defined around \eqref{e:content_vector}, and $\tilde n$, defined around \eqref{eq:su2ansatzintegral},
are related by $n=\tilde n -1$.}
These coset representatives are related to the scalar metric via
\begin{equation}\label{eq:cosetm}
(M_{MN}) = \cV\cV^T = \cV^m\cV^m + \cV^a\cV^a
\end{equation}
and have to fulfill
\begin{equation}\label{eq:coseteta}
(\eta_{MN}) = -\cV^m\cV^m + \cV^a\cV^a \,.
\end{equation}
Before we can determine \(\cV\), it is necessary to diagonalize \(g_{ij}\) and \(H_{IJ}\). First we observe that \(g_{ij}\) can be expressed as
\begin{equation}
g_{ij} = e^{-\rho_2}k_i^k k_j^l \delta_{kl} \,,
\end{equation}
where \(k = e^{\rho_2/2} (\I\,\tau)^{-1/2}(1,\tau)\).

In \eqref{eq:su2ansatzhodge} we have introduced \(H_{IJ}\) via \(\ast \omega^I = -{H^I}_J \omega^J \wedge \mathrm{vol}_2^{(0)}\), and as described in \cite{Louis:2009dq} it only depends on \(\zeta^a_I\)
\begin{equation}\label{eq:zetaH}
H_{IJ} = 2 \zeta^a_I \zeta^a_J + \eta_{IJ} \,.
\end{equation}
From \eqref{eq:su2forms} and \eqref{eq:su2formsexpansion} one sees that
\begin{equation}
\zeta^a_I \eta^{IJ} \zeta^b_J = -\delta^{ab} \,.
\end{equation}
Therefore \(\zeta^a_I {H^I}_J = -\zeta^a_J\), which means that the three \(\zeta^a_I\) are eigenvectors of \({H^I}_J\) with eigenvalue \(-1\). 
If we now introduce an orthonormal basis \(\xi^\alpha_I\) (\(\alpha = 1,\dots,\tilde n-3\)) of the subspace orthogonal to all \(\zeta^a_I\) (\textit{i.e.}~\(\xi^\alpha_I \eta^{IJ} \xi^\beta_J = \delta^{\alpha\beta}\) and \(\zeta^a_I \eta^{IJ} \xi^\beta_J = 0\)), we can write 
\begin{equation}
H_{IJ} = \zeta^a_I \zeta^a_J + \xi^\alpha_I \xi^\alpha_J
\end{equation}
since we can deduce from \eqref{eq:zetaH} that the \(\xi^\alpha_I\) are eigenvectors of \({H^I}_J\) with eigenvalue \(+1\).
Moreover it follows that \(\xi^\alpha_I \xi^\alpha_J = \zeta^a_I \zeta^a_J + \eta_{IJ}\) and so
\begin{equation}
\eta_{IJ} = - \zeta^a_I \zeta^a_J + \xi^\alpha_I \xi^\alpha_J \,.
\end{equation} 
We can shorten the notation by defining
\begin{equation}\label{eq:combinedzetaxi}
E^\cI_I = (\zeta^a_I, \xi^\alpha_J) \,, \quad \cI = (a, \alpha) \,,
\end{equation}
which allows us to write
\begin{equation}
H_{IJ} = E^\cI_I E^\cJ_J \delta_{\cI\cJ}
\quad\mathrm{and}\quad
\eta_{IJ} = E^\cI_I E^\cJ_J \eta_{\cI\cJ} \,,
\end{equation}
with \(\eta_{\cI\cJ} = \mathrm{diag}(-1,-1,-1;+1,\dots,+1)\).

After this preparation we are able to write down \(\cV\),
\begin{equation}\begin{aligned}\label{eq:su2cosetrepr}
{\cV_i}^j &= e^{\rho_4/2} k_i^j \,, \\
{\cV_i}^\bj &= e^{-\rho_4/2} \delta^{j\bj} (k^{-1})_j^k(\epsilon_{ki} \gamma + \tfrac{1}{2}c_{kI}c_i^I) \,, \\
{\cV_i}^\cI &= -{E_I}^\cI c_i^I \,, \\
{\cV_\bi}^\bj &= e^{-\rho_4/2}\delta^{j\jb} \delta_{i\ib} (k^{-1})_j^i \,, \\
{\cV_I}^\bi &= e^{-\rho_4/2}\delta^{i\ib} (k^{-1})_i^j c_{jI} \,, \\
{\cV_I}^\cI &= {E_I}^\cI \,,
\end{aligned}\end{equation}
such that
\begin{equation}\label{eq:cosetnondiagm}
M_{MN} = (\cV \cV^T)_{MN} = {\cV_M}^i {\cV_N}^i + {\cV_M}^\ib {\cV_N}^\ib + {\cV_M}^\cI {\cV_N}^\cI \,,
\end{equation}
and
\begin{equation}\label{eq:cosetnondiageta}
\eta_{MN} = 2 \delta_{i\ib} {\cV_M}^i {\cV_N}^{\ib} + \eta_{\cI\cJ}{\cV_M}^\cI {\cV_N}^\cJ \,.
\end{equation}

In the end it is necessary to split \eqref{eq:su2cosetrepr} into \(\cV^m\) and \(\cV^a\), which corresponds to bringing \eqref{eq:cosetnondiageta} 
into diagonal form.
The result reads
\begin{equation}
{\cV_M}^m =
\begin{pmatrix}
\tfrac{1}{\sqrt 2}\left(-{\cV_M}^1 + {\cV_M}^{\bar1}\right) \\
\tfrac{1}{\sqrt 2}\left(-{\cV_M}^2 + {\cV_M}^{\bar2}\right) \\
{\cV_M}^{\cI = 1,2,3}
\end{pmatrix} \,, \quad
{\cV_M}^a =
\begin{pmatrix}
\tfrac{1}{\sqrt 2}\left({\cV_M}^1 + {\cV_M}^{\bar1}\right) \\
\tfrac{1}{\sqrt 2}\left({\cV_M}^2 + {\cV_M}^{\bar2}\right) \\
{\cV_M}^{\cI \neq 1,2,3}
\end{pmatrix}\,.
\end{equation}
Using \eqref{eq:cosetnondiagm} and \eqref{eq:cosetnondiageta} one can easily check that these combinations fulfill \eqref{eq:cosetm} and \eqref{eq:coseteta}.

\section[The Contracted Embedding Tensors for Calabi-Yau Manifolds with \texorpdfstring{$\chi = 0$}{chi=0}]{The Contracted
Embedding Tensors for Calabi-Yau Manifolds
with \texorpdfstring{$\chi = 0$}{chi=0}\sectionmark{Calabi-Yau Manifolds with \texorpdfstring{$\chi = 0$}{chi=0}}}
\sectionmark{Calabi-Yau Manifolds with \texorpdfstring{$\chi = 0$}{chi=0}}
\label{app:contracted_gaugings}
Using the results from \eqref{eq:su2cosetrepr} we can compute the contractions of the embedding tensors \eqref{eq:su2embeddingtensors} with the coset representatives as introduced in \eqref{e:dressed_gaugings}.
Hereby we restrict to the special case of Calabi-Yau manifolds with vanishing Euler number and use the relevant relations from \eqref{eq:5dCYconditions} that follow to simplify the resulting expressions.
We also restrict to the case without four-form flux and set \(n = n_I = 0\).

For \( (\xi^{mn}) \) we find that it takes the general form
\begin{equation}
(\xi^{mn}) = 
\left( \begin{array}{@{}cc|ccc@{}}
\multicolumn{2}{c|}{\multirow{2}*{${\bf0}_{2\times2}$}} & - & \xi^{1n} & - \\
&& - & \xi^{2n} & - \\
\hline
| & | & \multicolumn{3}{c}{\multirow{3}*{${\bf0}_{3\times3}$}} \\
\xi^{m1} & \xi^{m2} && \\
| & | &&& \\
\end{array}\right) \,,
\end{equation}
where its non-vanishing components are given by
\begin{equation}\begin{aligned}\label{eq:xi12m}
\xi^{1,m=3,4,5} = -\xi^{m1} &= \frac{1}{\sqrt2}e^{-\frac{1}{2}(\rho_2+\rho_4)}\sqrt{\I\,\tau} \,t^{2I} \zeta^{a=1,2,3}_I \,,\\
\xi^{2,m=3,4,5} = -\xi^{m2} &= - \frac{1}{\sqrt2}e^{-\frac{1}{2}(\rho_2+\rho_4)}\frac{1}{\sqrt{\I\,\tau}} \left(t^{1I} + \R\,\tau \,t^{2I}\right) \zeta^{a=1,2,3}_I \,.
\end{aligned}\end{equation}
Similarly we have
\begin{equation}
\xi^{ab} = 
\left( \begin{array}{@{}cc|ccc@{}}
\multicolumn{2}{c|}{\multirow{2}*{${\bf0}_{2\times2}$}} & - & \xi^{6b} & - \\
&& - & \xi^{7b} & - \\
\hline
| & | & \multicolumn{3}{c}{\multirow{3}*{${\bf0}_{(n-2)\times(n-2)}$}} \\
\xi^{a6} & \xi^{a7} && \\
| & | &&& \\
\end{array}\right),
\end{equation}
with
\begin{equation}\begin{aligned}\label{eq:xi12a}
\xi^{6,a=8,\dots, 5+n} = -\xi^{a6} &= \frac{1}{\sqrt2}e^{-\frac{1}{2}(\rho_2+\rho_4)}\sqrt{\I\,\tau} \,t^{2I} \xi^{\alpha=1,\dots,\tilde n-3}_I \,,\\
\xi^{7,a=8,\dots, 5+n} = -\xi^{a7} &= - \frac{1}{\sqrt2}e^{-\frac{1}{2}(\rho_2+\rho_4)}\frac{1}{\sqrt{\I\,\tau}} \left(t^{1I} + \R\,\tau \,t^{2I}\right) \xi^{\alpha=1,\dots,\tilde n-3}_I \,,
\end{aligned}\end{equation}
and finally for the mixed-index part
\begin{equation}
\xi^{ma} = 
\left( \begin{array}{@{}cc|ccc@{}}
\multicolumn{2}{c|}{\multirow{2}*{${\bf0}_{2\times2}$}} & - & \xi^{1a} & - \\
&& - & \xi^{2a} & - \\
\hline
| & | & \multicolumn{3}{c}{\multirow{3}*{${\bf0}_{3\times( n-2)}$}} \\
\xi^{m6} & \xi^{m7} && \\
| & | &&& \\
\end{array}\right) \,,
\end{equation}
where its entries are again given by \eqref{eq:xi12m} and \eqref{eq:xi12a}.

Following the notation introduced in \eqref{eq:combinedzetaxi} we obtain for the non-vanishing components of the contracted \(f_{MNP}\)
\begin{equation}\begin{aligned}
f^{m=1,\cI\cJ} = f^{a=6,\cI\cJ} =& - \frac{1}{\sqrt2}e^{-\frac{1}{2}(\rho_2+\rho_4)}\sqrt{\I\,\tau} \Bigl(T^J_{1K} \eta^{KI} + \tfrac{1}{2} t^2 \eta^{IJ} \Bigr) E^\cI_I E^\cJ_J \,, \\
f^{m=2,\cI\cJ} = f^{a=7,\cI\cJ} =& \frac{1}{\sqrt2}e^{-\frac{1}{2}(\rho_2+\rho_4)}\frac{1}{\sqrt{\I\,\tau}} \Bigl(\left(T^J_{2I} - \R\,\tau\, T^J_{1I}\right) \eta^{IK} \\& + \tfrac{1}{2}\left(t^1 + \R\,\tau \,t^2\right)\eta^{IJ} \Bigr) E^\cI_K E^\cJ_J \,.
\end{aligned}\end{equation}
For completeness we also give the contracted versions of \(\xi_M\) although they vanish for the special case of the Enriques Calabi-Yau,
\begin{equation}\begin{aligned}
\xi^{m=1} = \xi^{a=6} &= -\frac{1}{\sqrt2}e^{-\frac{1}{2}(\rho_2+\rho_4)}\sqrt{\I\,\tau} \,t^2 \,,\\
\xi^{m=2} = \xi^{a=7} &= \frac{1}{\sqrt2}e^{-\frac{1}{2}(\rho_2+\rho_4)}\frac{1}{\sqrt{\I\,\tau}} \left(t^1 + \R\,\tau \,t^2\right) \,.
\end{aligned}\end{equation}

It is important to notice that these expression are still subject to a set of constraints since there are redundancies in the scalar sector.
One has to use the relations in \cite{KashaniPoor:2013en} in order to extract the proper unconstrained contracted embedding tensors. For the Enriques Calabi-Yau
we find\footnote{The geometrical analysis of the Enriques Calabi-Yau was also carried out in \cite{KashaniPoor:2013en}.}
\begin{align}
 &f_{1,6\,\,\, 3,8\,\,\, 5,10} = f_{2,7\,\,\, 4,9\,\,\, 5,10} = \frac{1}{2} \Sigma^3 \, e^{-\frac{1}{2}(\rho_2 + \rho_4)}\, \I \, \tau \nn \\
 &\xi_{1,6\,\,\, 3,8} = \xi_{2,7\,\,\, 4,9} = \frac{1}{\sqrt 2}\, e^{-\frac{1}{2}(\rho_2 + \rho_4)}\, \I \, \tau \, ,
\end{align}
where there are two options for each index position.
We explicitly inserted the quantities $t^i_I$, $T^I_{iJ}$ for the Enriques Calabi-Yau \cite{KashaniPoor:2013en}
\begin{align}\label{e:torsion_classes}
 (t^i_I)=&\begin{pmatrix}
 0 & 1 & 0 & 0 & -1 & 0 & \textbf{0}_{1\times 8} \\
 -1 & 0 & 0 & 1 & 0 & 0 & \textbf{0}_{1\times 8}
         \end{pmatrix} \, , \nn\\
 (T^I_{1J})=&\begin{pmatrix}
 0 & 0 & 1 & 0 & 0 & -1 & \textbf{0}_{1\times 8} \\
 0 & 0 & 0 & 0 & 0 & 0 &  \textbf{0}_{1\times 8}\\
 -1 & 0 & 0 & 1 & 0 & 0 & \textbf{0}_{1\times 8} \\
 0 & 0 & 1 & 0 & 0 & -1 & \textbf{0}_{1\times 8} \\
 0 & 0 & 0 & 0 & 0 & 0 & \textbf{0}_{1\times 8} \\
 -1 & 0 & 0 & 1 & 0 & 0 & \textbf{0}_{1\times 8} \\
 \textbf{0}_{8\times 1} & \textbf{0}_{8\times 1} & \textbf{0}_{8\times 1} & \textbf{0}_{8\times 1} & \textbf{0}_{8\times 1} & \textbf{0}_{8\times 1} & 
 \textbf{0}_{8\times 8}
             \end{pmatrix} \, , \nn\\
 (T^I_{2J})=&\begin{pmatrix}
 0 & 0 & 0 & 0 & 0 & 0 & \textbf{0}_{1\times 8} \\
 0 & 0 & 1 & 0 & 0 & -1 & \textbf{0}_{1\times 8}\\
 0 & -1 & 0 & 0 & 1 & 0 & \textbf{0}_{1\times 8} \\
 0 & 0 & 0 & 0 & 0 & 0 & \textbf{0}_{1\times 8} \\
 0 & 0 & 1 & 0 & 0 & -1 & \textbf{0}_{1\times 8} \\
 0 & -1 & 0 & 0 & 1 & 0 & \textbf{0}_{1\times 8} \\
 \textbf{0}_{8\times 1} & \textbf{0}_{8\times 1} & \textbf{0}_{8\times 1} & \textbf{0}_{8\times 1} & \textbf{0}_{8\times 1} & \textbf{0}_{8\times 1} & 
 \textbf{0}_{8\times 8}
             \end{pmatrix} \, .
\end{align}
Note that the general elimination of redundancies is far from being straightforward.

\chapter{Comparison with Type IIA Supergravity on \texorpdfstring{$SU(2)$}{SU(2)}-Structure Manifolds}
\chaptermark{Type IIA on \texorpdfstring{$SU(2)$}{SU(2)}-structure Manifolds}
\label{IIAreduction}
Another way of reproducing the results from \autoref{ch:m_su2} is to take the four-dimensional theory obtained in \cite{KashaniPoor:2013en} by reducing type IIA string theory on \(SU(2)\)-structure manifolds, and relate it to the five-dimensional case.
Since type IIA string theory can be obtained from M-theory by compactifying it on a circle, our results should be connected
to the four-dimensional theory in the same way.
Thus it is possible to take the dictionary from \cite{Schon:2006kz} where exactly the relevant compactification of \(\cN = 4\), \(d = 5\) supergravity is described, and uplift the existing results to five dimensions.

It has been worked out in \cite{Danckaert:2011ju} how to group the vectors in four dimensions into \(SO(6,\tilde n)\) representations
\begin{equation}\begin{aligned}
A^{\tilde M+} &= \left(G^i, \tilde{B}^{\bar\imath}, A, \tilde{C}_{12}, C^J \right), \\
A^{\tilde M-} &= \left(B^i, \tilde{G}^{\bar\imath}, C_{12}, \tilde{A}, \tilde{C}^J \right)\, ,
\end{aligned}\end{equation}
where \(A^{\tilde M-}\) is the magnetic dual of \(A^{\tilde M+}\).\footnote{
We use indices \(\tilde M, \tilde N, \dots = 1, \dots, 6 + \tilde n\) for the \(SO(6,\tilde n)\) to distinguish them from the \(SO(5,\tilde n - 1)\) indices \(M, N, \dots\).
Notice also that the \(d = 4\) theory contains one additional vector multiplet compared to \(d = 5\), so \(SO(5,\tilde n - 1)\) in five dimensions corresponds indeed to \(SO(6,\tilde n)\) in four dimensions.
} The \(SO(6,\tilde n)\) metric is given by
\begin{equation}\label{eq:d4metric}
\eta_{\tilde M\tilde N} =
\begin{pmatrix}
0 & \delta_{i\jb} & 0 & 0 & 0 \\
\delta_{\ib j} & 0 & 0 & 0 & 0 \\
0 & 0 & 0 & 1 & 0 \\
0 & 0 & 1 & 0 & 0 \\
0 & 0 & 0 & 0 & \eta_{IJ}
\end{pmatrix}\,.
\end{equation}
It is now necessary to determine how to break \(A^{\tilde M+}\) and \(A^{\tilde M-}\) into \(SO(5,\tilde n-1)\) representations.
Therefore we will write \(\tilde M = \{M, \oplus, \ominus\}\).
Obviously \(A\) does not appear in the five-dimensional case.
When tracing back its origin from the reduction of M-theory to IIA supergravity, it is clear that it is the Kaluza-Klein vector coming from reducing the five-dimensional metric to four dimensions. Thus according to \cite{Schon:2006kz} we have to identify it with \(A^{\ominus+}\) and its magnetic dual with \(A^{\oplus-}\).
This makes it furthermore possible to fix \(A^{\oplus+} = \tilde{C}_{12}\) and \(A^{\ominus-} = C_{12}\).
Lastly \(B^i\) and \(\tilde{B}^{\bar\imath}\) do not appear in the five-dimensional theory as well, but since they originate from \(C^i\) and \(\tilde{C}^{\bar\imath}\), they can simply be replaced by the latter.
Using this information, the correct identification of the five-dimensional vectors with \(A^M\) and \(A^0\) is
\begin{equation}\begin{aligned}
A^M &= A^{M+} = \left(G^i, \tilde{C}^{\bar\imath}, C^J \right)\,, \\ 
A^0 &= A^{\ominus-} = C_{12}\,,
\end{aligned}\end{equation}
which reproduces our original results. Furthermore we can obtain \eqref{eq:su2metric} by crossing out the fifth and sixth row and line from \eqref{eq:d4metric}.

Note that we can also get \(\Sigma\) and the scalar metric \(M_{MN}\) from the four-dimensional results in \cite{Danckaert:2011ju}.
Namely \eqref{eq:scalarmetric} can be obtained from the four-dimensional \(M_{\tilde M\tilde N}\) by replacing \(\beta\) with \(\gamma\) and removing all scalars that do not exist in the five-dimensional theory. \(\Sigma\) is related to the \(M_{66}\) component
in four dimensions whereby here the additional factor of \(\I\,\tau\) and the different Weyl rescalings of the metric in four and five dimensions have to be taken into account.

Furthermore in \cite{Schon:2006kz} formulae are provided for the reduction of the embedding tensors
which together with the expressions from \cite{KashaniPoor:2013en} yield
\begin{equation}\begin{aligned}
\xi_i &= 2 f_{+i\oplus\ominus} = -2 f_{+i56}= - \epsilon_{ij}t^j\,, \\
\xi_{iI} &= f_{-\ominus iI} = f_{-5IJ} = \epsilon_{ij}t^j_I\,, \\
f_{ij\bar\imath} &= f_{+ij\bar\imath} = \delta_{\bar\imath[i}\epsilon_{j]k}t^k\,, \\
f_{iIJ} &= f_{+iIJ} = - T^K_{iI} \eta_{KJ} - \tfrac{1}{2}\epsilon_{ij}t^j \eta_{IJ} \,.
\end{aligned}\end{equation}
For \(\xi_{i}\) one can equally well use the relation
\begin{equation}
\xi_i = \xi_{+i} = - \epsilon_{ij}t^j \,.
\end{equation}

\chapter{Lie Theory}\label{c:lie}

\section{Lie Theory Conventions}\label{s:lie_conv}

In this appendix we summarize our conventions for the Lie algebra theory used in 
this thesis.

We consider a simple Lie algebra $\mathfrak{g}$ associated to the Lie group $G$. 
The definition of a (preliminary) basis of Cartan generators $\lbrace\tilde T_I \rbrace$ with 
\begin{align}
 \tr_{f} ( \tilde T_I \tilde T_J ) = \delta_{IJ} \ ,
\end{align}
will allow us to fix the normalization of the root lattice. The trace $\tr_{f}$ is taken in the fundamental representation.
We denote the simple roots by $\boldsymbol{\alpha}_I$, $I=1,\dots ,\rk \mathfrak{g}$, the simple coroots
are denoted by $\boldsymbol{\alpha}_I^\vee := \frac{2 \boldsymbol{\alpha}_I}{\langle \boldsymbol{\alpha}_I , \boldsymbol{\alpha}_I \rangle}$.

In order to match with the geometric setup it is important to introduce a 
coroot-basis $\lbrace T_I \rbrace$ for the Cartan-subalgebra. It is defined by
\begin{align}
 T_I := \frac{2\,\boldsymbol{\alpha}_I^J \tilde T_J}{\langle\boldsymbol{\alpha}_I ,\boldsymbol{\alpha}_I \rangle}
\end{align}
with $\boldsymbol{\alpha}_I^J$ the components of the simple roots.
We furthermore define the (normalized)
coroot intersection matrix $\cC_{IJ}$ as
\begin{align}\label{e:def_coroot_int_mat}
 \cC_{IJ} = \lambda_{\mathfrak{g}}^{-1} \langle \boldsymbol{\alpha}^\vee_I , \boldsymbol{\alpha}^\vee_J \rangle \ ,
\end{align}
with
\begin{align}
 2 \lambda_{\mathfrak{g}}^{-1} = \langle \boldsymbol{\alpha}_{\textrm{max}}, \boldsymbol{\alpha}_{\textrm{max}} \rangle  \  ,
\end{align}
where $\boldsymbol{\alpha}_{\textrm{max}}$ is the root of maximal length.
The normalization of the Cartan generators $T_I$ (in the coroot basis) is then given by
\begin{align}
 \tr_{f} (T_I T_J) = \lambda_{\mathfrak{g}} \, \cC_{IJ} \, .
\end{align}
Furthermore for some weight $w$ the Dynkin labels are defined as
\begin{align}
 w_I := \langle \boldsymbol{\alpha}^\vee_I , w \rangle \, .
\end{align}
The Coxeter labels $a^I$ denote the components of the highest root $\theta$ in the expansion
\begin{align}\label{def:cox}
 \theta =: \sum_{I} a^I \boldsymbol{\alpha}_I \, .
\end{align}
Finally, in \autoref{t:lie_conventions} we display the numbering of the nodes in the Dynkin diagrams, the Coxeter labels and the definition of the fundamental
representations of all simple Lie algebras as well as the values for the normalization factors $\lambda_{\mathfrak{g}}$ in our conventions.
\begin{table}
\begin{center}
\resizebox{\linewidth}{!}{\begin{tabular}{|m{1.0cm}||m{5.3cm}|m{3.3cm}|m{3.3cm}|m{0.4cm}|}
\hline
Type & Dynkin diagram &  Coxeter labels & Fund. rep. & $\lambda_{\mathfrak{g}}$\\
\hline \hline
$A_n$ & \scalebox{3}{\begin{dynkin}\tikzstyle{every node}=[font=\tiny\tiny]
    \node[align=left, scale=0.5] at (\dynkinstep*1,-.15cm){1};
    \node[align=left, scale=0.5] at (\dynkinstep*2,-.15cm){2};
    \node[align=left, scale=0.5] at (\dynkinstep*3,-.15cm){3};
    \node[align=left, scale=0.5] at (\dynkinstep*6,-.15cm){n-1};
    \node[align=left, scale=0.5] at (\dynkinstep*7,-.162cm){n};
    \node[align=left, scale=0.5] at (\dynkinstep*1,.1cm){};
    \dynkinline{1}{0}{4}{0};
    \dynkindots{4}{0}{5}{0};
    \dynkinline{5}{0}{7}{0};
    \foreach \x in {1,2,3,6,7}
    {
       \dynkindot{\x}{0}
    }
  \end{dynkin}} & $(1,1,1,\dots,1,1)$ & $(1,0,0,\dots,0,0)$ & 1\\
\hline 
$B_n$ & \scalebox{3}{\begin{dynkin}\tikzstyle{every node}=[font=\tiny]
    \node[align=left, scale=0.5] at (\dynkinstep*1,-.15cm){1};
    \node[align=left, scale=0.5] at (\dynkinstep*2,-.15cm){2};
    \node[align=left, scale=0.5] at (\dynkinstep*3,-.15cm){3};
    \node[align=left, scale=0.5] at (\dynkinstep*6,-.15cm){n-1};
    \node[align=left, scale=0.5] at (\dynkinstep*7,-.162cm){n};
    \node[align=left, scale=0.5] at (\dynkinstep*1,.1cm){};
    \dynkinline{1}{0}{4}{0};
    \dynkindots{4}{0}{5}{0};
    \dynkinline{5}{0}{6}{0}
    \dynkindoubleline{6}{0}{7}{0};
    \foreach \x in {1,2,3,6,7}
    {
       \dynkindot{\x}{0}
    }
  \end{dynkin}} & $(1,2,2,\dots,2,2)$ & $(1,0,0,\dots,0,0)$ & 2\\ 
\hline 
$C_n$ & \scalebox{3}{\begin{dynkin}\tikzstyle{every node}=[font=\tiny]
    \node[align=left, scale=0.5] at (\dynkinstep*1,-.15cm){1};
    \node[align=left, scale=0.5] at (\dynkinstep*2,-.15cm){2};
    \node[align=left, scale=0.5] at (\dynkinstep*3,-.15cm){3};
    \node[align=left, scale=0.5] at (\dynkinstep*6,-.15cm){n-1};
    \node[align=left, scale=0.5] at (\dynkinstep*7,-.162cm){n};
    \node[align=left, scale=0.5] at (\dynkinstep*1,.1cm){};
    \dynkinline{1}{0}{4}{0};
    \dynkindots{4}{0}{5}{0};
    \dynkinline{5}{0}{6}{0}
    \dynkindoubleline{7}{0}{6}{0};
    \foreach \x in {1,2,3,6,7}
    {
       \dynkindot{\x}{0}
    }
  \end{dynkin}} & $(2,2,2,\dots,2,1)$ & $(1,0,0,\dots,0,0)$ & 1\\
\hline 
$D_n$ & \scalebox{3}{\begin{dynkin}\tikzstyle{every node}=[font=\tiny]
    \node[align=left, scale=0.5] at (\dynkinstep*1,-.15cm){1};
    \node[align=left, scale=0.5] at (\dynkinstep*2,-.15cm){2};
    \node[align=left, scale=0.5] at (\dynkinstep*3,-.15cm){3};
    \node[align=left, scale=0.5] at (\dynkinstep*6,-.15cm){n-2};
    \node[align=left, scale=0.5] at (\dynkinstep*7,-.35cm){n-1};
    \node[align=left, scale=0.5] at (\dynkinstep*7,.35cm){n};
    \dynkinline{1}{0}{4}{0};
    \dynkindots{4}{0}{5}{0};
    \dynkinline{5}{0}{6}{0}
    \foreach \x in {1,2,3,6}
    {
       \dynkindot{\x}{0}
    }
    \dynkindot{7}{.8}
    \dynkindot{7}{-.8}
    \dynkinline{6}{0}{7}{.8}
    \dynkinline{6}{0}{7}{-.8}
  \end{dynkin}} & $(1,2,2,\dots,2,1,1)$ & $(1,0,0,\dots,0,0,0)$ & 2\\
\hline 
$E_6$ & \scalebox{3}{\begin{dynkin}\tikzstyle{every node}=[font=\tiny]
    \node[align=left, scale=0.5] at (\dynkinstep*1,-.15cm){1};
    \node[align=left, scale=0.5] at (\dynkinstep*2,-.15cm){3};
    \node[align=left, scale=0.5] at (\dynkinstep*3,-.15cm){4};
    \node[align=left, scale=0.5] at (\dynkinstep*4,-.15cm){5};
    \node[align=left, scale=0.5] at (\dynkinstep*5,-.15cm){6};
    \node[align=left, scale=0.5] at (\dynkinstep*3.5,.25cm){2};
    \foreach \x in {1,...,5}
    {
        \dynkindot{\x}{0}
    }
    \dynkindot{3}{1}
    \dynkinline{1}{0}{5}{0}
    \dynkinline{3}{0}{3}{1}
  \end{dynkin}} & $(1,2,2,3,2,1)$ & $(0,0,0,0,0,1)$ & 6\\
\hline 
$E_7$ & \scalebox{3}{\begin{dynkin}\tikzstyle{every node}=[font=\tiny]
    \node[align=left, scale=0.5] at (\dynkinstep*1,-.15cm){1};
    \node[align=left, scale=0.5] at (\dynkinstep*2,-.15cm){3};
    \node[align=left, scale=0.5] at (\dynkinstep*3,-.15cm){4};
    \node[align=left, scale=0.5] at (\dynkinstep*4,-.15cm){5};
    \node[align=left, scale=0.5] at (\dynkinstep*5,-.15cm){6};
    \node[align=left, scale=0.5] at (\dynkinstep*6,-.15cm){7};
    \node[align=left, scale=0.5] at (\dynkinstep*3.5,.25cm){2};
    \foreach \x in {1,...,6}
    {
        \dynkindot{\x}{0}
    }
    \dynkindot{3}{1}
    \dynkinline{1}{0}{6}{0}
    \dynkinline{3}{0}{3}{1}
  \end{dynkin}}& $(2,2,3,4,3,2,1)$  & $(0,0,0,0,0,0,1)$ & 12\\
\hline 
$E_8$ & \scalebox{3}{\begin{dynkin}\tikzstyle{every node}=[font=\tiny]
    \node[align=left, scale=0.5] at (\dynkinstep*1,-.15cm){1};
    \node[align=left, scale=0.5] at (\dynkinstep*2,-.15cm){3};
    \node[align=left, scale=0.5] at (\dynkinstep*3,-.15cm){4};
    \node[align=left, scale=0.5] at (\dynkinstep*4,-.15cm){5};
    \node[align=left, scale=0.5] at (\dynkinstep*5,-.15cm){6};
    \node[align=left, scale=0.5] at (\dynkinstep*6,-.15cm){7};
    \node[align=left, scale=0.5] at (\dynkinstep*7,-.15cm){8};
    \node[align=left, scale=0.5] at (\dynkinstep*3.5,.25cm){2};
    \foreach \x in {1,...,7}
    {
        \dynkindot{\x}{0}
    }
    \dynkindot{3}{1}
    \dynkinline{1}{0}{7}{0}
    \dynkinline{3}{0}{3}{1}
  \end{dynkin}}& $(2,3,4,6,5,4,3,2)$  & $(0,0,0,0,0,0,0,1)$ & 60\\
\hline 
$F_4$ &  \scalebox{3}{\begin{dynkin}\tikzstyle{every node}=[font=\tiny]
    \node[align=left, scale=0.5] at (\dynkinstep*1,-.15cm){1};
    \node[align=left, scale=0.5] at (\dynkinstep*2,-.15cm){2};
    \node[align=left, scale=0.5] at (\dynkinstep*3,-.15cm){3};
    \node[align=left, scale=0.5] at (\dynkinstep*4,-.15cm){4};
    \node[align=left, scale=0.5] at (\dynkinstep*1,.1cm){};
    \dynkindoubleline{2}{0}{3}{0}
    \foreach \x in {1,...,4}
    {
        \dynkindot{\x}{0}
    }
    \dynkinline{1}{0}{2}{0}
    \dynkinline{3}{0}{4}{0}
  \end{dynkin}}& $(2,3,4,2)$   & $(0,0,0,1)$ & 6\\
\hline 
$G_2$ &  \scalebox{3}{\begin{dynkin}\tikzstyle{every node}=[font=\tiny]
    \node[align=left, scale=0.5] at (\dynkinstep*1,-.15cm){1};
    \node[align=left, scale=0.5] at (\dynkinstep*2,-.15cm){2};
    \node[align=left, scale=0.5] at (\dynkinstep*1,.1cm){};
    \dynkintripleline{2}{0}{1}{0}
    \foreach \x in {1,2}
    {
        \dynkindot{\x}{0}
    }
  \end{dynkin}} & $(3,2)$  & $(1,0)$ & 2 \\
\hline
\end{tabular}}
\end{center}
\caption{We display our conventions for the simple Lie algebras.}
\label{t:lie_conventions}
\end{table}

\section{Trace Identities}\label{app:traces}
In the following we show that the factors appearing in trace reductions of highest weight representations of simple Lie algebras can be related to
certain sums over the weights in that representation. This will allow us to relate one-loop Chern-Simons terms to non-Abelian anomaly
cancelation conditions
since sums over weights are evaluated in the former while factors of trace reductions appear in the latter. It also gives a general tool
to evaluate trace factors in a straightforward way.

\subsection{Quadratic Trace Identities}
We start with the evaluation of the quadratic trace identity, \textit{i.e.}~we relate the quantity $A_R$ defined by
\begin{align}\label{e:quadratic_trace}
 \tr_R \hat F^2 = A_R\, \tr_f \hat F^2 
\end{align}
to a sum over weights. This was done in \cite{Grimm:2013oga}, and in contrast to the cubic and quartic trace identities the result takes a very simple form
\begin{align}
 \sum_{w \in R}\,w_{I} w_{J} = A_R \lambda_{\mathfrak{g}} \cC_{IJ} \, .
\end{align}
This equation holds for all highest weight representations of any simple Lie algebra.

\subsection{Cubic Trace Identities}
We now show that the conditions \eqref{4d_anomaly_alt2} and \eqref{e:6d_alter_6}
\begin{subequations}\label{e:cubic_CS_matching}
\begin{align}
 \sum_{R,q} F_{\sfrac 1 2}(R,q)\sum_{w \in R}\,w_{I} w_{J} w_{K} &= 0  \, , \\
 \sum_{R,q} F_{\fe}(R,q) \sum_{w \in R}\, q_m w_{I} w_{J}  w_{K} &=0 \, ,
\end{align}
\end{subequations}
are, depending on the choice of indices, either trivially fulfilled (as a group theoretical identity)
or equivalent to the four- and six-dimensional
anomaly conditions \eqref{4d_nA_anomaly} and \eqref{e:6d_anom_5}
\begin{subequations}\label{e:cubic_anomaly}
\begin{align}
 \sum_{R,q} F_{\sfrac 1 2}(R,q) \, E_R &= 0 \, , \\
 \sum_{R,q} F_{\fe}(R,q) \  q_m E_R &= 0 \, ,
\end{align}
\end{subequations}
where $E_R$ appears in the trace reduction
\begin{align}\label{e:cubic_trace}
 \tr_R \hat F^3 = E_R\, \tr_f \hat F^3 \, .
\end{align}
Expanding the traces we can write \eqref{e:cubic_trace} as
\begin{align}
 \hat F^I \hat F^J \hat F^K \sum_{w \in R} w_I w_J w_K = \hat F^I \hat F^J \hat F^K \, E_R \, \sum_{w^f} w_I^f w_J^f w_K^f \, ,
\end{align}
where $\hat F=\hat F^I T_I$ and we sum over all weights, in particular $w^f$ denote the weights of the fundamental representation.
Considering this equation as a generating function we find
\begin{align}\label{e:cubic_weight_sum}
 \sum_{w \in R} w_I w_J w_K = E_R \, \sum_{w^f} w_I^f w_J^f w_K^f \, .
\end{align}
The key point is now to try to generally evaluate the sum over the fundamental weights on the right hand side.
This procedure indeed will allow us to relate the factor $E_R$ to a certain sum over the weights in the representation $R$ which appears
in the calculation of one-loop Chern-Simons terms. In the following we carry this out for all simple Lie algebras.

\vspace{2cm}

{\noindent\large\framebox{$\bf A_1$, $\bf B_{n\geq 3}$, $\bf C_{n\geq 2}$, $\bf D_{n\geq 4}$, $\bf E_6$, $\bf E_7$, $\bf E_8$, $\bf F_4$, $\bf G_2$}}
 
 \vspace{0.4cm}
\noindent
 For these algebras there exists no cubic Casimir operator which is why non-Abelian anomalies are always trivially absent and one
 therefore defines $E_R = 0$. Via \eqref{e:cubic_weight_sum} the conditions from the one-loop Chern-Simons matchings
 \eqref{e:cubic_CS_matching} are then equivalent to the anomaly cancelation conditions
 \eqref{e:cubic_anomaly}.\footnote{Note also
 that the condition $\sum_{w^f} w_I^f w_J^f w_K^f = 0 \, \, \, \, \forall \, I,J,K$ precisely means that
 there is no cubic Casimir and one then also has by definition $E_R =0$.}
 
 \vspace{2cm}
 
{\noindent\large\framebox{ $\bf A_{n \neq 1}$ }}
 
  \vspace{0.4cm}
\noindent
 For $A_{n \neq 1}$ there exists a cubic Casimir and we start by
 explicitly evaluating the traces over the fundamental weights for different index choices.
 
 \begin{enumerate}[label=(\alph*)]\setlength{\itemindent}{1.5cm}
 \item $I=J=K$
 \begin{itemize}[leftmargin=.6in]
  \item[] We calculate
 \begin{align}
  \sum_{w^f} ( w_I^f )^3 = 0 \, 
 \end{align}
such that we can conclude using \eqref{e:cubic_weight_sum}
 \begin{align}
  \sum_{w \in R} ( w_I )^3 = 0 \, .
 \end{align}
The corresponding
  Chern-Simons matchings \eqref{e:cubic_CS_matching} are therefore trivial and impose no restrictions on the spectrum.
 \end{itemize}

 \item $I=K\neq J$
 \begin{itemize}[leftmargin=.6in]
  \item[] 
 Now we evaluate
 \begin{align}
  \sum_{w^f} ( w_I^f )^2 w_J^f = (I-J)\, \cC_{IJ} \, \, .
 \end{align}
 With \eqref{e:cubic_weight_sum} the Chern-Simons matchings \eqref{e:cubic_CS_matching} in this case become
 \begin{subequations}
 \begin{align}
  \sum_{R,q} F_{\sfrac 1 2}(R,q) \, E_R \, (I-J)\, \cC_{IJ} = 0 \, , \\
  \sum_{R,q} F_{\sfrac 1 2}(R,q) \, q_m E_R \, (I-J)\, \cC_{IJ} = 0 \, ,
 \end{align}
  \end{subequations}
 which are equivalent to the anomaly conditions \eqref{e:cubic_anomaly}.
 \end{itemize}
 
 \item $I\neq J \neq K$
 \begin{itemize}[leftmargin=.6in]
  \item[] 
 Finally it turns out that
 \begin{align}
  \sum_{w^f} w_I^f w_J^f w_K^f = 0 \, ,
 \end{align}
 which is why the Chern-Simons matchings are again trivial like in the case $I=J=K$.
 \end{itemize} 
 \end{enumerate}

 \vspace{2cm}

To put it in a nutshell we have shown that the Chern-Simons matchings \eqref{e:cubic_CS_matching} are completely equivalent to the anomaly cancelation
conditions \eqref{e:cubic_anomaly} for all simple Lie algebras.

We stress that here and in the following writing down
the kind of expansions $\tr_R \hat F^3 = \hat F^I \hat F^J \hat F^K \sum_{w \in R} w_I w_J w_K$ would already suffice in order to
show that the consistent action of large gauge transformations on Chern-Simons couplings is equivalent to the cancelation of anomalies.
This is easy to see since $\tr_R \hat F^3$ appears in the anomaly polynomial and $\sum_{w \in R} w_I w_J w_K$ in the variation of
Chern-Simons terms. However, as usually these conditions are written down by using the Casimir $E_R$ and all traces
are transferred to the fundamental
representation, we take a little more effort and relate Casimir operators to sums over weights. This procedure also yields convenient
formulae for the latter which are quite useful.

\subsection{Quartic Trace Identities}\label{app:quart_tr}
Let us perform the same steps as in the last subsection now for quartic traces.
More precisely we show that the condition \eqref{e:6d_alter_5}
\begin{align}\label{e:6d_CS_match}
   \sum_{R,q}  F_{\fe}(R,q) \sum_{w \in R} \ w_{I} w_{J}  w_{K} w_{L} =  - 3 b^{\alpha} b^{\beta} \eta_{\alpha\beta} \cC_{(I J}
\cC_{KL)} 
\end{align}
is equivalent to the six-dimensional pure non-Abelian gauge anomalies \eqref{e:6d_anom_3} and \eqref{e:6d_anom_4}
\begin{subequations}\label{e:6d_anomalies_nA}
\begin{align}
 \sum_{R,q} F_{\fe}(R,q) B_R &= 0 \, , \\
  \sum_{R,q} F_{\fe}(R,q) C_R  &= -3 \frac{b^\alpha}{\lambda_{\mathfrak{g}}} \frac{b^\beta}{\lambda_{\mathfrak{g}}} \eta_{\alpha\beta}\, ,
  \end{align}
\end{subequations}
where the constants $B_R , C_R$ are defined as
\begin{align}\label{e:quartic_trace_red}
 \tr_R \hat F^4 = B_R \, \tr_f \hat F^4 + C_R \, (\tr_f \hat F^2)^2 \, .
\end{align}
Expanding the traces on both sides of \eqref{e:quartic_trace_red} and taking derivatives with respect to $\hat F^I$
we obtain in analogy to \eqref{e:cubic_weight_sum}
\begin{align}\label{e:quartic_trace_weight_sum}
 &\sum_{w \in R} w_I w_J w_K w_L \nn \\
 &\,\,\, = B_R \sum_{w^f} w^f_I w^f_J w^f_K w^f_L  
  +\frac{C_R}{3} \bigg [ \Big(\sum_{w^f} w_I^f w_J^f\Big ) \Big(\sum_{ w^{' f}} w_K^{' f} w_L^{' f}\Big ) + \Big(\sum_{w^f} w_I^f w_K^f\Big )
  \Big(\sum_{w^{' f}} w_J^{' f} w_L^{' f}\Big ) \nn\\
 &\hspace{175pt}+ \Big(\sum_{w^f} w_I^f w_L^f\Big ) \Big(\sum_{w^{' f}} w_J^{' f} w_K^{' f}\Big ) \bigg ] \, .
\end{align}

Like in the preceding subsection we now explicitly evaluate the sums over the fundamental weights in order to rewrite \eqref{e:quartic_trace_weight_sum}.
For the different simple Lie algebras and all possible choices of indices \eqref{e:quartic_trace_weight_sum} then becomes:

\vspace{2cm}

{\noindent\large\framebox{$\bf A_n \, , \quad n\geq 1$}}

 \begin{enumerate}[label=(\alph*)]\setlength{\itemindent}{1.5cm}
	\item $I=J=K=L$
	\begin{align}
	 \sum_{w \in R} (w_I)^4 = B_R \,\, \mathcal{C}_{II}\,\, \lambda_{\mathfrak{g}} + C_R \,\, \mathcal{C}_{II}^2 \,\, \lambda_{\mathfrak{g}}^2 \, ,
	\end{align}
	\item $I=K=L$, $I \neq J$
	\begin{align}
	 \sum_{w \in R} (w_I)^3 \,  w_J = B_R \,\, \mathcal{C}_{IJ}\,\, \lambda_{\mathfrak{g}} + C_R  \,\,
	 \mathcal{C}_{II} \,\,\mathcal{C}_{IJ} \,\, \lambda_{\mathfrak{g}}^2  \, ,
	\end{align}
	\item $I=L$, $I \neq J \neq K$
	\begin{align}
	 \sum_{w \in R} (w_I)^2 \,  w_J w_K = \frac{1}{3}\,\,C_R \,\big ( 2\,\, \mathcal{C}_{IJ} \,\,\mathcal{C}_{IK} + \mathcal{C}_{II} \,\,\mathcal{C}_{JK} \big )
	 \,\, \lambda_{\mathfrak{g}}^2  \, ,
	\end{align}
	\item $I \neq J \neq K \neq L$
	\begin{align}
	 \sum_{w \in R} w_I w_J w_K w_L = C_R \, \, \mathcal{C}_{(IJ} \,\,\mathcal{C}_{KL)} 
	 \, \, \lambda_{\mathfrak{g}}^2  \, .
	\end{align}
 \end{enumerate}
 We can now insert these equations into the Chern-Simons matching \eqref{e:6d_CS_match} and find two linearly independent equations
 \begin{align}
  \sum_{R,q} F_{\fe}(R,q) \bigg(\frac{1}{2}\,B_R + C_R \bigg)  &= -3\, \frac{b^\alpha}{\lambda_{\mathfrak{g}}}
  \,\,\frac{b^\beta}{\lambda_{\mathfrak{g}}} \eta_{\alpha\beta} \, , \\
   \sum_{R,q} F_{\fe}(R,q)\,\, C_R   &= -3\, \frac{b^\alpha}{\lambda_{\mathfrak{g}}}
  \,\,\frac{b^\beta}{\lambda_{\mathfrak{g}}} \eta_{\alpha\beta} \, . 
 \end{align}
These equations are in fact equivalent to the gauge anomaly conditions \eqref{e:6d_anomalies_nA}.

\vspace{2cm}

{\noindent\large\framebox{$\bf B_n \, , \quad n\geq 3$}}

 \begin{enumerate}[label=(\alph*)]\setlength{\itemindent}{1.5cm}
	\item $I=J=K=L$
	\begin{align}
	 \sum_{w \in R} (w_I)^4 = \frac{1}{4}\,B_R \,\,\mathcal{C}_{In}^2\,\, \mathcal{C}_{II}\,\, \lambda_{\mathfrak{g}}
	 + C_R \,\, \mathcal{C}_{II}^2 \,\, \lambda_{\mathfrak{g}}^2  \, ,
	\end{align}
	\item $I=K=L$, $I \neq J$
	\begin{align}
	 \sum_{w \in R} (w_I)^3 \,  w_J = \frac{1}{4}\,B_R \,\,\mathcal{C}_{In}^2\,\,\mathcal{C}_{IJ}\,\, \lambda_{\mathfrak{g}} + C_R  \,\,
	 \mathcal{C}_{II} \,\,\mathcal{C}_{IJ} \,\, \lambda_{\mathfrak{g}}^2  \, ,
	\end{align}
	\item $I=L$, $I \neq J \neq K$
	\begin{align}
	 \sum_{w \in R} (w_I)^2 \,  w_J w_K = 
	 \frac{1}{3}\,\,C_R \,\big ( 2\,\, \mathcal{C}_{IJ} \,\,\mathcal{C}_{IK} + \mathcal{C}_{II} \,\,\mathcal{C}_{JK} \big )
	 \,\, \lambda_{\mathfrak{g}}^2  \, ,
	\end{align}
	\item $I \neq J \neq K \neq L$
	\begin{align}
	 \sum_{w \in R} w_I w_J w_K w_L = C_R \,\, \mathcal{C}_{(IJ} \,\,\mathcal{C}_{KL)} 
	 \,\, \lambda_{\mathfrak{g}}^2  \, .
	\end{align}
 \end{enumerate}
 Insertion into \eqref{e:6d_CS_match} yields
 \begin{align}
  \sum_{R,q} F_{\fe}(R,q) \bigg(\frac{1}{4}\,B_R + C_R \bigg)  &= -3\, \frac{b^\alpha}{\lambda_{\mathfrak{g}}}
  \,\,\frac{b^\beta}{\lambda_{\mathfrak{g}}} \eta_{\alpha\beta} \, , \\
   \sum_{R,q} F_{\fe}(R,q)\,\, C_R   &= -3\, \frac{b^\alpha}{\lambda_{\mathfrak{g}}}
  \,\,\frac{b^\beta}{\lambda_{\mathfrak{g}}} \eta_{\alpha\beta} \, , 
 \end{align}
which is equivalent to \eqref{e:6d_anomalies_nA}.

\vspace{2cm}

{\noindent\large\framebox{$\bf C_n \, , \quad n\geq 2$}}
 
 \begin{enumerate}[label=(\alph*)]\setlength{\itemindent}{1.5cm}
	\item $I=J=K=L$
	\begin{align}
	 \sum_{w \in R} (w_I)^4 = B_R \,\, \mathcal{C}_{II}\,\, \lambda_{\mathfrak{g}} + C_R \,\, \mathcal{C}_{II}^2 \,\, \lambda_{\mathfrak{g}}^2  \, ,
	\end{align}
	\item $I=K=L$, $I \neq J$
	\begin{align}
	 \sum_{w \in R} (w_I)^3 \,  w_J = B_R \,\, \mathcal{C}_{IJ}\,\, \lambda_{\mathfrak{g}} + C_R  \,\,
	 \mathcal{C}_{II} \,\,\mathcal{C}_{IJ} \,\, \lambda_{\mathfrak{g}}^2 \, ,
	\end{align}
	\item $I=L$, $I \neq J \neq K$
	\begin{align}
	 \sum_{w \in R} (w_I)^2 \,  w_J w_K = 
	 \frac{1}{3}\,\,C_R \,\big ( 2\,\, \mathcal{C}_{IJ} \,\,\mathcal{C}_{IK} + \mathcal{C}_{II} \,\,\mathcal{C}_{JK} \big )
	 \,\, \lambda_{\mathfrak{g}}^2  \, ,
	\end{align}
	\item $I \neq J \neq K \neq L$
	\begin{align}
	 \sum_{w \in R} w_I w_J w_K w_L = C_R \,\,  \mathcal{C}_{(IJ} \,\,\mathcal{C}_{KL)} 
	 \,\, \lambda_{\mathfrak{g}}^2  \, ,
	\end{align}
 \end{enumerate}
 which can be inserted into \eqref{e:6d_CS_match}
 \begin{align}
  \sum_{R,q} F_{\fe}(R,q) \bigg(\frac{1}{4}\,B_R + C_R \bigg)  &= -3\, \frac{b^\alpha}{\lambda_{\mathfrak{g}}}
  \,\,\frac{b^\beta}{\lambda_{\mathfrak{g}}} \eta_{\alpha\beta} \, , \\
   \sum_{R,q} F_{\fe}(R,q)\,\, C_R   &= -3\, \frac{b^\alpha}{\lambda_{\mathfrak{g}}}
  \,\,\frac{b^\beta}{\lambda_{\mathfrak{g}}} \eta_{\alpha\beta} \, . 
 \end{align}
These equations are equivalent to the anomaly conditions \eqref{e:6d_anomalies_nA}.

\vspace{2cm}

{\noindent\large\framebox{$\bf D_n \, , \quad n\geq 4$}}
 
 \begin{enumerate}[label=(\alph*)]\setlength{\itemindent}{1.5cm}
	\item $I=J=K=L$
	\begin{align}
	 \sum_{w \in R} (w_I)^4 = B_R \,\, \mathcal{C}_{II}\,\, \lambda_{\mathfrak{g}} + C_R \,\, \mathcal{C}_{II}^2 \,\, \lambda_{\mathfrak{g}}^2  \, ,
	\end{align}
	\item $I=K=L$, $I \neq J$
	\begin{align}
	 \sum_{w \in R} (w_I)^3 \,  w_J = B_R \,\, \mathcal{C}_{IJ}\,\, \lambda_{\mathfrak{g}} + C_R  \,\,
	 \mathcal{C}_{II} \,\,\mathcal{C}_{IJ} \,\, \lambda_{\mathfrak{g}}^2 \, ,
	\end{align}
	\item $I=L$, $I \neq J \neq K$
	\begin{align}
	 \sum_{w \in R} (w_I)^2 \,  w_J w_K = \alpha_{IJK}\,B_R +
	 \frac{1}{3}\,\,C_R \,\big ( 2\,\, \mathcal{C}_{IJ} \,\,\mathcal{C}_{IK} + \mathcal{C}_{II} \,\,\mathcal{C}_{JK} \big )
	 \,\, \lambda_{\mathfrak{g}}^2  \, ,
	\end{align}
	\item $I \neq J \neq K \neq L$
	\begin{align}
	 \sum_{w \in R} w_I w_J w_K w_L = C_R \,\, \mathcal{C}_{(IJ} \,\,\mathcal{C}_{KL)} 
	 \,\, \lambda_{\mathfrak{g}}^2  \, ,
	\end{align}
 \end{enumerate}
 with the definition
 \begin{align}
  \alpha_{IJK} := 4\,\Big(\delta_{I,n-2}\,\delta_{(J,n}\,\delta_{K),n-1}-\delta_{I,n-1}\,\delta_{(J,n}\,\delta_{K),n-2}-\delta_{I,n}\,
 \delta_{(J,n-1}\,\delta_{K),n-2}\Big)
 \end{align}
 Inserting into \eqref{e:6d_CS_match} we obtain
 \begin{align}
  \sum_{R,q} F_{\fe}(R,q) \bigg(\frac{1}{4}\,B_R + C_R \bigg)  &= -3\, \frac{b^\alpha}{\lambda_{\mathfrak{g}}}
  \,\,\frac{b^\beta}{\lambda_{\mathfrak{g}}} \eta_{\alpha\beta} \, , \\
   \sum_{R,q} F_{\fe}(R,q)\,\, C_R   &= -3\, \frac{b^\alpha}{\lambda_{\mathfrak{g}}}
  \,\,\frac{b^\beta}{\lambda_{\mathfrak{g}}} \eta_{\alpha\beta} \, ,  
 \end{align}
  which is equivalent to the anomaly conditions \eqref{e:6d_anomalies_nA}.

\vspace{2cm}

{\noindent\large \framebox{$\bf E_6$, $\bf E_7$, $\bf E_8$, $\bf F_4$, $\bf G_2$}}

\vspace{0.4cm}
\noindent
For these algebras there is no fourth-order Casimir, therefore by definition $B_R = 0$ for all representations. We find by explicit calculation
  \begin{enumerate}[label=(\alph*)]\setlength{\itemindent}{1.5cm}
	\item $I=J=K=L$
	\begin{align}
	 \sum_{w \in R} (w_I)^4 =  C_R \,\, \mathcal{C}_{II}^2 \,\, \lambda_{\mathfrak{g}}^2 
	\end{align}
	\item $I=K=L$, $I \neq J$
	\begin{align}
	 \sum_{w \in R} (w_I)^3 \,  w_J =  C_R  \,\, \mathcal{C}_{II} \,\,\mathcal{C}_{IJ} \,\, \lambda_{\mathfrak{g}}^2 
	\end{align}
	\item $I=L$, $I \neq J \neq K$
	\begin{align}
	 \sum_{w \in R} (w_I)^2 \,  w_J w_K = \frac{1}{3}\,\,C_R \,\big ( 2\,\, \mathcal{C}_{IJ} \,\,\mathcal{C}_{IK} + \mathcal{C}_{II} \,\,\mathcal{C}_{JK} \big )
	 \,\, \lambda_{\mathfrak{g}}^2 
	\end{align}
	\item $I \neq J \neq K \neq L$
	\begin{align}
	 \sum_{w \in R} w_I w_J w_K w_L = C_R \,\, \mathcal{C}_{(IJ} \,\,\mathcal{C}_{KL)} 
	 \,\, \lambda_{\mathfrak{g}}^2 
	\end{align}
 \end{enumerate}
 Plugging this in into the Chern-Simons matching \eqref{e:6d_CS_match} we get
 \begin{align}
   \sum_{R,q} F_{\fe}(R,q)\,\, C_R   = -3\, \frac{b^\alpha}{\lambda_{\mathfrak{g}}}
  \,\,\frac{b^\beta}{\lambda_{\mathfrak{g}}} \eta_{\alpha\beta} \, ,
 \end{align}
which is again equivalent to the cancelation of anomalies since the first equation in \eqref{e:6d_anomalies_nA} is trivial due to the absence of
a fourth-order Casimir.

\vspace{1cm}

Thus we have shown that the matching condition from one-loop Chern-Simons terms \eqref{e:6d_CS_match} is fully equivalent
to the cancelation of non-Abelian gauge anomalies \eqref{e:6d_anomalies_nA} for all simple Lie algebras.

\chapter{Identities for Circle-Reduced Theories}
\section{One-Loop Chern-Simons Terms}\label{sec:one_loop_calc}
In this section we derive the special form of one-loop corrections to the Chern-Simons terms in three- and five-dimensional Abelian gauge
theories when the latter arise from a circle compactification of four- and six-dimensional theories, respectively.
We stress that generic VEVs for the Wilson line scalars are assumed. The following discussions and formulae do not hold
\textit{e.g.~}for integer Wilson line backgrounds.
This setup is described in \autoref{ch:circle_theories} and we will use the notation which is introduced there.
Before we go into the details let us in general introduce zeta function regularization which will be exploited in our calculations.

\subsection{Zeta Function Regularization}
Since we are analyzing circle-compactified theories in this thesis, the full contributions to the one-loop Chern-Simons terms are generically
infinite sums over Kaluza-Klein modes, which need to be regularized. In the following calculations four different
types of infinite sums do appear
\begin{align}\label{e:inf_sums}
\sum_{n = -\infty}^{+\infty}\sign (x+n)\, , \qquad &\sum_{n = -\infty}^{+\infty}n\sign (x+n)\, ,\\
\sum_{n = -\infty}^{+\infty}n^2 \sign (x+n)\, , \qquad &\sum_{n = -\infty}^{+\infty}n^3 \sign (x+n) \, , \nn
\end{align}
with some generic constant $x$ which is to be specified but is \textit{not integer}.
Note that the zeta function is defined as
\begin{align}
 \zeta(s) = \sum_{n = 1}^{\infty} n^{-s} \, , \quad \mathrm{Re}(s)>1 \, .
\end{align}
Analytic continuation of this expression yields in particular
\begin{align}
 \zeta(-1) &= - \frac{1}{12} \, , &
  \zeta(-3) &= \frac{1}{120} \, ,
\end{align}
such that we can use the following regularization scheme
\begin{align}
 \sum_{n = 1}^{\infty}n &\mapsto \zeta(-1) = - \frac{1}{12} \, , 
 &\sum_{n = 1}^{\infty}n^3 &\mapsto \zeta(-3) = \frac{1}{120} \, .
\end{align}
The sums in \eqref{e:inf_sums} then become
{\allowdisplaybreaks\begin{align}
 \sum_{n = -\infty}^{+\infty}\sign (x+n) &= 2 \Big( l + \frac{1}{2} \Big )\sign (x) \, , \\
 \sum_{n = -\infty}^{+\infty} n \sign (x+n) &= - \frac{1}{6} - l\big(l+1\big) \, , \\
 \sum_{n = -\infty}^{+\infty} n^2 \sign (x+n) &= \frac{2}{3}\, l\big(l+1\big)\Big(l + \frac{1}{2}\Big)\sign (x) \, , \\
 \sum_{n = -\infty}^{+\infty} n^3 \sign (x+n) &= \frac{1}{60} - \frac{1}{2}\, l^2 \big(l+1\big)^2 \, ,
\end{align}}
with the definition
$l := \Big \lfloor \vert x \vert \Big \rfloor \, ,$
where we make use of the floor function $\lfloor \cdot \rfloor$.

\subsection{Three Dimensions}
We are now in a position to evaluate the one-loop Chern-Simons terms of a four-dimensional gauge theory on a circle which is pushed to the
Coulomb branch (generic VEVs).
Recall from \autoref{ch:CS} that the general correction from a massive charged spin-$\sfrac{1}{2}$ Dirac fermion reads
 \begin{align}
 \Theta_{\Lambda\Sigma}^{\textrm{loop}} = \frac{1}{2} q_\Lambda q_\Sigma \sign (m) \, ,
\end{align}
and in \autoref{tab:massive_3d} we have already listed the spectrum of massive modes. We depict it here once more for convenience 
\begin{center}
\begin{tabular}{c|ccc}
\rule[-5pt]{0pt}{5pt} \bf 4d & \multicolumn{3}{c}{\bf 3d}\\
\hline
\rule{0pt}{15pt}
\multirow{2}{*}{Field}  & \multirow{2}{*}{KK-tower} & 
 \multirow{2}{*}{Mass} & $(A^0, A^I , A^m)$ \\
\rule[-5pt]{0pt}{15pt} &&& Charge \\
\hline
\rule[-10pt]{0pt}{30pt} $\hat \Bpsi^{\sfrac12} (w,q)$  & $\Bpsi^{\sfrac12}_{(n)} (w,q)$ & $m_{\rm CB}^{w,q} +
\frac{n}{\langle r\rangle}$ & $(-n, w_I , q_m)$
\end{tabular}
\end{center}
With the definition
\begin{align}
 l_{w,q} := \Bigg\lfloor \bigg\vert \frac{m^{w,q}_{\rm CB}}{m_{\rm KK}} \bigg\vert \Bigg\rfloor
\end{align}
the one-loop corrections now are evaluated by using zeta function regularization \cite{Grimm:2013oga,Cvetic:2013uta}
\allowdisplaybreaks \begin{subequations}\label{e:all_3d_loops}
\begin{align}
 \Theta_{00}& = \frac{1}{3} \sum_{R,q} F_{\sfrac 1 2}(R,q) \sum_{w \in R} \ l_{w,q} \ \big(l_{w,q} +1 \big) \ \Big(l_{w,q} +\frac{1}{2} \Big) \
 \sign  \big(m^{w,q}_{\rm CB}\big)\, ,\\
\Theta_{0I} &= \frac{1}{12}\sum_{R,q} F_{\sfrac 1 2}(R,q) \sum_{w \in R} \ \Big(1+ 6\ l_{w,q} \ \big(l_{w,q} +1 \big)\Big) \ w_I \nn \\
\label{e:3d_0I} &= \frac{1}{2}\sum_{R,q} F_{\sfrac 1 2}(R,q) \sum_{w \in R} \ l_{w,q} \ \big(l_{w,q} +1 \big) \ w_I\, , \\
\Theta_{0m} &= \frac{1}{12}\sum_{R,q}  F_{\sfrac 1 2}(R,q) \sum_{w \in R} \ \Big(1+ 6\ l_{w,q} \ \big(l_{w,q} +1 \big )\Big) \ q_m\, , \\
 \Theta_{IJ} &=  \sum_{R,q} F_{\sfrac 1 2}(R,q) \sum_{w \in R}  \ \Big(l_{w,q} +\frac{1}{2} \Big) \ w_I w_J \ \sign \big(m^{w,q}_{\rm CB}\big) \, , \\
 \Theta_{mn} &=  \sum_{R,q}  F_{\sfrac 1 2}(R,q) \sum_{w \in R} \ \Big(l_{w,q} +\frac{1}{2} \Big) \ q_m q_n \ \sign \big(m^{w,q}_{\rm CB}\big) \, \\
 \Theta_{Im} &=  \sum_{R,q} F_{\sfrac 1 2}(R,q) \sum_{w \in R}  \ \Big(l_{w,q} +\frac{1}{2} \Big) \ w_I q_m \ \sign \big(m^{w,q}_{\rm CB}\big) \, ,
\end{align}
\end{subequations}
where the sums are over all representations $R$ of the non-Abelian gauge group and the $U(1)$ charges $q$, as well as
all weights $w$ of a given representation $R$. In \eqref{e:3d_0I} we used the relation
\begin{align}\label{e:single_weight_sum}
 \sum_{w \in R} w_I = 0 \, ,
\end{align}
which holds for all highest weight representations $R$ and is proven in \cite{Grimm:2013oga}.

Finally, as an application, note that for deriving the effective action of F-theory compactifications on Calabi-Yau fourfolds one has to consider a circle-reduced four-dimensional
$\cN=1$ supergravity theory. Since the spin-$\sfrac 12$ fermions in these settings are provided by chiral multiplets, we find
\begin{align}
 F_{\sfrac 1 2}(R,q) =   C(R,q) \, ,
\end{align}
where $C(R,q)$ is defined as the number of chiral multiplets transforming in the representation $R$ and with $U(1)$ charges $q$.

\subsection{Five Dimensions}
Let us now turn to circle-compactified
six-dimensional theories on the Coulomb branch (generic VEVs).
Recall the general form of the one-loop correction to the Chern-Simons terms in five dimensions as already introduced in \autoref{ch:CS}
\begin{align}
k_{\Lambda\Sigma\Theta}^{\textrm{loop}} = c_{AFF}\, q_\Lambda q_\Sigma q_\Theta \, \sign (m) \, , \\
k_{\Lambda}^{\textrm{loop}} = c_{A\cR\cR}\, q_\Lambda \, \sign (m) \, ,
\end{align}
with $c_{AFF},c_{A\cR\cR}$ given by
\begin{center}
\begin{tabular}{c|ccc}
 & Spin-$\sfrac{1}{2}$ fermion & Self-dual tensor & Spin-$\sfrac{3}{2}$ fermion\\
 \hline
\rule{0pt}{15pt} $c_{AFF}$ & $\ \ \frac{1}{2}$ & $-2$ & $\frac{5}{2}$\\
\rule{0pt}{15pt} $c_{A\cR\cR}$ & $-1$ & $-8$ & $19$\\
\end{tabular}
\end{center}
We also display once more the spectrum of massive modes from \autoref{tab:massive_5d}
\begin{center}
\begin{tabular}{cc|cccc}
\multicolumn{2}{c|}{\rule[-5pt]{0pt}{5pt} \bf 6d} & \multicolumn{3}{c}{\bf 5d} \\
\hline
\rule{0pt}{15pt}
\multirow{2}{*}{Field} & \multirow{2}{*}{$\mathfrak{su}(2) \times \mathfrak{su}(2)$} & \multirow{2}{*}{KK-tower} & 
\multirow{2}{*}{$\mathfrak{su}(2) \times \mathfrak{su}(2)$} & \multirow{2}{*}{Mass} & $(A^0, A^I , A^m)$ \\
\rule[-5pt]{0pt}{15pt} &&&&& Charge \\
\hline
\rule[-10pt]{0pt}{30pt} $\hat \Bpsi^{\sfrac12} (w,q)$ & $(\frac{1}{2},0),(0,\frac{1}{2})$ & $\Bpsi^{\sfrac12}_{(n)} (w,q)$ &
$(\frac{1}{2},0),(0,\frac{1}{2})$ & $m_{\rm CB}^{w,q} + \frac{n}{\langle r\rangle}$ & $(-n,w_I,q_m)$ \\
\rule[-10pt]{0pt}{10pt}$\hat B^\alpha$ & $(1,0),(0,1)$ & $\BB^\alpha_{(n>0)}$ & $(1,0),(0,1)$ & $\frac{n}{\langle r\rangle}$ & $(-n,0,0)$ \\
\rule[-10pt]{0pt}{10pt}$\hat \Bpsi^{\sfrac32}_{\mu}$ & $(1,\frac{1}{2}),(\frac{1}{2},1)$ & $\Bpsi^{\sfrac32}_{\mu \, (n)}$
& $(1,\frac{1}{2}),(\frac{1}{2},1)$ & $\frac{n}{\langle r\rangle}$ & $(-n,0,0)$
\end{tabular}
\end{center}
As before we define
\begin{align}
 l_{w,q} := \Bigg\lfloor \bigg\vert \frac{m^{w,q}_{\rm CB}}{m_{\rm KK}} \bigg\vert \Bigg\rfloor
\end{align}
and use zeta function regularization in order to obtain \cite{Grimm:2013oga} 
{\allowdisplaybreaks\begin{subequations}\label{e:all_5d_loops}
\begin{align}
 k_{000} &= \frac{1}{120} \bigg (- \sum_{R,q} F_{\fe}(R,q)\sum_{w \in R} \Big( 1-30 \ l_{w,q}^2 \ \big (l_{w,q} +1 \big)^2 \Big )  
 +2 \mathfrak T -5 F_{\gr}\bigg ) \, , \\
k_{00I} &=  \frac{1}{3}\sum_{R,q} F_{\fe}(R,q) \sum_{w \in R} \ l_{w,q} \ \big(l_{w,q} +1 \big) \ \Big(l_{w,q} +\frac{1}{2}\Big) \ w_I \ \sign 
\big(m^{w,q}_{\rm CB}\big)\, ,\\
k_{00m} &=  \frac{1}{3}\sum_{R,q} F_{\fe}(R,q) \sum_{w \in R} \ l_{w,q} \ \big(l_{w,q} +1 \big) \ \Big(l_{w,q} +\frac{1}{2}
\Big) \ q_m \ \sign  \big(m^{w,q}_{\rm CB}\big)\, ,\\
k_{0IJ} &=  \frac{1}{12}\sum_{R,q} F_{\fe}(R,q) \sum_{w \in R} \ \Big(1+ 6\ l_{w,q} \ \big(l_{w,q} +1 \big)\Big) \ w_I w_J\, ,\\
k_{0mn} &=  \frac{1}{12}\sum_{R,q} F_{\fe}(R,q) \sum_{w \in R} \ \Big(1+ 6\ l_{w,q} \ \big(l_{w,q} +1 \big)\Big) \ q_m q_n\, ,\\
k_{0Im} &=  \frac{1}{12}\sum_{R,q} F_{\fe}(R,q) \sum_{w \in R} \ \Big(1+ 6\ l_{w,q} \ \big(l_{w,q} +1 \big)\Big) \ w_I q_m \nn \\
\label{e:5d_0Im} &=  \frac{1}{2}\sum_{R,q} F_{\fe}(R,q) \sum_{w \in R} \ l_{w,q} \ \big(l_{w,q} +1 \big) \ w_I q_m\, ,\\
 k_{IJK} &= \sum_{R,q} F_{\fe}(R,q) \sum_{w \in R} \ \Big(l_{w,q}  +\frac{1}{2} \Big) \ w_I w_J w_K \ \sign  \big(m^{w,q}_{\rm CB}\big)\, ,\\
 k_{mnp} &= \sum_{R,q} F_{\fe}(R,q) \sum_{w \in R} \ \Big(l_{w,q}  +\frac{1}{2}\Big) \ q_m q_n q_p \ \sign  \big(m^{w,q}_{\rm CB}\big)\, ,\\
 k_{IJm} &= \sum_{R,q} F_{\fe}(R,q) \sum_{w \in R} \ \Big(l_{w,q}  +\frac{1}{2} \Big) \ w_I w_J q_m \ \sign  \big(m^{w,q}_{\rm CB}\big)\, ,\\
k_{Imn} &= \sum_{R,q} F_{\fe}(R,q) \sum_{w \in R} \ \Big(l_{w,q} +\frac{1}{2} \Big) \ w_I q_m q_n \ \sign  \big(m^{w,q}_{\rm CB}\big)\, ,\\
k_{0} &= \frac{1}{6} \bigg (- \sum_{R,q} F_{\fe}(R,q) \sum_{w \in R} \Big( 1 + 6 \  l_{w,q} \ \big(l_{w,q} +1 \big) \Big ) 
-4 \mathfrak T + 19 F_{\gr} \bigg ) \, ,  \\
 k_{I} &= -2\sum_{R,q} F_{\fe}(R,q) \sum_{w \in R} \ \Big(l_{w,q} +\frac{1}{2} \Big) \ w_I \ \sign  \big(m^{w,q}_{\rm CB}\big) \, , \\
 k_{m} &= -2\sum_{R,q} F_{\fe}(R,q) \sum_{w \in R} \ \Big(l_{w,q} +\frac{1}{2} \Big) \ q_m \ \sign  \big(m^{w,q}_{\rm CB}\big) \, ,
\end{align}
\end{subequations}}
where we made use of \eqref{e:single_weight_sum} in \eqref{e:5d_0Im}.

Once more let us specify these general formulae to F-theory compactifications.
Putting the latter theory on Calabi-Yau threefolds one has to investigate a circle-reduced six-dimensional
$\cN=(1,0)$ supergravity theory in order to derive the effective action. We denote the number of six-dimensional
tensor multiplets by $T$, vector
multiplets by $V$ and hypermultiplets by $H$. We find
\begin{subequations}
\begin{align}
 F_{\fe} (R,q) &= (n_{U(1)}-T) \ \delta_{R, 1} \cdot \delta_{q,0} +  \delta_{R, \textrm{adj}} \cdot \delta_{q,0} - H(R,q)  \, , \\
 \mathfrak T  & = 1 - T \, , \\
 F_{\gr} &= 1 \, .
\end{align}
\end{subequations}
The factor $\delta_{R, 1} \cdot \delta_{q,0}$ is only non-vanishing for uncharged singlets and
implements the contribution of tensorini and $U(1)$ gaugini
while $\delta_{R, \textrm{adj}} \cdot \delta_{q,0}$ is different from zero for the adjoint representation, thus 
taking care of non-Abelian gaugini.

\section{Coulomb Branch Identities}\label{sec:cb_id}
In this section we show the central identity
\begin{align}
 \Big(\tilde l_{w,q} + \frac{1}{2}\Big) \sign \big(\tilde m^{w,q}_{\rm CB}\big)  - \Big(l_{w,q} + \frac{1}{2}\Big) \sign \big(m^{w,q}_{\rm CB}\big) 
 =    \fn^I w_{I} + \fn^m q_{m}
\end{align}
under the transformation \eqref{eq:LGTgeneral_1}. It again only holds for \textit{generic} Wilson line backgrounds, \textit{i.e.~}no
integer multiples of the radius.

Consider a massive mode with weight $w$ under a non-Abelian gauge group
and charges $q_m$ under Abelian gauge bosons $A^m$, Coulomb branch mass $m^{w,q}_{\rm CB}$ and KK-level $n$. 
We pick vectors $\fn^I,\fn^{m}$ and perform the basis change \eqref{eq:LGTgeneral_1}.
It is important to notice that this transformation leaves all VEVs invariant except of
\begin{align}
\langle \zeta^{I} \rangle &\mapsto \langle \zeta^{I} \rangle + \frac{\fn^I}{\langle r \rangle} \, , \\
 \langle \zeta^{m} \rangle &\mapsto \langle \zeta^{m} \rangle + \frac{\fn^m}{\langle r \rangle} \, .
\end{align}
One can then easily show that the sign function fulfills
\begin{align}\label{e:sign_trafo}
\sign \big(m^{w,q}_{\rm CB} + n \, m_{\rm KK}\big) = \sign \big(\tilde  m^{w,q}_{\rm CB} + ( n - \fn^I w_{I} - \fn^m q_{m})\, m_{\rm KK}\big) \, .
\end{align}
Depending on the sign of the Coulomb branch masses we have to investigate four different cases:
\begin{enumerate}
 \item[] \framebox{$\sign \big(m^{w,q}_{\rm CB}\big) > 0 $}
 
 The integer quantity $l_{w,q}$ is then defined via the following property
 \begin{align}
  \sign \big(m^{w,q}_{\rm CB} - l_{w,q} \, m_{\rm KK}\big) > 0 \quad \wedge \quad \sign \big(m^{w,q}_{\rm CB} - ( l_{w,q} + 1 ) \, m_{\rm KK}\big) < 0 \, .
 \end{align}
 Using \eqref{e:sign_trafo} we find
 \begin{align}
  &\sign \big(\tilde m^{w,q}_{\rm CB} - (l_{w,q} + \fn^I w_{I} + \fn^m q_{m}) \, m_{\rm KK}\big)&& > 0 \, ,  \\ & \sign 
  \big(\tilde m^{w,q}_{\rm CB} -
  (l_{w,q} + \fn^I w_{I} + \fn^m q_{m}+1) \, m_{\rm KK}\big)&& < 0 \, . \nn
 \end{align}
 Depending on the sign of $\tilde m^{w,q}_{\rm CB}$ we can now read off $\tilde l_{w,q}$
 \begin{subequations}\label{e:cases1} 
 \begin{align}
 &\tilde  l_{w,q} = l_{w,q} + \fn^I w_{I} + \fn^m q_{m}  && \textrm{for } \sign\big(\tilde m^{w,q}_{\rm CB}\big) > 0 \, , \\
& \tilde l_{w,q} = -l_{w,q} - \fn^I w_{I} - \fn^m q_{m}-1  && \textrm{for } \sign\big(\tilde m^{w,q}_{\rm CB}\big) < 0 \, .
 \end{align}
 \end{subequations}
\item[] \framebox{$\sign (m^{w,q}_{\rm CB}) < 0 $}

 Now $l_{w,q}$ is defined as
 \begin{align}
  \sign \big(m^{w,q}_{\rm CB} + ( l_{w,q} + 1 ) \, m_{\rm KK}\big) > 0 \quad \wedge \quad \sign \big(m^{w,q}_{\rm CB} + l_{w,q} 
  \, m_{\rm KK}\big) < 0 \, .
 \end{align}
 With \eqref{e:sign_trafo} we get
 \begin{align}
  &\sign \big(\tilde m^{w,q}_{\rm CB} + (l_{w,q} - \fn^I w_{I} - \fn^m q_{m} +1 ) \, m_{\rm KK}\big)&& > 0 \, , \\
   &\sign \big(\tilde m^{w,q}_{\rm CB} + (l_{w,q} - \fn^I w_{I} - \fn^m q_{m}) \, m_{\rm KK}\big)&& < 0 \, . \nn
 \end{align}
 From this we can again determine $\tilde l_{w,q}$
 \begin{subequations}\label{e:cases2}
 \begin{align}
& \tilde l_{w,q} = - l_{w,q} + \fn^I w_{I} + \fn^m q_{m} -1  && \textrm{for } \sign\big(\tilde m^{w,q}_{\rm CB}\big) > 0 \, , \\
&\tilde l_{w,q} = l_{w,q} - \fn^I w_{I} - \fn^m q_{m}  && \textrm{for } \sign\big(\tilde m^{w,q}_{\rm CB}\big) < 0 \, .
 \end{align}
 \end{subequations}
\end{enumerate}
It is now easy to check that the relations \eqref{e:cases1}, \eqref{e:cases2} are indeed summarized as
\begin{align}\label{e:CBiden}
\Big(\tilde l_{w,q} + \frac{1}{2}\Big) \sign \big(\tilde m^{w,q}_{\rm CB}\big) - \Big(l_{w,q} + \frac{1}{2}\Big)
 \sign \big(m^{w,q}_{\rm CB}\big) =   \fn^I w_{I} + \fn^m q_{m} \, .
\end{align}

As shown in \autoref{sec:anom_lgt}, this expression plays a crucial role
when one investigates the relation between one-loop Chern-Simons terms which have been calculated in different frames (related by
large gauge transformations). Indeed, the characteristic factor $\Big(l_{w,q} + \frac{1}{2}\Big)
 \sign \big(m^{w,q}_{\rm CB}\big)$ directly appears in the formulae for
 \begin{align}
  \delta \Theta_{IJ}, \,\delta \Theta_{mn}, \,\delta \Theta_{Im}, \,\delta k_{IJK}, \,\delta k_{mnp}, \,\delta k_{IJm}, 
  \,\delta k_{Imn}, \,\delta k_{I}, \,\delta k_{m} , \nn
 \end{align}
as one can see by looking at their explicit expressions in \autoref{sec:one_loop_calc}.
In order to perform a similar analysis for the other types of one-loop Chern-Simons terms we will need additional
identities which however can be derived from \eqref{e:CBiden} straightforwardly.
In particular, in order to relate 
  \begin{align}
  \delta \Theta_{0I}, \,\delta \Theta_{0m}, \,\delta k_{0IJ}, \,\delta k_{0mn}, \,\delta k_{0Im},  \,\delta k_{0} , \nn
 \end{align}
to anomalies
one has to use the relation
 \begin{align}\label{e:add_rel_1}
\tilde l_{w,q} \, \big(\tilde l_{w,q} + 1\big)  - l_{w,q} \, \big(l_{w,q} + 1\big) 
= & \, 2\ \big (\fn^I w_{I} + \fn^m q_{m} \big ) \big(l_{w,q} + \frac{1}{2}\big) \sign (m^{w,q}_{\rm CB}) \nn   \\
& + \big (\fn^I w_{I} + \fn^m q_{m} \big )^2 \, ,
\end{align}
since these are the type of factors which appear in the loop calculations \autoref{sec:one_loop_calc}.
Similarly for
  $\delta \Theta_{00},  \,\delta k_{00I}, \,\delta k_{00m}$
 we exploit
\begin{align}\label{e:add_rel_2}
\tilde l_{w,q} \, \big(\tilde l_{w,q} + 1\big) &\, \big(\tilde l_{w,q} + \frac{1}{2}\big) \sign (\tilde m^{w,q}_{\rm CB})
- l_{w,q} \, \big(l_{w,q} + 1\big)  \, \big(l_{w,q} + \frac{1}{2}\big) \sign ( m^{w,q}_{\rm CB}) \nn \\
= & \, \frac{1}{2} \big (\fn^I w_{I} + \fn^m q_{m} \big ) \ \Big(1 + 6 \ l_{w,q} \, \big(l_{w,q} + 1\big)\Big)  \\
& + 3 \ \big (\fn^I w_{I} + \fn^m q_{m} \big )^2 \big(l_{w,q} + \frac{1}{2}\big) \sign (m^{w,q}_{\rm CB})
+ \big (\fn^I w_{I} + \fn^m q_{m} \big )^3 \nn \, ,
\end{align}
and finally for $\delta k_{000}$
the relevant identity is
\begin{align}\label{e:add_rel_3}
\tilde l_{w,q}^2 \, \big(\tilde l_{w,q} + 1\big)^2 & - l_{w,q}^2 \, \big(l_{w,q} + 1\big) ^2 \\
= &\,  4 \ \big (\fn^I w_{I} + \fn^m q_{m} \big ) \ l_{w,q} \, \big(l_{w,q} + 1\big) 
\, \big(l_{w,q} + \frac{1}{2}\big) \sign ( m^{w,q}_{\rm CB}) \nn \\
& +  \big (\fn^I w_{I} + \fn^m q_{m} \big )^2 \ \Big(1 + 6 \ l_{w,q} \, \big(l_{w,q} + 1\big)\Big) \nn \\
& + 4 \ \big (\fn^I w_{I} + \fn^m q_{m} \big )^3 \big(l_{w,q} + \frac{1}{2}\big) \sign (m^{w,q}_{\rm CB})
+ \big (\fn^I w_{I} + \fn^m q_{m} \big )^4 \, . \nn
\end{align}
We will comment on the precise relation of the individual Chern-Simons couplings to anomalies in the upcoming section.

\section{Large Gauge Transformations of Chern-Simons Terms}
In \autoref{sec:anom_lgt} we conveyed how one can obtain all gauge anomaly cancelation conditions
in four and six dimensions (and also mixed gauge-gravitational anomalies in six dimensions) by considering two different ways
of evaluating large gauge transformations on one-loop induced Chern-Simons couplings and demanding consistency of both approaches.
We showed there that it suffices to consider
only a subset of all one-loop Chern-Simons terms in order to obtain all anomalies. In this section we provide the complete list
of which anomaly cancelation conditions one obtains from the large gauge transformations of all possible one-loop Chern-Simons terms.
These calculations are straightforward, however besides of the identity \eqref{e:CBiden}, which we used in \autoref{sec:anom_lgt}
in order to evaluate $\delta\tilde \Theta_{IJ}$, $\delta\tilde \Theta_{mn}$, $\delta\tilde \Theta_{Im}$,
$\delta\tilde k_{IJK}$, $\delta\tilde k_{mnp}$, $\delta\tilde k_{IJm}$, $\delta\tilde k_{Imn}$, $\delta\tilde k_{I}$, $\delta\tilde k_{m}$,
one now also
has to make use of the additional relations \eqref{e:add_rel_1}, \eqref{e:add_rel_2}, \eqref{e:add_rel_3} which
however can be derived from \eqref{e:CBiden} as we have just shown.
In four-dimensional theories on the circle we have
\begin{center}
\begin{tabular}{r|ccc}
\rule[-.3cm]{0cm}{.8cm} & $\partial_{\fn^L}\partial_{\fn^M}\dots$ & $\partial_{\fn^L}\dots\partial_{\fn^q}\dots$ 
 & $\partial_{\fn^q}\partial_{\fn^r}\dots$\\
 \hline
\rule[-.3cm]{0cm}{.8cm}$\delta\tilde \Theta_{00}\overset{!}{=}0$ &  \eqref{4d_anomaly_alt2}  &  \eqref{4d_anomaly_alt3} & \eqref{4d_anomaly_alt4}  \\
\rule[-.3cm]{0cm}{.8cm}$\delta\tilde \Theta_{0I}\overset{!}{=}0$ & \eqref{4d_anomaly_alt2} & \eqref{4d_anomaly_alt3} & 0 \\
\rule[-.3cm]{0cm}{.8cm}$\delta\tilde \Theta_{0m}\overset{!}{=}0$ & \eqref{4d_anomaly_alt3} & 0 & \eqref{4d_anomaly_alt4}  \\
\rule[-.3cm]{0cm}{.8cm}$\delta\tilde \Theta_{IJ}\overset{!}{=}0$ & \eqref{4d_anomaly_alt2} & 0 & \eqref{4d_anomaly_alt3} \\
\rule[-.3cm]{0cm}{.8cm}$\delta\tilde \Theta_{mn}\overset{!}{=}0$ & 0 & 0 & \eqref{4d_anomaly_alt4} \\
\rule[-.3cm]{0cm}{.8cm}$\delta\tilde \Theta_{Im}\overset{!}{=}0$ & \eqref{4d_anomaly_alt3} & 0 & 0
\end{tabular}
\end{center}
while in the circle-reduced six-dimensional settings the full list is
\begin{center}
\begin{tabular}{r|ccc}
\rule[-.3cm]{0cm}{.8cm} & $\partial_{\fn^L}\partial_{\fn^M}\dots$ & $\partial_{\fn^L}\dots\partial_{\fn^q}\dots$ 
 & $\partial_{\fn^q}\partial_{\fn^r}\dots$\\
 \hline
\rule[-.3cm]{0cm}{.8cm}$\delta\tilde k_{000}\overset{!}{=}0$&\eqref{e:6d_alter_5}& \eqref{e:6d_alter_6},\eqref{e:6d_alter_7} &  \eqref{e:6d_alter_8} \\
\rule[-.3cm]{0cm}{.8cm}$\delta\tilde k_{00I}\overset{!}{=}0$ & \eqref{e:6d_alter_5} &  \eqref{e:6d_alter_7} & 0\\
\rule[-.3cm]{0cm}{.8cm}$\delta\tilde k_{00m}\overset{!}{=}0$ & \eqref{e:6d_alter_6} & \eqref{e:6d_alter_7} &  \eqref{e:6d_alter_8} \\
\rule[-.3cm]{0cm}{.8cm}$\delta\tilde k_{0IJ}\overset{!}{=}0$ & \eqref{e:6d_alter_5} & \eqref{e:6d_alter_6} & \eqref{e:6d_alter_7} \\
\rule[-.3cm]{0cm}{.8cm}$\delta\tilde k_{0mn}\overset{!}{=}0$ & \eqref{e:6d_alter_7} & 0& \eqref{e:6d_alter_8} \\
\rule[-.3cm]{0cm}{.8cm}$\delta\tilde k_{0Im}\overset{!}{=}0$ & \eqref{e:6d_alter_6} & \eqref{e:6d_alter_7} & 0\\
\rule[-.3cm]{0cm}{.8cm}$\delta\tilde k_{IJK}\overset{!}{=}0$ & \eqref{e:6d_alter_5} & 0 & \eqref{e:6d_alter_6}\\
\rule[-.3cm]{0cm}{.8cm}$\delta\tilde k_{mnp}\overset{!}{=}0$ & 0 & 0 & \eqref{e:6d_alter_8}\\
\rule[-.3cm]{0cm}{.8cm}$\delta\tilde k_{IJm}\overset{!}{=}0$ & \eqref{e:6d_alter_6} & 0 & \eqref{e:6d_alter_7}\\
\rule[-.3cm]{0cm}{.8cm}$\delta\tilde k_{Imn}\overset{!}{=}0$ & \eqref{e:6d_alter_7} & 0 & 0\\
\rule[-.3cm]{0cm}{.8cm}$\delta\tilde k_{0}\overset{!}{=}0$ & \eqref{e:6d_alter_3} & 0 & \eqref{e:6d_alter_4}\\
\rule[-.3cm]{0cm}{.8cm}$\delta\tilde k_{I}\overset{!}{=}0$ & \eqref{e:6d_alter_3} & 0 & 0\\
\rule[-.3cm]{0cm}{.8cm}$\delta\tilde k_{m}\overset{!}{=}0$ & 0 & 0 & \eqref{e:6d_alter_4} 
\end{tabular}
\end{center}
Let us explain these tables: In order to obtain the indicated anomaly conditions one has to take an appropriate number of derivatives
of the equations $\delta\tilde \Theta_{\Lambda\Sigma}\overset{!}{=}0$, $\delta\tilde k_{\Lambda\Sigma\Theta}\overset{!}{=}0$,
$\delta\tilde k_{\Lambda}\overset{!}{=}0$ with respect to
$\fn^L$ and $\fn^q$, \textit{i.e.}~in the non-Abelian and Abelian directions, respectively. The total number of derivatives one has to take is given by
one plus the number
of 0-indices in $\delta\tilde \Theta_{\Lambda\Sigma}$, $\delta\tilde k_{\Lambda\Sigma\Theta}$,
$\delta\tilde k_{\Lambda}$, \textit{e.g.}~three derivatives for $k_{00I}$ and one derivative for $\Theta_{mn}$.
In the first column we only take derivatives in the Cartan directions of the non-Abelian gauge group, though possibly different 
directions, while
in the third column the derivatives are only with respect to $U(1)$ large gauge transformation parameters, also possibly different ones.
In the second column we assume derivatives with respect to both Cartan and $U(1)$ directions.

\chapter{Intersection Numbers}\label{app_inter}
In this chapter we list useful intersection numbers of elliptically-fibered Calabi-Yau four- and threefolds along with their matched
quantity in the M-theory to F-theory duality. Special emphasis is put on Chern-Simons couplings $\Theta_{\Lambda\Sigma}$,
$k_{\Lambda\Sigma\Theta}$, $k_{\Lambda}$.

For Calabi-Yau fourfolds we consider the specific intersections 
\begin{align}
 \pi(D_\Lambda \cdot D_\Sigma)^\alpha := ( D_\Lambda \cdot D_\Sigma \cdot \cC^\beta ) \ \tensor{\eta}{^{-1}_\beta^\alpha}  
\end{align}
 and the induced Chern-Simons couplings
 \begin{align}
 \Theta_{\Lambda\Sigma} = - \frac{1}{4}\ D_\Lambda \cdot D_\Sigma \cdot [G_4 ] \, , 
\end{align}
where $\tensor{\eta}{_\alpha^\beta}$ is the full-rank intersection matrix \eqref{e:def_metric}.
One finds
\begin{align}
 \pi(D_\alpha \cdot D_\beta)^\gamma & = 0 \, ,  
 &\pi(D_m \cdot D_\alpha)^\beta  & = 0 \, ,  & \pi(D_I \cdot D_\alpha)^\beta  & = 0 \, ,  \\
 \pi(D_0 \cdot D_0)^\alpha  & = 0 \, ,  &\pi(D_0 \cdot D_m)^\alpha  & = 0 \, , 
  &\pi(D_0 \cdot D_I)^\alpha  & = 0 \, ,  \nn \\
  \pi(D_0 \cdot D_\alpha)^\beta  & = \delta^\beta_\alpha \, , &\pi(D_I \cdot D_J)^\alpha & = - b^\alpha \ \cC_{IJ} \, ,
 &\pi(D_m \cdot D_n)^\alpha & = - b_{mn}^\alpha  \nn
\end{align}
 and
 \begin{align}
  D_\alpha \cdot D_\beta \cdot [G_4 ] &= 0 \, ,  &D_\alpha \cdot D_0 \cdot [G_4 ] &= 0 \, , \\
 D_\alpha \cdot D_I \cdot [G_4 ] &= 0 \, ,  & D_\alpha \cdot D_m \cdot [G_4 ] & = -2\theta_{\alpha m} \, . \nn
 \end{align}
The remaining Chern-Simons couplings are one-loop expressions which are evaluated field-theoretically in \autoref{sec:one_loop_calc}.

Finally for Calabi-Yau threefolds all relevant intersection numbers correspond to Chern-Simons couplings
\begin{align}
 k_{\Lambda\Sigma\Theta} = D_\Lambda \cdot D_\Sigma \cdot D_\Theta \, , \qquad
 k_{\Lambda} = D_\Lambda \cdot [c_2] \, ,
\end{align}
and one evaluates
\begin{align}
 D_\alpha \cdot D_\beta \cdot D_\gamma &=0 \ , &  D_0 \cdot D_\alpha \cdot D_\beta &= \eta_{\alpha\beta} \ , 
 &  D_I \cdot D_\alpha \cdot D_\beta &=0 \ , \nn \\
   D_m \cdot D_\alpha \cdot D_\beta &=0 \ , &  D_0 \cdot D_0 \cdot D_\alpha &=0 \ ,
   &  D_I \cdot D_J \cdot D_\alpha &= -\eta_{\alpha\beta} b^\beta \ \cC_{IJ} \ , \nn \\
  D_m \cdot D_n \cdot D_\alpha &= -\eta_{\alpha\beta} b^\beta_{mn} \ , &
   D_0 \cdot D_I \cdot D_\alpha &=0 \ , & D_0 \cdot D_m \cdot D_\alpha &=0 \ ,\nn \\ 
 D_\alpha \cdot [c_2] &= -12\ \eta_{\alpha\beta} a^\beta \ .
\end{align}

Finally let us collect some useful geometrical facts about elliptic fibrations. For any divisor which corresponds to a
rational section $S$ we have
\begin{align}
 \pi (S \cdot S) = K
\end{align}
where the map $\pi$ is defined in \eqref{def_proj} and $K$ is the canonical class of the base. The same
quantity appears in connection with the second Chern class for Calabi-Yau threefolds
\begin{align}
 \big( D_\alpha \cdot [c_2 ]\big ) \ \eta^{-1 \, \alpha \beta} \ D_\beta  = -12K \, .
\end{align}
Furthermore, the pullback of the divisor $S^{\rm b}$ over which the elliptic fiber degenerates fulfills
\begin{align}
 \pi (D_I \cdot D_J) = - \cC_{IJ} \ S \, .
\end{align}
Finally for the case of a \textit{holomorphic} zero-section one can explicitly evaluate some intersections
which are matched to one-loop Chern-Simons terms. One finds for Calabi-Yau fourfolds
\begin{align}
 D_0 \cdot D_m \cdot [G_4] &= -\frac{1}{2} K \cdot D_m \cdot [G_4] \, , &
 D_0 \cdot D_I \cdot [G_4] &= 0 \, , &
 D_0 \cdot D_0 \cdot [G_4] &= 0 \, ,
\end{align}
and for Calabi-Yau threefolds
\begin{align}
 D_0 \cdot D_m \cdot D_n &= -\frac{1}{2} K \cdot D_m \cdot D_n \, , & D_0 \cdot D_0 \cdot D_m &= 0 \, , & D_0 \cdot D_m \cdot D_I &= 0 \, , \nn\\
 D_0 \cdot D_I \cdot D_J &= -\frac{1}{2} K \cdot D_I \cdot D_J \, , &
  D_0 \cdot D_0 \cdot D_I &= 0 \, , &
 D_0 \cdot D_0 \cdot D_0 &= \frac{1}{4} K \cdot K  \cdot D_0 \, , \nn \\
 D_0 \cdot [c_2 ]  &= 52 - 4 h^{1,1} (B_2) \, .
\end{align}
Note that the condition of holomorphicity is absolutely crucial here.

 \chapter{Six-Dimensional Supermultiplets}\label{app:multi}
In the following we display the six-dimensional supermultiplets. The first two factors encode the representation under the little group in terms of spins whereas the second two entries
label the representation under the R-symmetry group $USp(2N_L)\times USp(2N_R)$ in terms of the dimension of the representation. The table is taken from \cite{deWit:2002vz} and adjusted to our
chirality conventions.
 \begin{center}
\begin{tabular}{lll}
\textbf{Multiplet} & \textbf{Bosons} & \textbf{Fermions} \\
\hline\hline
\rule[-.3cm]{0cm}{.8cm}(1,0) hyper & $\big(0,0;2,1\big)\oplus\rm h.c.$ & $\big(0,\frac{1}{2};1,1\big)\oplus\rm h.c.$\\ \hline
\rule[-.3cm]{0cm}{.8cm}(1,0) tensor & $\big(0,1;1,1\big)\oplus\big(0,0;1,1\big)$ & $\big(0,\frac{1}{2};2,1\big)$\\ \hline
\rule[-.3cm]{0cm}{.8cm}(1,0) vector & $\big(\frac{1}{2},\frac{1}{2};1,1\big)$ & $\big(\frac{1}{2},0;2,1\big)$\\ \hline
\rule[-.3cm]{0cm}{.8cm}(1,0) graviton & $\big(1,1;1,1\big)\oplus\big(1,0;1,1\big)$ & $\big(1,\frac{1}{2};2,1\big)$\\ \hline
\rule[-.3cm]{0cm}{.8cm}(2,0) tensor & $\big(0,1;1,1\big)\oplus\big(0,0;5,1\big)$ & $\big(0,\frac{1}{2};4,1\big)$\\ \hline
\rule[-.3cm]{0cm}{.8cm}(2,0) graviton & $\big(1,1;1,1\big)\oplus\big(1,0;5,1\big)$ & $\big(1,\frac{1}{2};4,1\big)$\\ \hline
\rule[-.3cm]{0cm}{.8cm}(1,1) vector & $\big(\frac{1}{2},\frac{1}{2};1,1\big)\oplus \big( 0,0;2,2 \big)$ & $\big(0,\frac{1}{2};1,2\big)\oplus \big(\frac{1}{2},0;2,1\big)$\\ \hline
\rule[-.3cm]{0cm}{.8cm}(1,1) graviton & $\big(1,1;1,1\big)\oplus\big(\frac{1}{2},\frac{1}{2};2,2\big)\oplus$ &
$\big(\frac{1}{2},1;1,2\big)\oplus\big(1,\frac{1}{2};2,1\big)\oplus$\\
\rule[-.3cm]{0cm}{.8cm}& $\big(1,0;1,1\big)\oplus\big(0,1;1,1\big)\oplus\big(0,0;1,1\big) $ & $\big(\frac{1}{2},0;1,2\big)\oplus \big(0,\frac{1}{2};2,1\big)$\\ \hline
\rule[-.3cm]{0cm}{.8cm}(2,1) graviton & $\big(1,1;1,1\big)\oplus\big(\frac{1}{2},\frac{1}{2};4,2\big)\oplus$ &
$\big(\frac{1}{2},1;1,2\big)\oplus\big(1,\frac{1}{2};4,1\big)\oplus$\\
\rule[-.3cm]{0cm}{.8cm}& $\big(1,0;5,1\big)\oplus\big(0,1;1,1\big)\oplus\big(0,0;5,1\big) $ & $\big(\frac{1}{2},0;5,2\big)\oplus \big(0,\frac{1}{2};4,1\big)$\\ \hline
\rule[-.3cm]{0cm}{.8cm}(2,2) graviton & $\big(1,1;1,1\big)\oplus\big(\frac{1}{2},\frac{1}{2};4,4\big)\oplus$ &
$\big(\frac{1}{2},1;1,4\big)\oplus\big(1,\frac{1}{2};4,1\big)\oplus$\\
\rule[-.3cm]{0cm}{.8cm}& $\big(1,0;5,1\big)\oplus\big(0,1;1,5\big)\oplus\big(0,0;5,5\big) $ & $\big(\frac{1}{2},0;5,4\big)\oplus \big(0,\frac{1}{2};4,5\big)$ \\  \hline
\end{tabular}
\end{center}

\bibliography{references}
\bibliographystyle{utcaps}

\end{document}